\documentclass[aps, prx, reprint,floatfix,superscriptaddress,onecolumn,nofootinbib]{revtex4-1}
\usepackage{comment}
\usepackage{graphicx}
\usepackage{indentfirst}
\usepackage{braket}
\usepackage{float}
\usepackage{amsmath}
\usepackage{amsthm}
\usepackage{wasysym}
\usepackage{mathrsfs}
\usepackage{bm}
\usepackage{epstopdf}
\usepackage{mathtools}
\usepackage{CJK}
\usepackage{esint}
\usepackage{color}
\usepackage[T1]{fontenc}
\usepackage{subfigure}
\usepackage{amsfonts}
\usepackage{footmisc}
\usepackage{scrextend}
\usepackage{multirow}
\usepackage{amssymb}
\usepackage[hyperfootnotes=false]{hyperref}
\usepackage{multirow}
\usepackage{diagbox}
\usepackage[english]{babel}
\usepackage{url}
\usepackage{siunitx}
\usepackage{bm}
\usepackage{hyperref}
\usepackage[dvipsnames]{xcolor}
\usepackage{soul}
\usepackage{physics}

\definecolor{darkblue}{rgb}{0,0,0.5}
\hypersetup{
colorlinks=true,
linkcolor=darkblue,
filecolor=blue,
citecolor=darkblue,  
urlcolor=darkblue,
}
\newcommand{\bs}{\bm s}

\newcommand{\bx}{\boldsymbol x}

\newcommand{\bI}{\boldsymbol{I}}

\newcommand*\diff{\mathop{}\!\mathrm{d}}

\newcommand{\calA}{{\cal A}}

\newcommand{\calL}{{\cal L}}
\newcommand{\calN}{{\cal N}}

\def\be{\begin{equation}}
\def\ee{\end{equation}}
\def\ba{\begin{eqnarray}}
\def\ea{\end{eqnarray}}

\newcommand{\QZ}[1]{{{\textcolor{black}{#1}}}}

\newtheorem{theorem}{Theorem}
\newtheorem*{conjecture*}{Conjecture}

\newtheorem{proposition}{Proposition}

\theoremstyle{definition}

\newtheorem{example}{Example}[section]

\begin{document}

\title{Advances in Bosonic Quantum Error Correction with Gottesman–Kitaev–Preskill Codes: Theory, Engineering and Applications}

\author{Anthony J. Brady}
\affiliation{
Ming Hsieh Department of Electrical and Computer Engineering, University of Southern California, Los
Angeles, California 90089, USA
}
\affiliation{
Department of Electrical and Computer Engineering, University of Arizona, Tucson, Arizona 85721, USA
}

\author{Alec Eickbusch}
\affiliation{Departments of Physics and Applied Physics, Yale University, New Haven, 06520, Connecticut, USA}
\affiliation{Yale Quantum Institute,
Yale University, 
New Haven,
06511, 
Connecticut,
USA}

\author{Shraddha Singh}
\affiliation{Departments of Physics and Applied Physics, Yale University, New Haven, 06520, Connecticut, USA}
\affiliation{Yale Quantum Institute,
Yale University, 
New Haven,
06511, 
Connecticut,
USA}

\author{Jing Wu}
\affiliation{
J. C. Wyant College of Optical Sciences, University of Arizona, Tucson, Arizona 85721, USA}

\author{Quntao Zhuang}
\email{qzhuang@usc.edu}
\affiliation{
Ming Hsieh Department of Electrical and Computer Engineering, University of Southern California, Los
Angeles, California 90089, USA
}

\affiliation{
Department of Electrical and Computer Engineering, University of Arizona, Tucson, Arizona 85721, USA
}
\affiliation{
J. C. Wyant College of Optical Sciences, University of Arizona, Tucson, Arizona 85721, USA}
\affiliation{
Department of Physics and Astronomy, University of Southern California, Los
Angeles, California 90089, USA
}

\begin{abstract}
 Encoding quantum information into a set of harmonic oscillators is considered a hardware efficient approach to mitigate noise for reliable quantum information processing. Various codes have been proposed to encode a qubit into an oscillator---including cat codes, binomial codes and Gottesman-Kitaev-Preskill (GKP) codes---and are among the first to reach a break-even point for quantum error correction. Though GKP codes are widely recognized for their promise in quantum computation, they also facilitate near-optimal quantum communication rates in bosonic channels and offer the ability to safeguard arbitrary quantum states of oscillators. This review focuses on the basic working mechanism, performance characterization, and the many applications of GKP codes---emphasizing recent experimental progress in superconducting circuit architectures and theoretical advancements in multimode GKP qubit codes and oscillators-to-oscillators (O2O) codes. We begin with a preliminary continuous-variable formalism needed for bosonic codes. We then proceed to the quantum engineering involved to physically realize GKP states.  We take a deep dive into GKP stabilization and preparation in superconducting architectures and examine proposals for realizing GKP states in the optical domain (along with a concise review of GKP realization in trapped-ion platforms). Finally, we present multimode GKP qubits and GKP-O2O codes, examine code performance and discuss applications of GKP codes in quantum information processing tasks such as computing, communication, and sensing.


\end{abstract}

\keywords{
Quantum error correction; Bosonic codes; Gottesman-Kitaev-Preskill codes.
}
\maketitle

\tableofcontents

\section{Introduction}

Quantum physics has brought new opportunities in information processing, which can be broadly classified into three categories: computing, communication, and sensing (see Fig.~\ref{fig:phys_systems}). Quantum computers can speed up the solution of classically hard problems such as factoring~\cite{shor1999polynomial} and unstructured search~\cite{grover1996fast}. Quantum communication allows unconditional security and quantum resource allocation to sites to complete distributed quantum information processing tasks~\cite{kimble2008quantum,wehner2018quantum,kozlowski2019towards,quics2021prxQuantum}. Quantum sensing protocols enable unprecedented precision in measurement of unknown parameters with the help of nonclassical resources such as squeezing and entanglement~\cite{giovannetti2011advances,zhang2020distributed}. Quantum information processing has led to paradigm shifts not only in fundamental science but also in engineering capabilities and has emerged as a multidisciplinary field of research. 

Various platforms have been proposed and demonstrated for quantum information processing, including optical and nanophotonic systems, microwave circuits and cavities, trapped ions and nuclear spins to name a few. Quantum information processing across all platforms is challenged by loss and environmental noise. Although much has been devoted to achieving quantum advantages in near-term devices, quantum error correction (QEC) is still necessary for scalable and robust quantum information processing. Moreover, as we move onto implementing QEC codes in real physical systems, the design of such codes needs to be tailored to the physical system at hand.

Regarding physical systems of quantum information processors, we can broadly classify them into two major types, discrete-variable (DV) and continuous-variable (CV) systems. DV systems have finite quantum degrees of freedom, i.e. the Hilbert space is finite-dimensional. The most common case of a system with dimension two naturally encodes a quantum bit or qubit. Bosonic CV systems, on the other hand, have infinite quantum degrees of freedom in principle, i.e. the Hilbert space is infinite dimensional.~\footnote{In practice, a system with finite energy does not have infinite Hilbert space dimension. However, it is often still convenient to approximate it as an infinite dimensional system.} A harmonic oscillator is a typical example, where the occupation number can in general be zero, one, $\dots$ up to infinity.

An important goal of QEC concerns the robust encoding of a qubit to a physical system. Regarding bosonic CV systems, a qubit can be naively encoded by making the otherwise equidistant gap between energy levels of an oscillator unequal, enabling the transition probability between the first two states, $\ket{0}$ and $\ket{1}$, controllable. This realizes an anharmonic oscillator focusing on the lowest two levels, as is the case of transmon qubits~\cite{Koch_2007,majer2007coupling,kjaergaard2020superconducting}. However, such a simple zero and one cut-off scheme is not making full use of the bosonic degree of freedom for CV systems, as pointed out by the breakthrough works of Chuang, Leung and Yamamoto~\cite{chuang1997bosonic}, and Gottesman, Kitaev and Preskill (GKP)~\cite{gkp2001}---the latter being the focus of this review. \QZ{Alternative approaches that rely on Gaussian states and Gaussian operations of bosonic systems have proven inadequate in achieving robust encoding~\cite{eisert2002nogodistillation,fiuravsek2002gaussian,giedke2002characterization,niset2009nogo} as universal quantum computing relies non-Gaussian elements~\cite{lloyd1999quantum}. In this sense, non-Gaussian states, such as GKP states, are a necessary ingredient in bosonic error correction.} Indeed, bosonic quantum error correction is widely recognized as a promising and hardware-efficient strategy for combating noise in futuristic quantum information processors. For example, bosonic QEC codes implemented with microwave circuit quantum electrodynamics (cQED) systems are among the first codes to surpass the QEC break-even point~\cite{ofek2016extending,campagne2020quantum}. 

\QZ{While numerous works and reviews have extensively documented the remarkable accomplishments in protecting a qubit with GKP codes~\cite{albert2018pra,terhal2020towards,puri2021rvw,cai2021bosonic,joshi2021quantum,albert2022bosonic}, a dedicated review on the engineering and diverse applications of GKP codes is currently missing.} Notably, the experimental capabilities of cavity QED systems have substantially improved over the years. Besides the increase of speed and quality in single-mode control and engineering~\cite{eickbusch2022fast}, multiple ocillators can now be engineered with high-fidelity fast beamsplitters~\cite{chapman2022high,lu2023microwaveBS} and universal controls~\cite{diringer2023conditional}.
Furthermore, as multiple oscillators can be manipulated in a physical system, there is much more one can do with bosonic QEC. Two emergent approaches of multi-oscillator codes have shed new light on how to exploit the power of more modes. The first one regards a direct encoding of an oscillator into multiple oscillators as proposed in Ref.~\cite{noh2020o2o} (or multiple oscillators into more oscillators as in Ref.~\cite{wu2022optimal}). Enabled by GKP states, such oscillators-to-oscillators (O2O) codes extend the applicability of QEC to protect continuous-variable quantum information such as squeezing and continuous-variable multi-partite entanglement that are crucial for suppressing noise in quantum sensors. At the same time, a second approach of encoding a qubit into multiple oscillators has also shown advantages in error correction performance for a geometrical (lattice) perspective~\cite{harrington2001rates,baptiste2022multiGKP,conrad2022lattice}. These new perspectives have brought new opportunities in quantum computing, quantum sensing and quantum communication.

This review provides a self-contained introduction and summary of recent advances in bosonic QEC using GKP codes---focusing on the engineering of GKP codes and the breadth of applications that they enable. When possible, we prioritize pedagogy by presenting simple examples and intuitive explanations of concepts, ensuring accessibility to a broad readership. We thoroughly examine recent developments on multimode codes, including both encoding qubits into many oscillators and encoding oscillators into more oscillators. We provide an extensive summary on the experimental systems to support these new codes---focusing particularly on microwave cQED systems for their multimode engineering capabilities and promising demonstration of QEC beyond break-even; we also discuss current proposals for generating optical (i.e., ``flying'') GKP states. The review also delves into applications of GKP codes beyond (but also including) the typical use-case of fault-tolerant quantum computing, such as quantum communication and quantum sensing. We conclude with open problems and future research directions, which we hope sparks further interest into this exciting field of bosonic QEC with GKP codes. The remainder of the review is organized as follows.

Chapter~\ref{sec:bosonic_qi} provides a brief overview of bosonic quantum information. We begin with introducing the relevant physical systems, and then proceed to the basic theoretical tools required for analysing bosonic quantum information. This chapter only assumes knowledge of basic quantum mechanics and provide the additional tools to analyze bosonic quantum systems. Readers with backgrounds in these areas can directly jump to later sections, while referring back to this section when necessary. 

Chapter~\ref{sec:noise_model} provides a unified description of common noise model adopted for bosonic quantum error correction analyses, including photon loss, additive noise and cQED auxiliary noise sources. We also include a summary on approaches of noise model conversions (e.g., from loss to additive noise).

Chapter~\ref{sec:gkp_lattice} delves into the heart of GKP states, offering a comprehensive mathematical description of multimode GKP states from a lattice perspective. We start with a clear presentation of simple single-mode square lattice GKP states and gradually progress to explore more complex lattices and multimode GKP states. This section serves as a solid foundation for readers to grasp the concepts of GKP QEC codes. By the end of this chapter, we hope that readers will be well-prepared to understand and work with GKP QEC codes effectively in cutting edge research.

Chapter~\ref{sec:GKP_eng} summarizes the recent advances in the quantum engineering of GKP states. We focus on microwave superconducting cQED systems where GKP memory experiments have achieved beyond break-even results and multimode engineering capability has been developed. We discuss stabilization using Hamiltonian and dissipation engineering, and the universal control of GKP qubits using auxiliary qubits. While a high-fidelity experimental realization of GKP states in the optical domain has not yet been achieved (although there is some preliminary work, as indicated in Ref.~\cite{konno2023propagatingGKP}), it is crucial to note that the optical domain stands out as the exclusive platform for long-distance quantum communication. We thus provide an overview of recent proposals for optical GKP state engineering along with the associated experimental challenges. Limited by length, we do not go into details about recent promising trapped ion experiments~\cite{fluhmann2019encoding,home2022QECgkp}. Finally, we offer a brief perspective on the potential scalability of GKP codes, which could be a resource-efficient path towards quantum computing.

Chapter~\ref{sec:QEC-multimode} introduces multimode GKP qubits. We begin with the motivation of multimode GKP codes, and then highlight comparison of error correction in the simplest case of two-mode GKP codes using two types of lattices, D4 lattice code and tesseract code. These lattices have different benefits when it comes to practical error correction. Understanding these differences is important for scaling up to a practical fault-tolerant quantum computation. Finally, the concatenation between GKP codes with discrete-variable qubit codes is summarized.

Chapter~\ref{sec:gkp_o2o_codes} introduces the general formulation of GKP oscillator-to-oscillator (O2O) codes. We start with the general encoding scheme, the stabilizer measurement and error syndromes and then describe decoding strategies. Then we review fundamental lower bounds of GKP code performance and discuss the ``no-threshold theorem'' for general O2O codes. To solidify concepts, we provide some simple examples of O2O codes and review recent results of O2O code reduction and code optimization.

Chapter~\ref{sec:applications} addresses the applications of bosonic QEC in quantum computing, communication and sensing. In terms of computing and communication, bosonic codes provide a resource efficient approach of QEC and can be adopted in concatenation with qubit codes. On the other hand, protecting CV quantum information in quantum sensing applications heavily rely on GKP O2O codes. By the end of this chapter, readers will gain insights into the versatile applications of bosonic QEC and how it may positively impact various quantum technologies, paving the way for advanced and reliable quantum information processing and sensing capabilities.

Chapter~\ref{sec:discussion} concludes the review with a colloquial summary of the key points covered throughout the document. We also discuss open problems in bosonic QEC with GKP codes, shedding light on the challenges that researchers may face and the exciting potential for further advancements in the field.

\section{Bosonic Quantum Information}
\label{sec:bosonic_qi}

 In this section, we provide a brief introduction to bosonic quantum information, starting from physical systems in Chapter~\ref{sec:physical_systems} followed by an overview of quantum harmonic oscillators to establish basic notation in Chapter~\ref{sec:basic_notation}. In Chs.~\ref{sec:gauss_evol} and \ref{sec:Gaussian_measurement}, we survey the special class of linear bosonic quantum systems which provides theoretical tools for analysing GKP codes; this survey is quick as much of it relies on techniques from Gaussian quantum information theory, a well-developed field. In Chapter~\ref{sec:NG_states}, we discuss non-Gaussian resources and channels that are pertinent for GKP codes. Finally, we provide details on quantum capacity in Chapter~\ref{sec:quantum_capacity}, which enables one to find bounds on error correction performance.

\subsection{Physical systems}
\label{sec:physical_systems}

\begin{figure}
    \centering
    \includegraphics[width=.75\linewidth]{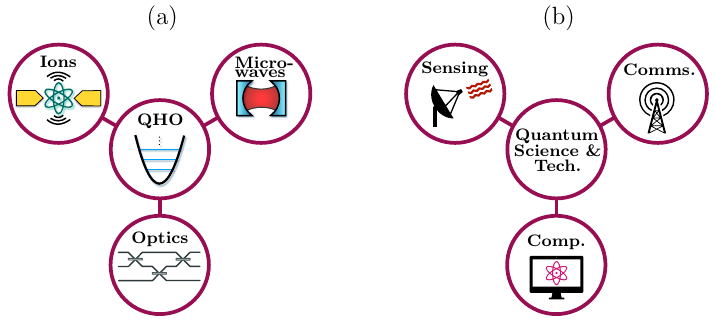}
    \caption{Quantum science with quantum oscillators. (a) Physical manifestations of a quantum harmonic oscillator (QHO): Motional degrees of freedom of a trapped ion, modes of a microwave resonator, electromagnetic waves at optical frequencies. (b) Quantum harmonic oscillators play a major role in quantum-specific applications such as quantum sensing, quantum communication, and quantum computing.}
    \label{fig:phys_systems}
\end{figure}

Bosonic quantum systems refer to systems of interacting quantum oscillators. There exists a wide spectrum of physical realizations of a quantum oscillator (see Figure~\ref{fig:phys_systems} for an illustration), such as electromagnetic waves at microwave and optical frequencies and the motional degrees of freedom of trapped ions. For instance, electromagnetic waves are important for various quantum information processing applications, including quantum sensing, computation and communication. When enhanced by cavities, they can couple to various solid state quantum systems and enable powerful quantum control for state engineering and error correction. Here we very briefly introduce a few physical realizations of bosonic quantum systems including optical systems, microwave resonators, and trapped ions. In this review we focus primarily on optical and microwave modes of light, but we include a succinct description of trapped ions below since GKP states have been recently demonstrated in these systems.

\subsubsection{Optical systems}

Optical systems refer to electromagnetic waves with high (visible or near-infrared) frequencies, such that the thermal noise at room temperature is negligible. In these systems, quantum effects such as coherence and entanglement can be maintained for a fairly long time at room temperature. Quantum information encoded in optical modes can propagate through the atmosphere and fibers, with low loss, e.g. 0.2 dB per kilometers in fibers at 1550nm. In this regard, optical photons are the ideal candidates for transmitting quantum information over long distances~\cite{Bennett20147,Ekert_1991,gisin2002,xu2020,pirandola2020advances}. On the other hand, the capability of state engineering and quantum operations is limited in optical systems due to weak nonlinearities that are otherwise required to couple optical photons. The available optical quantum states and operations are thus often restricted to the class of Gaussian states and operations; see Chs.~\ref{sec:gauss_evol} and
\ref{sec:Gaussian_measurement} for descriptions of Gaussian states and operations. 

State-of-the-art experimental systems in the optical domain mainly rely on photon counting to go beyond Gaussian operations and Gaussian states. For instance, there are many proposals for generating optical GKP states using linear optics, squeezing, and photon counting, \QZ{as well as a promising (though preliminary) demonstration~\cite{konno2023propagatingGKP}}; see Chapter~\ref{sec:GKP_eng} for further details. Photon counting requires photon-number-resolving detection which is currently one drawback in optical systems because high efficiency single photon counting devices typically require extensive cooling. On the other hand, photon detectors are usually the only component in optical setups that require cryogenics. We note that integrating the required optical components onto a nanophotonic chip is a promising route towards efficient and scalable quantum optical information processing~\cite{zhu2021integratedLiNrvw,moody2022roadmapQuPhotonics, lu2020toward1percent,yanagimoto2022tempTrapping}.

\subsubsection{Microwave cavities}
Microwave systems refer to electromagnetic waves with frequencies in the range $\sim\SI{1}{GHz}$ to $\sim\SI{100}{GHz}$. These frequencies enable strong nonlinear interaction at the single-quanta level, which can be readily realized with the Josephson effect in superconducting circuits~\cite{blais2021cqedRMP,wang2020efficient,elder2020high,heeres2017implementing}. Harmonic oscillators realized in 3D microwave cavities have longer lifetimes compared to their anharmonic counterparts like transmons and fluxoniums. Hence, interest in bosonic error correction in superconducting cavities has peaked in recent years. As we discuss thoroughly in Chapter~\ref{sec:GKP_eng}, GKP state engineering using microwave cavities relies on universal control via auxiliary resources (e.g., an auxiliary qubit) or a non-linear term in the Hamiltonian. 

One of the challenges in microwave quantum information processing is the abundant thermal noise at room temperature at microwave frequencies. To maintain quantum coherence, cooling to $\sim10$mK is required. Another drawback of microwave frequencies is the high attenuation in either the atmosphere or a waveguide, which prevents direct transmission of quantum states at microwave frequencies over long distances. Current realizations with an auxiliary control qubit are furthermore limited by auxiliary qubit errors. Nevertheless, superconducting circuits provide one of the best platforms for exquisite control of quantum systems~\cite{blais2021cqedRMP}. For instance, GKP and other non-Gaussian states have been realized in superconducting circuits~\cite{campagne2020quantum,eickbusch2022fast,sivak2022breakeven} via universal control of the qubit-oscillator coupled system. The potential for scaling up to multiple interacting modes is also promising~\cite{lu2023microwaveBS,chapman2022high,diringer2023conditional}. 

We point out that the complementary characteristics of microwave and optical photons suggests a hybrid quantum processing unit where quantum information may be processed by superconducting circuits and then transmitted with optical photons. Therefore, an efficient transduction scheme to interconvert quantum states between microwave and optical photons is also desired~\cite{lauk2020PerspTransdux,jiang2021OpticaTrdxRvw,vainsencher2016bi,balram2016coherent,fan2018superconducting,shao2019microwave,han2020cavity,zhong2020proposal,mirhosseini2020superconducting,forsch2020microwave,jiang2020efficient,fiaschi2021optomechanical} and will be a crucial technology in future quantum networks~\cite{kimble2008quantum,wehner2018quantum,kozlowski2019towards,quics2021prxQuantum}.

\subsubsection{Trapped ions}

Trapped-ion quantum systems, in principle, provide a ``clean'' approach to process quantum information, as the fundamental elements of such systems (ions) are designed by nature and thus do not suffer from inherent fabrication errors~\cite{bruzewicz2019trapped}. Nevertheless, the precise control of ion behavior in a collective and scalable manner presents its own challenges and complexities. For example, inhomogeneous trapping fields may cause variations in level splittings, leading to effective disparities between the ion qubits at the operational level. 

Many current quantum computing platforms based on trapped ions selectively choose two discrete atomic levels to serve as the basis states for a computational qubit~\cite{bruzewicz2019trapped}. However, in search for a more resource-efficient system of error correction, there is rising interest to encode quantum information into continuous-variable degrees of freedom, such as the motional degrees of freedom of the ions. For instance, Fl\"uhmann et al~\cite{fluhmann2019encoding} created a GKP qubit state in the axial motional mode of a single ion and, furthermore, implemented single-qubit logical operations with high fidelity. In a follow up work~\cite{home2022QECgkp}, de Neeve et al (of the same group) further demonstrated quantum error correction of both the square and hexagonal GKP codes through engineered dissipation. An enhancement in the coherence time of the logical states by a factor of three was also demonstrated, marking an important step towards break-even in trapped ion systems for bosonic error correction. For further details for trapped ion systems, please refer to dedicated reviews such as Refs~\cite{bruzewicz2019trapped,brown2021materials,monroe2021programmable} and others. In terms of experimental platforms, the main focus of our review is on superconducting circuit implementations of GKP codes, for which more literature (in terms of experimental realizations and proposals) has currently been published. 

\subsection{Basics and notation}
\label{sec:basic_notation}
In this section, we review the basic mathematical formalism to describe a simple quantum harmonic oscillator and then broaden the framework to encompass multiple oscillators. This extension is performed in a methodical and efficient way, allowing for the analysis of various properties---including the Wigner function and entropy of a bosonic quantum state consisting of $N$ modes.

Throughout this review, we primarily write a `hat' on operators only in places where operators can potentially be confused with numbers---e.g., for the canonical operators $\hat{q}$ and $\hat{p}$. We omit `hat' for unitary operators, density matrices, and others that are easily recognized as operators. For finite-dimensional matrices, vectors etc. associated with the phase-space of the modes (see below for more details), we generally use boldface.

\subsubsection{A single harmonic oscillator}
\label{sec:qho}

We begin with a textbook explanation of a simple quantum harmonic oscillator. Consider the (normalized) position and momentum operators $\hat{q}$ and $\hat{p}$ of the oscillator, which obey the canonical commutation relations
\begin{equation}
    \comm{\hat{q}}{\hat{p}}=i\hbar,
\end{equation}
where $\hbar$ is the reduced Planck's constant. 
A quantum harmonic oscillator of mass $m$ and frequency $\omega$ has a Hamiltonian,
\begin{align}
    \hat{H}_{\rm osc}= \frac{1}{2}\left(m\omega^2\hat{q}^2+\frac{1}{m}\hat{p}^2\right)=\hbar\omega(\hat{n}+1/2),
\end{align}
where the additive constant $\hbar\omega/2$ is the `zero-point energy'. We ignore it for the most part. The operator $\hat{n}\coloneqq\hat{a}^\dagger\hat{a}$ is the number operator and tells us how many quanta occupy an oscillator state. The annihilation and creation operators, $\hat{a}$ and $\hat{a}^\dagger$, are related to the canonical operators $\hat{q}$ and $\hat{p}$ via
\begin{equation}
    \hat{a}=\sqrt{\frac{m\omega}{2\hbar}}\left(\hat{q}+\frac{i}{m\omega}\hat{p}\right),
\end{equation}
which obey $\comm{\hat{a}}{\hat{a}^\dagger}=1$. Likewise, $\sqrt{2\hbar/m\omega} \Re{\hat{a}}=\hat{q}$ and $\sqrt{2\hbar m\omega}\Im{\hat{a}}=\hat{p}$. In this review, we often set $\hbar=1$ and $m\omega=1$ for convenience, so that $\hat{a}=(\hat{q}+i\hat{p})/\sqrt{2}$.

Physically, the annihilation and creation operators describe the destruction or creation of single oscillator quanta.
One can show that the following \textit{Fock states} are eigenstates of the number operator (and thus the oscillator Hamiltonian),
\begin{equation}
    \ket{n}\coloneqq\frac{\left(\hat{a}^{\dagger}\right)^n}{\sqrt{n!}}\ket{\rm vac}  \qq{and} \hat{n}\ket{n}=n\ket{n},
\end{equation}
where $n$ is a positive integer and $\ket{\rm vac}\coloneqq\ket{0}$ is the vacuum (i.e., lowest energy) state defined implicitly via $\hat{a}\ket{\rm vac}=0$. The set $\{\ket{n}\}_{n=0}^{\infty}$ forms an orthonormal basis in the bosonic Hilbert space of a single mode (of mode frequency $\omega$) $\mathscr{H}$, i.e.
$
    \innerproduct{m}{n}=\delta_{mn} $ and $
   \sum_{n=0}^\infty\dyad{n}=\hat{I},
$
where $\delta_{mn}$ is the Kronecker delta and $\hat{I}$ is the identity on $\mathscr{H}$. Finally, one can show that
$
    \hat{a}\ket{n}=\sqrt{n}\ket{n-1},
    \hat{a}^\dagger\ket{n}=\sqrt{n+1}\ket{n+1},
$
i.e. $\hat{a}$ annihilates one quanta and $\hat{a}^\dagger$ creates one quanta of the oscillator.

We may also ask, at what position is the oscillator, or what momentum does the oscillator have? In this case, we rely on the (non-normalizable) position and momentum eigenstates $\ket{q}$ and $\ket{p}$, where $\hat{q}\ket{q}=q\ket{q}$, $\hat{p}\ket{p}=p\ket{p}$ and $q,p\in\mathbb{R}$. The position and momentum eigenstates resolve to the identity in the continuum,
\begin{equation}
\int_{q\in\mathbb{R}}\dd{q}\dyad{q}= \int_{p\in\mathbb{R}}\dd{p}\dyad{p}=\hat{I},
\label{q_p_eigenstate}
\end{equation}
and are related via Fourier transform,
$
    \ket{q}=\frac{1}{\sqrt{2\pi}}\int_{p\in\mathbb{R}}\dd{p}{\rm e}^{-ipq}\ket{p},
$
under the convention that $\delta(x-x^\prime)=\int\dd{k}{\rm e}^{i(x-x^\prime)k}/2\pi$.

We can ``shift'' (or displace) the position of the oscillator by an amount $x$ via the unitary operator ${\rm e}^{-ix\hat{p}}$. Likewise we can ``boost'' the momentum by an amount $k$ via ${\rm e}^{ik\hat{q}}$. We can both shift and boost the oscillator using the general \emph{displacement operator}
\begin{equation}\label{eq:displace_1mode}
    D_{x,k}={\rm e}^{-i(x\hat{p}-k\hat{q})}.
\end{equation}
The general definition of displacement operator is given in terms of annihilation and creation operators. Let $\alpha=(x+ik)/\sqrt{2}$, then using the fact that $\hat{q}=\sqrt{2}\Re\hat{a}$ and $\hat{p}=\sqrt{2}\Im\hat{a}$, we can write the displacement operator as
\begin{equation}
    D_\alpha={\rm e}^{-\alpha^*\hat{a}+\alpha\hat{a}^\dagger}.
\end{equation}
Displacing the vacuum by an amount $\alpha$ leads to a so-called \emph{coherent state} 
\be 
\ket{\alpha}\coloneqq D_\alpha\ket{\rm vac}.
\ee 
The coherent state is an eigenstate of the annihilation operator $\hat{a}$ with eigenvalue $\alpha$. Coherent states form an (overcomplete) basis of states in $\mathscr{H}$ such that
$
\int_{\alpha\in\mathbb{C}}\frac{\dd{\alpha}}{\pi}\dyad{\alpha}=\hat{I}.
$
We note that, using the displacement operator, one can define the characteristic function and Wigner function of the quantum state, which we defer to the multimode case for a general description [see Eqs.~\eqref{eq:characteristic_fn} and~\eqref{Wigner_func}].

\subsubsection{Multimode bosonic system}
\label{sec:multi_mode_bosonic_system}
We can extend the previous discussion for a single harmonic oscillator to a set of $N$ harmonic oscillators. Consider an $N$-mode bosonic Hilbert space $\mathscr{H}^{\otimes N}$, where $\mathscr{H}$ is the Hilbert space of a single bosonic mode similar to Chapter~\ref{sec:qho}. More formally, we have an $N$ mode bosonic quantum system associated with a Hilbert space $\mathscr{H}^{\otimes N}$. The canonical variables of the system are the $\hat{q}$'s and $\hat{p}$'s of the 2N dimensional phase space $\mathbb{R}^{2N}$. The bosonic system has a symplectic structure on $\mathbb{R}^{2N}$ induced by the commutation relations of the canonical variables. We define a vector of canonical operators for the $N$ modes as,
\begin{equation}
    \hat{\bm r}^\top\coloneqq\left(\hat{q}_1,\hat{p}_1,\dots,\hat{q}_N,\hat{p}_N\right),
\end{equation}
such that,
\begin{equation}\label{eq:ccr}
    [\hat{\bm r}_k,\hat{\bm r}_j]=i \bm{\Omega}_{kj},
\end{equation}
where $\bm\Omega$ is a $2N\times2N$ matrix representing the $N$-mode symplectic form,
\begin{equation}\label{eq:symplectic_form}
    \bm\Omega = \bigoplus_{i=1}^N\bm\Omega_1 \qq{with} 
    \bm\Omega_1=
    \begin{pmatrix}
        0 & 1 \\
        -1 & 0
    \end{pmatrix}.
\end{equation}
Equation~\eqref{eq:ccr} is nothing but the standard canonical commutation relations between the various $\hat{q}$'s and $\hat{p}$'s of the modes. For later use, we define the mean $\bm\mu$ and covariance matrix $\bm\sigma$ for a quantum state $\rho$ as
\begin{align}
    \bm\mu&\coloneqq\Tr\big(\hat{\bm r}\rho\big),\label{eq:mu_def}\\
    \bm\sigma_{ij}&\coloneqq
    \frac{1}{2}\Tr\Big(\acomm{\hat{\bm r}_i-\bm\mu_i}{\hat{\bm r}_j-\bm\mu_j}\rho\Big),\label{eq:sigma_def}
\end{align}
where $\acomm*{\hat{A}}{\hat{B}}\coloneqq\hat{A}\hat{B}+\hat{B}\hat{A}$. For Gaussian quantum states (introduced in Chapter~\ref{sec:gauss_states}), the mean and covariance completely specify the state. In this sense, the way of describing a Gaussian state is analog to the classical description of a Gaussian distribution; nevertheless, Gaussian states can in general possess non-classical features such as squeezing and entanglement. In addition to quadrature operators, sometimes it is useful to consider annihilation operators $\hat{a}_1,\cdots,\hat{a}_N$, with each $\hat{a}_k=(\hat{q}_k+i\hat{p}_k)/\sqrt{2}$.

We can shift or displace the canonical coordinates by some constant amount $\hat{\bm r}\rightarrow\hat{\bm r}+\bm\zeta$, where $\bm\zeta\in\mathbb{R}^{2N}$. Displacements are generated by a Hamiltonian that is strictly linear in the canonical operators. The unitary transformation for a shift is given by the multimode displacement (or Weyl) operator,
\begin{equation}\label{eq:weyl_op}
    {D}_{\bm\zeta}\coloneqq\exp\left(-i{\bm\zeta}^\top\bm\Omega\hat{\bm r}\right),
\end{equation}
such that $D_{\bm\zeta}^\dagger\hat{\bm r}D_{\bm\zeta}=\hat{\bm r}+\bm\zeta$, which is a generalization of Eq.~\eqref{eq:displace_1mode}. The displacement operators form an operator basis for the space of bounded operators $\mathcal{B}(\mathscr{H}^{\otimes N})$, which include physical quantum states $\rho\in\mathcal{B}(\mathscr{H}^{\otimes N})$~\cite{serafini2017book}. The orthonormality condition satisfied by displacement operators is given by,
\begin{equation}\label{eq:weyl_opbasis}
    \Tr({D}_{\bm\zeta}{D}_{-\bm\nu})=(2\pi)^N\delta^{2N}(\bm\zeta-\bm\nu),
\end{equation}
where $\delta^{2N}(\cdot)$ is the $2N$-dimensional Dirac delta distribution. Therefore, we can expand a general operator function $f(\hat{\bm r})$ in the Weyl basis via
\begin{equation}\label{eq:weyl_transform}
    f(\hat{\bm r})=\frac{1}{(2\pi)^N}\int_{\bm\zeta\in\mathbb{R}^{2N}}\dd{\bm\zeta}\Tr(f(\hat{\bm r})D_{-\bm\zeta})D_{\bm\zeta}.
\end{equation}
Displacement operators can be shown to satisfy a composition rule, 
\begin{equation}
    {D}_{\bm\zeta+\bm\nu}={\rm e}^{i\omega(\bm\zeta,\bm\nu)/2}{D}_{\bm\zeta}{D}_{\bm\nu},\label{eq:weyl_comp}
\end{equation}
where $\omega(\bm\zeta,\bm\nu)\coloneqq\bm\zeta^\top\bm\Omega\bm\nu$ is the \textit{symplectic inner product} between $\bm\zeta$ and $\bm\nu$. The symplectic product obeys $\omega(\bm\nu,\bm\zeta)=-\omega(\bm\zeta,\bm\nu)$ and is invariant under symplectic transformations $\bm S$ where $\bm S^\top\bm\Omega \bm S=\bm \Omega$, that is, $\omega(\bm S\bm\zeta,\bm S\bm\nu)=\omega(\bm\zeta,\bm\nu)$. Symplectic transformations represent any basis change in the continuous variable systems and are discussed in greater detail in the next section. By swapping $\bm\zeta$ and $\bm\nu$, it follows that ${D}_{\bm\zeta}{D}_{\bm\nu}={\rm e}^{-i\omega(\bm\zeta,\bm\nu)}{D}_{\bm\nu}{D}_{\bm\zeta}$. If $\omega(\bm\zeta,\bm\nu)=2\pi n$ with $n\in\mathbb{Z}$, then the displacement operators commute; we come back to this point later when we discuss GKP states. Observe that displacements are always local in the sense that ${D}_{\bm\zeta}=\bigotimes_{i=1}^N{D}_{\bm\zeta_i}$, where $\bm\zeta_i\in\mathbb
{R}^2$ are single-mode displacements.

We can apply the operator basis expansion of Eq.~\eqref{eq:weyl_transform} to a density matrix $\rho$ to obtain,
\begin{equation}\label{eq:characteristic_fn}
    \rho=\frac{1}{(2\pi)^{N}}\int_{\bm\zeta\in\mathbb{R}^{2N}}\dd{\bm\zeta}\chi(\bm\zeta)D_{\bm\zeta},
\end{equation}
where $\chi(\bm\zeta)\coloneqq\Tr(\rho D_{-\bm\zeta})$. The function $\chi(\bm\zeta)$ is known as the \emph{characteristic function} of $\rho$. Since $\Tr\rho=1$ and $\rho^\dagger=\rho$, we have $\chi(\bm 0)=1$ and $\chi^*(-\bm\zeta)=\chi(\bm\zeta)$. By a Fourier transform of the characteristic function $\chi(\bm\zeta)$, we obtain the \emph{Wigner function} $W(\bm\zeta)$,
\begin{equation}
    W(\bm\zeta)\coloneqq\frac{1}{(2\pi)^{2N}} \int_{\bm\nu\in\mathbb{R}^{2N}}\dd{\bm\nu}{\rm e}^{i\bm\zeta^\top\bm\Omega\bm\nu}\chi(\bm\nu).
    \label{Wigner_func}
\end{equation}
The Wigner function is a quasi-probability distribution in phase space (quasi because it can be negative). The marginal of the Wigner function along the $q$ ($p$) direction gives the probability that the state has momentum $p$ (position $q$)---e.g., for a single mode, $\int\dd{q}W(q,p)=\expval{\rho}{p}$ where $\ket{p}$ is the (non-normalizable) eigenstate of the momentum operator $\hat{p}$ (see Chapter~\ref{sec:qho}).

We conclude this section with the (von Neumann) entropy of a quantum state $\rho$,
\begin{equation}
    S(\rho)\coloneqq-\Tr\left(\rho\log\rho\right),
\end{equation}
where $\log$ is base 2 here. The entropy is strictly 0 for pure states and, for an $N$ mode bosonic system with total mean quanta $\bar{N}$, the entropy per mode ($S/N$) is upper-bounded by the thermal entropy 
\begin{equation}\label{eq:thermal_entropy}
s_{\rm th}(\bar{n})\coloneqq (\bar{n}+1)\log(\bar{n}+1)-\bar{n}\log\bar{n},
\end{equation}
where $\bar{n}=\bar{N}/N$ is the number of quanta per mode. The entropy is invariant under unitaries, $S(U\rho U^\dagger)=S(\rho)$ and is sub-additive, such that $S(\rho_{AB})\leq S(\rho_A)+S(\rho_B)$. Finally, the entropy is concave---i.e., for an ensemble density of matrices $\{\lambda_i, \rho_i\}$, where $0\leq\lambda_i\leq1$ and $\sum_i\lambda_i=1$, $S(\sum_i\lambda_i\rho_i)\geq\sum_i\lambda_i S(\rho_i)$.

\subsection{Quantum channels and dynamics of open quantum systems}
\label{sec:channel_dynamics}

General quantum processes can be modeled as interaction between a quantum system and environment. In the following, we describe two theoretical tools that are often adopted in their modeling.

A quantum channel $\mathcal{N}$ models quantum dynamics in a finite time via input-output relations. Consider a quantum system $S$ and an environment $E$ that the system interacts with. The overall evolution of the system-environment can be modeled as a unitary evolution $ {U}_{SE}$, and the initial environment can always be taken as a pure state $\sigma_E=\dyad{e_0}_E$. 
Now, given a system described by the density matrix $\rho_S$, a set of basis vectors $\{\ket{e_k}_E\}$ of the environment, and the joint unitary interaction $ {U}_{SE}$, the input-output relations for a quantum channel can be generically written as  
\begin{align}
    \mathcal{N}(\rho_S)&=\Tr_E\Big\{ {U}_{SE}(\rho_S\otimes\sigma_E) {U}^\dagger_{SE}\Big\}\nonumber\\
    &=\sum_{k}\bra{e_k}_E\left( {U}_{SE}(\rho_S\otimes\dyad{e_0}_E) {U}^\dagger_{SE}\right)\ket{e_k}_E\nonumber\\
    &=\sum_k\left(\bra{e_k} {U}_{SE}\ket{e_0}\right)\rho_S\left(\bra{e_0} {U}^\dagger_{SE}\ket{e_k}\right)\nonumber\\
    &=\sum_k {K}_k\rho_S {K}_k^\dagger,\label{eq:general_dilation}
\end{align}
where $ {K}_k=\bra{e_k} {U}_{SE}\ket{e_0}$ are the Kraus operators that act on the system Hilbert space $\mathscr{H}_S$. By the completeness relation $\sum_k\dyad{e_k}_E= {I}_E$ and the normality condition $\innerproduct{e_0}=1$, the Kraus operators satisfy $\sum_k {K}_k^\dagger {K}_k= {I}_S$ as required of a good quantum channel. Observe that the Kraus decomposition for the channel is not unique, since an arbitrary unitary transformation on the environment basis vectors leads to a different set of Kraus operators $\{K_k\}$ but nevertheless describe the same channel $\mathcal{N}$.

In cases where one tries to control an open quantum system continuously, the simple input-output quantum channel model above is insufficient. Instead, it is more appropriate to adopt a dynamic approach to describe the open quantum system. We assume a Markov approximation for the open system dynamics, where the system is thought to interact with a memoryless environment/bath. In simpler terms, the state of the system (after tracing over the bath degrees of freedome) at time $t+\delta t$ only depends on its state at time $t$, without considering the complete history of the system.  Under the Markov approximation, one can show that the density matrix $\rho$ obeys the master equation in Lindblad form,\footnote{A concise and pedagogical introduction to the master equation can be found in Section 3.5 of John Preskill's lecture notes~\cite{preskill_LectureNotes}.}
\begin{equation}
    \dot\rho=-i [H,\rho]+\sum_i\Gamma_i\mathbb{D}[L_i]\rho,\label{eq:master_eq} 
\end{equation}
where $\mathbb{D}[O]\rho=O\rho O^\dagger-\{O^\dagger O,\rho\}/2$ is the standard dissipation superoperator, $L_i$ is the (dimensionless) Lindbladian jump operator, and $\Gamma_i$ is a damping rate. The open system dynamics may lead to a fixed point (or steady-state) solution $\rho_{\infty}$ such that $\dot{\rho}_\infty=0$. We give some examples of open system dynamics in Chapter~\ref{sec:noise_model} that are important for modeling certain bosonic platforms.

\subsection{Gaussian evolution and Gaussian states} 
\label{sec:gauss_evol}
Gaussian states and unitaries are a fundamental class of quantum resources that enable many quantum information processing tasks. They play a crucial role in bosonic QEC and offer significant advantages in terms of feasibility, particularly in the optical domain. In the context of infinite-energy GKP qubit codes (refer to Chs.~\ref{sec:gkp_lattice} and~\ref{sec:QEC-multimode}), Gaussian operations serve as the foundation for basic logical operations. Additionally, GKP oscillators-to-oscillators codes (as discussed in Chapter~\ref{sec:gkp_o2o_codes}) rely on Gaussian unitaries during the encoding and decoding processes. Here we briefly review Gaussian unitaries, Gaussian channels, and Gaussian states. 
For a more complete exposition, we refer the reader to Refs.~\cite{weedbrook2012rmp,serafini2017book}. In superconducting cavities (see Chapter~\ref{sec:GKP_eng}), displacement operations are relatively easy, while the high-fidelity realization of other Gaussian unitaries (such as beamsplitting and squeezing) without inducing unwanted interactions/loss is an active topic of research \cite{chapman2022high,lu2023microwaveBS}.

\subsubsection{Gaussian unitaries}\label{sec:gauss_unitaries}
We first provide a formal explanation of Gaussian unitaries---corresponding to symplectic transformations in the $2N$-dimensional phase space of $N$ modes---as well as some typical examples of single- and two-mode transformations. We also make connections between the mathematical formalism and physical realizations throughout, for example, parametric amplification and linear-optical networks correspond to squeezing and orthogonal transformations, respectively.

Consider a set of $N$ bosonic modes with canonical operators $\hat{\bm r}$. We assume the modes interact via quadratic interactions---i.e., the Hamiltonian is at most quadratic in $\hat q$'s and $\hat p$'s. Since the Hamiltonian is quadratic, the evolution of the canonical operators is linear such that $\hat{\bm r}\rightarrow\bm S\hat{\bm r}$, where $\bm S$ is some $2N\times 2N$ matrix; we ignore possible shifts (displacements) in the coordinates $\hat{\bm r}$ for now. In order to preserve the canonical commutation relations~\eqref{eq:ccr}, we must have that
\begin{equation}\label{eq:symplectic_cond}
    \bm S\bm\Omega\bm S^\top=\bm\Omega.
\end{equation}
That is, the matrix $\bm S$ must preserve the (real) symplectic form $\bm\Omega$. Any matrix which satisfies this condition is called a \emph{symplectic matrix} and $\bm S\in{\rm Sp}(2N,\mathbb{R})$, where ${\rm Sp}(2N,\mathbb{R})$ is a $2N\times2N$ matrix representation of the real symplectic group (which has dimension $\abs{{\rm Sp}(2N,\mathbb{R})}=2N^2+N$). From the symplectic matrix $\bm S$, we can find a unitary representation $U_{\bm S}$, which encodes the symplectic transformation $\bm S$ (or vice versa) and produces an input-output relation for the canonical operators as,
\begin{equation}\label{eq:r_S}
    U_{\bm S}^\dagger\hat{\bm r}U_{\bm S} = \bm S \hat{\bm r}.
\end{equation}
Using this equation and the general conjugation formula $U^\dagger f(\hat{g})U=f(U^\dagger\hat{g}U)$, a symplectic transformation $\bm S$ transforms the displacement operators $D_{\bm \mu}$ of Eq.~\eqref{eq:weyl_op} as
\begin{equation}\label{eq:weyl_S}
    U_{\bm S}^\dagger{D}_{\bm\zeta}U_{\bm S}={D}_{\bm S^{-1}\bm\zeta}.
\end{equation}

For an arbitrary symplectic matrix $\bm S$, it turns out that we can decompose $\bm S$ into simple single-mode and two-mode elements by the so-called \emph{Bloch-Messiah decomposition}~\cite{braunstein05}
\begin{equation}\label{eq:bloch_messiah}
    \bm S = \bm B^\prime\cdot\left(\bigoplus_{i=1}^N{\rm e}^{-r_i\bm Z}\right)\cdot\bm B,
\end{equation}
where $\bm Z$ is the $2\times2$ Pauli-Z matrix and ${\rm e}^{-r_i\bm Z}$ is a single mode squeezing transformation on the $i$th mode with squeezing strength $r_i$. Here $\bm B^\prime,\bm B\in{\rm Sp}(2N,\mathbb{R})\cap{\rm SO}(2N)\simeq U(N)$, where $U(N)$ is the unitary group of dimension $\abs{U(N)}=N^2$. In other words, $\bm B^\prime$ and $\bm B$ are passive (e.g., linear optical) transformations ($\bm B\bm\Omega\bm B^\top=\bm\Omega$ and $\bm B\bm B^\top=\bm I$). 

\begin{figure}
    \centering
    \includegraphics[width=\linewidth]{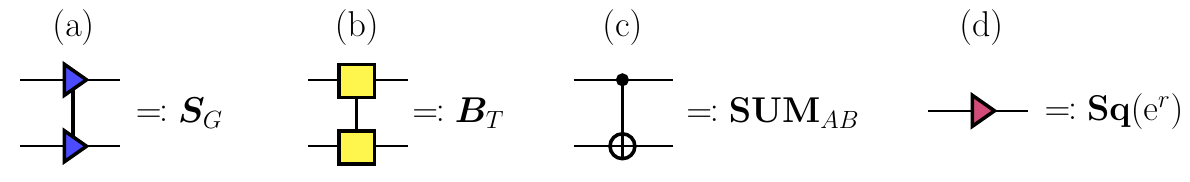}
    \caption{Diagrammatic representations of common single- and two-mode transformations: (a) two-mode squeezer, (b) beamsplitter, (c) SUM gate, and (g) single-mode squeezer. Squeezer notation taken in analogy with classical circuit representations of linear amplifiers.}
    \label{fig:circuit_dictionary}
\end{figure}

Below, we provide some examples of Gaussian unitaries in the single- and two-mode cases. Simple diagrams representing these basic elements, which we can compose to form multimode CV quantum circuits, are shown in Figure~\ref{fig:circuit_dictionary}. We also give an explicit example of the Bloch-Messiah decomposition for a TMS interaction in Example~\ref{example:tms} below.

\begin{example}[Displacements]
Although we introduced displacement as the basis operators in Eq.~\eqref{eq:weyl_op}, they are physical operations that can be experimentally realized. The sympleclic transformation is the trivial identity $\bm I$, as displacement only leads to constant shifts of quadrature operators.
In linear optics, a displacement operation is realized by interacting two modes on a beamsplitter where one of the modes is in a strong coherent state (e.g., high-power laser). A vanishing portion of the strong coherent state is then mixed into the the mode, leading to a constant displacement. For microwave cavities, a displacement is realized by coupling the oscillator to a transmission line (either capacitive or inductive coupling) and driving the oscillator on-resonance with a microwave signal (see \cite{blais2021cqedRMP}).
\end{example}

\begin{example}[Single-mode transformations]\label{example:single_transform}
Applying the Bloch-Messiah decomposition~\eqref{eq:bloch_messiah} to a single mode, we have that general single-mode symplectic transformation reduces to a set of phase-shifts and a single-mode-squeezing, $\bm S=\bm R(\phi_2)\textbf{Sq}({\rm e}^r)\bm R(\phi_1)$.
The single-mode squeezer $\textbf{Sq}({\rm e}^r)$ with squeezing strength $r$ has a symplectic matrix representation,
\begin{equation}
    \textbf{Sq}({\rm e}^r)= {\rm e}^{-r\bm Z},
    \label{eq:Sq_r}
\end{equation}
whereas a phase rotation $\bm R(\phi)$ is a $2\times2$ rotation matrix,
\begin{equation}
    \textbf{R}(\phi)=
    \begin{pmatrix}
    \cos\phi & \sin\phi \\
    -\sin\phi & \cos\phi
    \end{pmatrix},
    \label{eq:R_phi}
\end{equation}
which, for instance, takes $\hat{a}\rightarrow\hat{a}e^{-i\phi}$. The single mode squeezer is generated by the Hamiltonian $\hat{a}^2+h.c.$, and the phase rotation is generated by the Hamiltonian $\hat{a}^\dagger \hat{a}$, where $h.c.$ stands for Hermitian conjugate.
\end{example}

\begin{example}[Beam-splitter]
    The two-mode beamsplitter has a symplectic representation
\begin{equation}\label{eq:bs_matrix}
    \bm B = 
    \begin{pmatrix}
        \cos\theta\bm I & \sin\theta\bm I\\
        -\sin\theta\bm I & \cos\theta\bm I
    \end{pmatrix},
\end{equation}
where $\cos^2\theta$ is the transmittance of the beamsplitter. A beamplitter between two modes $\hat{a}_1$ and $\hat{a}_2$ is generated by the Hamiltonian $\hat{a}_1^\dagger \hat{a}_2+h.c.$. For a 50:50 beamsplitter ($\theta=\pi/4$), we may sometimes express the symplectic matrix as $\bm B_{1/2}$, or for a beamsplitter with transmittance $T=\cos^2\theta$, we may write $\bm B_{T}$. Practically, we can concatenate phase-shifts and beamsplitters to, e.g., construct a Mach-Zehnder interferometer or an arbitrary linear-optical network on $N$ modes~\cite{reck94}.
\end{example}

\begin{example}[Two-mode squeezing]\label{example:tms}
Two-mode squeezing, ubiquitous in CV quantum information processing, is used to generate entangled photon pairs. In optics, this is achieved through a process called (non-degenerate) spontaneous parametric down conversion~\cite{Jeff_notes_20}. For superconducting circuits, two-mode squeezing can be achieved by coupling cavities to nonlinear \QZ{Josephson} mixers and driving three- or four-wave processes (see, for example, \cite{Chien_2020}). A two-mode squeezing transformation with gain $G$ has a symplectic matrix representation,
\begin{equation}\label{eq:tms_matrix}
    \bm S_{G}=
    \begin{pmatrix}
        \sqrt{G}\bm I & \sqrt{G-1}\bm Z\\
        \sqrt{G-1}\bm Z & \sqrt{G}\bm I
    \end{pmatrix}.
\end{equation}
As an interesting example of the Bloch-Messiah decomposition~\eqref{eq:bloch_messiah}, we can reparameterize the gain $G$ in terms of the squeezing strength $r$ via $G=\cosh^2r$ and write the two-mode squeezer in terms of 50/50 beamsplitters and single-mode squeezers,
\begin{equation}\label{eq:tms_decomp}
    \bm S_{G}= \bm B_{1/2}^{\top}\cdot\Big(\textbf{Sq}({\rm e}^{-r})\oplus\textbf{Sq}({\rm e}^{r})\Big)\cdot\bm B_{1/2}.
\end{equation}
\end{example}

\begin{example}[SUM-gate or ``CV CNOT'']
We introduce the SUM-gate, which has a unitary representation ${\rm SUM}_{AB}\coloneqq{\rm e}^{-i \hat{q}_A\otimes\hat{p}_B}$. For many (infinite-energy) GKP encodings, the SUM-gate is the CV equivalent of a CNOT gate~\cite{gkp2001} and can be used for ancilla-assisted measurement protocols in the CV domain. The SUM gate transforms the canonical operators as,
\begin{align}
\begin{aligned}
     \hat{q}_A&\rightarrow\hat{q}_A,& \qq{} \hat{p}_A&\rightarrow \hat{p}_A-\hat{p}_B,&\\
    \hat{q}_{B}&\rightarrow\hat{q}_{B}+\hat{q}_A,&\qq{}\hat{p}_{B}&\rightarrow\hat{p}_{B}.&
\end{aligned}
\end{align}
and has symplectic representation,
\begin{equation}
    \textbf{SUM}_{AB}=
    \begin{pmatrix}
    \bm{I}_2 & -\bm\Pi_p\\
    \bm\Pi_q & \bm{I}_2
    \end{pmatrix}, \label{eq:sum_matrix}
\end{equation}
where $\bm\Pi_q={\rm diag}(1, 0)$ and $\bm\Pi_p={\rm diag}(0, 1)$ represent projections along the $q$ and $p$ directions of the modes. See Ref.~\cite{tzitrin2020ProgressGKP} for a Bloch-Messiah decomposition of the SUM-gate.
\end{example}

Before moving forward to general Gaussian channels, we consider a useful property of unitary Gaussian transformations:
\begin{theorem}[Modewise entanglement theorem~\cite{reznik2003modewise}]\label{thm:modewise}
Consider a subsystem $A$ of $N$ modes and a subsystem $B$ of $M$ modes, such that the joint system $AB$ consists of $K=M+N$ modes, and define the positive definite matrix $\bm{\tau}_{AB}=\tau\bm S\bm S^\top$, where $\tau\in(0,\infty)$ and $\bm S\in{\rm Sp}(2K,\mathbb{R})$. Then, there exists local symplectic matrices $\bm S_A\in{\rm Sp}(2N,\mathbb{R})$ and $\bm S_B\in{\rm Sp}(2M,\mathbb{R})$ such that,
\begin{equation}
\left(\bm S_A\oplus\bm S_B\right)\bm{\tau}_{AB}\left(\bm S_A^\top\oplus\bm S_B^\top\right)=\tau\left(\bigoplus_{i=1}^{N}\bm S_{G_i}\bm S_{G_i}^{\top}\right)\oplus\bm I_{2(M-N)},
\end{equation}
where $\bm S_{G_i}$ is a TMS squeezing operation of gain $G_i$ [see, e.g., Eq.~\eqref{eq:tms_matrix}] between the $i$th mode in $A$ and the $(N+i)$th mode in $B$ and $\bm I_{2(M-N)}$ is the identity on the remaining $M-N$ modes.
\end{theorem}
We refer the reader to Refs.~\cite{serafini2017book,reznik2003modewise,adesso2005localizable,adesso2007thesis} for detailed proofs and more in depth discussion. We make use of this theorem in Chapter~\ref{sec:code_redux} to significantly reduce the complexity of generic O2O codes that are used to battle additive Gaussian noise (AGN)~\cite{wu2022optimal}.


\subsubsection{Gaussian channels}\label{sec:gauss_channels}

As alluded to in Chapter~\ref{sec:channel_dynamics}, we can describe the \emph{non-unitary} evolution of a bosonic quantum system $Q$ of $N$ modes (prepared in an initial state $\Psi_Q$) through a quantum channel $\mathcal{C}:\mathscr{H}^{\otimes N}\rightarrow\mathscr{H}^{\otimes N}$ by a larger \emph{unitary} evolution $U:\mathscr{H}^{\otimes N}\otimes\mathscr{H}^{\otimes M}\rightarrow\mathscr{H}^{\otimes N}\otimes\mathscr{H}^{\otimes M}$ which couples the $N$-mode system $Q$ to a $M$-mode environment $E$ (in the quantum state $\rho_E$).\footnote{For a system of $N$ modes, it is sufficient to choose an environment of $M\leq2N$ modes~\cite{weedbrook2012rmp}.} For a Gaussian channel $\mathcal{G}$, the environment state $\rho_E$ is a Gaussian state (see Chapter~\ref{sec:gauss_states} for details about Gaussian states), and the unitary is determined by a symplectic matrix $\bm S$. Formally, we can write the Gaussian channel as $\mathcal{G}({\Psi})=\Tr_E\left[U_{\bm S}\left({\Psi}_Q\otimes{\rho}_E\right)U_{\bm S}^\dagger\right]$, which is a specific instance of Eq.~\eqref{eq:general_dilation}.

Below we introduce some common single-mode Gaussian quantum channels, which will be useful throughout this review. We note that all single-mode (non-unitary) Gaussian channels fall within 6 channel classes~\cite{holevo2007classification} (see also Table 1 of Ref.~\cite{weedbrook2012rmp} and surrounding discussion), but we focus primarily on four such channels--- the unitary displacement channel, thermal loss channel, amplifier channel, and additive Gaussian noise channel ---because of their prevalence in bosonic quantum information processing. Due to the simplicity of Gaussian channels, we find it convenient to directly describe them with the evolution of the annihilation operator $\hat{a}$; the corresponding quadrature evolution can be obtained from the real and imaginary parts. 

\begin{example}[Displacement channel]
Consider a single-mode and the displacement operator $D_\alpha=\exp(\alpha\hat{a}^\dagger-\alpha^*\hat{a})$, where $\alpha\in\mathbb{C}$. It acts on the annihilation operator $\hat{a}$ as
\begin{equation}\label{eq:displacement}
D^\dagger_\alpha\hat{a}D_\alpha=\hat{a}+\alpha.
\end{equation}
We define the unitary displacement channel, $\mathcal{D}_{\alpha}$, as a unitary conjugation by $D_\alpha$. For a quantum state $\Psi$, $\mathcal{D}_{\alpha}(\Psi)=D_\alpha\Psi D^\dagger_\alpha$, which leads to Eq.~\eqref{eq:displacement} in the Heisenberg picture. For $N$ modes with canonical operators $\hat{\bm r}$, the displacement operator is $D_{\bm\zeta}$ per Eq.~\eqref{eq:weyl_op}, where $\bm\zeta\in\mathbb{R}^{2N}$. We define the displacement channel $\mathcal{D}_{\bm\zeta}$ which acts in the Heisenberg picture as 
$\mathcal{D}_{\bm\zeta}: \hat{\bm r}{\,\longrightarrow\,} \hat{\bm r} +\bm\zeta$. 
\end{example}

\begin{example}[Thermal loss channel or Photon loss channel]
\label{example:thermal_loss}
Consider a thermal loss channel $\calL_{\eta,\bar{n}}$ which obeys the following input-output relation on the annihilation operators,
\be 
\hat{a}'=\sqrt{\eta}\hat{a}+\sqrt{1-\eta} \hat{e},
\ee 
where $0\leq\eta\leq1$ is the transmittance of the channel and $\hat{e}$ represents an environment mode in a (Gaussian) thermal state with $\expval{\hat{e}^\dagger \hat{e}}=\bar{n}$ mean number of quanta. 
The intensity of the input is reduced by $\eta$ and thermal noise $(1-\eta)(1/2+\bar{n})$ is added to the quadratures; the factor of $1/2$ originates from vacuum noise. For $\bar{n}=0$, the channel is often referred to as a pure-loss channel. Physically, the thermal-loss channel stems from interacting the system mode $\hat{a}$ with a thermal environment mode $\hat{e}$ by a beamsplitter-like interaction.
\end{example}

\begin{example}[Thermal amplifier channel]
Consider an amplifier channel $\calA_{G, \bar{n}}$ which obeys the following input-output relation,
\be 
\hat{a}'=\sqrt{G}\hat{a}+\sqrt{G-1} \hat{e}^\dagger,
\ee 
where $G\geq1$ is the gain of the channel and $\hat{e}$ represents an environment mode in a (Gaussian) thermal state with $\expval{\hat{e}^\dagger \hat{e}}=\bar{n}$ mean number of quanta. 
The intensity of the input is amplified by $G$ and thermal noise $(G-1)(1/2+\bar{n})$ is added to the quadratures. For $\bar{n}=0$, the channel is often referred to as a quantum-limited amplifier. Physically, the thermal amplifier channel stems from interacting the system mode $\hat{a}$ with a thermal environment mode $\hat{e}$ by a two-mode squeezing-like interaction.
\end{example}

\begin{example}[Additive Gaussian noise channel]
One can define an additive Gaussian noise (AGN) channel formally via 
\be 
\calN_{\sigma}=\lim_{\eta\to 1}\calL_{\eta, \bar{n}/(1-\eta)},
\ee
where $\sigma^2\coloneqq\bar{n}$ is the quadrature variance of the AGN channel. Equivalently, we can view the AGN channel as applying Gaussian random displacements $\mathcal{D}_{\bm\xi}$ to a quantum state $\Psi$, where $\bm\xi\in\mathbb
{R}^2\sim\mathcal{N}(0,\sigma^2\bm I_2)$ and $\mathcal{N}(0,\sigma^2\bm I_2)$ is a bi-variate normal distribution with variance $\sigma^2$,
\begin{equation}\label{eq:displacement_AGN}
    \mathcal{N}_{\sigma}(\Psi)=\frac{1}{2\pi\sigma^2}\int_{\xi\in\mathbb{R}^2}\dd{\bm\xi}\,{\rm e}^{-\frac{\abs{\bm\xi}^2}{2\sigma^2}}\mathcal{D}_{\bm\xi}\left(\Psi\right).
\end{equation}
We can easily generalize the single-mode AGN channel to a $N$-mode AGN channel with (a generally correlated) noise matrix $\bm Y\geq0$,
\begin{equation}
    \mathcal{N}_{\bm Y}(\Psi)=\frac{1}{(2\pi)^N\sqrt{\det\bm Y}}\int_{\xi\in\mathbb{R}^{2N}}\dd{\bm\xi}\,{\rm e}^{-\frac{1}{2}\bm\xi^\top\bm Y^{-1}\bm\xi}\mathcal{D}_{\bm\xi}\left(\Psi\right).
    \label{N_Gaussian_multi}
\end{equation}
Furthermore, through symplectic diagonalization, we may decompose the $N$-mode AGN channel into $N$ independent AGN channels with different variances~\cite{wu2021continuous}. Explicitly, there exists a symplectic transformation $\bm S$ such that $\bm S\bm Y\bm S^\top=\bigoplus_{i=1}^{N}\sigma^2_i\bm I_2$, where $\sigma^2_i$ are the independent variances, also known as symplectic eigenvalues of $\bm Y$. The correlated AGN channel and the independent AGN channels are related via conjugation by the unitary channel $\mathcal{U}_{\bm S}$,  $\mathcal{U}_{\bm S}\circ\mathcal{N}_{\bm Y}\circ\mathcal{U}_{\bm S}^{-1}=\bigotimes_{i=1}^N\mathcal{N}_{\sigma_i}$. For example, $\mathcal{U}_{\bm S}$ could be a linear-optical network, potentially with some squeezing if there is $q$ or $p$ bias.
\end{example}

It is often useful to concatenate Gaussian channels to simplify analyses, prove bounds on information theoretical quantities, or use concatenated channels as part of a decoding strategy. For instance, a thermal loss channel can be converted to an AGN channel by pre- or post-amplification---making them amenable to quantum error correction with GKP states as GKP encoding can combat AGN error quite well~\cite{gkp2001,albert2018pra}. 

Interestingly, we can produce less noise by pre-amplification compared to post-amplification\footnote{As a practical example, in microwave electronics, low-noise amplifiers are often used as the first step to combat noise later in the chain. For superconducting circuits, this is often achieved by using a quantum-limited amplifier as the first step in a readout chain.}~\cite{sharma2018bounding,rosati2018narrow,noh2019quantumcapacity}. See the example below:
\begin{example}[Concatenation of Gaussian channels: Amplification then loss = less AGN]\label{example:ampthenloss}
Consider a loss channel $\mathcal{L}_{\eta,\bar{n}_{\mathcal{L}}}$ with transmittance $0<\eta\leq 1$ and noisy quanta $\bar{n}_{\mathcal{L}}\geq0$ and an amplifier channel $\mathcal{A}_{G,\bar{n}_{\mathcal{A}}}$ with gain $G=1/\eta$ and noisy quanta $\bar{n}_{\mathcal{A}}\geq0$. Amplification prior to loss, $ \mathcal{L}_{\eta,\bar{n}_{\mathcal{L}}}\circ\mathcal{A}_{G,\bar{n}_{\mathcal{A}}}$, results in an AGN channel $\mathcal{N}_{\sigma_{\mathcal{L}\mathcal{A}}}$ with
\begin{equation}\label{eq:ampthenloss_sigma}
    \sigma^2_{\mathcal{L}\mathcal{A}}=(1-\eta)(1+2\bar{n}),
\end{equation}
where $\bar{n}\coloneqq(\bar{n}_{\mathcal{L}}+\bar{n}_{\mathcal{A}})/2$. Furthermore, for the same values of $\eta$ and $\bar{n}$, pre-amplification introduces less AGN than post-amplification, i.e. $\sigma_{\mathcal{L}\mathcal{A}}^2\leq\sigma^2_{\mathcal{A}\mathcal{L}}=\left(\frac{1-\eta}{\eta}\right)(1+2\bar{n})$.
\end{example}


\subsubsection{Gaussian states}\label{sec:gauss_states}
Gaussian states refer to the class of states with Gaussian Wigner functions of Eq.~\eqref{Wigner_func}. A Gaussian state $\rho_G$ is thus completely determined by its first and second moments, $\bm\mu_G$ and $\bm\sigma_G$ [defined generally in Eqs.~\eqref{eq:mu_def} and~\eqref{eq:sigma_def}]. Some examples of Gaussian states are coherent states, with non-trivial mean and vacuum noise $\bm\sigma_{\rm vac}=\bm I/2$; thermal states, with zero mean and $\bm\sigma_{\rm th}=(1/2+\bar{n})\bm I$; and single-mode squeezed states, with zero mean and $\bm\sigma_{\rm sqz}={\rm e}^{-2r\bm Z}/2$. GKP states, which are the focus of this review, are examples of non-Gaussian states; we elaborate on non-Gaussianity and GKP states in particular in later sections.

Any Gaussian state can be reduced to a product of thermal states via a Gaussian unitary. The resulting thermal quanta in each mode $\bar{n}_{{\rm th},j}=\nu_j-1/2$ is related to the the symplectic eigenvalues $\nu_j$ of the covariance matrix ${\bm \sigma}_G$. It then follows that the entropy of a Gaussian state is given by the sum of the thermal entropies for the individual modes.\footnote{Recall that unitaries leave the entropy invariant. Hence the displacements $D_{\bm\mu}$ and symplectic transformation $\bm S$ do not contribute to the entropy of the global state.} Written in terms of the symplectic eigenvalues $\nu_j$ of $\rho_G$,
\begin{equation}
S(\rho_G)=\sum_{j=1}^N s_{\rm th}\left(\nu_j-\frac{1}{2}\right),
\end{equation}
where the function $s_{\rm th}(x)$ is defined in Eq.~\eqref{eq:thermal_entropy}. Gaussian states are particularly special when it comes to additive, unitary-invariant quantities like entropy in that Gaussian input states extremize such quantities (\emph{Gaussian extremality})~\cite{holevo1999gauss_channels,wolf2006extremality}. For example, given a quantum state $\rho$ and a Gaussian quantum state $\rho_G$ with the same first and second moments, the entropy is maximized by $\rho_G$,
\begin{equation}\label{eq:gauss_extreme}
    S(\rho)\leq S(\rho_G).
\end{equation}


\subsection{Gaussian measurement: homodyne and heterodyne}
\label{sec:Gaussian_measurement}
An ideal Gaussian measurement (sometimes referred to as a general-dyne measurement) of a state $\rho$ is a projection onto a pure Gaussian state $\ket{\psi_G}$ with some probability $\propto\expval{\rho}{\psi_G}$. We can think of this at the level of positive operator-valued measurements (POVMs) by the resolution of the identity,
\begin{equation}\label{eq:generaldyne}
    \frac{1}{(2\pi)^N}\int_{\bm\mu\in\mathbb{R}^{2N}}\dd{\bm\mu}D_{\bm\mu}U_{\bm S}\dyad{\rm vac}U_{\bm S}^\dagger D_{\bm\mu}^\dagger =\mathbb{I},
\end{equation}
where $\ket{\psi_G(\bm\mu,\bm S)}=D_{\bm\mu}U_{\bm S}\ket{\rm vac}$ defines a (parameterized) pure Gaussian state that we project onto; the POVM is thus $E_{\bm\mu,\bm S}=\dyad{\psi_G}/(2\pi)^N$, with the constant factor $1/(2\pi)^N$ due to overcompleteness of the coherent state basis. Typical homodyne and heterodyne measurements fall into this general class of Gaussian measurements in certain limits as we discuss below.

\subsubsection{Homodyne measurement}
Mathematically, \emph{homodyne detection} of a single mode follows from Eq.~\eqref{eq:generaldyne} by setting $U_{\bm S}$ as a squeezing operator---with squeezing along the $q$ direction such that $\bm S={\rm diag}({\rm e}^{-r},{\rm e}^{r})$---and taking the infinite squeezing limit ($r\rightarrow\infty$). This results in a projection along the $q$ quadrature of the mode. 

In the optical domain, if we want to perform homodyne detection on a quantum state $\rho_S$, we can mix the system $S$ with a strong coherent state at 50/50 beam-splitter and measure the intensity difference at the output by photodetection, however this is a destructive measurement. To measure the $q$ quadrature of $\rho_S$ non-destructively, we can couple the system $S$ via a SUM-gate [see Eq.~\eqref{eq:sum_matrix}] to a measurement-ancilla $M$ prepared in a position eigenstate, $\ket{q}_M$, and then perform a projective homodyne measurement on the ancilla to infer the position of the system $S$. Later, we discuss a similar non-destructive detection strategy to measure the stabilizers of GKP states.

Homodyne detection is ``off-the-shelf'' in the optical domain and commonly applied in telecom systems.
Typically, efficiency of optical homodyne detection is given by the product of the mode-mixing efficiency and the photodetector efficiency. Overall efficiency $>90\%$ can be routinely achieved, not taking into account any additional mode matching efficiency or fiber coupling efficiency, e.g. as studied in Ref.~\cite{grandi2017experimental}. In addition, excess noise is usually small, as thermal noise is low at optical frequencies.

Homodyne detection in the microwave domain is more involved, often requiring phase-sensitive quantum-limited amplification; see, e.g., Ref.~\cite{blais2021cqedRMP} and references therein. As discussed in \cite{puri2021rvw}, due to the low measurement efficiency in the microwave domain, repeated phase-estimation measurements of GKP logicals and stabilizers using an auxiliary qubit can lead to lower error probability than direct homodyne detection given current measurement efficiencies for microwave cavities. See Chapter \ref{ssec:logical readout} for a discussion of GKP logical readout. 

\subsubsection{Heterodyne measurement}
We primarily focus on homodyne measurement in this review, but we mention heterodyne measurements for completeness. \emph{Heterodyne detection} corresponds to a projection onto a coherent state (e.g., $D_{\alpha}\ket{\rm vac}$); thus, we take $U_{\bm S}=\mathbb{I}$ in Eq.~\eqref{eq:generaldyne} to realize a heterodyne measurement. Theoretically, a heterodyne setup can be seen as a double-homodyne scheme, where a system $S$ is split in two and then each output undergoes homodyne measurement (albeit with a $\pi/2$ phase difference between the homodyne detectors). Similar to the homodyne case, optical heterodyne detection is off-the-shelf and efficiency of $>90\%$ can be achieved experimentally. For microwave domain, heterodyne detection often involves phase-insensitive (also called phase-preserving) amplification before detection to suppress noise at the detector. Loss during the measurement 
chain and noise added during amplification can limit the fidelity of these measurements in the microwave domain \cite{blais2021cqedRMP}.

\subsection{Non-Gaussian states and channels}
\label{sec:NG_states}

General unitary transforms in an infinite dimension system are quite complicated. An arbitrary unitary can be generated by Hamiltonians that are polynomial in operators $\hat{q}_k$'s and $\hat{p}_k$'s. A set of operations is considered universal if by a finite number of applications of operations in the set, one can approach arbitrarily close to any unitary evolution generated by such Hamiltonians~\cite{lloyd1999quantum}. Under this definition of universality, Ref.~\cite{lloyd1999quantum} shows that Gaussian operations alone are not universal, since Gaussian unitaries corresponds to generators of second order polynomials in $\hat{q}_k$'s and $\hat{p}_k$'s. However, an arbitrary extra unitary with generators of higher order than two, in addition to Gaussian operations, is universal. In this regard, non-Gaussian operations can be considered as a resource for quantum information processing~\cite{takagi2018,Walschaers2021NGaussPRXQuantum}. \QZ{In fact, Gaussian operations and Gaussian states alone are efficiently simulatable on a classical computer~\cite{bartlett2002efficient}; thus non-Gaussian elements are prerequisite for realizing a quantum computational advantage with bosons~\cite{Chabaud2023BosonicQAdvantage}.}

Based on this finding of universality, ref.~\cite{sefi2011decompose} developed a systematic way of performing the decomposition of any unitary generated by polynomial Hamiltonians to a basic set of Gaussian unitaries $\left\{e^{i\pi\left(\hat{p}^2+\hat{q}^2\right)/2}, e^{i t_1\hat{q}}, e^{i t_2\hat{q}^2} \right\}$ and the cubic phase gate
$
\hat{V}(\gamma)=e^{i\gamma \hat{q}^3}.
$
The choice of the non-Gaussian unitary is not unique; the cubic phase gate, generated by $\hat{q}^3$, is chosen since it is one of the most simple non-Gaussian unitaries. 

There are a number of experimental proposals of realizing the cubic phase gate in the optical domain, involving genuine non-Gaussian resource states and Gaussian operations combined with feed-forward~\cite{gkp2001,ghose2007non,marek2011deterministic,sabapathy2018states}. The cubic phase gate can be realized by consuming the cubic phase state
$
 \ket{\gamma} = \hat{V}(\gamma)\ket{0}_p = \int dq e^{i\gamma q^3}\ket{q}
$
as the resource state, where $\ket{0}_p$ is the zero-momentum state at the infinite squeezing limit and unnormalizable~\cite{gkp2001}. Normalized version of cubic phase state can be defined by having a finite squeezing to begin with. While optical realization is still challenging, microwave engineering of cubic phase gate and states have been achieved~\cite{hillmann2020universal, Kudra_2022}. However, for the realization of non-Gaussian control for microwave superconducting cavities, the cubic phase gate as a resource is likely not the most practical option, as \QZ{Josephson} junctions readily gives the necessary nonlinearity to be universal \cite{heeres2017implementing, ma2021quantum,eickbusch2022fast}. The difficulty instead comes with mitigating the propagation of errors. 

Other types of non-Gaussian states include Fock states, the ON state 
$
\ket{ON}\propto \left(\ket{0}+a\ket{N}\right)
$~\cite{sabapathy2018states} and GKP states~\cite{gkp2001} that are most relevant to this review. We discuss universality of GKP states plus Gaussian operations in Chapter~\ref{sec:gkp_universality}.

To help understand the special role of non-Gaussian states and operations, here we briefly review the resource theory of non-Gaussianity. A more detailed review can be found in Ref.~\cite{lachman2022quantum}. To quantify non-Gaussianity, one can first observe that any Gaussian state has a positive Wigner function, which is just a multivariate Gaussian distribution~\cite{weedbrook2012rmp,serafini2017book}. For pure states, a state is non-Gaussian if and only if its Wigner function has negative values~\cite{hudson1974wigner,soto1983wigner}. Starting from this observation, Refs.~\cite{takagi2018,albarelli2018resource} proposed to utilize logarithmic negativity of the Wigner function to quantify the amount of non-Gaussianity, which has been found useful in resource distillation~\cite{takagi2018,jee2021resource}. \QZ{Negativity as a resource is also consistent with the fact that a positive Wigner function renders classical simulation of the system efficient~\cite{mari2012positive} and also serves as a resource for computing in discrete-variable systems~\cite{veitch2012negative}.}
More recently, the stellar rank---the number of zeros of Husimi Q function -- is introduced to characterize non-Gaussianity, which equals the minimal number of single-photon additions needed to engineer them~\cite{chabaud2020stellar}, \QZ{and, furthermore, can be related to the complexity of quantum computations with bosons~\cite{Chabaud2023BosonicQAdvantage}}. Besides developing quantifiers, experimentally a more feasible witness of Wigner negativity or non-Gaussianity is also important~\cite{chabaud2021witnessing,Chabaud2021certification}. In terms of resource theory for non-Gaussian operations, Ref.~\cite{zhuang2018resource} proposed quantifiers based on generation power of non-Gaussianity in a non-convex fashion. In general, quantifying non-Gaussianity and connecting them to operational tasks is an active research direction.


\subsubsection{No-go theorem for Gaussian error correction}

The fact that universality of quantum computation cannot be done by Gaussian operations alone~\cite{lloyd1999quantum} indicates that error correction is probably impossible just with Gaussian operations as well. References~\cite{eisert2002nogodistillation,fiuravsek2002gaussian,giedke2002characterization,niset2009nogo} further made this intuition rigorous. In Ref.~\cite{eisert2002nogodistillation}, the authors show that Gaussian entanglement cannot be distilled by only local Gaussian operations and Gaussian measurement. For example, two identical copies of less squeezed two-mode squeezed vacua cannot be transformed by Gaussian operations into a single two-mode squeezed vacuum with higher squeezing. Reference~\cite{fiuravsek2002gaussian} shows that probabilistic distillation of entanglement at a single copy level is impossible, simply because conditional Gaussian maps changes the covariance matrix in a deterministic fashion. Reference~\cite{giedke2002characterization} generalizes the conclusion to an arbitrary number of modes. 
Ref.~\cite{niset2009nogo} shows that Gaussian operations cannot protect Gaussian states against Gaussian errors. The proof considers degradation of entanglement under Gaussian channels and shows that Gaussian operations on top of the channel cannot improve the quality of entanglement. 

From these observations, we see that non-Gaussian resource is the key ingredient for quantum error correction. Since in-line non-Gaussian operations (where the gates need to be applied on general quantum input) are more challenging to implement, ideally we want to have Gaussian encoding and decoding operations, while relying on non-Gaussian ancilla---which is exactly the case for the GKP codes discussed in this review.

\subsubsection{General additive noise channel}
We previously discussed Gaussian quantum channels, however in some scenarios, non-Gaussian additive noise channels naturally arise---e.g., in the input-output relations of multimode GKP-O2O (oscillators-to-oscillators) codes. Generally, a $N$-mode non-Gaussian additive noise channel $\widetilde{\mathcal{N}}$ acts on an input quantum state $\rho$ as
\begin{equation}\label{eq:ANGN_channel}
\rho^\prime=\widetilde{\mathcal{N}}\left(\rho\right)=\int_{\bm\zeta\in\mathbb{R}^{2N}}\dd{\bm\zeta}P(\bm \zeta)D_{\bm\zeta}\rho D_{\bm\zeta}^\dagger,
\end{equation}
where $P(\bm\zeta)$ is a multivariate non-Gaussian pdf. We shall generally assume that the pdf $P(\bm\zeta)$ has zero mean $\expval{\bm \zeta}=0$ and covariance noise matrix $\bm{\widetilde{Y}}_{ij}=\expval{\bm\zeta_i\bm\zeta_j}$. Though higher-order moments beyond the mean and covariance are needed to fully specify the channel, we can still give a Gaussian approximation to the channel $\rho^\prime_G=\mathcal{N}_{\bm{\widetilde{Y}}}(\rho_G)$, where $\rho_G$ is a Gaussian approximation to the input $\rho$. The Gaussian approximation $\mathcal{N}_{\bm{\widetilde{Y}}}$ (see Eq.~\eqref{N_Gaussian_multi}) is amenable to analysis with Gaussian techniques and actually allows us to place a lower bound on the quantum capacity of the corresponding non-Gaussian channel $\widetilde{\mathcal{N}}$ (see below), which is relevant for quantum information studies of GKP-type QEC codes.

\subsection{Quantum communication capacity}
\label{sec:quantum_capacity}

A fundamental task in the quantum information sciences is quantum communication. Quantum communication revolves around the transmission of quantum information in the form of coherent quantum states, such as qubits or CV states, across a noisy channel. A naturally arising question, how much quantum information can we send through a noisy quantum channel per channel use in principle? This is quantified by the so-called \emph{quantum capacity}~\cite{lloyd1997capacity,shor2002quantum,devetak2005capacity} of the channel. We give a pedagogical explanation for the quantum capacity and refer the reader to, e.g., Refs.~\cite{wilde2013qit,gyongyosi2018survey} for rigorous details.

\begin{figure}
    \centering
    \includegraphics[width=.75\linewidth]{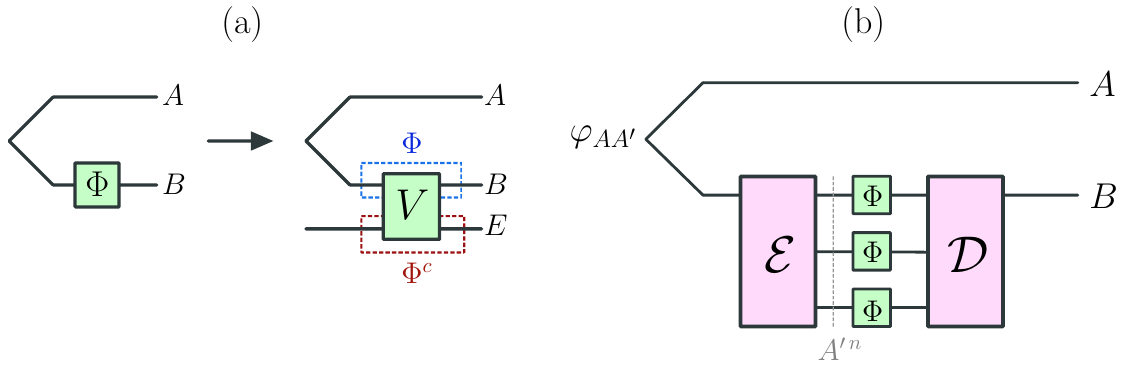}
    \caption{Channel schematics: (a) Unitary extension of a channel $\Phi$ with complementary channel $\Phi^c$; (b) General QEC circuit with encoder $\mathcal{E}$ and decoder $\mathcal{D}$.}
    \label{fig:qec_schematic}
\end{figure}

Suppose that a sender, Alice ($A$), has an entangled state $\varphi_{AA^\prime}$ that she wants to share with a receiver, Bob ($B$), where the subsystems $A$, $A^\prime$, and $B$ have Hilbert space dimension $\abs{A}=\abs{A^\prime}=\abs{B}$, and that there is a quantum channel $\Phi\coloneqq\Phi_{A^\prime\rightarrow B}$ that Alice can use to communicate with Bob. Alice knows that the channel is noisy, so she devises an encoding scheme $\mathcal{E}\coloneqq{\rm id}_{A}\otimes\mathcal{E}_{A^\prime\rightarrow A^{\prime\,n}}$ that encodes the sub-system $A^\prime$ into a larger system $A^{\prime\,n}$ to protect the information from noise in the channel. Bob is aware of the encoding procedure and possesses a corresponding decoder $\mathcal{D}\coloneqq{\rm id}_{A}\otimes\mathcal{D}_{A^{\prime\,n}\rightarrow B}$ that he uses to decode the information and retrieve the intended message from his share of entanglement, if the channel is not ``too noisy''. The whole transmission process from Alice to Bob---encoding, noisy transmission, and decoding---results in the quantum state $\tilde{\varphi}_{A B}=(\mathcal{D}\circ\Phi^{\otimes n}\circ\mathcal{E})(\varphi_{A A^\prime})$ at Bob's end. A schematic of this setup is shown in Figure~\ref{fig:qec_schematic}(b). The \emph{rate} of the communication process is $R=\log\abs{A}/n$, and we say that the rate is achievable for the channel $\Phi$ if the transmitted state $\tilde{\varphi}_{A B}$ is $\epsilon$-close to the intended state ${\varphi}_{A B}$, where $\epsilon$ is an arbitrarily small number. The \emph{quantum capacity} $C_{\mathcal{Q}}(\Phi)$ is then the supremum of all achievable rates $R$. 

A landmark result in quantum information theory is the \emph{quantum capacity theorem}, which states that the quantum capacity is equal to the regularized coherent information of the channel~\cite{lloyd1997capacity,devetak2005capacity},
\begin{equation}\label{eq:qcapacity}
    C_{\mathcal{Q}}(\Phi)=\mathcal{Q}_{\rm reg}(\Phi),
\end{equation}
where $\mathcal{Q}_{\rm reg}=\lim_{k\rightarrow\infty}\mathcal{Q}(\Phi^{\otimes k})/k$. The channel coherent information $\mathcal{Q}(\Phi)$ is
\begin{equation}
    \mathcal{Q}(\Phi)=\max_{\varphi} \big[S(\rho_B)-S(\rho_{AB})\big],
\end{equation}
where $\rho_{AB}=\Phi(\varphi_{A A^\prime})$ and $\rho_B=\Tr_{A}(\rho_{AB})$ and the maximization is over all bi-partite pure states $\varphi_{AA^\prime}$.\footnote{We can just as well maximize over all input density matrices $\varphi_{A^\prime}=\Tr_A(\varphi_{AA^\prime})$.} By purification, we can introduce an environment $E$ such that the joint system $ABE$ is described by a pure state; see Figure~\ref{fig:qec_schematic}(a) for a schematic. Then $S(\rho_{AB})=S(\rho_E)$ where $\rho_E=\Tr_{A}\big(\Phi^c(\varphi_{A^\prime A})\big)$ is the output state to the environment via the complementary channel $\Phi^c:A^\prime\rightarrow E$ of $\Phi$. Thus, the quantum capacity directly relates to the amount of information Bob receives $S(\rho_B)$ minus the amount of information the environment receives $S(\rho_E)$. We note that $\mathcal{Q}(\Phi)\leq C_{\mathcal{Q}}(\Phi)$.

The regularized coherent information appearing in Eq.~\eqref{eq:qcapacity} makes evaluating the quantum capacity generally quite challenging. In principle, an infinite number of channel uses might be needed due to superadditivity~\cite{smith2008quantum,hastings2009superadditivity,smith2011quantum,zhu2017superadditivity,zhu2018superadditivity}. However, for \emph{degradable} quantum channels, the problem simplifies since $\mathcal{Q}(\Phi^{\otimes k})=k\mathcal{Q}(\Phi)$ for such channels~\cite{devetakshor2005degradable}; in other words, the quantum capacity for degradable quantum channels is equivalent to the single-shot quantum capacity, $\mathcal{Q}(\Phi)$. Furthermore, for anti-degradable channels, $C_{\mathcal{Q}}(\Phi)=0$. A channel $\Phi$ is said to be degradable if there exists a channel $\mathcal{D}$ such that $\Phi^c=\mathcal{D}\circ\Phi$; i.e., Bob can simulate the environment channel $\Phi^c$ by ``degrading'' his channel $\Phi$ via $\mathcal{D}$. Contrariwise, $\Phi$ is anti-degradable if there exists a channel $\mathcal{A}$ such that $\Phi=\mathcal{A}\circ\Phi^c$; i.e., the environment can simulate Bob's channel.

\subsubsection{Capacity of Gaussian channels}

Due to the infinite energies involved in an infinite-dimensional Hilbert space of bosonic systems, a bit of care has to be taken when referring to the quantum capacity of Gaussian bosonic quantum channels. As pointed out in Ref.~\cite{wilde2018energyconst}, it is physically meaningful to place a photon-number constraint $\Tr(\rho\hat{n})\leq\bar{n}$ on input states $\rho$, and then evaluate physically meaningful quantities (such as capacities) under said constraint. For instance, in reality, we always deal with a finite amount of squeezing, a finite number of thermal quanta etc.; so such a constraint is implicitly present in practice. This has led to a technical separation between energy-constrained (finite $\bar{n}$) and energy-unconstrained ($\bar{n}\rightarrow\infty$) capacities. For simplicity though, we focus on the energy-unconstrained setting in this review.

Special cases of degradable, single-mode Gaussian bosonic channels exist, and their quantum capacities have been characterized. Examples of single-mode degradable bosonic channels are the quantum-limited amplifier channel $\mathcal{A}_{G,0}$ and the bosonic pure-loss channel $\mathcal{L}_{\eta,0}$ for $\eta>1/2$. For $\eta<1/2$, the bosonic pure-loss channel is anti-degradable and thus $C_{\mathcal{Q}}(\mathcal{L}_{\eta,0})=0$. The quantum capacity of the pure-loss channel $\mathcal{L}_{\eta,0}$ has been known for years~\cite{wolf2007PRLcapacities,HW2001capacities} and is given explicitly by,
\begin{equation}
    C_{\mathcal{Q}}(\mathcal{L}_{\eta,0})=\log\left(\frac{\eta}{1-\eta}\right),
\end{equation}
which is non-zero iff $\eta>1/2$. \QZ{Here and what follows, it is understood that the quantum capacity is bounded from below by zero. Thus, for brevity, we do not include maximization, i.e. $\max[0,\cdot]$, in the corresponding expressions.}

General Gaussian bosonic quantum channels are neither degradable nor anti-degradable, and explicitly finding quantum capacity of such channels is an open problem. On the other hand, there exists upper and lower bounds for many channels of interest. For instance, lower~\cite{HW2001capacities} and upper bounds~\cite{rosati2018narrow,noh2019quantumcapacity} on the quantum capacity of a thermal loss channel $\mathcal{L}_{\eta, \bar{n}}$ are known,
\begin{equation}\label{eq:bounds_thermalLoss}
    \log\left(\frac{\eta}{1-\eta}\right)-s_{\rm th}(\bar{n})\leq C_{\mathcal{Q}}(\mathcal{L}_{\eta,\bar{n}})\leq\log\left(\frac{\eta-(1-\eta)\bar{n}}{(1-\eta)(1+\bar{n})}\right).
\end{equation}
The lower bound is obtained from an input thermal state~\cite{HW2001capacities} (a Gaussian state), whereas the upper bound follows from a data-processing argument~\cite{rosati2018narrow,noh2019quantumcapacity}. 

Since $\mathcal{N}_{\sigma^2}=\lim_{\eta\rightarrow1}\mathcal{L}_{\eta,\bar{n}/(1-\eta)}$ with $\sigma^2=\bar{n}$, we can use the results above to bound the quantum capacity of the single-mode AGN channel $\mathcal{N}_\sigma$,\footnote{The upper bound follows directly from substitution of $\bar{n}\rightarrow\bar{n}/(1-\eta)$ into Eq.~\eqref{eq:bounds_thermalLoss}. For the lower bound, let $x=\bar{n}/(1-\eta)$ and use $s_{\rm th}(x)\approx\log(ex)$ for $x\gg1$.}
\begin{equation}\label{eq:agn_bounds}
    \log\left(\frac{1}{e\sigma^2}\right)\leq C_{\mathcal{Q}}(\mathcal{N}_{\sigma})\leq\log\left(\frac{1-\sigma^2}{\sigma^2}\right).
\end{equation}
These results can be easily extended to a $N$-mode AGN channel $\mathcal{N}_{\bm Y}$ since we can interpret the $N$-mode channel as a set of independent AGN channels via $\mathcal{U}_{\bm S}\circ\mathcal{N}_{\bm Y}\circ\mathcal{U}_{\bm S}^{-1}=\bigotimes_{i=1}^N\mathcal{N}_{\sigma_i}$, where $\bm S\bm Y\bm S^\top=\bigoplus_{i=1}^{N}\sigma^2_i\bm I_2$. From which one can show that,
\begin{equation}\label{eq:bounds_NmodeAGN}
    \log\left(\frac{1}{e^N\sqrt{\det\bm Y}}\right)\leq C_{\mathcal{Q}}(\mathcal{N}_{\bm Y})\leq\log\left(\frac{1-\sqrt{\det\bm Y}}{\sqrt{\det\bm Y}}\right).
\end{equation}
The lower bound follows by treating the $N$-modes of the joint channel independently and using uncorrelated thermal input states, while the upper bound follows from a data processing argument~\cite{wu2021continuous}. 

\subsubsection{Capacity of general additive noise channel}
Using results from the previous section, we can actually lower-bound the quantum capacity of an arbitrary non-Gaussian additive noise channel $\widetilde{\mathcal{N}}$~\eqref{eq:ANGN_channel}. Consider the noise covariance matrix $\widetilde{\bm Y}$ of $\widetilde{\mathcal{N}}$. By Gaussian extremality\footnote{In particular, Gaussian states minimize the coherent information~\cite{eisert2005GaussChannels,wolf2006extremality}; i.e., $S(\rho_B)-S(\rho_{AB})\geq S[(\rho_{B})_G]-S[(\rho_{AB})_G]$ where $(\rho_{AB})_G$ is a Gaussian state with equal first and second moments as $\rho_{AB}$.} and using similar arguments that led to the lower bound of Eq.~\eqref{eq:bounds_NmodeAGN}, one can show that,
\begin{equation}\label{eq:capacity_nonAGN}
    \log\left(\frac{1}{e^N\sqrt{\det\widetilde{\bm Y}}}\right)\leq C_{\mathcal{Q}}(\widetilde{\mathcal{N}}).
\end{equation}
Thus we can estimate the capacity of a general---not necessarily Gaussian---additive noise channel by considering only the additive noise covariance matrix $\widetilde{\bm Y}$ of the channel, without the need for higher-order moments. Later, we use the lower bound in Eq.~\eqref{eq:capacity_nonAGN} to place a lower bound on the QEC properties of GKP-O2O codes (Chapter~\ref{sec:output_bound}).

\section{Noise Models}
\label{sec:noise_model}

\begin{figure}
    \centering
    \includegraphics[width=.6\linewidth]{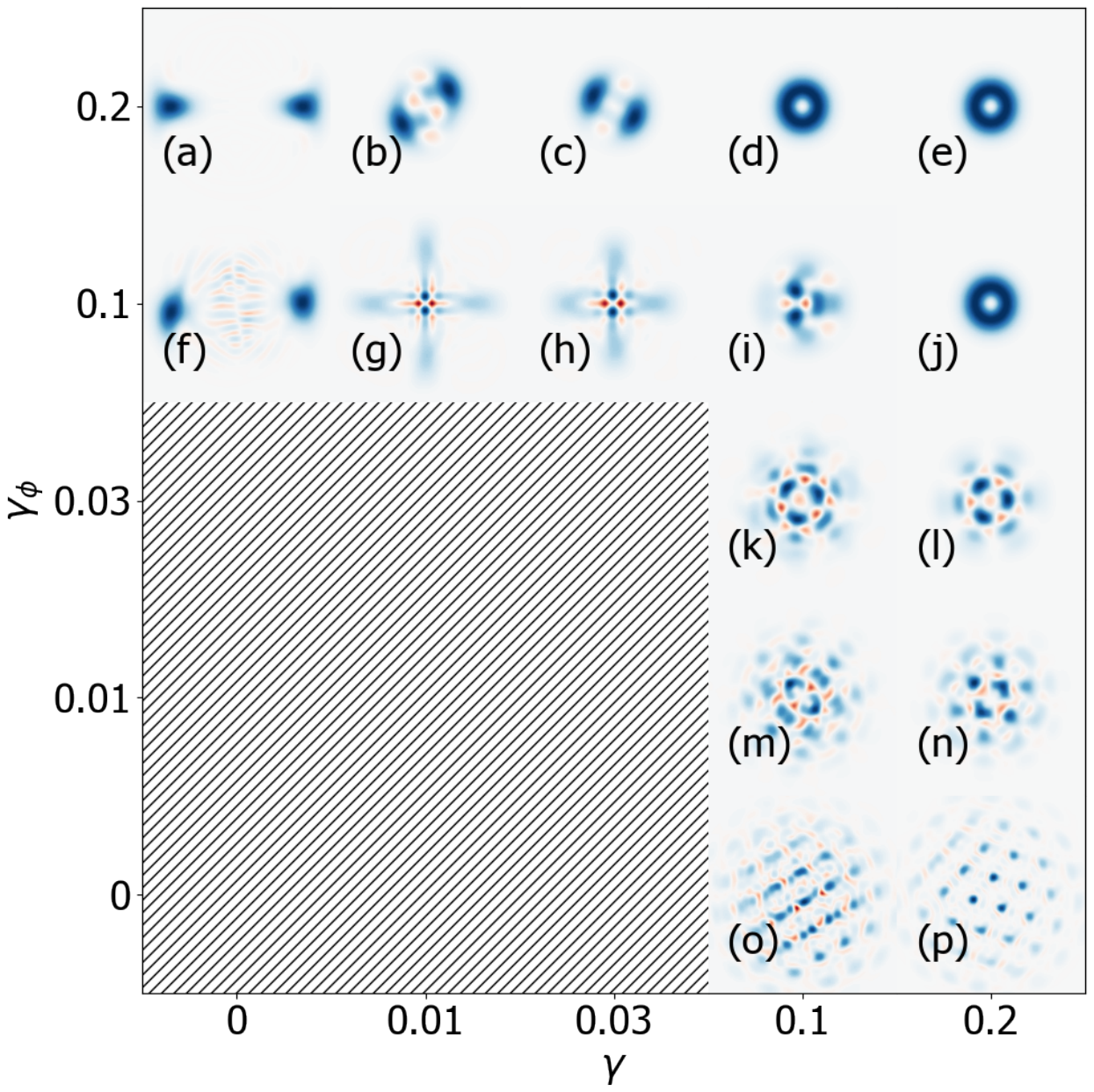}
    \caption{\QZ{Wigner functions of bosonic codes that approach the quantum communication capacity of the (single-mode, pure) loss-dephasing channel~\cite{leviant2022quantum}}. For pure loss ($\gamma_\phi=0$), the hexagonal GKP code emerges as optimal (see also Ref.~\cite{noh2019quantumcapacity} for similar results on thermal loss channels). For non-zero dephasing, having rotation symmetry in the code becomes favorable and GKP lattice codes are no longer optimal. Hashed region represents a degeneracy of codes with similar performance. Figure adapted from Ref.~\cite{leviant2022quantum} where the authors used semi-definite programming methods for optimal recovery to obtain this figure.}
    \label{fig:convergence2gkp}
\end{figure}

In this chapter, we review common noise in bosonic quantum systems---including photon loss, AGN, phase noise, and cQED auxiliary noise sources---and their models to facilitate quantum error correction analyses. Two theoretical frameworks are often adopted for noise modeling: a quantum channel model or an open quantum system dynamics approach, as introduced in Chapter~\ref{sec:channel_dynamics}. The former is often relevant for propagating bosonic modes, such as in quantum communication. While the latter is often relevant in, e.g., cQED analyses, where continuous quantum control can be applied. We discuss noise from both perspectives below.

Before diving into details of noise models,  let us first provide some rationale for why GKP codes (extensively reviewed in the forthcoming chapters) are effective in mitigating such noises. The original work by Gottesman, Kitaev, and Preskill~\cite{gkp2001} highlighted the effectiveness of GKP lattice codes in dealing with displacement errors because---as we explain in more detail in Chapter~\ref{sec:GKP_simple_part}---such codes allow simultaneous estimation of displacements along both position and momentum quadratures (modulo the lattice spacing). This concept can be extended in a heuristic manner to other types of noise using the following observation: Since the set of displacement operators is complete [see Eq.~\eqref{eq:weyl_op}], one can always decompose \textit{any} operation in the basis of displacement operators, similar to the case of Pauli errors in the DV setting. This displacement basis decomposition gives some credence as to why GKP codes might be capable of protecting more than just random displacements (additive noises), as was also highlighted in the original work of Ref.~\cite{gkp2001} for photon loss. Furthermore, we note that it is theoretically possible to degrade any noise channel to a general additive noise channel  [Eq.~\eqref{eq:ANGN_channel}], $\widetilde{\mathcal{N}}\left(\rho\right)=\int_{\bm\zeta\in\mathbb{R}^{2N}}\dd{\bm\zeta}P(\bm \zeta)D_{\bm\zeta}\rho D_{\bm\zeta}^\dagger$, via channel twirling~\cite{conrad2021twirling}, though twirling might not be practically feasible. For Gaussian noise sources, $P(\bm\zeta)$ is a multivariate Gaussian distribution. Using similar arguments regarding QEC capabilities of GKP states for random displacements, we infer that GKP codes are likewise ideal for mitigating the noise of the twirled channel $\tilde{\mathcal{N}}$.\footnote{We emphasize that, although twirling to an additive noise channel is neat for heuristic arguments (and often convenient for analytical assessments), this is not the way one corrects errors in practice and, furthermore, degrading a quantum channel to an AGN channel via twirling is not optimal for QEC purposes in principle.}

To further reinforce the QEC capabilities of GKP codes, it is crucial to highlight that GKP codes arguably provide the optimal encoding scheme for safeguarding quantum information against Gaussian noise sources, such as additive noise (mentioned above) and thermal loss. This is evidenced by the many works that discuss achieving quantum capacities of Gaussian noise channels via GKP codes~\cite{harrington2001rates,albert2018pra,noh2019quantumcapacity,leviant2022quantum} (see also Chapter~\ref{subsec:apps_comms} of this review and Ref.~\cite{albert2018pra} for comparisons among a zoo of bosonic codes). On the other hand, the situation changes when non-Gaussianity enters into the noise model---a prime example of non-Gaussian noise being dephasing. For instance, GKP codes are known to be inadequate in combating pure dephasing noise~\cite{albert2018pra,puri2021rvw,leviant2022quantum}, whereas codes with rotation symmetry become preferable method of encoding~\cite{Grimsmo_2020,leviant2022quantum}; see Fig.~\ref{fig:convergence2gkp} for an illustration of this point. Though, from a practical perspective, Gaussian noise sources generally dominate the noise budget, as can be seen by comparing the loss and dephasing rates in cQED architectures---with typical intrinsic energy relaxation rates for superconducting 3D cavities on the order of $\kappa \approx 1/(\SI{1}{ms})$ down to $\kappa \approx 1/(\SI{30}{ms})$\footnote{State-of-the art niobium superconducting cavities have even smaller relaxation rates as low as $\kappa = 1/(\SI{1}{s})$ \cite{Romanenko_2020}. To the best of our knowledge, single-photon level intrinsic dephasing rates for these cavities has not yet been measured.} \cite{reagor_2016,milul2023superconducting} and upper bounds on intrinsic dephasing rates on the order of $\kappa_\phi \lesssim 1/(\SI{10}{ms})$ \cite{Rosenblum_2018,reagor_2016} down to $\kappa_\phi \lesssim 1/(\SI{500}{ms})$ \cite{milul2023superconducting,campagne2020quantum,eickbusch2022fast}.  In optics (e.g., integrated photonic structures~\cite{zhu2021integratedLiNrvw,moody2022roadmapQuPhotonics}), photon loss dominates, and dephasing is usually negligible, at least when active phase stabilization (more important for long distance quantum communication) is present.

\subsection{Photon loss}

Loss of encoded quanta plagues all bosonic quantum information processors and is arguably the most prominent source of noise. As a quantum channel model, loss can be described by a pure-loss channel $\calL_{\eta,0}$ in Example~\ref{example:thermal_loss}. Promisingly, Noh et al~\cite{noh2019quantumcapacity} demonstrated that GKP codes are optimal for combating loss by proving that GKP codes approach the quantum capacity of loss channels (up to a constant gap), with numerical support from an optimized encoding/decoding strategy; see also Ref.~\cite{albert2018pra} for a prior comparison of many bosonic codes. Unfortunately, the encoding/decoding strategy presented in that paper was not constructive, and therefore, at present, there is no known concrete and provably optimal encoding/decoding strategy to handle loss with GKP codes. Nevertheless, there exist several (sub-optimal) loss-mitigation strategies that are practically relevant. 

In terms of open system dynamics picture, loss can be described with a single dissipation superoperator $\mathbb{D}[\sqrt{\kappa}\hat{a}]$; see Eq.~\eqref{sec:channel_dynamics}. As we shall see in Chapter~\ref{sec:GKP_eng}, one way to combat loss in cQED architectures is via dissipation engineering, where loss and coherent Hamiltonian dynamics work in unison to stabilize the microwave modes onto the GKP manifold. Another strategy is to convert loss to AGN via pre- or post-amplification (see the following subsection), which can be accomplished in the channel model or in the continuous time model, and then correct for the AGN directly via SUM-gates plus GKP measurement ancillae~\cite{gkp2001}. The loss-conversion technique is often used as a neat mathematical trick to simplify calculations, however loss conversion may be practically pertinent to optical platforms that do not have active stabilization methods.

\subsection{Additive Gaussian noise via loss conversion}

\begin{figure}
    \centering
    \includegraphics[width=.65\linewidth]{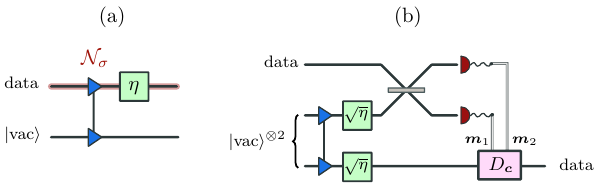}
    \caption{Noise conversion from loss to AGN. (a) Pre-amplification (via two-mode squeezing with an ancillary mode) with gain $G=1/\eta$. (b) Teleportation-based noise conversion where only ancillary modes undergo noise. Here a two-mode squeezed vacuum source is placed ``in the middle'' of a transmission line with total transmittance $\eta$.}
    \label{fig:amp_loss_protocol}
\end{figure}

For the quantum channel model, a pure-loss channel, $\calL_{\eta,0}$, as defined in Example~\ref{example:thermal_loss}, can also be converted to an AGN channel. Intuitively, amplification can counter the decay of quantum states amplitudes from loss, but at the cost of increased noise. As shown in Figure~\ref{fig:amp_loss_protocol}(a) and discussed in Examples~\ref{example:ampthenloss}, a simple approach is to append a quantum-limited amplifier $\calA_{G,0}$ before the loss channel. From 
Eq.~\eqref{eq:ampthenloss_sigma}, we have the effective channel as AGN, $\calL_{\eta,0}\circ \calA_{G=1/\eta,0}=\calN_{\sigma_{\mathcal{LA}}^2}$, with noise\footnote{Similar conversion to AGN also holds for thermal-loss channel, as discussed in Example~\ref{example:ampthenloss}.}
\be 
\sigma_{\mathcal{LA}}^2=1-\eta.
\label{pre_amp}
\ee 
It is known that this conversion strategy is not optimal for quantum error correction performance~\cite{albert2018pra,noh2019quantumcapacity} but a better strategy is currently unknown. To this end, we note that quantum-limited amplification has an equivalent representation as coupling the data to an ancillary vacuum mode via two-mode squeezing; see Fig~\ref{fig:amp_loss_protocol}(a). It thus seems plausible that intelligently utilizing the ancillary mode may lead to improved performance but how to explicitly do so is an open problem.

An alternative approach to noise conversion is a CV teleportation-based approach. When it is possible to break the loss channel into two parts, Ref.~\cite{wu2022continuous} proposed to put a two-mode squeezed vacuum source in between each lossy path, then distribute the two-mode squeezed vacuum along each path and perform CV teleportation. The effective AGN channel in this scheme is $\calN_{\sigma^2_{\rm center-TP}}$ with variance
\be 
\sigma^2_{\rm center-TP}=\sqrt{\eta}e^{-2r}+1-\sqrt{\eta},
\label{eq:sigma_center_TP}
\ee 
where $r$ is the two-mode squeezing strength defined in Eq.~\eqref{eq:tms_matrix}.

Within the framework of the open system dynamics described by Eq.~\eqref{eq:master_eq}, a continuous-time model of additive Gaussian noise can be represented by two dissipation superoperators, $\mathbb{D}[\sqrt{\kappa}\hat{q}]$ and $\mathbb{D}[\sqrt{\kappa}\hat{p}]$. Similar to loss conversion in the channel model, AGN in continuous time can likewise be generated from a loss dissipator by subsequently amplifying the quadratures over time to counteract the loss.




\subsection{Phase noise} 

Phase noise models the uncertainty in phase stabilization or optical path alignment. Likewise, phase noise can appear in, e.g., frequency fluctuations of a microwave mode in cQED. In terms of quantum channel model, where phase noise is commonly referred to as a dephasing channel~\cite{lami2023exactDephasing,leviant2022quantum}, it corresponds to randomly applying phase rotations which are described by the symplectic transform in Eq.~\eqref{eq:R_phi}. In terms of open system dynamics picture of Eq.~\eqref{eq:master_eq}, phase noise can be  described by the dissipation superoperator $\mathbb{D}[\sqrt{2\kappa_\phi}\hat{a}^\dagger\hat{a}]$ with jump operator $L=\hat{a}^\dagger\hat{a}$. Since the jump operator appearing in the dissipator is non-linear in the quadrature operators, the dephasing channel is intrinsically a non-Gaussian quantum channel. We note that this dissipator is a simplified model of dephasing, since some physical systems (such as superconducting oscillators) can have phase noise with a non-uniform spectral density~\cite{Niepce_2021}.

As noted in the original GKP paper, when decomposed to displacement operators, phase noise involves displacements of arbitrarily large amplitudes. For this reason, the qubit GKP code is known to be less robust against phase noise compared to, e.g., rotation symmetric bosonic codes~\cite{albert2018pra,Grimsmo_2020,leviant2022quantum}. Indeed, in the pure dephasing case, it is found that two-legged cat codes and squeezed two-legged cat codes are preferred in terms of quantum communication rates~\cite{leviant2022quantum}.

\subsection{Auxiliary noise sources}\label{ssec:auxiliary-noise}
Apart from these dominant oscillator noise channels, there are some tertiary noise sources which could harm the encoded quantum information. Such noise sources may arise from unavoidable higher order terms or couplings with auxiliary qubits. Below, we divide these sources into two categories depending on their effect on the oscillator state, coherent and incoherent errors.
\paragraph{Coherent errors}
To realize non-Gaussian states such as GKP codes, a nonlinearity is required. However, the nonlinearity can also introduce unwanted coherent Hamiltonian terms that distort GKP states. One example is discussed in Chapter~\ref{sec:exp_arch} in the context of realizing GKP codes in superconducting circuits through interactions with an auxiliary qubit in the dispersive regime. In such a system, the oscillator inherits a Kerr-type nonlinearity, with a Hamiltonian given by $H=K\hat{a}^{\dag 2} \hat{a}^2$. Although Kerr could be used to perform non-Clifford gates \cite{baptiste2022multiGKP}, it is generally harmful when idling \cite{albert2018pra,campagne2020quantum,eickbusch2022fast}. 

Other types of coherent errors emerge when scaling to multiple oscillators. For example, in superconducting circuits, couplers that could be used for realizing multi-oscillator gates as discussed in Chapter~\ref{sec:scaling_up} can also give rise to unwanted cross-Kerr interactions of the form $H=\chi_{ab}\hat{a}^{\dag} \hat{a}\hat{b}^\dag b$. Realizing fast gates while suppressing these unwanted interactions is the topic of engineering a large on-off ratio \cite{chapman2022high}. 

\paragraph{Incoherent errors}
When GKP code preparation and error correction is realized by coupling the oscillator to an auxiliary qubit (Chapter~\ref{sec:exp_arch}), errors of the auxiliary qubit can propagate to the oscillator in various ways. For superconducting circuit architectures in the dispersive regime, thermal jumps of the auxiliary qubit's state can lead to additional dephasing of the oscillator \cite{reagor_2016} and coupling to the qubit can decrease the $T_1$ of the oscillator through the \textit{reverse Purcell effect} if the qubit's bare lifetime is lower than the oscillator  \cite{blais2021cqedRMP}. These effects can also be present for couplings between superconducting oscillators. Additionally, during gates between the oscillator and qubit, bit flips and phase flips of the qubit can propagate to the oscillator state, depending on the exact circuit being performed, as described in Chapter~\ref{sec:qubit-dissipation} and experimentally measured in Chapter~\ref{sec:exp_arch}.

\section{Mathematical Description of GKP Lattice States}\label{sec:gkp_lattice}

So far, we have maintained a broad view on topics in bosonic quantum information processing---discussing bosonic quantum channels, unitaries, and common bosonic noise processes along the way---without delving much into the specifics of GKP codes and states, which are the central focus of this review. In this section, we take our first deep dive into this fascinating topic. Here we present the formalism to describe the mathematical properties of GKP lattice states, which is useful for theoretical analyses of multimode GKP codes. For a gentle introduction to GKP states, we begin with the single-mode square lattice GKP states in Chapter~\ref{sec:GKP_simple_part}, where we explicitly write down the wave functions and Wigner functions. We then continue towards a generic geometrical description of multimode GKP lattice states, which provides an efficient way of mathematically handling multimode states. In Chs.~\ref{sec:canonical_lattice} and~\ref{sec:comp_lattice}, we review canonical GKP states and computational GKP states, respectively. We provide some discussion on stabilizer measurements and error syndromes in Chapter~\ref{sec:syndromes} and finish in Chapter~\ref{sec:finite_GKP} by discussing the impacts of deforming these states to realistic finite-energy GKP codes which occupy a finite volume in phase space.

\subsection{Introducing GKP states: The square lattice}
\label{sec:GKP_simple_part}
\begin{figure}
    \centering
    \includegraphics[width=.5\linewidth]{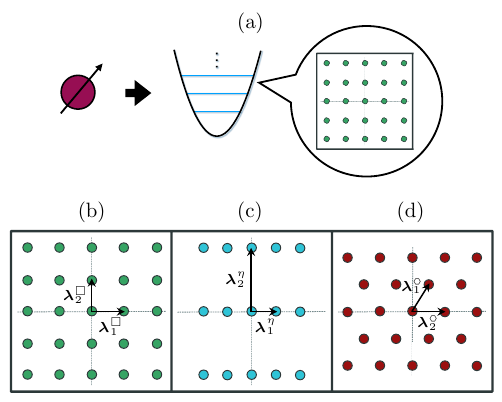}
    \caption{Illustrating GKP states. (a) Encoding a qubit into the infinite dimensional Hilbert space of an oscillator that is in a GKP lattice state~\cite{gkp2001}. Common two-dimensional lattices: (b) square lattice, (c) rectangular lattice, (d) hexagonal lattice. Basis vectors for the lattices are also shown.}
    \label{fig:lattice_schematic}
\end{figure}

GKP states are a class of non-Gaussian states with non-trivial Wigner functions that are highly concentrated at points on a rigid lattice in phase space~\cite{gkp2001}; see Figure~\ref{fig:lattice_schematic} for an illustration and Figure~\ref{fig:GKP_evolution}(d) of Chapter~\ref{sec:GKP_eng}. To motivate the need of GKP states for error correction~\cite{terhal2016encoding,duivenvoorden2017}, we begin with the uncertainty principle that all states need to obey: $\Delta^2 q \Delta^2 p \ge \frac{1}{4}$ (choosing $\hbar=1$). The uncertainty relations imply that the Wigner functions of valid quantum states have to spread in phase space. For error correction, the ideal case of a single-mode Wigner function is a delta function at origin, for which any random displacement can then be reverted back to the origin. According to the uncertainty principle though, having a delta-distribution as a Wigner function is non-physical as it implies $\Delta^2 p \Delta^2 q=0$. Gottesman, Kitaev, and Preskill introduced an ingenuous approach to work around this implausibility~\cite{gkp2001} by constructing quantum states described by grids of delta functions, popularly acronymed \emph{the GKP states}. In this manner, the variance in both quadratures can be large; while if one focuses on the individual lattice points, say origin, there is a delta-like distribution about that point. Thus, if a small displacement error happens, one can correct it by reverting the shifted lattice back to the original lattice positions~\cite{terhal2016encoding,duivenvoorden2017}.\footnote{We note, however, that displacement errors on the GKP state are uncorrectable when the shifts of the grid are larger than half of the lattice spacing because the direction of the shift error is ambiguous in this case.}. 

In Ref.~\cite{gkp2001}, the authors constructed general qudit encoding for $d$ quantum states into $N$ harmonic oscillators via what we now broadly refer to as GKP states. Since the inception GKP states, researchers have developed different applications and realizations of these states. However, in this context, we concentrate on two fundamental versions of GKP states, which have been extensively employed in the scientific literature: the \emph{canonical GKP state} ($d=1$)\footnote{Sometimes referred to as a qunaught state or sensor state.}---adopted for CV quantum information processing---and the \emph{computational GKP state} or the \emph{GKP qubit} ($d=2$)---widely adopted for digital quantum information processing. The canonical GKP state can be written in the position or momentum basis as 
\begin{equation}    
\qq{Canonical GKP state:}\ket{\square}=\sum_{n\in\mathbb{Z}}\ket{n\sqrt{2\pi}}_q=\sum_{n\in\mathbb{Z}}\ket{n\sqrt{2\pi}}_p,
\label{eq:canonicalGKP}
\end{equation}
where we use the simplified notation $\ket{\cdot}_q$ to represent the infinite-energy position eigenstates (likewise for the momentum eigenstate $\ket{\cdot}_p$) represented by Eq.~\eqref{q_p_eigenstate}. The canonical GKP state is a $+1$ eigenstate of the canonical GKP stabilizers,
\begin{equation}\label{eq:canonical_stabilizers}
   \qq{Canonical stabilizers:} S_1^{\square}={\rm e}^{i\sqrt{2\pi}\hat{q}} \qq{and} S_2^{\square}={\rm e}^{-i\sqrt{2\pi}\hat{p}}
\end{equation}
The computational GKP state, corresponding to a logically encoded qubit with bit value $\mathsf{b}\in\{0,1\}$ [see Figure~\ref{fig:lattice_schematic}(a) for an illustration], is given by
\begin{equation}
    \qq{Computational GKP state:} \ket{\mathsf{b}_{\square}}=\sum_{n\in\mathbb{Z}}\ket{(2n+\mathsf{b})\sqrt{\pi}}_q=\sum_{n\in\mathbb{Z}}{\rm e}^{-in\pi\mathsf{b}}\ket{n\sqrt{\pi}}_p.\label{eq:logicalGKP}
\end{equation}
The computational GKP state is a $+1$ eigenstate of the computational stabilizers
\begin{equation}\label{eq:computational_stabilizers}
   \qq{Computational stabilizers:} S_X={\rm e}^{-i2\sqrt{\pi}\hat{p}} \qq{and} S_Z={\rm e}^{i2\sqrt{\pi}\hat{q}},
\end{equation}
where the subscripts $Z$ and $X$ refer to the logical Pauli operators, $Z=\sqrt{S_Z}$ and $X=\sqrt{S_X}$, such that $Z\ket{\mathsf{b}_\square}=(-1)^{\mathsf{b}}\ket{\mathsf{b}_\square}$ and $X\ket{\mathsf{b}_\square}=\ket{(\mathsf{b}\oplus1)_\square}$. The GKP states described above correspond to a square lattice in phase space, and other lattices can be created from the square GKP state by symplectic transformations. Some common lattices are the square lattice just discussed, the rectangular lattice, and the hexagonal lattice; see Figure~\ref{fig:lattice_schematic}(b-d) for a schematic and the following sections for more details.

While ideal GKP states are not normalizable---and therefore non-physical---the original authors in Ref.~\cite{gkp2001} presented the finite-squeezed GKP state $\ket{\square(\Delta)}$, where $\Delta$ is an effective squeezing parameter. Writing out a canonical finite-energy canonical GKP state in the position and momentum bases, we obtain
\begin{align}
&\ket{\square(\Delta)}\propto 
\sum_{t=-\infty}^\infty e^{-\pi \Delta^2 t^2 }
\int e^{-(q-\ell t)^2/2\Delta^2} \ket{q} dq
\propto 
\sum_{t=-\infty}^\infty  \int e^{-\Delta^2p^2/2} e^{-(p-\ell t)^2/2\Delta^2}\ket{p}dp,
\label{GKP_finte_Delta}
\end{align}  
where $\ell=\sqrt{2\pi}$ is often used throughout this review for brevity. For $\Delta\ll1$, this complicated expression can be written succinctly as $\ket{\square(\Delta)}\propto{\rm e}^{-\Delta^2\hat{n}}\ket{\square}$, where ${\rm e}^{-\Delta^2\hat{n}}$ is an envelope operator that restricts the GKP lattice to a ball of radius $\sim\Delta^{-1}$ in phase space; see Chapter~\ref{sec:finite_GKP} (and also Refs.~\cite{royer2020stabilization,rojkov2023twoqubitStabiliz}) for more details about finite-energy GKP states and extension to multiple modes. Also see Ref.~\cite{matsuura2020EquivApproxGKP}, where the authors show the equivalence between different mathematical descriptions of finite-energy GKP states.

We can formally solve for the Wigner function of the finite-energy GKP state,
\begin{equation}
W(p,q;\square(\Delta))=
\frac{1}{N_\Delta}\sum_{t_1,t_2=-\infty}^\infty e^{-\Delta^2 p^2+i\ell  p\left(t_1-t_2\right)-\pi \Delta^2(t_1^2+t_2^2) -\frac{1}{\Delta^2}\left[q-\left(t_1+t_2\right)\ell/2\right]^2},
\end{equation}
where the normalization constant $N_\Delta=\sum_{t_1,t_2=-\infty}^\infty \pi \exp\left(-\frac{\pi \left(t_1-t_2\right)^2}{2\Delta^2}-\pi\Delta^2(t_1^2+t_2^2)\right)$. When $\Delta\ll1$, the Wigner function of $\ket{\square(\Delta)}$ is peaked around a square grid of spacing $\sqrt{2\pi}$ for a canonical square GKP, with positive and negative peaks alternating, for example, see Figure~\ref{fig:GKP_evolution}(d) for the Wigner function of a computational GKP state. Hence, for the canonical GKP, displacement errors must be less than $\sqrt{2\pi}/2$ for unambiguous characterization and correction. Taking a cut of the Wigner function along the $q$ (or $p$) direction, the alternating peaks cancel, and the position (or momentum) wavefunction is invariant under shifts of $\sqrt{2\pi}$. The quadrature variances are $\expval{\hat{q}^2}\simeq \expval{\hat{p}^2}\simeq 1/2\Delta^2$ and equal the mean photon number $N_S$ of the state. However, if we consider only the phase space region close to a single peak, the variances in position \textit{and} momentum around the peak are $\Delta^2/2\simeq 1/4N_S\ll 1$, which is twice the variance of a squeezed-vacuum state with the same number of photons as the GKP state. Due to the reduced variance around lattice points, the (canonical) GKP state has been considered for sensing applications (hence, the ``sensor state'' moniker~\cite{duivenvoorden2017}); see Chapter~\ref{subsec:gkp_sensor} for more discussion on this.


\begin{figure}
    \centering
    \includegraphics[width=.4\linewidth]{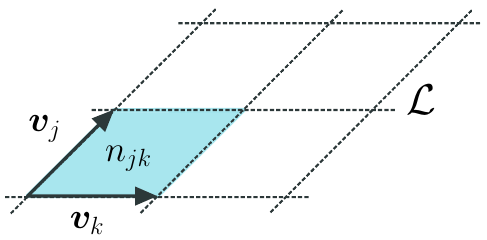}
    \caption{Area between lattice vectors is an integer $n_{jk}$ for a GKP lattice state $\ket{\mathcal{L}}$.}
    \label{fig:integral_condition}
\end{figure}

\subsection{GKP lattice states}
\label{sec:stabilizer_lattice}

In the previous discussion, we introduced the canonical and computational square GKP states, which are characterized by simple square grids with varying spacings in a two-dimensional phase space. We now aim to expand these concepts to encompass multimode GKP states---also known as \emph{GKP lattice states}---by adopting a geometrical approach to describe them. These GKP lattice states exhibit a direct correspondence with classical lattices in a $2N$-dimensional real space, as demonstrated in the original works of~\cite{gkp2001,harrington2001rates} and considered in more detail in the recent works of~\cite{baptiste2022multiGKP, conrad2022lattice,wu2022optimal}. Consequently, we delve a bit into the formalism required to describe useful classical lattices.

Consider a rigid, $2N$-dimensional lattice $\mathcal{L}$ defined by some set of lattice basis vectors $\bm{v}_j\in\mathbb{R}^{2N}$ where $j=1,\dots,2N$. We assume that
\begin{equation}\label{eq:basis_norm}
\omega(\bm{v}_j,\bm{v}_k)=\bm{v}_j^{\top}\bm\Omega\bm{v}_k=n_{jk},
\end{equation}
where $n_{jk}\in\mathbb{Z}$. In other words, the symplectic inner product between basis vectors---which is equal to the area between the two vectors in phase space---is an integer; see Figure~\ref{fig:integral_condition} for an illustration of this principle. Any lattice with basis vectors satisfying this integral condition is called a \textit{symplectically integral lattice}. From the basis vectors, we can build a \emph{generator matrix} $\bm M$, 
\begin{equation}
    \bm M=(\bm{v}_1, \bm{v}_2,\dots,\bm{v}_N).
\end{equation}
With this notation, we can more properly define the classical lattice $\mathcal{L}$ as the set of points $\mathcal{L}\coloneqq\{\bm M \bm a\, |\, \bm a\in\mathbb{Z}^{2N}\}$. Moreover, the integral conditions~\eqref{eq:basis_norm} can be packaged in a compact form,
\begin{equation}
    \bm M^\top\bm\Omega\bm M=\bm A,
\end{equation}
where the \emph{symplectic Gram matrix} $\bm A$ is an anti-symmetric matrix with integer entries. As we discuss later in more detail, in order to encode a $d$-level system (a qudit) into the classical lattice $\mathcal{L}$, the lattice must satisfy $\det\bm A=d^2$~\cite{gkp2001,harrington2001rates,baptiste2022multiGKP, conrad2022lattice}, where $d$ is the \emph{code dimension}.

We can provide an extension of the lattice $\mathcal{L}$ to the \emph{dual lattice} $\mathcal{L}^*$. The dual lattice consists of all points $\bm v^*$ which have integer symplectic inner product with the lattice basis vectors, i.e.
$\mathcal{L}^*=\{\bm v^*|\bm M^\top\bm\Omega \bm v^*\in\mathbb{Z}^{2N}\}$. It follows that $\mathcal{L}\subseteq\mathcal{L}^*$ since integer combinations of the basis vectors are also included in $\mathcal{L}^*$. One can relate the generator matrix of the dual lattice $\bm M^*$ to the generator matrix of the lattice $\bm M$ via the simple relation $\bm M^{*}=\bm M \bm A^{-\top}$. A (symplectic) \emph{self-dual lattice} is a lattice for which $\mathcal{L}=\mathcal{L}^*$ and $\det\bm A=1$~\cite{gkp2001,harrington2001rates, conrad2022lattice}.

For some insight into code construction from classical lattices, we give examples of valid classical lattices in dimensions of $2,4,8,24$ obtained from~\cite{conway2013sphere} and mention the corresponding code dimensions as well as associated lengths of logical Pauli operators (discussed in more detail in the following section). We then show how one can build a \emph{GKP lattice quantum state} $\ket{\mathcal{L}}$ from a classical lattice $\mathcal{L}$. For a detailed discussion into why we might care about higher dimensional lattices for quantum error correction, see Chapter~\ref{sec:QEC-multimode}.

\begin{example}[Hypercube]
    The simplest example of a classical lattice in $2N$ dimensions is a scaled hypercube with generator matrix $\bm M(\square^N)=a\bm I_{2N}$, such that $\det\bm A(\square^N)=a^{4N}$. Since the determinant of the Gram matrix determines the code dimension $d$ via $\det\bm A= d^2$, we can encode an ensemble of $N$ qubits by taking $a=\sqrt{2}$.\footnote{We could try to encode $n$ qubits into an $2N$-dimensional hypercube by taking $a=2^{n/2N}$, such that $\bm A=a^{n/N}\bm I_{2N}$ and $\det\bm A=d^2=(2^n)^2$. However, for this code to correspond to a symplectically integral lattice, we must have that $a^2\in\mathbb{Z}$, which is not satisfied for $a^2=2^{n/N}$ unless $n=N$.} With $N$ qubits encoded into an $2N$-dimensional hypercube, the minimal length of logical Pauli operators is $1/\sqrt{2}$ (in units $\ell=\sqrt{2\pi}$). 
\end{example}

\begin{example}[D-Type Lattice or Checkerboard Lattice]
The $D_4$ lattice has the densest packing in 4 dimensions and could therefore be used to efficiently encode a qubit into two modes. The generator matrix is given by,
 \begin{equation}   
 \bm M({D_{2N}})=\begin{pmatrix}
-1&-1&0&0&.&.&0&0\\
1&-1&0&0&.&.&0&0\\
0&0&1&-1&.&.&0&0\\
..&...&..&..&..&..&..&..\\
0&0&0&0&0&1&-1&0\\
0&0&0&0&0&0&1&-1\\
\end{pmatrix}
\end{equation}
The determinant of the Gram matrix for the $D_4$ lattice is $\det\bm A({D_{2N}})=4$ and hence, it can be used to encode a qubit in $N$ modes. One can show that the length of all logical operators is 1 (in units $\ell=\sqrt{2\pi}$)~\cite{baptiste2022multiGKP}.
\end{example}

\begin{example}[E-type Lattice]
The $E_8$ lattice has the densest packing in $8$ dimensions~\cite{viazovska2018sharp}. One can choose a canonical generator matrix for the $E_8$ lattice as,
 \begin{equation}  
 \bm M({E_{8}})=\begin{pmatrix}
2&0&0&0&0&0&0&0\\
-1&1&0&0&0&0&0&0\\
0&-1&1&0&0&0&0&0\\
0&0&-1&1&0&0&0&0\\
0&0&0&-1&1&0&0&0\\
0&0&0&0&-1&1&0&0\\
0&0&0&0&0&-1&1&0\\
1/2&1/2&1/2&1/2&1/2&1/2&1/2&1/2\\
\end{pmatrix},
\end{equation}
with $\det \bm M({E_8})=1$. Hence, scaling of the $E_8$ lattice [$\bm M({E_8})\rightarrow a\bm M({E_8})$] \QZ{yields a code dimension $d=a^8$---i.e., 4 qudits of dimension $\sqrt{a}$}. The length of the logical operators is 1 (in units $\ell=\sqrt{2\pi}$) for the $E_8$ lattice, which is larger than the ensemble of 4 qubits constructed from square lattices~\cite{baptiste2022multiGKP}.
\end{example}

\begin{example}[Leech Lattice] 
The Leech lattice is the optimal lattice in $24$ dimensions as proved in Ref.~\cite{cohn2017sphere}. The generator matrix for the Leech lattice is a unimodular matrix with determinant $\det\bm M_{\rm Leech}=1$, and \QZ{hence is valid for code dimensions $d=a^{24}$ upon scaling $\bm M_{\rm Leech}\rightarrow a\bm M_{\rm Leech}$. For $a=\sqrt{2}$}, it can yield an ensemble of 12 qubits. In this case, the length of logical operators ($=\sqrt{2}$) is twice as large as the square code.
\end{example}

We now sketch how to construct a GKP lattice quantum state $\ket{\mathcal{L}}$ from a classical lattice $\mathcal{L}$. From some set of lattice basis vectors $\{\bm v_j\}$, we define a corresponding set of \emph{stabilizer generators} via, 
\begin{equation}
    S_{j}\coloneqq D_{\ell\bm{v}_j},
\end{equation}
where we have introduced the canonical spacing $\ell\coloneqq\sqrt{2\pi}$. Combining the integral conditions~\eqref{eq:basis_norm} with the Weyl commutation relation~\eqref{eq:weyl_comp}, it follows that,
\begin{equation}
    \omega(\bm v_j,\bm v_k)=n_{jk}\implies\comm{S_{j}}{S_{k}}=0\quad\forall\,j,k.
\end{equation}
The stabilizers therefore generate a $2N$-dimensional stabilizer group $\mathcal{S}(\mathcal{L})\coloneqq\langle S_{1},S_{2},\dots,S_{2N}\rangle$. A \emph{GKP lattice state} $\ket{\mathcal{L}}$ is then defined as a simultaneous $+1$ eigenstate of all the commuting stabilizers, i.e. $S_j\ket{\mathcal{L}}=\ket{\mathcal{L}}\,\forall j$. From here, it should be clear that the GKP lattice state $\ket{\mathcal{L}}$ has the same symmetries in quantum phase space as the classical lattice $\mathcal{L}$ does in the classical real space. Before providing some concrete examples of GKP lattice states, we quote a fact about all single-mode lattice states.  
\begin{proposition}\label{prop:1mode_lattice}
Any single-mode GKP lattice state $\ket{\mathcal{L}}$ can be generated from a square GKP state by a symplectic transformation $\bm\Lambda$, i.e. $\ket{\mathcal{L}}=U_{\bm \Lambda}\ket{\square}$. 
\end{proposition}
This result follows from the fact that the Gram matrix $\bm A$ for any two-dimensional (symplectically integral) lattice is a $2\times2$ anti-symmetric matrix with degenerate matrix elements and is thus proportional to the symplectic form $\bm\Omega_1$, such that $\bm M^\top\bm\Omega_1\bm M\propto\bm\Omega_1$. Therefore, the generator matrix $\bm M$ must be proportional to a symplectic transformation; see Refs.~\cite{gkp2001,harrington2001rates,baptiste2022multiGKP,conrad2022lattice} for formal proofs and extensions. Hexagonal and rectangular GKP states are prototypical examples of this.


\subsubsection{Canonical lattice states}
\label{sec:canonical_lattice}

\begin{figure}
    \centering
    \includegraphics[width=.9\linewidth]{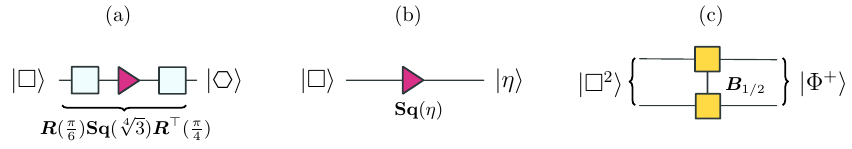}
    \caption{Circuit schematic of various GKP lattice states: (a) hexagonal GKP $\ket{\hexagon}$, (b) rectangular GKP $\ket{\eta}$, (c) GKP Bell state $\ket{\Phi^{+}_{\rm GKP}}$.}
    \label{fig:lattice_state_schematic}
\end{figure}

Here we restrict to lattices $\mathcal{L}_C$ that have generator matrices $\bm M_C$ satisfying $|\det\bm M_C|=1$, i.e. with trivial code dimension ($d=1$); we will call such lattice states \emph{canonical lattice states} (or canonical GKP states). Though one cannot encode digital information into canonical lattice states, these states can nevertheless be useful for other purposes, such as protecting arbitrary CV quantum states via oscillators-to-oscillators codes~\cite{noh2020o2o,wu2022optimal}.

We first consider the single-mode canonical GKP state discussed in Chapter~\ref{sec:GKP_simple_part}.
Using lattice notation, we can write the stabilizers of the canonical square GKP state $\ket{\square}$ as $ {S}^{\square}_{j}=\exp(i\ell{\bm{v}}^{\square\,\top}_j\bm\Omega_1\hat{{\bm r}})$, where
\begin{equation}
    {\bm{v}}^{\square}_1=\left(1,0\right)^\top \qq{and}
    {\bm{v}}^{\square}_2=\left(0,1\right)^\top,
\end{equation}
are the canonical square lattice vectors. The generator matrix of the canonical square lattice is then,
\begin{equation}
    \bm M_C(\square)=\left({\bm{v}}^{\square}_1\hspace{.6em}{\bm{v}}^{\square}_2\right)
    =\bm I_2,
\end{equation}
where $C$ refers to canonical. We define an $N$-mode canonical GKP hypercube as $\ket{\square^N}\coloneqq\ket{\square}^{\otimes N}$, with a generator matrix $\bm M_C(\square^N)\coloneqq\bigoplus_{i=1}^N\bm{M}_C(\square)=\bm I_{2N}$.

It turns out that we can generate any canonical lattice state $\ket{\Lambda}$ by applying symplectic transformations on the canonical hypercube~\cite{gkp2001,harrington2001rates,conrad2022lattice}. Indeed it is easy to show that such states correspond to a valid symplectically integral lattice $\Lambda$. To see this, define $\Lambda$ by the basis vectors $\bm{v}^{\Lambda}_j=\bm\Lambda\bm{v}^\square_{j}$. It follows that $\omega(\bm{v}^{\Lambda}_j,\bm{v}^{\Lambda}_k)=\omega(\bm{v}^\square_{j},\bm{v}^\square_{k})$ because the symplectic inner product is invariant under symplectic transformations. Thus, we can construct the stabilizers ${S}_j^{\Lambda}\coloneqq{D}_{\ell\bm{v}^{\Lambda}_j}$, which generate the stabilizer group $\mathcal{S}(\Lambda)=\langle{S}_1^{\Lambda}, {S}_2^{\Lambda},\dots,{S}_{2N}^{\Lambda}\rangle$, where e.g.
\begin{equation}
    {S}_j^{\Lambda} = U_{\bm\Lambda}  {S}_j^{\square} U_{\bm\Lambda}^\dagger.
\end{equation}
Hence, the lattice state $\ket{\Lambda}= U_{\bm\Lambda}\ket{\square^N}$ is a $+1$ eigenstate of $\mathcal{S}(\Lambda)$ by construction. Furthermore, it is straightforward to show that a generator matrix $\bm M_C(\Lambda)$ of the lattice $\Lambda$ and the symplectic transformation $\bm \Lambda$ can be related via
\begin{equation}\label{eq:M_lattice}
    \bm M_C(\Lambda)=\bm\Lambda.
\end{equation}
Therefore, $\bm M^{\top}_C(\Lambda)\bm\Omega\bm M_C(\Lambda)=\bm I_{2N}$. We show some examples of canonical lattice states in Figure~\ref{fig:lattice_state_schematic} and below.

\begin{example}[GKP Bell state]
    A GKP Bell state $\ket{\Phi^+_{\square}}=(\ket{\mathsf{0}_{\square}}\otimes\ket{\mathsf{0}_{\square}}+\ket{\mathsf{1}_{\square}}\otimes \ket{\mathsf{1}_{\square}})/\sqrt{2}$ can be created by interacting two canonical square GKP states on a 50:50 beamsplitter~\cite{walshe2020CVteleport}. The GKP Bell state has the same lattice spacing as the two-mode canonical GKP state and is thus not very useful for O2O codes. However, GKP Bell states are a necessary resource for, e.g., robust DV quantum teleportation~\cite{walshe2020CVteleport,vanLoock2022gkp}. A generator matrix for the GKP Bell state is simply $\bm M_C(\Phi^+)=\bm B_{1/2}$.
\end{example}

\begin{example}[Canonical $D_4$ GKP]
    We can generate a canonical $D_4$ from a canonical hypercube via the following two-mode symplectic transformation, 
\begin{equation}\label{eq:SD4}
    \bm\Lambda_{D_4}=\sqrt[4]{2}
    \begin{pmatrix}
        \frac{1}{2} & -\frac{1}{\sqrt{2}} & \frac{1}{2} & 0\\
        0 & \frac{1}{\sqrt{2}} & 0 & \frac{1}{\sqrt{2}}\\
        0 & \frac{1}{\sqrt{2}} & 0 & -\frac{1}{\sqrt{2}} \\
        -\frac{1}{2} & 0 & \frac{1}{2} & \frac{1}{\sqrt{2}}
    \end{pmatrix},
\end{equation}
such that $\bm M_C(D_4)=\bm\Lambda_{D_4}$. The canonical $D_4$ GKP state can be written as $\ket{{D_4}}= U_{\bm\Lambda_{D4}}\ket{\square^2}$. Since $\bm\Lambda_{D4}\bm\Lambda_{D4}^{\top}\neq\bm I$, the $D_4$ transformation has squeezing. We consider the QEC properties of the canonical $D_4$ state in oscillators-to-oscillators codes in Chapter~\ref{sec:gkp_o2o_codes}. 
\end{example}

\subsubsection{Computational lattice states}
\label{sec:comp_lattice}

Here we consider lattice states with non-trivial code dimension $d\neq1$. As previously mentioned, a lattice $\mathcal{L}$ can support a $d$-level quantum system if the determinant of the Gram matrix satisfies $\det(\bm A)=d^2$~\cite{gkp2001,harrington2001rates,baptiste2022multiGKP,conrad2022lattice}. For instance, a computational square GKP state [introduced in Eq.~\eqref{eq:logicalGKP}] can be represented by $\bm M_L(\square)=\sqrt{2}\bm I_2$ ($L$ for logical information) and thus $\det\bm A_L(\square)=2^2$, indicating that the computational GKP state can support a single qubit as advertised. Conversely, any canonical GKP state has $\det\bm A_C=1$, meaning that the encoded Hilbert space is trivial ($d=1$). Below, we give a few recipes to find generator matrices for multimode GKP qubits from classical lattices~\cite{baptiste2022multiGKP}. We then discuss logical Pauli operators of a multimode GKP qubit and introduce the Pauli displacement vectors. The main takeaways from this section are summarized in Table~\ref{tab:lattice-code}.

First, let us consider a general, $2N$-dimensional (symplectically integral) lattice $\mathcal{L}$ with generator matrix $\bm M(\mathcal{L})$ and Gram matrix $\bm A$, with code dimension $d$ given by $\det\bm A=d^2$. We can scale the lattice by $a\in\mathbb{R}$, such that,
    \begin{equation}
        \bm M'=a\bm M \implies
         \bm A'=a^{2}\bm A \implies
    \det\bm A'=a^{4N}\det\bm A.
    \end{equation}
    Since $\det\bm A=d^2$, we see that the code dimension can perhaps be increased by a factor  $a^{2N}$ via scaling.
We apply this intuition to two cases:
\begin{itemize}
    \item[(1)] \emph{Integral Lattices:} If we scale a $2N$-dimensional canonical lattice $\mathcal{L}_C$ with Gram matrix $\bm A_C$ (such that $\det\bm A_C=1$) to a lattice $\mathcal{L}^\prime$ with $\bm A^\prime=a^2\bm A_C$, then we must have that $a^2\in\mathbb{N}$ for the lattice $\mathcal{L}^\prime$ to be symplectically integral. For a single mode ($N=1$), we can use the canonical square lattice, as described in Chapter~\ref{sec:canonical_lattice}, such that any constant $a=\sqrt{d}$ yields a code of dimension $d$. Extending this example to $N$ modes, a canonical hypercubic lattice with generator matrix $\bm M_C(\square^N)=\bm I_{2N}$ can only be scaled to obtain an ensemble of $N$ qubits.
     
     \item[(2)] \textit{Non-Integral Lattices}: Another strategy is to scale a lattice to an invalid code dimension such that  the lattice is not symplectically integral. Then we can search for a non-symplectic orthogonal transformation ($\bm O^\top\bm\Omega \bm O\neq \bm\Omega$)---which does not change the code dimension since $\det\bm O=1$---but can change the Gram matrix as $\bm A=\bm M^\top \bm\Omega \bm M\rightarrow \bm A'=\bm M^\top \bm O^\top\bm\Omega \bm O\bm M$. If the transformation matrix $\bm O$ yields an integral Gram matrix $\bm A^\prime$, then we have a valid lattice with the desired code dimension, which could not be achieved by simply scaling.
\end{itemize}

\begin{example}[Tesseract lattice]
    A two-mode \emph{Tesseract} qubit code is an example of how one can scale then rotate a classical lattice to find an integral lattice that can support a qubit. Consider a scaled 4-dimensional hypercube with generator matrix $\sqrt[4]{2}\bm I_{4}$. This alone is not a symplectically integral lattice. However, by performing a $\pi/4$ rotation of the $p_1$ and $p_2$ quadratures $\tilde{\bm O}$, one can show that the generator matrix,\footnote{We use the notation $\widetilde{\square}^2$ since the tesseract is just a rotated hypercube.}
    \begin{equation}\label{eq:tess_M}
        \bm M({\widetilde{\square}^2})=\sqrt[4]{2}\tilde{\bm O}=\sqrt[4]{2}\begin{pmatrix}
             1&0&0&0\\
             0&\frac{1}{\sqrt{2}}&0&\frac{1}{\sqrt{2}}\\
             0&0&1&0\\
             0&\frac{1}{\sqrt{2}}&0&-\frac{1}{\sqrt{2}},
         \end{pmatrix}
    \end{equation}
    corresponds to a valid symplectically integral lattice with code dimension $d=2$.
\end{example}

\begin{table}[t]
\centering

\renewcommand{\arraystretch}{1.0}

\begin{tabular}{ c | c  c  c }
\hline\hline
    &&\\
    \quad & Ideal GKP Codespace ($\mathcal{S}$) &Matrix form ($\bm M$)\\
    \hline\hline
    &&\\
    Stabilizer Generators & $S_i=D_{\ell\bm v_i}$ & $\bm M= (\bm v_1,\bm v_2,..,\bm v_{2N})$  \\
    \hline
    &&\\
    Commutation Relations & $[S_i,S_j]=0\quad \forall i, j$ & $\bm v_i^\top\bm\Omega \bm v_j\in 2\mathbb{Z}\quad\forall i, j$ \\
    \hline 
    &&\\
    Code Dimension & $d$ & $\det\left(\bm M^\top \bm\Omega \bm M\right)= d^2$ \\
    \hline
    &&\\
    Logical Pauli Operators &  $[D_{\bm p},S_i]=0 \quad\forall i$ & $\bm p^{\top}\bm \Omega \bm v_i\in 2\mathbb{Z}+1\quad\forall i$ \\
    \hline\hline
\end{tabular}
\label{tab:lattice-code}
\caption{Correspondence between a classical lattice $\mathcal{L}$ with generator matrix $\bm M$ and the associated multimode GKP qudit, with code dimension $d$, defined by the stabilizer group $\mathcal{S}\coloneqq\mathcal{S}(\mathcal{L})$. Basis vectors $\{\bm v_j\}$ generate the lattice $\mathcal{L}$. Pauli displacement vectors $\bm p$ are taken as the shortest vectors of the dual lattice $\mathcal{L}^*$ and have (odd) integer symplectic inner product with $\{\bm v_j\}$.}
\end{table}

Interestingly, it turns out that any $N$-mode GKP qubit---i.e., a $2N$-dimensional lattice with $|\det\bm M|=2$---can be generated by acting on 1 single-mode GKP square qubit plus $(N-1)$ canonical square GKP states with a (multimode) symplectic transformation $\bm\Lambda$ (see Corollary 2 of Ref.~\cite{conrad2022lattice}). One can thus write down a generator matrix for the resulting GKP lattice state as 
\begin{equation}
    \bm M_L=\bm \Lambda\left(\sqrt{2}\bm I_2\oplus \bm I_{2(N-1)}\right).
\end{equation}
We provide some common examples of this for the simple case of encoding one qubit into one mode below.

 
\begin{example}[Hexagonal GKP]
    A hexagonal GKP qubit can be generated from the computational square GKP by the following symplectic transformation,
\begin{equation}\label{eq:lambda_hex}
    \bm\Lambda_{\hexagon}=
    \frac{\ell_{\hexagon}}{\ell}\begin{pmatrix}
        1 & -\frac{1}{2} \\
        0 & \frac{\sqrt{3}}{2}
    \end{pmatrix},
\end{equation}
where $\ell_{\hexagon}/\ell=\sqrt{2}/3^{1/4}\approx1.07$. Since $\bm\Lambda_{\hexagon}$ is a single-mode transformation, it admits a decomposition in terms of a single-mode squeezer sandwiched between two phase shifts; see Figure~\ref{fig:lattice_state_schematic}(a) for values of the phase shifts and squeezing strength. This encoding has logical Pauli vectors that are 1.07 times longer than the square code and, thus, in principle, is better for QEC.
\end{example}  

\begin{example}[Rectangular GKP]
    A rectangular GKP qubit can be created by squeezing the computational square by the symplectic squeezing matrix $\textbf{Sq}(\eta)$ with squeezing strength $\eta={\rm e}^r$. The generator matrix for the rectangular lattice is, 
    \begin{equation}
        \bm M_L(\eta)=\sqrt{2}~{\rm diag}(1/\eta,\eta).
    \end{equation}
    Computational rectangular GKP states have been recently studied for bias-enhanced QEC~\cite{hanggli2020pra,Zhang2022GKPxzzx,stafford2022GKPbiasedREP}.
\end{example}


So far, we have discussed about encoding $d$-level systems into a lattice but have yet to discuss, e.g., the logical Pauli operators associated with the code. We now make this notion concrete. For simplicity, we restrict to qubit encodings such that $d=2$. In Chapter~\ref{sec:stabilizer_lattice}, we introduced the dual lattice $\mathcal{L}^*$ as the set of all vectors $\{\bm v^*\}$ that have an integral symplectic inner product with the lattice basis vectors $\{\bm v_j\}$. The elements $\bm v^*\in\mathcal{L}^*/\mathcal{L}$ (i.e., $\bm v^*\in\mathcal{L}^*$ but $\bm v^*\notin\mathcal{L}$) correspond to (integer multiples of) logical \emph{Pauli displacement vectors} $\bm p\in\{\bm x, \bm y, \bm z\}$ that we associate with translations $D_{\bm p}$~\cite{harrington2001rates}. These Pauli translations implement logical Pauli operations---e.g., $X_L\coloneqq D_{\bm x}$. Here we equate the logical Pauli vectors $\bm p$ with the smallest vectors of the dual lattice $\mathcal{L}^*$, such that the norms $\norm{\bm x}$, $\norm{\bm y}$, and $\norm{\bm z}$  represent the \textit{Pauli distances} of the GKP code. The Pauli distances are significant because they roughly tell us how much random displacement noise a GKP qubit code can tolerate before a logical Pauli error occurs. Recall that the generator of matrix of the dual lattice can be found from $\bm M^*=\bm M\bm A^{-\top}$. Then the Pauli displacements can, more or less, be associated with the columns (or rows) of $\bm M^*$. A code is said to be \emph{balanced} if $\norm{\bm x}=\norm{\bm y}=\norm{\bm z}$, which is the case for the hexagonal lattice and $D_4$ lattice. Two examples for a GKP qubit are provided below to solidify some of these concepts.

\begin{example}[Square qubit]\label{example:square_code}
    For the square GKP qubit code, we have that $\bm M_L(\square)=\sqrt{2}\bm I_2$ and $\bm A_L(\square)=2\bm\Omega$, such that $\bm M^*_L(\square)=\bm \Omega/\sqrt{2}$. The Pauli vectors (i.e., shortest vectors of the dual lattice) are just the columns of $\bm M^*_L(\square)$. We thus have that $\norm{\bm x_\square}=\norm{\bm z_\square}=1/\sqrt{2}$ and $\norm{\bm y_\square}=1$.
\end{example}

\begin{example}[Tesseract qubit]
    The generator matrix for the tesseract qubit code can be written as $\bm M({\widetilde{\square}^2})=\sqrt[4]{2}\tilde{\bm O}$, where $\tilde{\bm O}$ is a non-symplectic orthogonal transformation; see Eq.~\eqref{eq:tess_M}. It follows that $\bm A({\widetilde{\square}^2})^{-\top}=\tilde{\bm O}^{\top}\bm\Omega\tilde{\bm O}/\sqrt{2}$, and one can write down a generator matrix for the dual lattice simply as, $\bm M({\widetilde{\square}^2})^*=\bm\Omega\tilde{\bm O}/\sqrt[4]{2}$. This is enough to show that $\norm{\bm x_{\widetilde{\square}^2}}=\norm{\bm z_{\widetilde{\square}^2}}=1/\sqrt[4]{2}$ and $\norm{\bm y_{\widetilde{\square}^2}}=\sqrt[4]{2}$, which are $\sqrt[4]{2}$ larger than the Pauli displacements of the single-mode square code.
\end{example}


\subsection{Error syndromes and stabilizer measurements}
\label{sec:syndromes}

As we summarize in Chapter~\ref{sec:noise_model}, general noise models can be reduced to displacement errors.
We briefly consider errors induced by random displacements on GKP lattice states and present an operational approach to extract error syndromes via stabilizer measurements of the lattice.

\subsubsection{Syndrome vector}
Consider a lattice state $\ket{\mathcal{L}}$, with stabilizers $\{S_j\}$, and a displacement error $\bm e$. Then define the error state $\ket{\mathcal{L}; \bm e}={D}_{\bm e}\ket{\mathcal{L}}$. By the Weyl commutation relation~\eqref{eq:weyl_comp}, it is easy to show that $\ket{\mathcal{L};\bm e}$ is an eigenstate of the stabilizer $S_j$ with eigenvalue $\exp[-i\omega\left(\ell\bm{v}_j,\bm e\right)]$. The set of quantities $\{\omega(\ell\bm{v}_j,\bm e)\mod{2\pi}\}_{j=1}^{2N}$ (one for each stabilizer) are the \emph{error syndromes}, and we can package them into a syndrome vector $\bm s$ via,
\begin{equation}\label{eq:syndrome}
    \bm s \coloneqq \bm s(\bm e)=\bm M^{\top}\bm\Omega\bm e\mod\sqrt{2\pi},
\end{equation}
where the modulo operation acts elementwise and we have factored out the canonical spacing $\ell=\sqrt{2\pi}$ for convenience. Given the syndrome information, we can perform syndrome-informed counter-displacements to the error state $\ket{\mathcal{L};\bm e}$ to correct for the error $\bm e$. If the error is small enough, then the error can be corrected with high probability. A simple example is the case where the lattice state is a canonical GKP state, such that $\bm M=\bm\Lambda$. Then,
\begin{equation}\label{eq:syndrome_symplectic}
    \bm s = \bm\Lambda^{-1}\bm e \mod\sqrt{2\pi}.
\end{equation}
An operational approach to measure the error syndromes is discussed below; see also Figure~\ref{fig:measurement} for a depiction of the measurement scheme for canonical GKP states.

\subsubsection{Stabilizer measurements: An operational approach}

\begin{figure}
    \centering
    \includegraphics[width=0.6\linewidth]{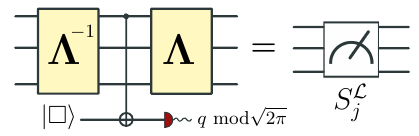}
    \caption{GKP-assisted stabilizer measurements for the canonical GKP state $\ket{\Lambda}=U_{\bm\Lambda}\ket{\square^N}$. Similar measurement schemes can be be constructed for multimode GKP qubits.}
    \label{fig:measurement}
\end{figure}

To measure the stabilizers (and thus extract the syndrome $\bm s$) of an error state $\ket{\mathcal{L}; \bm e}$, we could perform direct homodyne measurements along the effective quadratures, $\hat{\bm g}=\ell\bm M^\top\bm\Omega\bm{\hat{r}}\mod{2\pi}$, which is equivalent to measuring the stabilizers $\{S_j\}$. However, direct homodyne measurements are destructive; an ancilla-assisted measurement scheme is thus desired. One approach to handle this is to perform syndrome measurements with the help of SUM gates and GKP measurement ancillae~\cite{gkp2001}. This is an operational approach and may not be the go-to method for specific platforms [see, e.g., Chapter~\ref{sec:GKP_eng} for measurement methods in a cQED architecture where an auxiliary qubit is used to realize stabilizer measurements in a bit-wise fashion], however it illustrates the main features of stabilizer measurements and syndrome extraction.

For concreteness, we first discuss measuring the stabilizers of a square GKP state. Imagine trying to measure the stabilizers $S_j^\square$ [Eq.~\eqref{eq:canonical_stabilizers}] of a system $A$ prepared in a canonical GKP state $\ket{\square}_A$. To do so, we can couple the system $A$ to a GKP measurement ancilla $\ket{\square}_B$ via the SUM gate $\textbf{SUM}_{AB}$ and subsequently perform a $q$-homodyne measurement on $B$ to realize a stabilizer measurement of $S_1^{\square}$. Likewise, we can couple the system $A$ to another ancilla $B^\prime$ via $\textbf{SUM}_{B^\prime A}^{-1}$ and perform a $p$-homodyne measurement on $B^\prime$ to realize $S_2^{\square}$. Or if the system $A$ can be discarded after the syndrome measurements, we can directly perform a $p$-homodyne on $A$ following the $S_1^\square$ measurement, in order to save GKP resources; a similar measurement strategy holds for measuring stabilizers of computational GKP states. This stabilizer measurement strategy can be easily extended to measure the stabilizers of any canonical GKP state, i.e. states of the form $\ket{\Lambda}=U_{\bm\Lambda}\ket{\square^N}$. Indeed, we can simply conjugate the aforementioned measurement strategy to measure the stabilizers $\mathcal{S}(\Lambda)$; see Figure~\ref{fig:measurement} for a depiction. With this scheme, we can perform homodyne measurements in parallel on the individual modes. A slightly augmented setup can be constructed to measure the stabilizers of a multimode GKP qubit. 

A few comments about resources for stabilizer measurements are in order. There are generally $2N$ stabilizers---one for each quadrature---that need to be measured for a $N$-mode GKP lattice state. This holds for any GKP lattice state, even when, e.g., GKP codes are concatenated with DV qubit codes.\footnote{For a concatenated $[[n,k,d]]$ GKP qubit code (see Chapter~\ref{sec:QEC-multimode}), one often thinks about measuring the $2n$ stabilizers of the GKP code and then measuring the $n-k$ stabilizers of the outer qubit code, with a total of $3n-k$ measurements to be performed. In principle, the same can be accomplished with less ($2n$ measurements to be precise)~\cite{baptiste2022multiGKP,conrad2022lattice}.} Observe that, if the lattice state can be discarded after the stabilizers have been measured then only $N$ additional GKP ancillae (rather than $2N$) are required, thus reducing the GKP resources by half. Furthermore, in principle, we can perform these measurements in parallel on the individual modes by pushing the complexity of the measurement to the multimode symplectic transformation $\bm\Lambda$ that generates the lattice state from single-mode square GKP states.

Finally, we point out that the measurement schemes described above rely on the SUM gate, however the SUM-gate necessarily requires some amount of inline squeezing, as the SUM-gate is not a passive transformation. An equivalent measurement strategy, which moves squeezing offline, has been proposed in Ref.~\cite{vanLoock2022gkp}.

\subsection{Finite-energy GKP states}\label{sec:finite_GKP}

One practical issue or limitation with GKP states is the finite amount of squeezing to generate them in the lab~\cite{fluhmann2019encoding,campagne2020quantum}. The finite squeezing introduces non-idealities which can spoil QEC properties that are derived from the infinite-energy code---related to, for instance, fault-tolerance thresholds for quantum computing architectures based on GKP codes (Chapter~\ref{sec:app_ftqc})---if the squeezing is not beyond a certain threshold. We briefly covered some aspects of single-mode finitely squeezed GKP states in Chapter~\ref{sec:GKP_simple_part}, but we elaborate further on these matters here and extend to multiple modes. By a simple twirling argument, we also heuristically illustrate how the effective GKP noise from finite squeezing can affect the QEC performance of GKP codes.

The version of the GKP lattice state $\ket{\mathcal{L}}$ discussed so far has an infinite amount of energy and is thus an unphysical (non-normalizable) state.  We will now focus on its finite-energy counterpart~\cite{gkp2001} and regularize the state as follows. Consider a positive semi-definite matrix ${\bm\Delta}$, which quantifies the finite amount of squeezing in a GKP state. We then define a generic finite-energy GKP state via 
\begin{equation}\label{eq:envelope_lattice}
    \ket{\mathcal{L}(\bm\Delta)}=N_{\Delta}\underbrace{\exp\left(-\frac{\hat{\bm r}^\top\bm\Delta^\top\bm\Delta\hat{\bm r}}{2}\right)}_{\coloneqq E_{\Delta}}\ket{\mathcal{L}},
\end{equation}
where $N_{\Delta}$ is a normalization constant, and we have introduced the \emph{envelope operator}, $E_\Delta = \exp(-\frac{\hat{\bm r}^\top\bm\Delta^\top\bm\Delta\hat{\bm r}}{2})$. By introducing finite squeezing in this generic way, we allow for different levels of regularization for different modes and along different quadratures. However, if we assume the same amount of squeezing for all modes and all quadratures (isotropic squeezing), then $\bm\Delta=\Delta \bm I_{2N}$ where $\Delta$ is a positive constant, and we recover typical regularization considered in, e.g., Refs.~\cite{noh2019quantumcapacity,noh2020fault,noh2020o2o,royer2020stabilization}. In this case, finite squeezing has a direct interpretation as restricting the lattice to an $N$-dimensional ball of radius $\Delta^{-1}$. We note that an alternative definition of finite-energy GKP states is given by the quasi-degenerate ground state manifold of the finite-energy GKP Hamiltonian as discussed in Chapter~\ref{sec:GKP_Hamiltonian}. This alternative definition does not rely on non-unitary operators.

We can gain some insight as to why a finite-energy GKP state can, for instance, add to the noise budget for quantum information processing by a simple twirling model introduced in Ref.~\cite{noh2020fault}. We emphasize that the following twirling argument is heuristic and appropriate consideration of finite-energy states and operations need to be considered before strong conclusions are made. To simplify matters, we consider the scenario of isotropic squeezing, $\bm\Delta=\Delta\bm I_{2N}$. In this case, we can write the envelope operator in the displacement operator basis as,
\begin{equation}
    E_{\Delta}\propto\int\dd{\bm \mu}\exp\left(-\frac{\bm\mu^2}{4\tanh(\Delta^2/2)}\right)D_{\bm \mu}.
\end{equation}
We can then think about the regularization procedure as coherently applying displacements with a Gaussian envelope to a GKP state. To gain further intuition about the effects of finite-energy GKP states, we twirl the finite-energy state~\cite{conrad2021twirling,noh2020fault}, effectively applying incoherent displacements to the ideal GKP state via
\begin{equation}\label{eq:twirling_gkp}
    \ket{\mathcal{L}(\Delta)}\xrightarrow{\text{Twirling}} \bigotimes_{i=1}^N\mathcal{N}_{\sigma_{\rm GKP}}(\dyad{\mathcal{L}}),
\end{equation}
where the GKP noise per mode is defined as $\sigma_{\rm GKP}^2=\tanh(\Delta^2/2)$. We emphasize that the twirling here is a conceptual tool to simplify analyses of finite-energy GKP states and does not correspond to a realistic operation. In this twirling model, the lattice points of an ideal GKP state have an intrinsic jiggle with standard deviation $\sigma_{\rm GKP}$, which one can quantify by the effective squeezing in dB via $s_{\rm{GKP}}=-10\log_{10}(2\sigma_{\rm{GKP}}^2)$. This simple heuristic analysis indicates that finite squeezing presents an intrinsic noise that can ultimately constrain the performance of GKP codes. This effective ``GKP noise'' can enter in through the encoded GKP lattice as well as in stabilizer measurements that utilize ancillary GKP states. Each layer of GKP noise may effect QEC performance in a different manner. Furthermore, such noise can be a limiting factor for low levels of squeezing; see Chapter~\ref{sec:applications} where the effects of GKP noise on fault-tolerant quantum computing are assessed. High quality GKP states are thus crucial for high-fidelity quantum information processing with GKP codes.

One further dilemma with the finite-energy GKP state  $\ket{\mathcal{L}(\Delta)}$ is that it is not a $+1$ eigenstate of the ideal GKP stabilizer group $\mathcal{S}(\mathcal{L})$---unless the generators of $\mathcal{S}$ happen to commute with the envelop operator $E_{\Delta}$---which is not guaranteed. To alleviate this difficulty, one can introduce the regularized stabilizer generators~\cite{royer2020stabilization},
\begin{equation}\label{eq:envelope_stabilizer}
    S_j(\Delta)\coloneqq E_{\Delta} S_j E_{\Delta}^{-1},
\end{equation}
such that $S_j(\Delta)\ket{\mathcal{L}(\Delta)}=+1\ket{\mathcal{L}(\Delta)}\,\forall\,j$ by construction. The caveat is that $S_j(\Delta)$ is no longer a unitary operation simply because $E_{\Delta}$ does not correspond to unitary evolution. On the other hand, the envelope map $\mathcal{E}_\Delta(\bullet)\coloneqq E_{\Delta} \bullet E_{\Delta}^{-1}$ is technically a quantum channel ($\bullet$ is a placeholder for any linear operator). Thus, in principle, there exists a physical scheme to realize finite-energy stabilizer measurements. Indeed, such a scheme was introduced in Ref.~\cite{royer2020stabilization} where the authors considered coupling to an auxiliary qubit (transmon) to effectively realize the envelope map; a similar scheme was considered in Ref.~\cite{rojkov2023twoqubitStabiliz} to realize finite-energy two-qubit gates. In Chapter~\ref{sec:GKP_eng}, we discuss details regarding physical realizations of finite-energy stabilization, with a special focus on finite-energy logical operations in Chapter~\ref{sssec:GKP-control}.

\subsection{Modular variables and the Zak basis}
\label{ssec:Zak}

\QZ{In this section we discuss one final approach for the representation of GKP states and operators, called the modular variables formalism or the Zak basis~\cite{Zak_1967, Aharonov_1969, Ketterer_2016,Pantaleoni_Modular_2020,Pantaleoni_subsystem_2021,Mensen_phase_space_2021,shaw2022stabilizer, pantaleoni2023zak}}, which is useful for representing oscillator wavefunctions and operators on a restricted domain of two variables instead of an unrestricted domain of one variable. In particular, the modular wavefunction serves as an efficient basis for representing (and simulating) GKP-like states, or more generally, oscillator states that are close to periodic such as finite-energy GKP states. The formalism has been used in various other contexts in physics, such as in the analysis of spatial interference patterns and in quantum-Hall-effect literature.

The (single-mode) Zak basis has been used extensively in GKP literature, with applications in analysis of the GKP Hamiltonian (see Chapter~\ref{sec:GKP_Hamiltonian}), engineering superconducting circuits for realizing passive or active error correction (see Chapter~\ref{sec:GKP_circuits}), and an understanding of GKP error correction through the lens of a modular variable subsystem decomposition  \cite{, pantaleoni2023zak}. In this decomposition, the oscillator's Hilbert space is divided into two subsystems: one that stores logical qubit information and a second continuous-variable gauge subsystem that carries no logical information. In this subsection, we will briefly review the single-mode Zak basis and the Zak transform, using the notation and conventions from Ref.~\cite{pantaleoni2023zak}.

Given a position periodicity $a$ (in GKP, typically chosen to be the length of a stabilizer translation), the Zak kets form a complete single-mode basis and are given in the position and momentum representation by
\begin{align}
&\ket{u,v} \coloneqq \sqrt{\frac{a}{2\pi}}\sum_{m\in \mathbb{Z}} e^{iamv}\ket{u + am}_q =\sqrt{\frac{1}{a}}e^{-iuv}\sum_{m\in \mathbb{Z}} e^{-i2\pi mu/a}\ket{v + 2\pi m/a}_p.
\end{align}
They are orthonormal in the Dirac-comb sense, $\bra{u,v}\ket{u',v'} = \sum_m \delta\left(u-u' + am\right) \sum_n \delta\left(v-v' + (2\pi/a)n\right)$. The Zak kets are eigenstates of displacement operators,
\begin{align}
&e^{-ia\hat p}\ket{u,v} = e^{-iav}\ket{u,v} \qq{and} e^{i \frac{2\pi}{a} \hat q} \ket{u,v} = e^{i \frac{2\pi}{a} u}\ket{u,v},
\end{align}
with quasi-periodicity in the first variable and periodicity in the second, satisfying
\begin{align}
&e^{-it\hat{p}}\ket{u,v} = \ket{u+t,v},
&\ket{u+a,v} = e^{-iav}\ket{u,v}, \\
&e^{it\hat{q}}\ket{u,v} = e^{iut}
\ket{u,v+t},
&\ket{u,v+2\pi/a} = \ket{u,v}.
\end{align}
This leads to a restricted domain for $u$ (of width $a$) and $v$ (of width $2\pi/a$), giving a total of domain of area of $2\pi$ called Zak patch $\mathcal{P}$ with a center that we are free to choose. One choice of domain that is convenient for representing computational GKP states is $u \in [-a/4, 3a/4)$ and $v \in [-\pi/a, \pi/a)$. For example, using $a = 2\sqrt{\pi}$, the (single-mode) infinite-energy square GKP computational states are represented as $\ket{0_\square} = \ket{0,0}$ and $\ket{1_\square} = \ket{a/2,0}$. Arbitrary oscillator states can be represented in the basis of modular wavefunctions living on a torus $\ket{\psi} = \int_\mathcal{P} du dv\, \psi(u,v) \ket{u,v}$ with boundary conditions $\psi(u+a, v) = e^{iav}\psi(u,v)\,,\,\psi(u, v+2\pi/a) = \psi(u,v)$.

The power of the Zak basis is that square-integrable wavefunctions of an unbounded variable (such as position $q$) can be mapped onto quasi-periodic modular wavefunctions of two real variables with a bounded domain. This mapping is given by the Zak transform \cite{Zak_1967, Zak_1972, pantaleoni2023zak}
\begin{align}
\left( Z\psi \right) \left(u,v\right) = \sqrt{\frac{a}{2\pi}}\sum_{m\in \mathbb{Z}} e^{-iamv}\psi(u + am).
\end{align}

Finally, operators can also be described in the context of the Zak basis and modular wavefunctions. Explicitly, the position and momentum operators can be broken up into a modular part and a remainder part, 
\begin{align}
&\hat{q} = \hat{u} + a \hat{m} & \hat{p} = \hat{v} + \frac{2\pi}{a} \hat{n}
\end{align}
where $\hat{n}$ ($\hat{m}$) have integer eigenvalues (interpreted as the ``which-bin'' information), and $\hat{u} = (\hat{q} + c_q)\text{mod}\left[a\right] - c_q$ ($\hat{v} = (\hat{p} + c_p)\text{mod}\left[2\pi/a\right] - c_p$) are the modular position (momentum) operators (interpreted as the relative position or momentum within the specified bin). Here, $c_q$ ($c_p$) are constants used to center the Zak patch \cite{Ketterer_2016}. In the modular wavefunction representation these variables have a differential form,
\begin{align}
&\bra{u,v}\hat{u} = u \bra{u,v} & \bra{u,v}\hat{v} = v\bra{u,v} \\
& \bra{u,v} a\hat{m} = i\frac{\partial}{\partial v}\bra{u,v} &
\bra{u,v} \frac{2\pi}{a}\hat{n} = -\left(i \frac{\partial}{\partial u} + v\right) \bra{u,v},
\end{align}
giving a differential form for the oscillator position and momentum variables,
\begin{align}
\label{eq:zak_differential_ops}
& \bra{u,v} \hat q = \left(u + i \frac{\partial}{\partial v}\right) \bra{u,v} & \bra{u,v} \hat{p} = -i \frac{\partial}{\partial u} \bra{u,v} 
\end{align}
used to analyze the action of operators and Hamiltonians on modular wavefunctions \cite{Ganeshan_formalism}, as is done in Chapter~\ref{sec:GKP_Hamiltonian}. So far the Zak basis has only been used in the context of single-mode GKP states and GKP error correction. Also, a multimode Zak basis and modular variable approach could be useful in the
development of error correction strategies for multi-oscillator GKP codes~\cite{shaw2022stabilizer}. 

\section{Quantum Engineering with GKP codes}
\label{sec:GKP_eng}

The preceding section introduced a mathematical framework for defining and describing GKP lattice states, but it did not address practical methods for their engineering. We now bridge this gap by presenting an overview of recent advancements in quantum engineering with GKP codes. Our primary focus will be on progress made in the field of superconducting circuits, although we will also touch upon a few proposals related to optical systems. 

By \textit{quantum engineering}, we are referring to the design, development, and optimization of quantum systems and quantum control techniques to realize desired properties, such as protection against environmental noise, non-Gaussian evolution, and dissipative dynamics. We classify quantum engineering with GKP codes into two broad sections: (1) In Chapter~\ref{sec:GKP_Hamiltonian}, we discuss \textit{Hamiltonian engineering}, in which various methods---both driven (\textit{active}) and non-driven (\textit{passive})---have been proposed to realize the GKP Hamiltonian for which the GKP states are ground states; (2) in Chapter~\ref{dissipation-engineering}, we discuss \textit{dissipation engineering} for GKP codes. Dissipation engineering includes both stroboscopic methods---for which repeated interactions with an auxiliary qubit is used---and a recently proposed continuous method for direct engineering of the GKP dissipators in a driven superconducting circuit. 
Next, in Chapter~\ref{sssec:GKP-control} we review additional techniques for control of finite-energy GKP states using an auxiliary qubit. Lastly, in Chapter~\ref{sec:optical_gkp} we describe some proposals for realizing GKP codes in optical platforms, and in Chapter~\ref{sec:scaling_up}, we conclude with comments on code concatenation and next steps for scaling up towards practical quantum computing.

\subsection{GKP Hamiltonian engineering}
\label{sec:GKP_Hamiltonian}
Quantum error correction requires dissipation to correct errors using measurements or bath engineering \cite{terhal_QEC_review, lidar_brun_2013, Fowler_review}. A complimentary approach is realizing protection at the Hamiltonian level by encoding logical information in a manifold of eigenstates that is robust to changes in environmental variables coupling as local perturbations \cite{Kitaev_anyons_2003}. Such \textit{protected} qubits often gain error suppression through a combination of related factors, including a degenerate gound state manifold with a large energy gap to higher excited states, wavefunction delocalization leading to flat energy bands, and wavefunctions with explicit disjoint support so that localized noise only weakly affects the qubit's eigenstates \cite{Doucot_2012, Gyenis_protected_2021}.

Disjoint wavefunction support is also at the heart of realizing a protected GKP codespace. As a result, many protected superconducting qubits have wavefunctions resembling GKP states \cite{Doucot_2012, Gyenis_protected_2021, Groszkowski_2018, Paolo_2019, Gyenis_2021, Manucharyan_fluxonium, Pechenezhskiy_2020, Smith_2020, Smith_2022}. Additionally, a number of circuits have been proposed to engineer the GKP Hamiltonian directly \cite{Rymarz_2021, Le_2019} or through Floquet engineering \cite{Liang_2018, conrad2021twirling, sellem2023gkp, kolesnikow_2023, guo2023engineering, wang2023quantum}. We discuss these proposals in Chapter~\ref{sec:GKP_circuits}. 

In this section, we focus on the single-mode GKP Hamiltonian and proposals for its direct realization in superconducting circuits. Later, in Chapter~\ref{sec:qubit-dissipation}, we discuss proposals and experimental realizations of dissipative QEC with the GKP code.

\subsubsection{The GKP Hamiltonian}

As we defined in Sec.~\ref{sec:qho}, $\hat{q} = \left(a + a^\dag\right)/\sqrt{2}$ and $\hat{p} = -i\left(a - a^\dag\right)/\sqrt{2}$ are the position and momentum operators of a quantum harmonic oscillator with $\left[a, a^\dag\right] = 1$ (unit is chosen as $\hbar = 1$).  The infinite-energy GKP Hamiltonian for a single-mode rectangular-lattice can be written as \cite{gkp2001}
\begin{equation}
\label{eq:H_GKP}
H_\text{GKP} = -E_p\cos\left(\frac{\sqrt{2\pi d}}{\eta}\, \hat{p} \right) - E_q \cos\left( \eta \sqrt{2\pi d}\, \hat{q}\right).
\end{equation}

$H_\text{GKP}$ has a $d$-fold degenerate ground state manifold encoding the infinite-energy rectangular $d$-dimensional GKP code where $\eta$ sets the lattice aspect ratio in phase space; here $d$ is the code dimension (introduced in Chapter~\ref{sec:stabilizer_lattice}) of the single-mode GKP code.\footnote{In this section, we often restrict to the qubit ($d=2$ case) and use the nomenclature ``2D GKP'' state, Hamiltonian etc. to refer to a \emph{single-mode} GKP with $d=2$ code dimension.} Analogous Hamiltonians exist for other lattices, such as the hexagonal lattice. The spectrum of $H_\text{GKP}$ is continuous with eigenstates given by Zak basis states \cite{Zak_1967,Le_2019, Rymarz_2021, pantaleoni2023zak, Ganeshan_formalism, Liang_2018} (see Chapter~\ref{ssec:Zak}) and $E_q/E_p$ can be used to tune the relative dispersion along the $q$ and $p$ directions. We note that the GKP Hamiltonian is closely related to Harper's equation \cite{Harper_1955}, which is a tight binding model for motional dynamics of noninteracting electrons in the presence of a 2D periodic potential and uniform magnetic field, as discussed more below \cite{gkp2001,Liang_2018, Rymarz_2021, Hofstadter}.

The usual argument behind passive QEC is to construct a Hamiltonian for which the codespace is the degenerate ground state manifold, protected by an energy gap \cite{Kitaev_anyons_2003}.
However, passive QEC for continuous-variable systems is different than in multi-qubit systems such as the toric or surface code. In both cases, the protection Hamiltonian is of the form $H=-\sum_k E_k S_k$, where $\left\{S_k\right\}$ are the stabilizer generators and $E_k$ are the energy scales which should be positive and very large compared with typical couplings to the bath. For multi-qubit codes, the discrete spectrum of Pauli stabilizers gives rise to a spectral gap. For the infinite-energy GKP code, the spectrum of displacement stabilizers is continuous, rendering $H_\text{GKP}$ gapless.
As a result, the perturbation theory argument relying on a gap, such that local perturbations of the Hamiltonain give rise to small variations in energy, does not apply to $H_\text{GKP}$ directly \cite{Doucot_2012, Rymarz_2021}.

Depending on the physical realization of $H_\text{GKP}$, it is unclear if a continuous spectrum is really an issue, as proper thermalization to a cold bath could still prevent uncorrectable errors; this is highly context dependent and an active topic of research. Nonetheless, a gapped spectrum can be engineered by including a weak confinement potential, giving rise to the finite-energy GKP Hamiltonian $H_{\text{GKP},\Delta}$ \cite{gkp2001,Doucot_2012, Rymarz_2021, Ganeshan_formalism}.
The most straightforward is a harmonic confinement\footnote{A general harmonic confinement of $H = \hat{p}^2/2m + k\hat{q}^2/2$ can cast into the form of Eq.~\ref{eq:finite_energy_H} by scaling position and momentum as $\hat{q} \rightarrow \hat{q}/\sqrt{Z}, \hat{p} \rightarrow \hat{p}\sqrt{Z}$, where $Z = 1/\sqrt{km}$ is the impedance of the harmonic confinement},
\begin{equation}
\label{eq:finite_energy_H}
H_{\text{GKP},\Delta} = \frac{\omega_0}{2}\left(\hat{p}^2 + \hat{q}^2 \right) -E_p\cos\left(\frac{\sqrt{2\pi d}}{\eta}\, \hat{p} \right) - E_q \cos\left( \eta \sqrt{2\pi d}\, \hat{q}\right). 
\end{equation}
The $\Delta$ notation here indicates the quasi-degenerate ground states are now finite-energy GKP states (see, for instance, Chapter~\ref{sec:gkp_lattice} for an introduction to finite-energy GKP). The ground state manifold of the $H_{\text{GKP},\Delta}$ serves as an alternative definition to the single-mode finite-energy code manifold. We anticipate this definition can be extended to multimode GKP encodings.

To analyze the Hamiltonian, the Zak basis introduced in Chapter~\ref{ssec:Zak} can be used \cite{Ganeshan_formalism}. \QZ{In particular, using a Zak basis period of $a = \sqrt{2\pi d}/\eta$} and a Zak patch\footnote{Any patch of of width $a$ in $u$ and $2\pi/a$ in $v$ will do, however we find this patch to be the most convenient for representing GKP qudit states} of $\mathcal{P} = u \in [-a/2d,a(2d-1)/2d),\, v \in [-\pi/a,\pi/a )$, the Hamiltonian can be written as a differential operator in the Zak-wavefunction basis $H\ket{\psi} = \int_\mathcal{P} du dv\,  H(u,v)\psi(u,v)\ket{u,v}$ using Eq. \ref{eq:zak_differential_ops} as
\begin{equation}
\label{eq:H_GKP_ZAK}
H_{\text{GKP},\Delta}(u,v) = \frac{\omega_0}{2}\left(p_u^2 + \left(u - p_v\right)^2\right) - E_p\cos\left(\frac{\sqrt{2\pi d}}{\eta}\, v \right) - E_q \cos\left( \eta \sqrt{2\pi d}\, u\right)
\end{equation}
where we have defined effective momenta in the $u,v$ directions as $p_u = -i \frac{\partial}{\partial u}$ and $p_v = -i \frac{\partial}{\partial v}$ \cite{Ganeshan_formalism}. From this, the finite-energy GKP Hamiltonian $H_{\text{GKP,}\Delta}$ can be thought of as describing a particle of mass $1/\omega_0$ on a torus parameterized by $u,v$, coupled to a vector potential and two cosine potentials.

In the limit of weak harmonic confinement, $E_q, E_p \gg \omega_0$, the tunneling between cosine minima is suppressed. In this regime, $H_{\text{GKP,}\Delta}$ has a set of $d$ nearly degenerate ground states, each localized in different minima at positions $\left(u, v \right) = \left(an/d,0\right)$ with $n = 0,1,...,d-1$. The cosine terms can be expanded to quadratic order around a minima, leading to a local effective Hamiltonian that is well approximated by two uncoupled harmonic oscillators with frequencies $\omega_{u} = \left(\eta \sqrt{2\pi d}\right)\left(\sqrt{\omega_0 E_q}\right)$ and $\omega_{v} = \left(\sqrt{2\pi d}/\eta\right)\left(\sqrt{\omega_0 E_p}\right)$ and a spectral gap between the quasi-degenerate ground state manifold and the first excited states of approximately $E \approx \text{min}\left(\omega_u, \omega_v\right)$. 

Expanding the cosine potentials in Eq. \ref{eq:H_GKP_ZAK} to quadratic order can also be used to estimate the GKP squeezing parameters. GKP states can have different squeezings along the position and momentum directions in cases where $\eta \neq 1$ or $E_q \neq E_p$, and these can be estimated from the zero-point motion associated with the effective oscillator impedances of $u$ and $v$, \QZ{given by $Z_u = \left(\eta \sqrt{2\pi d}\right)^{-1} \sqrt{\omega_0/E_q}$ and $Z_v = \left( \sqrt{2\pi d}/\eta\right)^{-1} \sqrt{\omega_0/E_p}$.} The zero-point motion leads to GKP squeezing parameters of $\Delta_q \approx 1/\sqrt{Z_u}$ and $\Delta_p \approx 1/\sqrt{Z_p}$. Other quantities, such as the tunnel splitting of the ground state manifold, can be estimated by applying different perturbation techniques to the Zak-basis Hamiltonian, see for example Refs.~\cite{Rymarz_2021,Ganeshan_formalism}.

\begin{figure}[t]
    \centering
    \includegraphics[width=\linewidth]{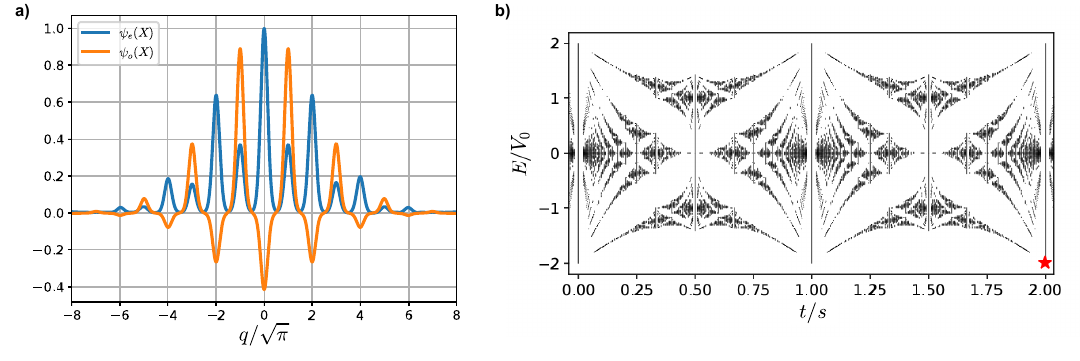}
    \caption{ a) Position wavefunctions for quasi-degenerate lowest energy states (Eq.~\eqref{eq:ground_states}) of the 2-D GKP Hamiltonian with a weak Harmonic confinement of $\Delta = 0.25$. b) The spectrum of $H_\text{LLL}$ (Eq.~\eqref{eq:H_LLL}) plotted as a function of inverse magnetic flux ratio $t/s$, resulting in Hofstadter’s butterfly. The 2D GKP manifold corresponding to $s/t = 1/2$ is marked by the red star. Figures reproduced with permission from \cite{Rymarz_2021}.}
    \label{fig:GKP_experiment}
\end{figure}

In the case of a square lattice with equal energy scales in $q$ and $p$ ($\eta = 1, E_q = E_p$), the ground states are well approximated by the finite-energy Hadamard eigenstates
\begin{align}
\label{eq:ground_states}
&\psi_{H+}\left(q\right) \approx \cos\left(\frac{\pi}{8}\right)\psi_{\Delta,0}(q) + \sin\left(\pi/8\right) \psi_{\Delta,1}\left(q\right), \\
&\psi_{H-}\left(q\right) \approx -\sin\left(\frac{\pi}{8}\right)\psi_{\Delta,0}(q) + \cos\left(\pi/8\right) \psi_{\Delta,1}\left(q\right),
\end{align}
where $\psi_{\Delta,{_0^1}}(q)$ are the wavefunctions for the finite-energy $0$ and $1$ logical states. This can be seen from the Hamiltonian's invariance with respect to the Fourier transform (a $\pi/2$ rotation in phase space). As an example from Ref.~\cite{Rymarz_2021}, the numerically obtained lowest-energy eigenstates are plotted in Figure \ref{fig:GKP_experiment}a. Modifying the impedance of the confinement potential can break this symmetry and lead to other ground state manifolds. Varying from one confinement potential to another could be used to engineer transformations of the finite-energy GKP states \cite{Doucot_2012}.

The GKP Hamiltonian can also arise as the low-energy Hamiltonian of a single electron confined to a two-dimensional plane with a periodic potential and a perpendicular magnetic field \cite{gkp2001, Rymarz_2021}. Although this system is likely not practical to implement, requiring unrealistically large magnetic fields, it is useful for a theoretical understanding of the GKP code. In Ref.~\cite{Rymarz_2021}, the authors show that the effective low energy Hamiltonian [the lowest Landau level (LLL) Hamiltonian] for this system in the weak Landau-level coupling limit is
\begin{equation}
\label{eq:H_LLL}
H_\text{LLL} = - V_0 \left[ \cos\left(2\sqrt{\pi}\hat{q}\right) + \cos\left(\frac{t}{s} \sqrt{\pi} \hat{p}\right)\right].
\end{equation}
Here $\left\{t,s\right\}$ are coprime natural numbers describing the rational multiple of flux quantum contained in a loop enclosed by magnetic translation operators, $\Phi = \left(s/t\right) \Phi_0$ where $\Phi_0$ is the (non-superconducting) flux quantum. The eigenvalue equation associated with $H_\text{LLL}$ is the Harper equation, resulting in an energy spectrum in the form of a Hofstadter butterfly \cite{Hofstadter}. We plot the spectrum  in Figure \ref{fig:GKP_experiment}b as a function of $t/s$ \cite{Liang_2018, Rymarz_2021}. This spectrum has $s$ bands that are $t$-fold degenerate, and the two-dimensional GKP code space corresponds to the ground state manifold at $s/t=1/2$ (equivalently $t/s=2$) as shown by the red star in the figure. In Ref.~\cite{Liang_2018}, the authors discuss this model in the context of topological order and introduce the notion of a phase-space interaction potential, connecting to the dynamics of some many-body systems.

\subsubsection{Proposals for realizing the GKP Hamiltonian in superconducting circuits}

\begin{figure}[ht]
    \centering
    \includegraphics[width=\linewidth]{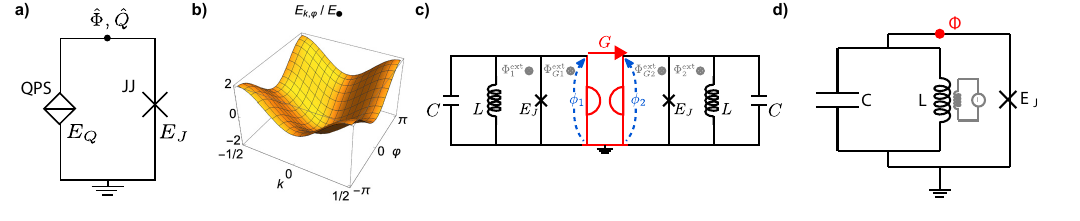}
    \caption{Realizing the GKP Hamiltonian in cQED. a) Dualmon circuit consisting of a quantum phase slip (QPS) element and Josephson junction (JJ) in parallel \cite{Le_2019}. b) Spectrum of the dualmon Hamiltonian \cite{Le_2019}. c) Circuit for realizing a hardware-encoded confined GKP Hamiltonian, consisting of two fluxonium modes coupled by a gyrator \cite{Rymarz_2021}. d) An oscillator with impedance $Z=2R_Q$ is periodically displaced via an inductive coupler. When coupled to a junction, GKP states are stabilized through the effective Floquet Hamiltonian \cite{conrad2021twirling}.} 
    \label{fig:GKP_circuits}
\end{figure}

\label{sec:GKP_circuits}

There are a few proposals for realizing $H_\text{GKP}$ in superconducting circuits, with a selection shown in Figure \ref{fig:GKP_circuits}. The most direct implementation of $H_\text{GKP}$ was analyzed in Ref.~\cite{Le_2019}, where the authors proposed the \textit{dualmon} circuit consisting of a \QZ{Josephson} junction (JJ) in parallel with a quantum phase slip (QPS) element as shown in Figure \ref{fig:GKP_circuits}a. The JJ is a superconducting circuit element in which particles of charge $2e$ coherently tunnel across an insulating barrier between two superconducting films with a current ($I_J$) and phase ($\Phi$) relation $I_J = I_C \sin\left(2\pi \Phi / \Phi_{0,s}\right)$; here, $\Phi = \int_{-\infty}^{\tau}V(\tau)d\tau$ is the flux linked by the JJ, $V(\tau)$ is the voltage across the junction, \QZ{and $\Phi_{0,s} = 2e/h$ is the superconducting magnetic flux quantum}.
The QPS is the dual element with a constitutive voltage ($V_Q$) charge ($Q$) relation $V_Q = V_c \sin\left(2\pi Q/(2e)\right)$, where $Q$ is the charge that has flowed through the QPS \cite{Mooij_2005, Mooij_2006,Astafiev_2012,Graaf_2018}.\footnote{For comprehensive introductions to superconducting circuits, we refer the reader the following reviews~\cite{Vool_2017,Krantz_2019,blais2021cqedRMP}.} The quantized Hamiltonian of this circuit is given by
\begin{equation}
H_\text{dualmon} = -E_Q \cos\left(2\pi \hat{n}\right) - E_J \cos\left(\hat{\phi}\right)
\end{equation}
where $\hat{\phi} = 2\pi\hat{\Phi}/\Phi_{0,s}$ is the reduced phase across the junction and $\hat{n}=Q/2e$ is the conjugate number of Cooper pairs that have tunneled across the junction, such that $\comm*{\hat{\phi}}{\hat{n}} = i$. The dualmon Hamiltonian corresponds to the $\left\{d=1, \eta = 1/\sqrt{2\pi}\right\}$ GKP Hamiltonian in Eq.~\eqref{eq:H_GKP}, indicating that it has a single ground state given by the canonical ($d=1$) square GKP state $\ket{\square}$; see Chapter~\ref{sec:GKP_simple_part} for an introduction. The circuit could be promoted to the $d=2$ GKP Hamiltonian---so that the degenerate ground state manifold encodes the GKP qubit (again see Chapter~\ref{sec:GKP_simple_part} for an introduction)---by replacing the Josephson junction with a $\pi$-periodic junction, also called the $\cos(2\phi)$ circuit element, which allows coherent tunneling of pairs of Cooper-pairs \cite{Blatter_2001, Gladchenko_2009, Doucot_2012, Le_2019, Smith_2020}. We note that $\hat{\phi}$ is treated here as an unbounded variable instead of as a compact variable. For a discussion of such subtleties, we refer the reader to Ref.~\cite{Devoret_phase_difference}.

The eigenstates of the dualmon $\ket{k, \varphi}$ are Zak states (see Sec. \ref{ssec:Zak}) and characterized by Bloch quantum numbers in the rectangular Zak patch $k \in (-1/2, 1/2]$ and $\varphi \in (-\pi, \pi]$.
The spectrum is given by $E_{k, \varphi} = -E_Q \cos\left(2\pi k\right) - E_J \cos\left(\varphi\right)$ and plotted in Figure \ref{fig:GKP_circuits}b. The spectrum has four critical points, $(k, \varphi) = \left\{ (0,0), (0,\pi), (1/2,0), (1/2,\pi)\right\}$ where $\Delta E_{k,\varphi} = \left(\partial_k E_{k,\varphi}, \partial_\varphi E_{k,\varphi}\right)=0$. As a result, the sensitivity to charge and flux noise vanishes at these charge- \textit{and} flux- insensitive points to linear order. Additionally, small-amplitude flux and charge noise will commute with the Hamiltonian and does not cause transitions between eigenstates to first order. The authors in Ref.~\cite{Le_2019} proposed using the states $\ket{0,0}$ and $\ket{0,\pi}$ as a logical qubit, corresponding to the canonical GKP states with $\pm 1$ eigenvalues of the canonical GKP stabilizers. By introducing realistic circuit elements into the dualmon circuit, including a linear inductance in series with the QPS element and capacitance shunting the junction, a second high-frequency oscillator mode is added, and the dualmon energy dispersion is found as the effective low-energy band of the circuit when projected onto the oscillator's ground state. Importantly, the first excited band could facilitate addressability of the logical states.

In Ref.~\cite{Rymarz_2021}, the authors proposed a different superconducting circuit to realize the GKP Hamiltonian, shown in Figure \ref{fig:GKP_circuits}c. The circuit consists of two fluxonium modes coupled by a gyrator (shown in red). The gyrator is a two-port non-reciprocal circuit element with the current voltage relations
\begin{equation}
\begin{pmatrix}
    I_1\\
    I_2
\end{pmatrix} 
= 
\begin{pmatrix}
    0 & -G\\
    G & 0
\end{pmatrix}
\begin{pmatrix}
    V_1\\
    V_2
\end{pmatrix},
\end{equation}
where $G$ is the gyration conductance \cite{Hogan_1952}. The gyrator can be included in the circuit Lagrangian as $\mathcal{L}_G = \left(G/2\right) \phi_1 \dot{\phi}_2 - \dot{\phi}_1 \phi_2$ where $\phi_i = \int_{-\infty}^t V_i(\tau) d\tau$ are the node fluxes---which is similar to a homogeneous magnetic field of strength $B=G/e$ in the $\phi_1 \phi_2$ plane. As a result, the circuit Hamiltonian can be mapped exactly to the Hamiltonian of an electron in a periodic potential with a perpendicular magnetic field, as discussed in Chapter~\ref{sec:GKP_Hamiltonian}, for which the effective low-energy Hamiltonian takes the form of the LLL Hamiltonian in Eq.~\eqref{eq:H_LLL}. The exact mapping of the magnetic field to the gyration conductance is $s/t = G/G_0$ where $G_0 = (2e)^2/h$ is the superconducting conductance quantum. As a result, to obtain the 2D GKP Hamiltonian with $s/t = 1/2$, we require $G=G_0/2$. This value of $G\approx 1/\left(\SI{13}{kOhm}\right)$ is unrealistic for superconducting gyrators \cite{Rosenthal_2017, Chapman_2017, Lecocq_2017}, however it could be easily reached using quantum (anomalous) Hall effect devices \cite{Viola_2014, Bosco_2017,Bosco_2017_2,Mahoney_2017}.

Finally, it was shown in Refs.~\cite{conrad2021twirling},\cite{kolesnikow_2023}, and \cite{sellem2023gkp} that the GKP Hamiltonian can emerge as the effective Floquet Hamiltonian of a driven superconducting circuit. An example from Ref.~\cite{conrad2021twirling} is shown in Figure \ref{fig:GKP_circuits}, consisting of a high-impedance quantum harmonic oscillator shunted by a Josephson junction. The oscillator must be tuned to have impedance $Z = 2R_Q$, where $R_Q = \frac{h}{4e^2} \approx \SI{6.5}{kOhm}$ is the resistance quantum, such that the circuit's Hamiltonian (in the characteristic function representation) is well supported on points corresponding to the 2D GKP stabilizers (i.e., the GKP qubit stabilizers of Eq.~\eqref{eq:computational_stabilizers}). By periodically displacing the oscillator corresponding to multiples of logical displacements in a ``bang-bang'' approach, the dynamical decoupling sequence approximately filters out Hamiltonian terms that are not close to stabilizer shifts, and the effective Floquet Hamiltonian is the 2D GKP Hamiltonian \cite{conrad2021twirling}. 

A similar GKP-Floquet scheme was introduced in Refs.~\cite{kolesnikow_2023} and \cite{sellem2023gkp}, consisting of a high-impedance harmonic oscillator coupled to a switchable Josephson element, which could be implemented by a flux-driven SQUID. The switchable junction allows controllable tunneling of a charge $2e$ which can displace the oscillator. By tunneling every quarter period of the oscillator's frequency, an effective cosine potential can be engineered in both the position and momentum. As shown in Ref.~\cite{sellem2023gkp}, as long as the oscillator's impedance is greater than or equal to twice the resistance quantum, the driving can be tuned so that an effective Hamiltonian for the 2D GKP code can be realized by using a diamond-shaped lattice in phase space. These methods are superconducting circuit realizations of the \textit{kicked Harper model}, a special case of the \textit{kicked harmonic oscillator} (see Ref.~\cite{Liang_2018} and references therein).

Without dissipation, the Floquet approach can be used for preparing GKP states, but they do not stabilize GKP states against noise, for which colored dissipation is likely needed. In superconducting circuits, the thermalization of effective Floquet Hamiltonians is an active topic of research. The extension of this method to full dissipative QEC is discussed in Chapter~\ref{sec:dissipation_continuous}. We note that other similar Floquet engineering proposals have been introduced for realizing protected circuits and Hamiltonians, such as the \textit{Kapitzonium} (a Floquet $0-\pi$ qubit) \cite{wang2023quantum}, or the Floquet engineering of other lattices in phase space \cite{Guo_2022, guo2023engineering}. 

Finally, other protected superconducting circuits realize ground state manifolds of GKP states (or GKP-like states) since they are designed using wavefunction delocalization and disjoint support for protection against environmental noise \cite{Doucot_2012, Gyenis_protected_2021}. For example, the $0-\pi$ qubit \cite{BKP_paper, Groszkowski_2018,Paolo_2019,Gyenis_2021} combines two widely used protected qubits, a high-impedance \textit{fluxonium} mode \cite{Manucharyan_fluxonium, Pechenezhskiy_2020} and a \textit{transmon} mode \cite{Koch_2007} to realize a quasi-degenerate ground state manifold that is spanned by a pair of two-mode GKP states \cite{conrad2022lattice}. Coupling the $0-\pi$ qubit to an oscillator leads to encoded GKP states of the oscillator that could be used to realize protected gates \cite{BKP_paper}. Another example is the $\cos(2\varphi)$ qubit \cite{Doucot_2012, larsen2021, Smith_2020, Smith_2022}, with GKP-like states as the effective ground state like manifold. Additionally, GKP states can arise as the ground state of the RF-SQUID\footnote{RF-SQUID stands for radio-frequency superconducting quantum interference device.} in special parameter regimes \cite{Devoret_phase_difference} where effective phase-slip dynamics can emerge. Realizing superconducting circuits with emergent dynamics, such as coherent phase slip dynamics or pair-wise Cooper pair tunneling, is an active topic of research.

\subsection{GKP dissipation engineering}\label{dissipation-engineering}
    Quantum error correction through stabilizer measurements could be thought of as \textit{dissipation engineering}. An alternative to measuring the stabilizers of the quantum codes is engineering a system-bath interaction,
    \begin{equation}
        H=\sqrt{\Gamma} (\hat{d}\hat{b}(t)^\dagger+\hat{d}^\dagger \hat{b}(t)),
    \end{equation}
    which relaxes the system to states satisfying $\hat{d}\ket{\psi}=0$, where $\hat{d}$ is known as the dissipator. 
    Any excitation in the system due to $\hat{d}^\dagger$ are transferred to the zero-temperature bath, autonomously cooling the system to the desired state $\ket{\psi}$. A Markovian model of dissipation is realized by the above Hamiltonian where the field operators (bath) obey $[\hat{b}(t),\hat{b}(t')^\dagger]=\delta(t-t')$, with $\delta(t)$ being the Dirac-delta distribution. 
    
    There are multiple ways to design dissipators into the codespace. In Ref.~\cite{royer2020stabilization}, the authors defined dissipators to the GKP codespace as the natural logarithm of the stabilizers $S$, since
    $\ln{S}\ket{\psi}=0$.\footnote{We note an alternative definition of the dissipators was introduced in Ref.~\cite{sellem2023gkp}, which we discuss in Chapter~\ref{sec:dissipation_continuous}.} Thus, in order to find equations for GKP dissipators, we analyze the finite-energy GKP stabilizers.  As discussed in Chapter~\ref{sec:finite_GKP} [see, e.g., Eq.~\eqref{eq:envelope_stabilizer}], the finite-energy GKP stabilizers can be obtained by the following deformation of an arbitrary ideal GKP stabilizers $S$,
\begin{align}
    S({\Delta})&=E_\Delta S E_\Delta^{-1}\nonumber\\
    &=E_{\Delta} e^{i\hat{v}} E_\Delta^{-1}\nonumber\\
&=e^{i\left[\cosh(\Delta^2)\hat{v}+i\sinh(\Delta^2)\hat{v}_{\perp}\right]}
\end{align}
where $\hat{v}=\alpha\hat{q}+\beta\hat{p}$ and $\hat{v}_{\perp}=\alpha\hat{p}-\beta\hat{q}$.
It can be easily checked that,
\begin{align}
    [E_\Delta S_iE_\Delta^{-1},E_\Delta S_jE_\Delta^{-1}]=E_\Delta[S_i,S_j]E_\Delta^{-1}= 0,
\end{align}
and thus, the new stabilizers and logical operators commute in the same way as the ideal stabilizers and logical operators, satisfying the minimum requirements for stabilizer-based error correction.\footnote{The non-hermiticity of $S_x, S_p$ or $\hat{d}_x, \hat{d}_p$ is not a problem here because we do not intend to measure these operators. Instead we want to build them into the dissipation Hamiltonian which will be Hermitian. In the next section, we discuss the engineering of this dissipation using an auxiliary qubit.} The dissipator corresponding to each stabilizer subspace becomes $\hat{d}=-\frac{i}{m\sqrt{2\cosh{\Delta^2}\sinh{\Delta^2}}}\ln S$ where $\ln S=(v_{[m/2\cosh{\Delta^2}]}/\sqrt{\tanh{\Delta^2}}+iv_\perp\sqrt{\tanh{\Delta^2}})/\sqrt{2}$. Here $v_{[l]}$ denotes symmetric version of the modular quadrature $v\quad mod\quad l$ also known as the Zak-basis described in Ch.\ref{ssec:Zak}. These modular quadratures are obtained from the multi-valued complex logarithm of the stabilizers $S_x,S_p$ such that $v_{[l]}\in(-l/2,l/2]$.

\begin{figure}
    \centering
    \includegraphics[width=\linewidth]{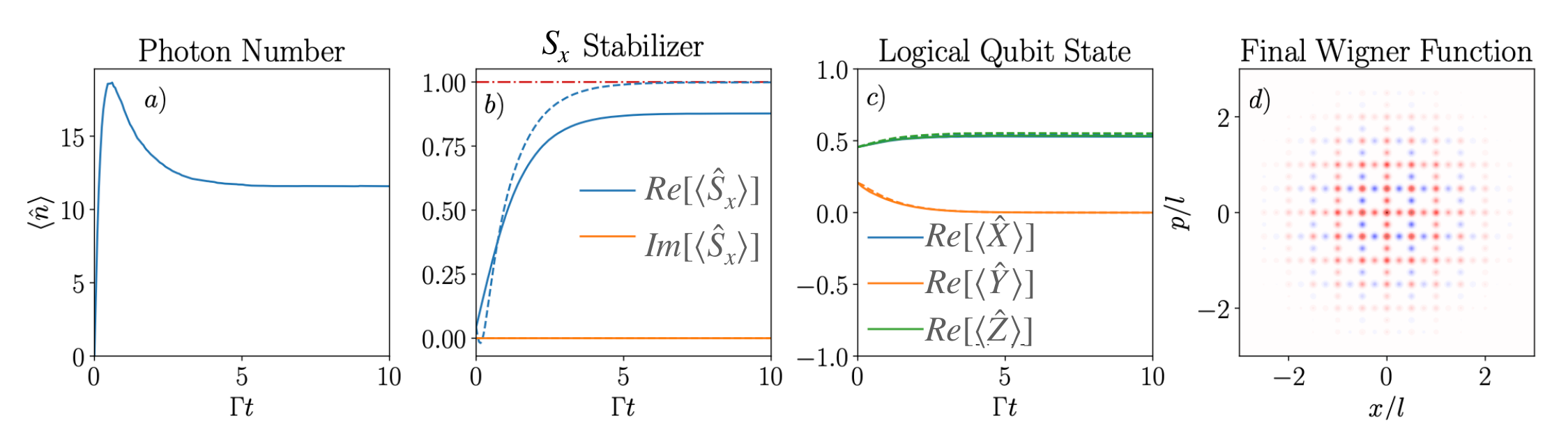}
    \caption{Evolution of system under the master equation given by Eq.~\eqref{ME} starting from vacuum with $\Gamma_1=\Gamma_2=\Gamma$ and $\Delta=0.2$. As a function of time, the figure shows, a) Excitation number of the system, b) Expectation value of the finite-energy stabilizer $S_x$ and, c) Real parts of the logical Pauli operators of the finite energy code. Here, $l=2\sqrt{\pi}$ is the lattice constant for square GKP codes. (d) Shows the Wigner function at $\Gamma t=10$. Figure adapted from~\cite{royer2020stabilization}}
    \label{fig:GKP_evolution}
\end{figure}

Focusing the discussion specifically to the single-mode square GKP states, we see that the two stabilizers generators and corresponding dissipators of square GKP code stabilizers, using the approximations $\cosh(\Delta^2)\simeq 1$ and $\sinh(\Delta^2)\simeq \Delta^2$, are given by,
\begin{align}
    S_{x}&=e^{i2\sqrt{\pi}(\hat{q}+i\Delta^2\hat{p})}\implies
    \hat{d}_x=(\hat{q}_{[\sqrt{\pi}]}/\Delta+i\hat{p}\Delta)/\sqrt{2},\\
    S_{p}&=e^{-i2\sqrt{\pi}(\hat{p}-i\Delta^2\hat{q})}\implies
    \hat{d}_p=-(\hat{p}_{[\sqrt{\pi}]}/\Delta-i\hat{q}\Delta)/\sqrt{2}.
\end{align}
Figure~\ref{fig:GKP_evolution} shows the evolution of the system to the GKP codespace generated by the master equation (see Eq.~\eqref{eq:master_eq}), 
\begin{equation}
    \dot\rho=\Gamma_1\mathbb{D}[\hat{d}_{x}]\rho+\Gamma_2\mathbb{D}[\hat{d}_{p}]\rho,\label{ME} 
\end{equation}
where $\mathbb{D}[O]\rho=O\rho O^\dagger-\{O^\dagger O,\rho\}/2$ is the standard dissipation superoperator; see Chapter~\ref{sec:channel_dynamics}. The photon number plot in Fig.~\ref{fig:GKP_evolution}a shows an instability in choosing these dissipators, as seen by the bump in initial photon number. The instabilties are taken care of by an alternative choice of dissipator (see Ref.~\cite{sellem2023gkp}) which uses $\hat{d}=S-I$. We discuss these alternative dissipators further in section \ref{sec:dissipation_continuous}.

\subsubsection{Engineered dissipation using an auxiliary qubit }\label{sec:qubit-dissipation}

One way to realize the non-local dissipators introduced just above is to use an auxiliary qubit coupled to the oscillator as a means for dissipation engineering. By preparing the auxiliary qubit in a known state, entangling the qubit and oscillator via a unitary operation, then resetting the auxiliary qubit, an effective dissipation can be realized. This method is sometimes called \textit{stroboscopic} dissipation engineering. 

\begin{figure}
    \centering
    \includegraphics[width=\linewidth]{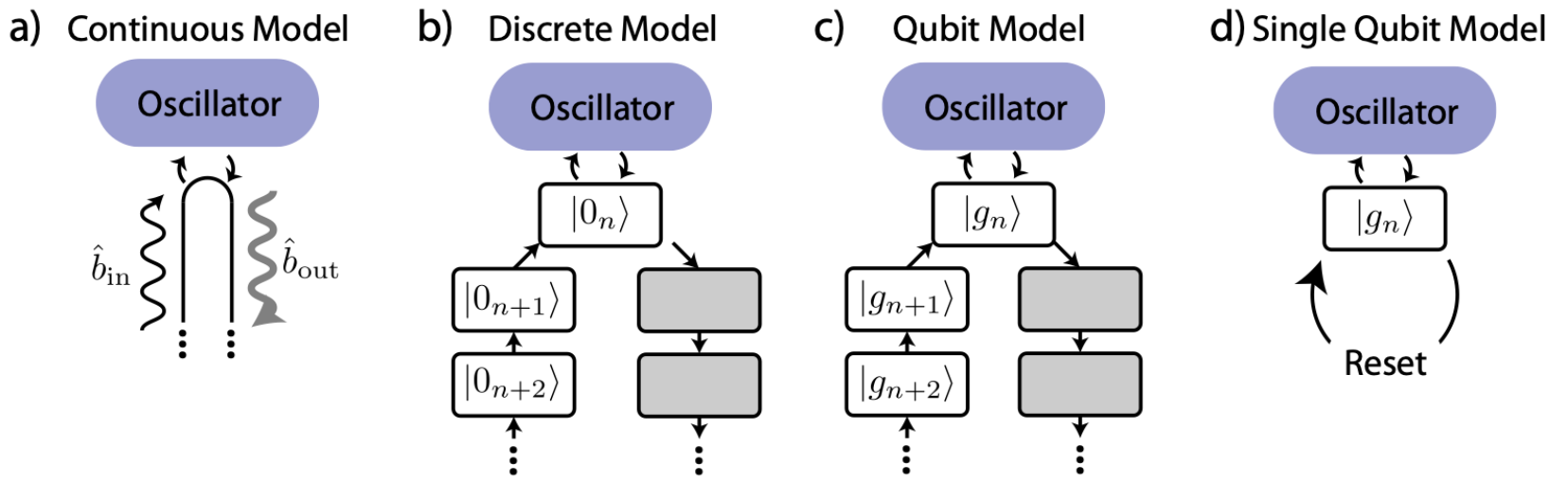}
    \caption{Cartoon of the discretization of the dissipation procedure. a) Continuous dissipation with an arbitrary input field and an output field carrying away excitations of the GKP state. The combined action of $b_{in}$ and $b_{out}$ can be envisioned as a conveyer belt carrying excess entropy of the oscillator. b) Discretization of the bath mode into slices of $\delta t$. c) The bath modes can be replaced by qubits in the limit that excitations $\propto\Gamma t\rightarrow 1$ is smaller than 1. d) Sunce the qubits are not re-used post this interaction, the same qubit can resetted to the state $\ket{g}$ and used over and over again in the same interaction with the oscillator. Figure adapted from~\cite{royer2020stabilization}}
    \label{fig:qubit-dissipation}
\end{figure}

As shown in Ref.~\cite{royer2020stabilization}, the continuous evolution under the Hamiltonian interaction, 
\begin{equation}
    H(t)=\sqrt{\Gamma} (\hat{d}\hat{b}^\dagger_t+\hat{d}^\dagger \hat{b}_t)
\end{equation}
can be discretized as if the system interacts with a different bath at every time step t, i.e.
\begin{align}
        U(t,t_0)&=\mathcal{T}e^{-i\int_{t_0}^t d\tau H(\tau) }\\
        &\approx\prod_{n=0}^T e^{-i\sqrt{\Gamma \delta t} (\hat{d}\hat{b}^\dagger_n+\hat{d}^\dagger \hat{b}_n) }\\
        &=\prod_{n=0}^T U_n,
\end{align}
where $t-t_0=T\delta t$ and $T\in \mathbb{Z}$. In the limit $\delta t\rightarrow 0$, we approach the continuous model. The excitation number, proportional to $\Gamma\delta t$, as shown in Ref.~\cite{royer2020stabilization}, needs to be small enough such that the $n$th bath mode contains less than one excitation. In this case, the bath mode can be realized using a qubit such that $\hat{b}_n\rightarrow \frac{\hat{\sigma}_{x,n}+i\hat{\sigma}_{y,n}}{2}$, where $\hat{\sigma}_{x,n},\hat{\sigma}_{y,n},\hat{\sigma}_{z,n}$ denote the Pauli matrices of $n$th qubit mode. The commutation relation between the bath operators $[\hat{b}_n,\hat{b}^\dagger_n]= 1$ is transformed as $\frac{1}{4}[\hat{\sigma}_{x,n}+i\hat{\sigma}_{y,n}, \hat{\sigma}_{x,n}-i\hat{\sigma}_{y,n}]=\hat{\sigma}_{z,n}$. For weakly populated qubits $\langle \hat{\sigma}_{z,n}\rangle\approx 1$, we retrieve the original commutation relation. In this qubit model, the time evolution is replaced by, 
\begin{equation}
    U(t,t_0)=\prod_{n=0}^Te^{-i\sqrt{\frac{\Gamma\delta t}{2\tanh(\Delta^2)}}\left(\hat{v}_{[\sqrt{\pi}]}\hat{\sigma}_{x,n}+\hat{v}_{\perp}\hat{\sigma}_{y,n}\tanh(\Delta^2)\right)}.
\end{equation}
Here, the qubits extract entropy from the oscillator and are left unused after. In other words, the ensemble of qubits can be replaced by a single qubit being reset after each time step, i.e.
\begin{align}
    U(t,t_0)&=\prod_{n=0}^Te^{-i\sqrt{\frac{\Gamma\delta t}{2\tanh(\Delta^2)}}\left(\hat{v}_{[\sqrt{\pi}]}\hat{\sigma}_x+\hat{v}_{\perp}\hat{\sigma}_y\tanh(\Delta^2)\right)}=\prod_{n=0}^T U_{\rm target}\label{eq:dissipation}
\end{align}
The correspondence between continuous and discrete models is depicted in Fig.~\ref{fig:qubit-dissipation}.

\begin{figure}
     \centering
     \includegraphics[width=\linewidth]{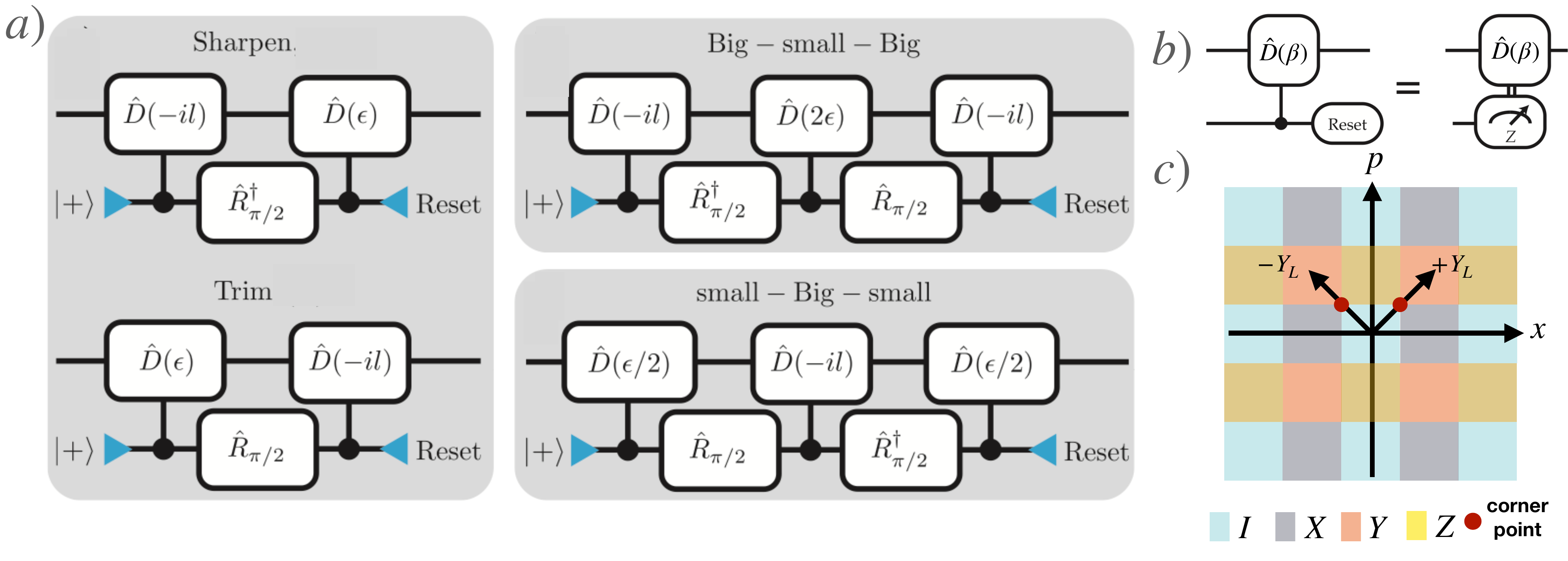}
     \caption{Dissipation engineering circuits that can be autonomous or measurement-based feedback circuits. a) Circuits obtained from trotterization of $U_{\rm target}$ in Eq.~\eqref{eq:dissipation}. The gates are conditional displacements, $\hat{D}(\alpha)=e^{(\alpha\hat{a}-\alpha^* \hat{a}^\dagger)\otimes\hat{\sigma}_Z}$ while $R_{\pi/2}$ are rotations about the $Y$-axis of the ancilla Bloch sphere by $\pi/2$. Here $l=\sqrt{\pi/2}$ and $\epsilon=l\Delta^2$. b) Circuit-equation to convert an autonomous circuit to measurement-based feedback circuit. The top rail is the oscillator and the bottom rail denotes the qubits. c) Effect of displacement errors on a GKP state. If the state is displaced to a point in the \{blue, gray, orange, yellow\}, stabilization corrects it back to the state with a $\{I,X_L,Y_L,Z_L\}$ error, respectively. If stabilizers along the arrows are measured in order to stabilize a lattice representing a $\ket{\pm Y}$ state, then ancilla decay will only cause logical errors if it occurs at the red \emph{corner points}. Figures a,b adapted from Ref.~\cite{royer2020stabilization}. Figure c adapted from Ref.~\cite{baptiste2022multiGKP}.}
     \label{fig:sBs}
 \end{figure}

The final task is to derive oscillator-qubit circuits which realize the Hamiltonian $\hat{v}_{[m]}\hat{\sigma}_x+\hat{v}_\perp\hat{\sigma}_y\tanh(\Delta^2)$ for $\hat{v}\in\{2\sqrt{\pi}\hat{q},2\sqrt{\pi}\hat{p}\}$ via trotterization. In Ref.~\cite{royer2020stabilization}, authors specify three different circuits using first-order and second-order trotterization given by Figs.~\ref{fig:sBs}a-c. All the circuits shown here are autonomous, but they can be easily converted into measurement-based feedback circuits using the circuit equation shown in Figure~\ref{fig:sBs}b.\footnote{The measurement-based version of the \textit{sharpen-trim} protocol was first introduced and experimentally demonstrated in \cite{campagne2020quantum}, as discussed in Chapter~\ref{GKP-exp}} We discuss one of the circuits in detail called the \emph{small-Big-small} or \emph{sBs} which was used to achieve beyond break-even quantum error-corrected memory in Ref.~\cite{sivak2022breakeven}; it was independently introduced and realized in trapped-ion system~\cite{home2022QECgkp}. The sBs circuit was obtained from using first order trotterization of $U_{\rm target}$ for $X,Z$ stabilization in Eq.~\eqref{eq:dissipation} as,
 \begin{align}     
 \label{eq:sBs unitary}
U_{sBs}^X&=e^{i\epsilon_q\hat{q}\hat{\sigma}_y}e^{-i\sqrt{\pi}\hat{p}\hat{\sigma}_x}e^{i\epsilon_q\hat{q}\hat{\sigma}_y}, \qq{where} \epsilon_q=\frac{\sqrt{\pi}}{2}\Delta_q^2,\\
 \qq{and} U_{sBs}^Z&=e^{-i\epsilon_p\hat{p}\sigma_y}e^{-i\sqrt{\pi}\hat{q}\hat{\sigma}_x}e^{-i\epsilon_p\hat{p}\hat{\sigma}_y}, \qq{where} \epsilon_p=\frac{\sqrt{\pi}}{2}\Delta_p^2.
 \end{align}
 The condition on modular quadratures is replaced by conditioning the whole unitary post trotterization to remain unchanged under translation $\hat{x}\rightarrow \hat{x}+m$, permitting $\hat{x}_{[m]}\rightarrow\hat{x}$. This condition is enforced by leveraging the modularity of the qubit by choosing $\Gamma\delta t$ such that the translation $\hat{x}\rightarrow \hat{x}+m$ leads to a trivial qubit operation after time $T$.

For a rectangular GKP code, we have $\Delta_q\neq \Delta_p$. We can thus generalize the sBs circuit to any lattice with arbitrary envelope shapes and sizes by using different stabilizer vectors for $\hat{v}$. It should be noted that, prior to Ref.~\cite{royer2020stabilization}, a similar circuit was obtained for GKP error correction via adaptive phase estimation using single-qubit ancillas by Terhal et al~\cite{terhal2016encoding}. While $sBs$ circuit was interpreted as another explanation for the phase estimation protocol, it is different in the sense that in the ideal GKP limit $\Delta\rightarrow 0$, the $sBs$ protocol comes down to only applying the big conditional displacement $B=e^{i\sqrt{\pi}\hat{x}\sigma_x}$ without any correction . The scheme can be seen as an amplitude amplification scheme when the ancilla at the end of both rounds $U_{sBs}^X$ and $U_{sBs}^Z$ is in the ground state ($g$); while if it is in the excited state ($e$) for either round, the circuit generates a corrective back action. See Figure~\ref{fig:Krauss} in Chapter~\ref{sec:exp_arch} for more details on the action of the $sBs$ map on the GKP code states and error states upon different ancillary measurement outcomes and Ref.~\cite{singh2023composite} for a complete description of the corrective back action.

 An ancilla decay during the larger conditional displacement could yield displacement errors larger than the distance of the code, and hence the logical error of the code depends linearly on ancilla decay. For example, if we define $CD(\sqrt{\pi/2})=e^{i\sqrt{\pi}\hat{p}\otimes\sigma_z}$, then the effect of an ancilla decay during this conditional displacement corresponds to,
\begin{equation}
    CD(\sqrt{\pi/2}-\alpha/2)(I\otimes \text{Err})CD(\alpha/2)=D(\sqrt{\pi/2}-\alpha)\otimes R_Y(\pi/2),
\end{equation}
where $\text{Err}=\ket{g}\bra{e}$ corresponds to an ancilla decay event. The displacement $\alpha$ is determined by the time at which the ancilla decay happened. Thus, an ancilla error during the course of conditional displacement can disrupt the displacement, leading to an error. Here, a displacement in position by $|x|=\sqrt{\pi}-2\alpha\in[-\sqrt{\pi},\sqrt{\pi}]$ can cause a logical error in the region where $\alpha\in[\sqrt{\pi}/4,3\sqrt{\pi}/4]$. Thus, the probability that an ancilla decay event causes a logical error is $50\%$, following this heuristic argument. 

The echoed conditional displacements used in superconducting circuits~\cite{campagne2020quantum,sivak2022breakeven} (further discussed in Chapter~\ref{sec:exp_arch}) will result in an equivalent probability of logical error rate on the GKP codewords due to ancilla decay. Ancilla dephasing on the other hand causes small displacement errors or measurement errors; the small displacement errors occur due to the dephasing errors which occur in between two conditional displacements of the $sBs$ circuit. These effects are correctable for the GKP encoding. Dependencies on ancilla errors have been demonstrated experimentally in Ref.~\cite{sivak2022breakeven}. Thus, circuits can be modified to ensure fault-tolerance with biased-noise ancilla such as Kerr-cats, fluxonium, squeezed cats, dissipatively stabilized cats, additional flag qubits.~\cite{puri2019stabilized,Shi_2019,grimm2020stabilization}. Another approach for suppression of ancilla errors is to use a GKP ancilla for error correction as discussed in Refs.~~\cite{terhal2020towards, siegele2023robust}.

Error correction of multimode GKP codes (discussed in Chs.~\ref{sec:gkp_lattice} and~\ref{sec:QEC-multimode}) using sBs has another source of improvement over the limitations caused by ancilla decay using the \emph{isthumus property}~\cite{baptiste2022multiGKP}.  The same idea can be used to stabilize a qubit for reduction in logical errors if there exist stabilizers which are not parallel to either of the logical operators of the code. This trick uses the extra degrees of freedom available when leveraging multiple modes for the encoding. The \emph{isthmus property} is discussed in detail in Ch.\ref{sssec:GKP-control}. 

Finally, in the Supplementary Material of Ref.~\cite{royer2020stabilization}, the authors discuss a qutrit model for the dissipation process. One major issue with this protocol is the presence of a super lattice which could lead to tunneling in sites with higher energy where the GKP state is more susceptible to errors coming from Kerr-nonlinearity (see Chapter~\ref{ssec:auxiliary-noise}). Even though the qutrit circuit equivalent to the sBs protocol yields a faster correction rate, the tunneling effect is higher in this case due to the larger conditional displacements required by the qutrit model. This is not good for ancilla noise nor for tunneling, since the tunneling probability to other cosine wells increases as the photon number increases. A protocol that overrides the issues caused by tunneling can benefit from faster error correction using qutrits.

\subsubsection{Engineered dissipation with continuous driving}
\label{sec:dissipation_continuous}
Chapter~\ref{sec:qubit-dissipation} focused on using repeated interactions with an auxiliary qubit to realize quantum error correction of GKP codes via engineered dissipation. However, as discussed, errors of the auxiliary qubit can spoil the QEC performance. An alternative approach to realizing engineered dissipation in superconducting circuits is to engineer an interaction between the target oscillator (or qubit) and a bath mode such that, after elimination of the bath mode, the desired dissipators are realized on the target mode \cite{Mirrahimi_2014, Kapit_2017}. Such continuous engineered dissipation has been used to realize experimental stabilization of the cat code \cite{Leghtas_2015} and the truncated 4-component cat (T4C) code \cite{Gertler_2021} in superconducting circuits. In the proposal from Ref.~\cite{sellem2023gkp}, this is done without an auxiliary qubit, eliminating the propagation of auxiliary qubit errors. 

\begin{figure}
    \centering
    \includegraphics[width=\linewidth]{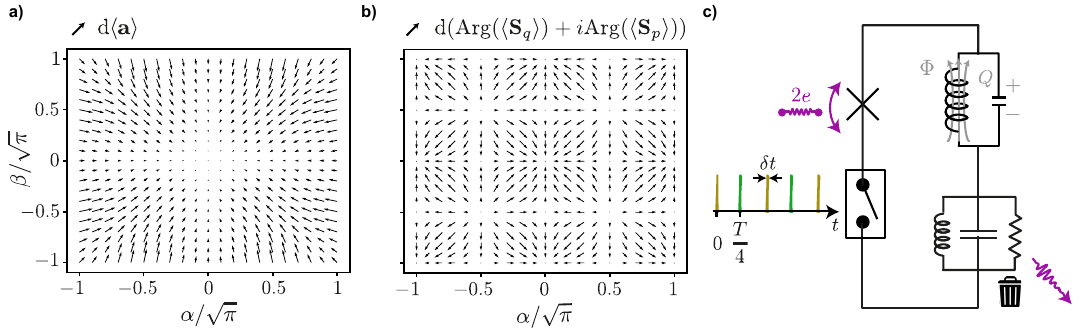}
    \caption{\label{fig:continuous_gkp_dissipation} a,b) For a finite-energy GKP code state with $\sinh(\Delta) = 0.2/\xi$ displaced by $\alpha + i\beta$, evolution of the state's center of mass (a) and modular coordinates (b) entailed by the Lindblad operators in Eq.~\eqref{eq:approximate_dissipators} is shown for a short time step $dt \ll 1/\Gamma$. c) Proposed circuit for realizing driven dissipation of the GKP code. A switch is closed for a short time $\delta t$ every quarter period of the target oscillator (top) in series with a cold bath oscillator (bottom). Figures reproduced with permission from \cite{sellem2023gkp}.}
\end{figure}

Similar to the dissipators introduced in Chapter~\ref{sec:qubit-dissipation}, the authors of Ref.\cite{sellem2023gkp} introduced a set of dissipators that continuously cool the oscillator's state towards the finite-energy GKP code manifold. Using the Lindblad equation $\dot{\rho} = \Gamma \sum_{k=0}^3 \mathbb{D}\left[M_k\right] \rho$ [see Eq.~\eqref{eq:master_eq}], they define a set of four dissipators $M_k = R_{k\pi/2}\left(S_{q}(\Delta) - I\right) R_{k\pi/2}^\dag$ where $R_\theta = \exp\left({i\theta \hat{a}^\dag \hat{a}}\right)$ performs a rotation by $\theta$ in phase space and $\Gamma$ is the dissipation rate.
Here, $S_q(\Delta) = E_\Delta S_q E_\Delta^{-1}$ is the finite-energy position stabilizer for the 2D square GKP code introduced in Eq.~\eqref{eq:envelope_stabilizer}.
As is evident from the dissipators, each Lindblad operator cancels the finite-energy code manifold. However, these operators are challenging to engineer in a continuous manner. Instead, the authors use the BCH formula to derive approximations to these dissipators via
\begin{equation}
\label{eq:approximate_dissipators}
L_k = \mathcal{A} R_{k\pi/2}e^{i\xi \hat q}\left(I - \epsilon \hat p\right) R_{k\pi/2}^\dag - I,
\end{equation}
where $\epsilon = \xi \sinh\left(\Delta^2\right)$, $\mathcal{A} = \exp\left(-\xi \epsilon/2\right)$, and $\xi  = 2\sqrt{\pi}$ is the stabilizer displacement length. The phase-space dynamics of these modular Lindblad operators is shown in Figure \ref{fig:continuous_gkp_dissipation}a and b, displaying convergence towards the finite energy code manifold. 

To understand the effectiveness of the dissipators, the authors analyzed the evolution of the generalized GKP Pauli operators, defined as $Z = \text{sgn}\left[\cos(\xi \hat{q}/2)\right]$, $X = \text{sgn}\left[\cos(\xi \hat{p}/2)\right]$ and $Y = iXZ$, where the function ${\rm sgn}(x)$ equals $+1$ if $x>0$, 0 if $x=0$, and $-1$ if $x<0$. These operators respect the Pauli algebra composition rules throughout the oscillator Hilbert space and coincide with the logical qubit Pauli operators inside the code manifold. With these four dissipators, the infinite-energy code stabilizers converge to their steady state value at a rate $\Gamma_c \gtrsim \mathcal{A}\epsilon \xi \Gamma$.
As shown in Refs.~\cite{sellem2023stability,sellem2023gkp}, under weak quadrature noise $\dot\rho = \mathbb{D}\left[\sqrt{\kappa}q\right]\rho + \mathbb{D}\left[\sqrt{\kappa}p\right]\rho$, such that $\kappa \ll \Gamma$, the generalized $X$ and $Z$ Pauli operators decay at a rate $\Gamma_L \simeq \frac{4\xi}{\pi}\sqrt{\frac{\kappa \Gamma}{2}}\exp\left(-\sqrt{8\Gamma/\xi^2 \kappa}\right)$ for an optimal choice in the parameter $\epsilon$. Importantly, this exponential scaling ensures logical errors can be heavily suppressed with a modest ratio of $\Gamma/\kappa$.

The authors of Ref.~\cite{sellem2023gkp} proposed a way to engineer the aforementioned dissipators using a high-impedance superconducting circuit driven with a frequency comb as shown in Figure \ref{fig:continuous_gkp_dissipation}c. Here, a switch controls the coherent tunneling of Cooper pairs across a Josephson junction placed in parallel in a two-mode circuit. The target GKP mode (top oscillator) has a large impedance such that each tunneling event translates its state by $\pm 2\sqrt{\pi}$ along the charge axis. By driving with a train of sharp pulses that activate tunneling every quarter period, shifts are generated along the charge axis and phase axis in phase-space, matching a GKP lattice. The second lower impedance auxiliary mode (bottom) is coupled to a cold load such that the target mode dynamics is irreversible, realizing the proposed dissipators and stabilizing the GKP code. This circuit can be seen as an extension of Floquet circuits discussed in Ch.~\ref{sec:GKP_circuits}. Without dissipation, the circuits for realizing a Floquet engineered GKP Hamiltonian do not correct errors, and the GKP states are not stabilized against noise. On the other hand, here the engineered dissipation provides a means of correcting errors, with exponential convergence to the code manifold in the strength of dissipation.

\subsection{Realization of GKP codes in superconducting circuits}
\label{sec:experiment}

The past few years have seen rapid advancement in the experimental realization of GKP state preparation and quantum error correction in superconducting circuit~\cite{campagne2020quantum,eickbusch2022fast,Kudra_2022, sivak2022breakeven} and trapped ion~\cite{Fluhmann_2018, fluhmann2019encoding,home2022QECgkp} platforms. In this section, we review the architecture for superconducting circuit experiments leading ultimately to the realization of an error corrected quantum memory with coherence beyond \textit{break-even} [as defined in Chapter~\ref{GKP-exp} around Eq.~\eqref{eq:break_even}]. 

\subsubsection{Experimental architecture and state preparation}
\label{sec:exp_arch}
\begin{figure}[t]
    \centering
    \includegraphics[width=\linewidth]{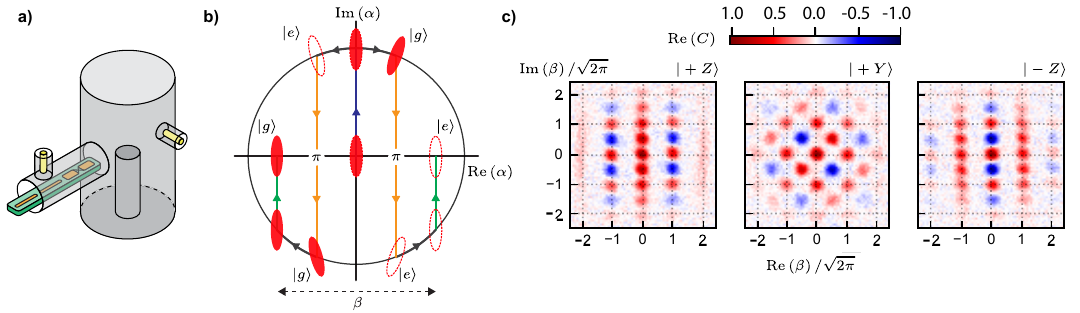}
    \caption{\label{fig:gkp_exp_1} a) Typical 3-D cQED setup consisting of a superconducting cavity (here, a post acts as a $\lambda/4$ resonator where $\lambda$ is the wavelength of the fundamental microwave mode) coupled to a transmon with an on-chip readout resonator and Purcell filter. Figure reproduced from \cite{sivak2022breakeven}. b) Phase-space displacement sequence used to realize the Echoed Conditional Displacement (ECD) gate; see \cite{eickbusch2022fast} for full pulse sequence. c) Measured characteristic functions for GKP states prepared with conditional displacements starting from vacuum. Figure reproduced from \cite{eickbusch2022fast}.}
\end{figure}

To date, all explicit realizations of the GKP code in superconducting circuits have been performed using a cQED architecture similar to that shown in Figure \ref{fig:gkp_exp_1}a~\cite{blais2021cqedRMP, Krantz_2019}. Here, a high quality-factor 3D superconducting cavity is coupled to an auxiliary superconducting qubit \cite{Paik_2011,reagor_2016,Axline_2016} and anchored at the base stage of a dilution refrigerator at a typical temperature of $\SI{20}{mK}$. The GKP state is encoded in the quantized electromagnetic field of the fundamental mode of the microwave cavity. Such an architecture has also been used to realize quantum error correction of other bosonic codes---including rotation-symmetric bosonic codes \cite{Grimsmo_2020} such as the cat code \cite{ofek2016extending} and binomial code \cite{Hu_2019, Zhongchu_2023} and also including fully autonomous realizations \cite{Gertler_2021}. Typical high-purity aluminum cavity single-photon lifetimes are on the order of $\SI{1}{ms}$, with state of the art niobium cavities reaching lifetimes of tens of milliseconds \cite{milul2023superconducting} up to seconds \cite{Romanenko_2020}. For the GKP experimental realizations, a transmon \cite{Koch_2007} has been used as the auxiliary control element, with current typical lifetimes on the order of $\SI{100}{us}$---of lower quality than the oscillator. As we see later, transmon errors during stabilization is a main limiting factor for reaching longer logical lifetimes in current cQED GKP experiments. Exploration of other types auxiliary qubits for GKP error correction is an ongoing topic of investigation in the field. 

The transmon-cavity system is operated in the dispersive regime of cQED \cite{blais2021cqedRMP} described by an effective Hamiltonian within in the rotating frame of the transmon and cavity as
\begin{equation}
H = \chi \hat{a}^\dag \hat{a} \frac{\hat{\sigma}_Z}{2} + \varepsilon^*(t) \hat{a} + \varepsilon(t) \hat{a}^\dag + \Omega^*(t) \hat{\sigma}_- + \Omega(t) \hat{\sigma}_+ + H_\text{spurious},
\end{equation}
where we have made the rotating wave approximation and two-level system approximation \cite{blais2021cqedRMP}. Both of these approximations are broken in GKP error correction experiments, however we use this Hamiltonian to explain the basic operating principles of the control. In the Hamiltonian above, $\hat{a}$ is the annihilation operator of the oscillator with $\left[\hat{a}, \hat{a}^\dag\right]=1$, $\hat{\sigma}_Z$ is the $Z$ Pauli operator for the auxiliary qubit (realized as the ground and first excited states of the transmon), and $\left\{\varepsilon(t), \Omega(t)\right\}$ are the complex-valued drives at the cavity and transmon frequency, respectively. These drives are delivered as microwave pulses sent through transmission lines that are coupled to the cavity mode and transmon. $H_\text{spurious}$ includes higher-order interactions that are generally harmful for GKP state preparation and error correction. For example, the oscillator Kerr nonlinearity $H_K = K a^{\dagger 2} a^2$ is a necessary byproduct of the cavity's dispersive coupling to the transmon, however it introduces an unwanted evolution of the oscillator that distorts GKP states \cite{albert2018pra, campagne2020quantum}. 

In other bosonic quantum error correction experiments using a similar architecture, the dispersive shift is engineered to be roughly $\chi/2\pi = \SI{1}{MHz}$, chosen so parity measurements that occur on a timescale $\tau_\text{parity} = \pi/\chi$ can be made fast relative to relaxation times of transmons and cavities \cite{ofek2016extending, Ma_2021}. This leads to typical values of the inherited Kerr in these other experiments on the order of $K/2\pi \approx \SI{1}{kHz}$. On the other hand, for GKP experiments, the Kerr coefficient must be engineered to be much smaller (on the order of $K/2\pi \approx \SI{50}{Hz}$ or less) in order for the Kerr effect not to reduce QEC performance relative to other error channels given current relaxation rates \cite{campagne2020quantum}. One approach to suppress Kerr is to reduce the coupling between the transmon and oscillator. Since the Kerr coefficient is proportional to the square of the dispersive shift to first order, $K \propto \chi^2$ \cite{blais2021cqedRMP}, the Kerr can be strongly suppressed this way. Such a \textit{low-$\chi$} approach was first demonstrated in Ref.~\cite{campagne2020quantum} with a dispersive shift of $\chi/2\pi = \SI{28}{kHz}$ and a suppressed Kerr coefficient on the order of $K/2\pi \approx \SI{1}{Hz}$.

To control the oscillator with a weak dispersive shift, an effective interaction is used that can be turned on and off \textit{in situ} using a microwave drive. In this approach, the oscillator is driven with a resonant tone activating a large displacement of the mean field, $\alpha(t)$. In the time-dependent displaced frame of the oscillator $\hat{a} \rightarrow \hat{a} + \alpha(t)$, the dispersive interaction $H = \chi \hat{a}^\dag \hat{a} \hat{\sigma}_Z/2$ becomes
\begin{equation}
	\label{eq:displaced H}
	\tilde{H} = \chi\left(\alpha(t) \hat{a}^\dagger + \alpha^*(t) \hat{a}\right) \frac{\hat{\sigma}_Z}{2} + \chi \hat{a}^\dagger \hat{a} \frac{\hat{\sigma}_Z}{2} + \chi |\alpha(t)|^2 \frac{\hat{\sigma}_Z}{2}.
\end{equation}
Here $d \alpha(t)/dt = -i\varepsilon(t)- \left(\kappa/2\right)\alpha(t)$ is the classical response to a resonant drive \cite{campagne2020quantum, eickbusch2022fast} \QZ{and $\kappa$ is the photon loss rate of the oscillator}. With a large displacement, the first term in $\tilde{H}$ dominates, and the effective interaction between the oscillator and qubit becomes a qubit-state-dependent force with an enhanced interaction strength $g_\text{eff}(t) = \chi |\alpha(t)|$. By using large oscillator displacements, the weak dispersive shift can be overcome.

As shown in Chapter~\ref{sec:qubit-dissipation}, the key entangling gate needed between the oscillator and auxiliary qubit is a conditional displacement. As introduced in Ref.~\cite{campagne2020quantum} and later refined in Ref.~\cite{eickbusch2022fast}, a conditional displacement can be constructed from $\tilde{H}$ using a suitable echo sequence as illustrated in Figure \ref{fig:gkp_exp_1}b. This sequence is called the \textit{Echoed Conditional Displacement} (ECD) gate, and the associated unitary is $\text{ECD}(\beta) = D(\beta/2)\ket{e}\bra{g} + D(-\beta/2)\ket{g}\bra{e}$. A similar conditional displacement gate was also realized in Ref.~\cite{diringer2023conditional} with a modified trajectory through phase space.

When the $\text{ECD}\left(\beta\right)$ gate is interleaved with rotations of the auxiliary qubit, a universal gate set is obtained, meaning any unitary on the oscillator and auxiliary qubit Hilbert space can be realized with enough applications of ECD gates and qubit rotations \cite{eickbusch2022fast}. One particular application of this universality is the unitary state preparation of finite-energy logical states. As shown in Figure \ref{fig:gkp_exp_1}, the $\left\{\ket{+Z},\ket{+Y}, \ket{-Z}\right\}$ logical GKP qubit states are attained in this manner with a target squeezing of $\Delta = 0.31$ and an achieved squeezing $\Delta = 0.35$ using $N=9$ layers of ECD control for the $Z$ eigenstates (and $N=10$ for $\ket{+Y}$). In particular, the measured data is an example of the characteristic function, $C\left(\beta\right) = \Tr\left(\rho D(\beta)\right)$ as introduced in Chapter~\ref{sec:qho}. Here the experimental state preparation fidelity ranges from $\mathcal{F} = 0.8$ to $\mathcal{F} = 0.85$, limited by errors of the auxiliary qubit and cavity errors during the ECD gates. For these pulses, the intermediate large displacement used was $|\alpha| = 30$, corresponding to an average of $900$ photons during the ECD gates. 

\subsubsection{Experimental realization of GKP quantum error correction}
\label{GKP-exp}

\begin{figure}
    \centering
    \includegraphics[width=\linewidth]{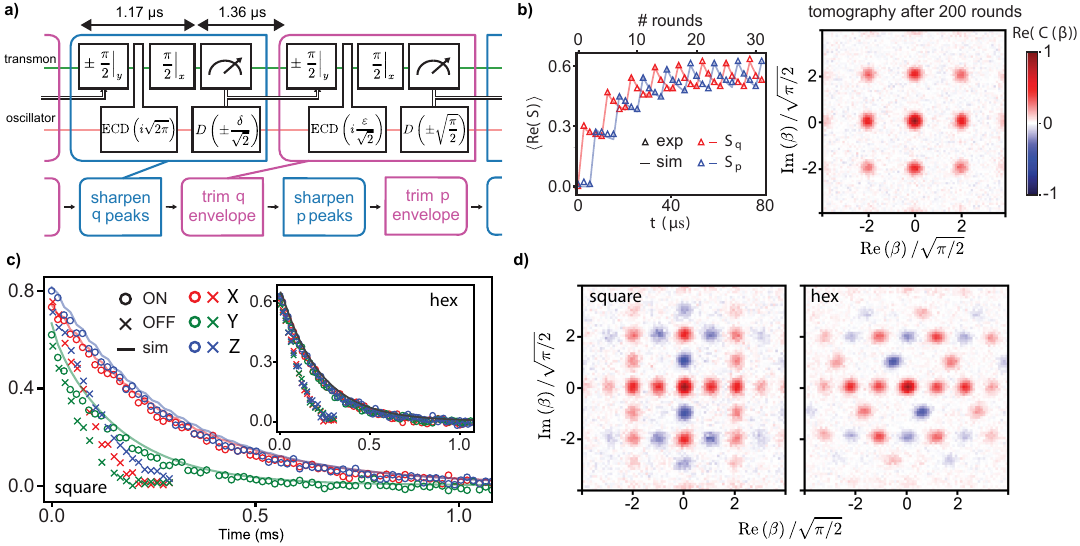}
    \caption{\label{fig:gkp_exp_2020}   a) Measurement based sharpen-trim protocol. b) Left: Evolution of the square-code position and momentum stabilizers under repeated QEC cycles starting from vacuum. Right: Measured characteristic function after 200 rounds.  c) Lifetimes of the uncorrected (crosses) and corrected (circles) square GKP qubit. Inset: Hexagonal lifetimes. d) Measured characteristic functions for Hadamard eigenstates prepared with a measurement-based gate teleportation protocol. Figures reproduced from \cite{campagne2020quantum}.}
\end{figure}

The first experiment to demonstrate quantum error correction of the GKP code was published in 2020 \cite{campagne2020quantum}. In this experiment, the authors introduced the measurement-based sharpen-trim protocol shown in Figure \ref{fig:gkp_exp_2020}a. This circuit implements engineered dissipation to the GKP code manifold as derived in Chapter~\ref{sec:qubit-dissipation}. By utilizing fast FPGA electronics with a latency on the order of $\SI{200}{ns}$, the transmon was measured and fast feedback displacements were employed to correct for small shift errors in real time in a four-round QEC cycle. As a verification experiment, the QEC cycle was repeated starting from the vacuum state of the oscillator, and the expectation values of stabilizers were measured after each step. The results of this experiment are shown in Figure \ref{fig:gkp_exp_2020}b, with a convergence to the quasi-steady state after about $20$ rounds (here, a round is defined as a single sharpen or trim step). The pattern in the measured stabilizers is clear: during each four round cycle, the measured position (or momentum) stabilizer value increases after the corresponding sharpen round, and slightly decreases during the next 3 rounds, matching simulation. Also shown in Fig.~\ref{fig:gkp_exp_2020}b is the convergence of the oscillator to a mixed state in the code manifold after 200 rounds of stabilization, longer than the decay constant of logical information.

To measure the performance of the quantum error correction protocol, Pauli eigenstates of the GKP code were prepared with a measurement-based protocol \cite{campagne2020quantum,fluhmann2019encoding}. In this protocol, the code is first stabilized to an arbitrary (mixed) state in the code manifold by many rounds of stabilization. Next, the GKP code is projected into one of the $\ket{\pm Z}$ logical eigenstates by an infinite-energy $Z$ logical measurement. For this measurement, the transmon is initialized in the $\ket{+x}$ state, a conditional displacement corresponding to a $Z$ logical Pauli is applied, and the transmon is measured in the $x$ basis. A real-time feedback displacement conditioned on the measurement result was then applied to prepare the desired logical state. The same procedure was used to prepare $X$ and $Y$ logical states by measuring the corresponding logical displacements. After state preparation, decay of the Pauli expectation values are measured in two cases: during free evolution and with repeated QEC cycles applied via the sharpen-trim protocol. \QZ{The results for the square GKP code are shown in Figure \ref{fig:gkp_exp_2020}c.} The lifetime of the error-corrected GKP Pauli expectation values is longer than the unstabilized counterparts by about a factor of two. The square code has a natural noise bias: Because the displacement corresponding to the $Y$ logical stabilizer is $\sqrt{2}$ times larger in phase-space than the $X$ and $Z$ stabilizers (see Example~\ref{example:square_code}), the code distance along the $Y$ direction is reduced, leading to a reduction in the lifetime of $Y$ eigenstates. 

\begin{figure}
    \centering
    \includegraphics[width=\linewidth]{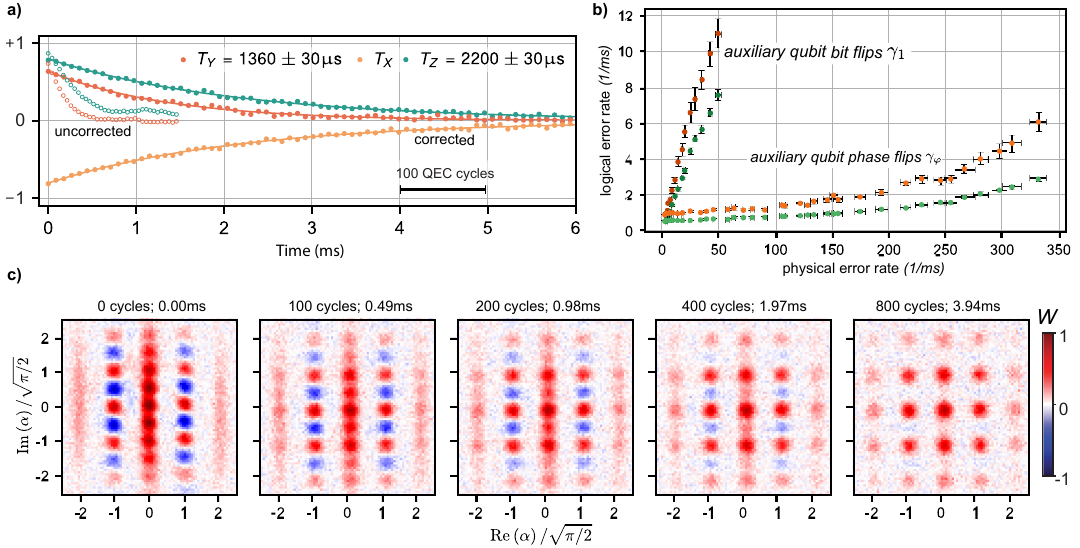}
    \caption{\label{fig:gkp_exp_2023} Figures reproduced from \cite{sivak2022breakeven}. a) Lifetimes of the uncorrected (open circles) and corrected (filled circles) square GKP code. b) Measured logical error rate sensitivity to increasing auxiliary qubit bit-flip or phase-flip rate. c) Measured Wigner functions of an error corrected $\ket{+Z}$ logical state, taken as snapshots after varying number of rounds.}
\end{figure}

To realize a logical qubit with a depolarizing error channel, a similar 6-round sharpen-trim protocol was used to stabilize the hexagonal GKP code, with details given in Ref.~\cite{campagne2020quantum} and results shown in the inset of Figure \ref{fig:gkp_exp_2023}c.  As expected, for the hexagonal code, the lifetimes of $X,Y$ and $Z$ are equal, due to the equal lengths of each displacement stablizer (and hence equal code distance). Finally, arbitrary rotations within the code manifold can be performed by a measurement-based gate teleportation protocol \cite{fluhmann2019encoding,campagne2020quantum}. With details given in Ref.\cite{campagne2020quantum}, the gate-teleportation protocol is similar to the logical measurement protocol, except the transmon is prepared in an arbitrary state on the $x-y$ plane of the Bloch sphere to perform a logical rotation within the manifold. This procedure was used to prepare Hadamard eigenstates in both the square and hexagonal codes, as shown in Figure \ref{fig:gkp_exp_2020}d. 

Although the authors of Ref.\cite{campagne2020quantum} realized stabilized GKP states with lifetimes longer than their unstabilized counterparts, the experiment did not realize error correction beyond \textit{break-even}. In particular, for a quantum channel $\mathcal{E}: \rho \rightarrow \mathcal{E} (\rho)$, we can define the average channel fidelity relative to a target unitary channel $U : \rho \rightarrow U\rho U^\dagger$ given by $\bar{\mathcal{F}} = \int d \psi \bra{\psi}U^\dagger \mathcal{E} \left(\ket{\psi}\bra{\psi}\right)U\ket{\psi}$ where the integral is over the uniform measure in state space. Using the Pauli transfer matrix approach, the experimental decay of Pauli expectation values can be used to compute the channel fidelity to the target identity channel $U=I$ \cite{Nielsen_2002}. At small times $\delta t$, this fidelity can be expanded as $\bar{\mathcal{F}}\left(\delta t\right) = 1 - \frac{1}{2} \Gamma \delta t$, allowing the short-time decay rate of different qubits and channels (which generally have non-exponential decay curves) to be compared through a single decay rate, $\Gamma$ \cite{sivak2022breakeven}. With this, the \textit{quantum error correction gain} is defined as
\begin{equation}
G = \frac{\Gamma_{\text{physical}}}{\Gamma_\text{logical}}
\label{eq:break_even}
\end{equation}
where $\Gamma_{\text{physical}}$ is the fidelity decay constant of the best physical qubit in an experiment and $\Gamma_{\text{logical}}$ is the decay constant of the error-corrected logical qubit. $G=1$ corresponds to the break-even point. To the best of our knowledge, only three experiments to date have achieved beyond break-even QEC ($G>1$) of a quantum memory given this metric, all using bosonic codes \cite{ofek2016extending,Zhongchu_2023,sivak2022breakeven}. We note that this definition is in the context of a \textit{quantum memory} experiment, where a long-lived stabilized manifold is the target application, i.e. the identity is the target unitary $U=I$. A more careful definition involving SPAM (state preparation and measurement) errors as well as logical gate errors will be needed in future works to compare error-corrected logical qubits for use in quantum computation.  

Using these definitions, for the 2020 GKP error correction experiment \cite{campagne2020quantum}, the average channel lifetimes of the stabilized logical qubits were $\Gamma_\text{square, 2020}^{-1} = \SI{222}{us}$ for the square encoding and $\Gamma_\text{hex, 2020}^{-1} = \SI{205}{us}$ for the hexagonal. This should be compared to the the best physical qubit in the system, the $\left\{\ket{0},\ket{1}\right\}$ Fock encoding of the high quality-factor microwave cavity, with decay constant $\Gamma_\text{Fock, 2020}^{-1} = \SI{368}{us}$ (for the cavity, $T_1 = \SI{245}{us}$, and we are assuming here that $T_2 = 2 T_1$, as the measured cavity intrinsic dephasing rate was negligible). From this, the realized QEC gain was $G_\text{square, 2020} = 0.6$ and $G_\text{hex, 2020} = 0.56$. As confirmed by master equation simulations, a limiting factor in realizing a larger gain was bit flips of the auxiliary transmon during the large conditional displacements. Since bit flips do not commute with the interaction Hamiltonian Eq.~\eqref{eq:displaced H}, they can lead to large displacement errors of the GKP state, causing logical errors, as is also discussed in Chapter~\ref{sec:qubit-dissipation}. The stabilization rounds were spaced optimally so the contribution from auxiliary qubit errors was on par with the contribution from cavity errors.

In a more recent 2023 experiment \cite{sivak2022breakeven}, advancements were made to reach quantum error correction beyond break-even. In particular, the incorporation of three main innovations led to this improvement in gain compared to the 2020 experiment. First, by using recently development fabrication techniques for realizing a 3D transmon with tantalum pads \cite{Place_2021}, a relatively long-lived auxiliary transmon was used with an average lifetime of $T_1 = \SI{280}{us}$, close to a six-time improvement over the auxiliary transmon used in 2020 \cite{campagne2020quantum}. Secondly, the measurement-based sharpen trim protocol used in 2020 was replaced with a \textit{semi-autonomous} version of the small-Big-small (sBs) protocol, as introduced and derived in Chapter~\ref{sec:qubit-dissipation}. Here, we call the protocol semi-autonomous since the final oscillator displacement in each sBs round is performed with a conditional displacement, however the auxiliary qubit and oscillator phase was still reset between QEC rounds using measurement and feedback (now incorporating reset of the $\ket{f}$ transmon state). The reset could be replaced with an autonomous transmon reset to make the protocol fully autonomous. The third major advancement was incorporating online optimization of the QEC protocol using model-free reinforcement learning \cite{Sivak_2022}. In particular, the sBs protocol was used as an ansatz for the QEC cycle, and a proximal policy optimization (PPO) reinforcement learning algorithm was used to train 45 real-valued parameters of the QEC cycle \textit{in-situ} to optimize the lifetime. Such training was essential to realizing the large QEC gain.

Together, these improvements led to the square GKP code logical decay curves shown in Figure \ref{fig:gkp_exp_2023}a, with logical lifetimes of $T_{X\text{, 2023}} = T_{Z\text{, 2023}} = \SI{2.20}{ms}$ and $T_{Y\text{, 2023}} = \SI{1.36}{ms}$ with decay time constant of the average channel fidelity of $1/\Gamma_\text{square, 2023} = \SI{1.82}{ms}$. Here, the GKP states were prepared using ECD control \cite{eickbusch2022fast}. Given the Fock encoding $\left\{\ket{0},\ket{1}\right\}$ lifetime of $\Gamma_{\left\{\ket{0},\ket{1}\right\}\text{,2023}} = \SI{0.8}{ms}$, the achieved QEC gain was $G_\text{square, 2023} = \Gamma_{\left\{\ket{0},\ket{1}\right\}\text{,2023}}/\Gamma_\text{square, 2023} = 2.27 \pm 0.07$, well beyond break-even. Snapshots of the Wigner functions at different points in the logical decay curve with QEC starting from $\ket{+Z}$ is shown in Figure \ref{fig:gkp_exp_2023}c, indicating the decay of the logical information encoded in the interference fringes. Finally, we note that the logical Pauli measurements used in Refs.~\cite{sivak2022breakeven} and \cite{campagne2020quantum} were the infinite-energy versions, which lowers the contrast of the decay curves. In future applications, these measurements could be replaced with measurements of the finite-energy Pauli operators, as discussed in Chapter~\ref{sssec:GKP-control}, to increase the contrast.

The authors of Ref.~\cite{sivak2022breakeven} also experimentally studied the stabilized GKP qubit's sensitivity to auxiliary qubit noise. For this, they injected noise to selectively increase the auxiliary qubit's bit-flip rate $\gamma_1$ or phase-flip rate $\gamma_\varphi$. The results of this experiment are shown in Figure \ref{fig:gkp_exp_2023}b. Fitting these slopes in the low physical error rate regime, the authors found that the QEC logical error rate is $65$-times more sensitive to auxiliary qubit bit flips than auxiliary phase flip, as expected from the discussion in Chapter~\ref{sec:qubit-dissipation}. 

To better understand the error correction protocol, the authors of Ref.~\cite{sivak2022breakeven} analyzed the Kraus maps for the small-Big-small protocol. In particular, the sBs protocol repeatedly implements a composite channel $\mathcal{R}_\Delta\left(\rho\right) = \left(\mathcal{R}_\Delta^Z \circ \mathcal{R}_\Delta^X\right)\left(\rho\right)$ where $\rho$ is the oscillator's density matrix and $\left\{\mathcal{R}_\Delta^X, \mathcal{R}_\Delta^Z\right\}$ are the rank-2 channels associated with sBs rounds in the position and momentum directions. 
These channels can be written as Kraus maps, $\mathcal{R}_\Delta^{X} \left(\rho\right) = \sum_{i=\left\{g,e\right\}} {K_i^X}^\dag \rho K^X_i$ where $\sum_{i} {K_i^{X}}^\dag K^X_i = I$ and a similar definition for $Z$.
Here, $K_{\left\{g,e\right\}}^X$ are the Kraus operators corresponding to measuring the auxiliary qubit in $\left\{\ket{g},\ket{e}\right\}$, explicitly given by 
\begin{align}
&K_g^X = \bra{g}U_{sBs}^X\ket{g}\\
&K_e^X = \bra{e}U_{sBs}^X\ket{g},
\end{align}
where $U_{sBs}^X$ is the unitary corresponding to an $X$- sBs round given explicitly in Eq.~\eqref{eq:sBs unitary}; see Chapter~\ref{sec:channel_dynamics} for general discussion about quantum channels and Kraus operators. 

\begin{figure}
    \centering
    \includegraphics[width=\linewidth]{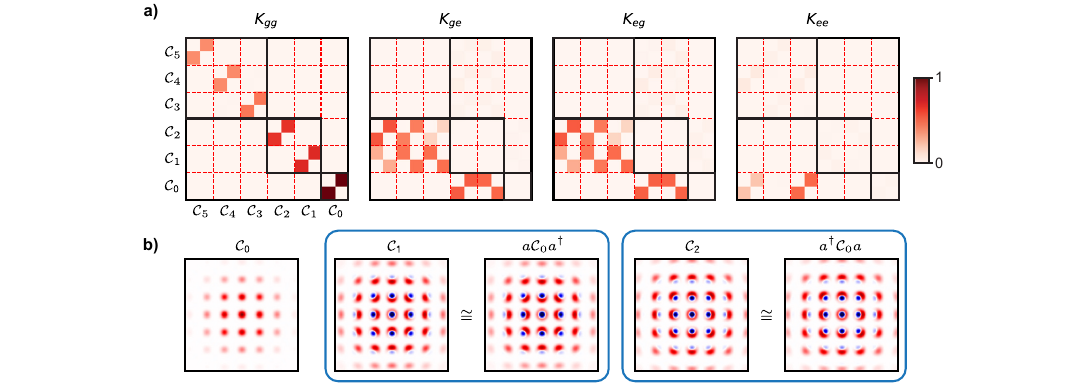}
    \caption{\label{fig:Krauss} Figures reproduced from \cite{sivak2022breakeven}. a) Kraus operators for the composite QEC channel, plotted in the basis for which $K_{gg}^\dag K_{gg}$ is diagonal (absolute values of matrix entries is shown). The code and error spaces are labeled, each consisting of two states forming a logical Bloch sphere. b) Wigner functions of projectors onto the code and error spaces, along with comparisons to a single photon loss or gain error applied to $\mathcal{C}_0$.}
\end{figure}

To understand these Kraus maps, the authors of Ref.~\cite{sivak2022breakeven} analyzed the four Kraus operators for the composite channel $\mathcal{R}_\Delta$ given by $K_{gg} = K_g^Z K_g^X$, $K_{ge} = K_g^Z K_e^X$, $K_{eg} = K_e^Z K_g^X$, and $K_{ee} = K_e^Z K_e^X$ corresponding to the two auxiliary qubit measurement outcomes of each full QEC cycle consisting of $X$ and $Z$ QEC rounds. For $\Delta = 0.34$, these Kraus operators are plotted as matrices in the truncated eigenbasis of $K_{gg}^\dag K_{gg}$ in Figure~\ref{fig:Krauss}. The eigenbasis splits into pairs of states $\mathcal{C}_i$ that define the various error spaces and are each orthogonal to the code space $\mathcal{C}_0$. These matrices shed a new light onto the sBs protocol by revealing a \textit{trickle-down} approach to error correction: \QZ{measuring $\ket{e}$ in either the $X$ or $Z$ sBs round} signals that an error has been corrected, and the corresponding Kraus matrix applied shifts the state down to the next lower error space. Similar code and error spaces can be defined as the quasi-degenerate pairs of eigenstates of the confined GKP Hamiltonian discussed in Chapter~\ref{sec:GKP_Hamiltonian}.

The authors of Ref.~\cite{sivak2022breakeven} also plotted these spaces explicitly, as reproduced in Figure~\ref{fig:Krauss}b, displaying the Wigner functions of the projectors $\Pi_0$, $\Pi_1$, and $\Pi_2$ onto the codespace and first two error spaces. As shown in the figure, the first two error spaces are well approximated by $\Pi_1 \approx a \Pi_0 a^\dag$ and $\Pi_2 \approx a^\dag \Pi_0 a$, indicating that the hierarchy closely resembles photon loss $a$ and photon gain $a^\dag$ type errors. The higher error spaces can also be studied this way, with more details in Ref.~\cite{sivak2022breakeven}. The Kraus operators can also be written in the position or momentum basis, as done in Ref.~\cite{campagne2020quantum} for the case of the sharpen-trim protocol. 

Besides the approaches to mitigate auxiliary qubit noise discussed in Chapter~\ref{sec:qubit-dissipation}, using the error corrected GKP qubits for quantum computation and other applications will require high-fidelity single-qubit and multi-qubit gates. For realistic GKP qubits, the gates must be engineered to suppress leakage out of the finite energy manifold. For Clifford gates, one simple approach is to apply the infinite-energy version of the gate (a Gaussian unitary) followed by many cycles of finite-energy QEC to project the states back onto the finite-energy manifold. An alternative is to engineer gates that directly respect the finite-energy condition, with one approach proposed in \cite{rojkov2023twoqubitStabiliz}. In superconducting circuits, purpose-built couplers must be used to achieve these Gaussian operations without introducing spurious nonlinearities, as discussed in Chapter~\ref{sec:exp_arch}. Some promising approaches include using Kerr-free parametric three-wave mixing with a SNAIL (Superconducting Nonlinear Asymmetric Inductive eLement) mixer \cite{Frattini_2018,Sivak_2019, chapman2022high} or other couplers that could be engineered in a Kerr-free regime \cite{lu2023microwaveBS, Ye_2021, Chien_2020}.


\subsection{Additional proposals for control of finite-energy GKP states using auxiliary qubits}
\label{sssec:GKP-control}

As introduced in Ref.~\cite{gkp2001}, Gaussian unitaries can be employed to realize Clifford operations on infinite-energy GKP states. However, for realistic finite-energy states as defined in Chapter~\ref{sec:finite_GKP}, Gaussian unitaries do not suffice.\footnote{In particular, any Gaussian unitary that does not commute with the number operator $a^\dag a$ (and hence does not commute with the envelope operator $E_\Delta$ as introduced in Chapter~\ref{sec:finite_GKP}) will lead to leakage outside of the finite-energy code manifold. For the computational single-mode GKP states, most Clifford operations fall into this category.} In addition, state preparation and measurement of finite-energy GKP states is challenging, requiring a non-linear resource to realize the non-Gaussian states.

In Chs.~\ref{sec:GKP_circuits}, \ref{dissipation-engineering}, and \ref{sec:experiment}, a few methods for state preparation and control of finite-energy GKP states was discussed in-depth. In this section, we extend this discussion to some other finite-energy control methods in the literature using auxiliary qubits. As these methods are not well-suited for, e.g., optical platforms, we relegate engineering optical GKP in to Ch.~\ref{sec:optical_gkp}.

\subsubsection{State preparation}

Reducing State Preparation And Measurement (SPAM) errors is key to achieving practical quantum computing \cite{Google_SC_35}. In spite of the recent beyond break-even results for error correction of a quantum memory described in Chapter~\ref{GKP-exp}, the contrast of measured decay curves (see Figs. \ref{fig:gkp_exp_2020} and \ref{fig:gkp_exp_2023}) as well as the relatively low state preparation fidelity (see Ref.~\cite{eickbusch2022fast}) suggest that SPAM errors need improvement to be competitive with other quantum computing architectures. 
Here, we give a brief summary of additional results and proposals for arbitrary finite-energy GKP state preparation and attempts to improve upon the fidelity by keeping auxiliary qubit errors in check, which is one of the main limitations of the methods presented in the previous sections.

As discussed in Ch.~\ref{sec:exp_arch},  experimentalists have used numerically optimized circuits to prepare GKP states with conditional displacements realized in a weak dispersive coupling regime~\cite{eickbusch2022fast}.
These gates are simpler and more favourable to GKP states. A measurement-free logical GKP state teleportation scheme was introduced in Ref.~\cite{Hastrup_2021_measurement_free} using numerical optimization with the same gates. Here, the authors find circuits composed of multiple conditional displacements to teleport arbitrary states from the single qubit ancilla to the corresponding logical GKP state in the oscillator, without any measurements on the ancilla. Ideally, post selection on the ancilla being in the expected state at the end of circuit can yield some protection against ancilla noise. However, both numerically optimized schemes yield circuits which are relatively long; thus the methods will still be restricted by multiple ancilla-error events. Due to auxiliary qubit errors during large-depth measurement-free circuits, measurement-based approaches or engineered dissipation based approaches might be a better option for high-fidelity state preparation of GKP states. For example, the states prepared experimentally using engineered dissipation, measurement, and feedback in Ref.~\cite{campagne2020quantum} (see also Fig. \ref{fig:gkp_exp_2020}) were visually of higher quality than those prepared using measurement-free approaches in Refs.~\cite{eickbusch2022fast, sivak2022breakeven} (see Figs. \ref{fig:gkp_exp_1} and \ref{fig:gkp_exp_2023}).

To prevent propagation of auxiliary qubit errors during state preparation, a fault-tolerant version of the phase estimation protocol introduced in Ref.~\cite{terhal2016encoding} was developed in Ref.~\cite{Shi_2019}. The authors of the latter use the non-adaptive phase estimation with one ancilla and one flag qubit. The flag qubit is used to detect errors during the conditional displacements by performing a CNOT gate between the auxiliary qubit and flag qubit before and after each conditional displacement. 
Due to the dependence of phase estimation protocols on measurement outcomes, the acceptance chance of this protocol is $ <50\%$. Also, in Ref.~\cite{singh2023composite}, the authors introduce an analytic framework from which one can derive deterministic generation of Pauli logical GKP states. The authors propose that auxiliary qubit resets in the deterministic scheme can be replaced by post-selection to detect auxiliary qubit errors. 
   
In addition to the above methods, Ref.~\cite{baptiste2022multiGKP} introduced a new feature, called the \emph{isthmus property}, to mitigate effects of auxiliary qubit errors using two-mode GKP codes (see Ch.\ref{ssec:twomode_GKP_qec}). This technique can also be understood in terms of single-mode GKP codes via preparation of single-mode $\ket{Y}_L$ states. Logical GKP $\ket{\pm Y}$ states can be prepared by measuring the stabilizers which are parallel to the $\pm Y_L$ operators as marked by the arrows along the diagonals in Figure~\ref{fig:sBs}c. These stabilizers are valid if we stabilize the +1 eigenspace of $+Y_L$ (right diagonal) and the $-1$ eigenspace of $-Y_L$ (left diagonal). The path that the oscillator state follows (during dissipative stabilization using dissipators engineered from these stabilizers) go along the marked black arrows in the figure. We denote the various regions of phase space displacements by colors which indicate whether, after several rounds of error correction, the state would be mapped to a logical $\{\text{no error}, X_L, Y_L, Z_L\}$. The black arrows cross the regions of $X_L, Z_L$ errors only at corners marked in red. If ancilla decay occurs anywhere apart from these points on the paths marked by the black arrows then the state suffers with no error since a $Y_L$ does not affect the logical information contained in a $\ket{Y}_L$ state. These points can be seen as the isthmus with harmful error regions on either sides. The probability of falling into either region decreases with decreasing $\Delta$. This is the isthmus property referred to above in the context of multimode GKP stabilization. Due to the extra degrees of freedom in two-mode GKP codes, all stabilizers can be measured using a path involving an \emph{isthmus} between erroneous regions; see Ch.\ref{ssec:twomode_GKP_qec} for further details.

\subsubsection{Unitary operations}
As we have discussed, when logical gates designed for the infinite-energy GKP codespace are performed on finite-energy GKP states, finite-squeezing effects can sometimes be thought of as additional noise. However, an alternative approach that is an active topic of research is to consider the finite-energy GKP manifold as a valid code space and modify all gates to account for finite-squeezing effects. As a result, in some representations, gates can become non-unitary, requiring dissipation.

Code deformation to finite-energy GKP states by the envelope operator $E_\Delta=e^{-\Delta^2\hat n}$, as discussed in Chapter~\ref{sec:finite_GKP}, transforms all logical operators to non-unitary operations with Hamiltonians of the same order. Gates that commute with $E_\Delta$, such as the Fourier gate, are unchanged. In Ref.~\cite{baptiste2022multiGKP}, it was shown that the square-root Hadamard gate for the single-mode computational square code is $\sqrt{\text{Had}} = e^{i \frac{\pi}{8} n^2}$ which also commutes with the envelope operator. Interestingly, as a result, this is a non-Clifford gate which could be performed without finite-squeezing modification. However, we do not know a valid teleportation and distillation scheme for the resource state which applies this gate. 

One method to find the finite-energy representation of infinite-energy gates is to apply the map $\mathcal{E}_\Delta(\bullet)\coloneqq E_{\Delta} \bullet E_{\Delta}^{-1}$. As an example, for single-mode infinite-energy computational GKP states, a phase gate and controlled-Z gate can be achieved using $P_{\mathrm{ideal}} = e^{iq^2}$ and $CZ_{\mathrm{ideal}} =e^{i\hat q\otimes \hat q}$, respectively~\cite{gkp2001}. These gates do not commute with the envelope operator and must be modified to account for finite-energy effects. They are transformed into the analogous finite-energy operations by conjugating them with the envelope operator $E_\Delta$ according to 
    \begin{align}
EP_{\mathrm{ideal}}E^{-1}=Ee^{i\hat{q}^2}E^{-1}&=e^{i(\cosh{\Delta^2}\hat{q}+i\sinh{\Delta^2}\hat{p})^2}\\
    &\approx e^{i(\hat{q}^2-\Delta^4\hat{p}^2+i\Delta^2\{\hat{q},\hat{p}\})}\quad\text{(Finite-energy Phase gate)}\\    ECZ_{\mathrm{ideal}}E^{-1}=Ee^{i\hat{q}\otimes\hat{q}}E^{-1}&=e^{i(\cosh{\Delta^2}\hat{q}+i\sinh{\Delta^2}\hat{p})\otimes(\cosh{\Delta^2}\hat{q}+i\sinh{\Delta^2}\hat{p})} \\    
    &\approx e^{i(\hat{q}\otimes\hat{q}-\Delta^4\hat{p}\otimes\hat{p}+i\Delta^2(\hat{q}\otimes\hat{p}+\hat{p}\otimes\hat{q}))}\quad \text{(Finite-energy CZ gate)}
    \end{align}
where the approximations hold in the small $\Delta$ limit such that $\cosh{\Delta^2}\approx 1$ and $\sinh{\Delta^2}\approx \Delta^2$. In Ref.~\cite{rojkov2023twoqubitStabiliz}, a circuit was proposed to perform these non-unitary gates using an auxiliary qubit through the dissipation model described in Ref.~\cite{royer2020stabilization}. The fidelity for the circuit obtained with first-order trotterization is reported to be $60\%$. The authors have shown that further trotterization can improve the circuit, however this would entail the use of ancilla for a longer time, subjecting the gate fidelity to be limited by ancilla errors.

In Ref.~\cite{singh2023composite}, the authors utilized an analytic framework to derive circuits for finite-energy GKP codespace using an auxiliary qubit. The authors present a measurement-free gate teleportation circuit using a single-qubit ancilla which corrects on the codespace while applying a logical operation. This unique property follows from the fact that the teleportation circuit is derived from the sBs protocol used for dissipative stabilization, as discussed in Chapter~\ref{sec:qubit-dissipation}. While the oscillator errors are taken care of in the circuits referred to in Ref.~\cite{singh2023composite}, the teleportation circuits are not fault-tolerant to ancilla errors. In order to achieve high fidelity in the presence of biased-noise ancilla, the authors construct piece-wise circuits that significantly mitigate the effects of ancilla dephasing. 

\begin{figure}
    \centering
\includegraphics[width=\linewidth]{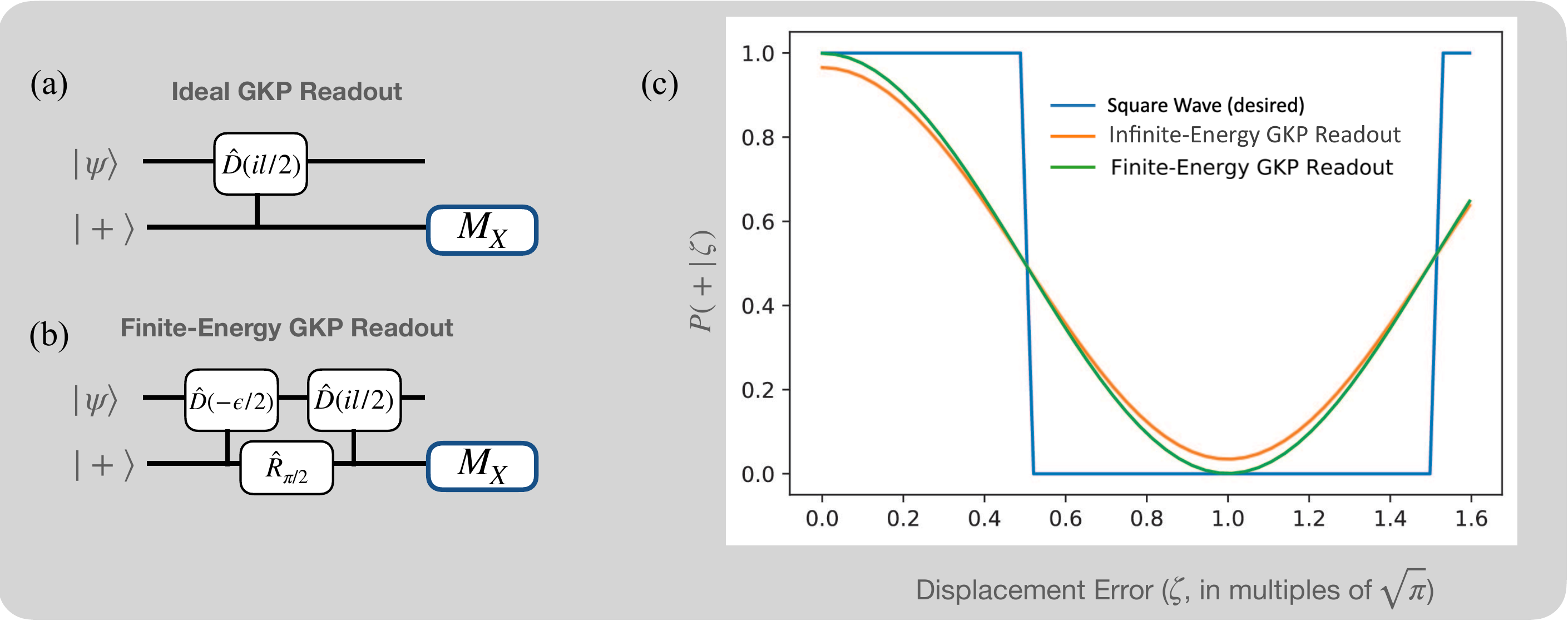}
\caption{\label{fig:GKP_readout}Readout fidelity of displaced GKP states~\cite{singh2023composite}. Here, $l=\sqrt{\pi/2}$. (a)Infinite-Energy Readout GKP Readout circuit (b) Finite-Energy GKP readout circuit~\cite{hastrup2021improved}. The circuit was optimized for $\epsilon=2l\Delta^2$, where $\Delta$ is the envelope-size of the concerned GKP state (see Sec.~\ref{sec:finite_GKP}). (c) Readout fidelity as a function of translation error. The graph plots the probability of getting the ancilla outcome $=+1$ upon measurement using the readout circuits shown in (a) and (b) versus the translation error in the logical state $\ket{\psi}=\ket{0_L}$. Ideally we would like the fidelity to be constant for all states with correctable displacement errors, see the blue curve. Finite-energy readout is better approximation to finite energy readout circuits compared to ideal readout fidelity, see the peak of the green and orange curves at zero translation errors. With increase in the translation error, the fidelities drop as a cosine function of $\zeta$, amount of translation error in the state.}
    
\end{figure}

It was shown in Ref.~\cite{baragiola2019GKPuniversal} that GKP states can achieve universality with the availability of Gaussian operations and logical Pauli eigenstates, modulo finite-energy corrections. The authors suggest preparing the GKP magic state using only beam-splitters and Pauli logical states. This direction is quite promising and at the heart of solving the issue of resource overhead for practical quantum computing~\cite{fowler2013surface}, however, it requires some additional analyses of how the finite-energy envelope affect the fidelity of the final state and what kind of corrections it entails.

\subsubsection{Logical readout}
\label{ssec:logical readout}
For optical systems, it is favorable to use Homodyne detection for logical readout of GKP states. However, this readout scheme is not well-suited to superconducting systems due to low measurement efficiency~\cite{puri2021rvw} (see details in Section.~\ref{sec:Gaussian_measurement}). Using the circuit shown in Fig. \ref{fig:GKP_readout}a, logical measurement of infinite-energy GKP operators
can be done using conditional displacements, which has been used in Refs.~\cite{campagne2020quantum,sivak2022breakeven}, however these measurements are not optimal for finite-energy GKP since the readout fidelity depends on the GKP squeezing. 

In Ref.~\cite{hastrup2021improved}, the authors appended a small conditional displacement on the ideal logical $X_L$ measurement circuit and found the readout fidelity to increase by an order of magnitude, as shown in Fig.~\ref{fig:GKP_readout}b. The magnitude of the added conditional displacement was optimized and found to be optimal when equal to $\frac{\sqrt{\pi}}{2}\Delta^2$, similar to the small displacement in the big-small-big protocol. This suggests that the circuit is closer to the logical measurement circuit for finite-energy GKP states. 

Finally, we would like to point out that the readout fidelity drops in the presence of correctable displacement errors following a cosine with the maxima at zero displacement error (see Figure.~\ref{fig:GKP_readout}c).
This issue requires one to perform a phase estimation type measurement using multiple qubits which might have worse back action on the state and also exposes the state to ancilla errors for a longer time. All these considerations indicate that there is still need for improvement in the universal control of logical GKP operations beyond achieving Gaussian control.

\subsection{Optical GKP: Proposals and challenges}\label{sec:optical_gkp}

\begin{figure}
    \centering
    \includegraphics[width=.85\linewidth]{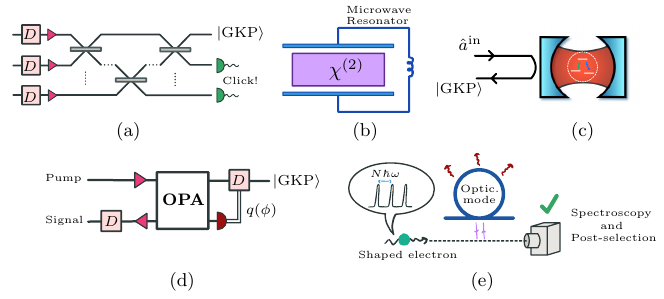}
    \caption{Schematic of proposed optical GKP state-generation devices. (a) Gaussian boson sampling (GBS) device~\cite{sabapathy2019QMLNonGauss,su2019pnrNonGauss,tzitrin2020ProgressGKP,bourassa2021blueprint}. (b) Microwave-to-optical transducer---e.g., a $\chi^{(2)}$ non-linear material embedded in a microwave resonator~\cite{lauk2020PerspTransdux,quics2021prxQuantum,jiang2021OpticaTrdxRvw}. (c) Reflection of an optical squeezed state off an optical cavity hosting a three-level system~\cite{hastrup2022cQED}. (d) Non-linear dynamics of a optical parametric amplifier (OPA)~\cite{yanagimoto2023GKPviaOPA}. \QZ{(e) Shaped, free electron interacting with an optical mode~\cite{dahan2023prxElectronGKP}}.}
    \label{fig:optical_GKP}
\end{figure}

To generate non-Gaussian states---such as GKP states---strong non-linearities are generally required. In the previous sections, we saw that, for microwave resonators, coupling a microwave mode to a nearby auxiliary qubit facilitates the necessary non-linearity. The situation is a bit more challenging in optics because the non-linearities are produced via third or higher-order photon-photon interactions in a non-linear material which are typically quite weak. \QZ{Recent developments---as evidenced by preliminary work in Ref.~\cite{konno2023propagatingGKP}---reveal initial glimpses of success, however these efforts currently exhibit relatively low fidelity and squeezing around 2.5 dB, below that of other platforms, such as microwave circuits and trapped ions, and what is required for fault-tolerant quantum computing with GKP states.} Therefore, it is imperative to maintain an open-minded approach and investigate various proposals that could facilitate the realization of high-fidelity, highly squeezed GKP states within the optical domain. In what follows, we discuss five proposals for optical GKP state generation; see Figure~\ref{fig:optical_GKP} for an illustration.

A popular approach to circumvent the weak non-linearities in optics is to probabilistically generate GKP (and other non-Gaussian) states via Gaussian boson sampling (GBS) devices~\cite{sabapathy2019QMLNonGauss,su2019pnrNonGauss,quesada2019NGaussPrep,tzitrin2020ProgressGKP,bourassa2021blueprint,takase2023GKP_synthesizer,konno2023propagatingGKP} (see also Refs.~\cite{Eaton2019catalysis,eaton2022MBQCgkpCreation} for similar but distinct protocols about PNR measurement-based GKP state-generation). The GBS device works as follows [see Figure~\ref{fig:optical_GKP}(a)]: Displaced squeezed vacua are injected into a multimode linear-optical network consisting of beamsplitters and phase-shifters. At the output, all modes but one are projected onto photon-number states via photon number resolving (PNR) detectors. Depending on the outcomes of the PNR measurements, the remaining mode collapses to the target (e.g., GKP) state with some probability of success. This scheme is probabilistic but heralded. Many GBS devices can then be multiplexed together to increase the success probability, however this comes with a large resource overhead~\cite{bourassa2021blueprint}. For instance, if $p_{\rm GBS}$ is the probability to successfully generate a (high-fidelity) GKP state from one GBS device,\footnote{We further note that there is a tradeoff between the fidelity of the target GKP state and the success probability $p_{\rm GBS}$~\cite{bourassa2021blueprint}.} then the probability for successfully generating a GKP state with $N_{\rm GBS}$ multiplexed devices is $p_{N-\rm GBS}\sim 1-(1-p_{\rm GBS})^{N_{\rm GBS}}$. Moreover, this GKP-via-GBS protocol relies on PNR detectors that must be operated at cryogenic temperatures. We note, however, that PNR detectors are the only elements in these optical quantum information processors that require such low temperatures.

An alternative but fairly ``straightforward'' way to generate optical GKP states is to directly transduce them from the microwave domain via an integrated optical-microwave device; see Figure~\ref{fig:optical_GKP}(b). For instance, one could have a cQED setup that creates GKP states in the microwave domain (something similar to architectures discussed in the previous section) but with the added functionality of quantum transduction~\cite{lauk2020PerspTransdux,quics2021prxQuantum,jiang2021OpticaTrdxRvw}. Transduction can be achieved by, for instance, embedding a non-linear (e.g., $\chi^{(2)}$) optical material in a microwave resonator and driving the optical material with a strong pump field~\cite{wangZorzetti2022transdux}. From the non-linear interactions induced by the pump, one can realize a beamsplitter-like interaction between the microwave and optical modes, thereby enabling one to route photons from microwave to optical frequencies (or vice versa); this is known as a direct conversion scheme. The caveat here is the low-conversion efficiencies ($\ll 50\%$) associated with such devices which can severely degrade the fidelity of non-classical output states and render them useless for quantum information processing~\cite{jiang2021OpticaTrdxRvw}. Recently, it has been proposed that squeezing can help to bypass the efficiency bottleneck of direct-conversion schemes~\cite{zhong2022SqueezTransdux}. Another workaround is to utilize two-mode squeezing interactions, as opposed to beamsplitter-like interactions, followed by a CV quantum teleporation protocol to effectively transduce the quantum state from microwave to optical frequencies~\cite{wu2021transduction}; it has been argued that this teleporation-based transduction scheme can outperform direct conversion schemes in large regions of parameter space.

Hastrup and Anderson~\cite{hastrup2022cQED} proposed an auxiliary cavity QED setup to generate optical GKP states. In their protocol, an optical squeezed state is iteratively displaced then reflected off an optical cavity hosting a three-level ``atom''~\cite{hastrup2022cQED}; see also Figure~\ref{fig:optical_GKP}(c). More iterations result in a GKP state with more peaks, however the protocol is ultimately limited by the cooperativity of the cavity (similar to transduction methods), suggesting that generating highly squeezed ($>10$ dB) GKP states with this method may be challenging in the near term.

Yanagimoto et al~\cite{yanagimoto2023GKPviaOPA} proposed an entirely optical GKP state-generation setup based on the non-linear dynamics of an optical parametric amplifier (OPA); see Figure~\ref{fig:optical_GKP}(d) for an illustration. In this proposal, a pump mode initialized in a squeezed vacuum and a signal mode initialized in a displace-squeezed state interact via three-wave mixing through the OPA.  Performing general-dyne measurements on the signal mode realizes a quantum non-demolition (QND) measurement on the pump (or vice versa), enabling deterministic generation of non-Gaussian states. Specifically, by projecting the signal mode onto a displaced squeezed state, a QND modular homodyne measurement is performed on the pump, resulting in the pump mode being projected onto an approximate GKP state (up to feed-forward displacements). One advantage of this scheme is that it does not require the integration of microwave and optical quantum technologies. In addition, the use of homodyne detectors instead of PNR detectors eliminates the need for cryogenic temperatures. The caveat though is the use of (generally weak) three-wave mixing, however the authors argue that their scheme appears promising in the near-term with non-linear nanophotonic structures~\cite{zhu2021integratedLiNrvw,moody2022roadmapQuPhotonics, lu2020toward1percent,yanagimoto2022tempTrapping}.

\QZ{Finally, Dahan et al~\cite{dahan2023prxElectronGKP} (see also Ref.~\cite{baranes2023ElectronGKPcontrol}) proposed a unique method for generating optical GKP states by coupling shaped, free-electrons to an optical mode. Their study demonstrates that the electron-photon coupling facilitates a conditional displacement on the optical mode, analogous to circuit QED setups. They also quantify the amount of squeezing (in dB) of the process, showing that the squeezing scales logarithmically with the number of free electrons, while the probability of successfully heralding a GKP state (of certain amount of squeezing) scales inversely with the number of free electrons. Projections indicate a GKP state with 10dB of squeezing is possible at $10\%$ success probability, assuming initial vacuum for the optical mode. Higher success rates ($\sim30\%$) can be achieved by initializing in a squeezed vacuum state.}

\subsection{Scaling up GKP codes}
\label{sec:scaling_up}

\begin{figure}
    \centering
    \includegraphics[width=.85\linewidth]{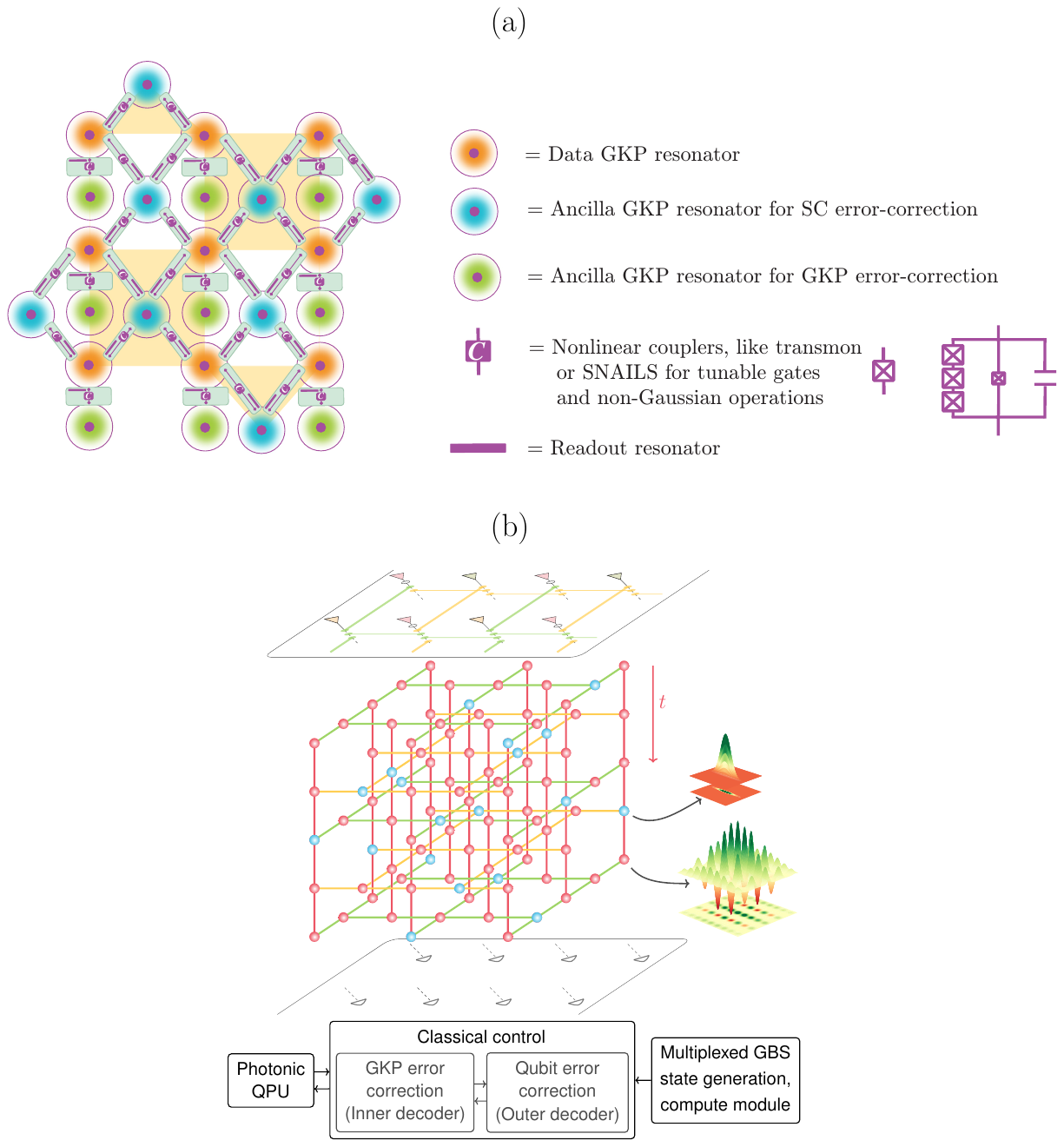}
    \caption{Scaling up with GKP codes. (a) A schematic of a low-level chip architecture for the all-GKP surface code~\cite{puri2021rvw}. Yellow and white regions indicate $X$ and $Z$ stabilizers for the surface code. The architecture includes nonlinear couplers, which could be implemented as three-wave mixers \cite{Frattini_2018, Sivak_2019,Ye_2021} or four-wave mixers, such as a transmon \cite{Koch_2007}. The ancilla GKP resonators and measure GKP qubits could be replaced with a biased-noise qubit, such as a Kerr-cat \cite{Puri_kerr_cat, grimm2020stabilization}, or an auxiliary two-level system, such as a transmon. Additionally, as shown here, each coupler includes an on-chip readout resonator. Such a GKP surface code could also be realized in a planar (2D) superconducting circuit architecture. (b) Proposed architecture for GKP cluster-state generation for MBQC with a photonic chip~\cite{bourassa2021blueprint}. The chip (top) generates a resource state---a GKP cluster state---in the time domain via switchable beam-splitters, controllable phase shifters, in-line squeezing, and delay lines. The GKP cluster state serves as the basis for fault-tolerant computation, which can be achieved through a measurement-based version of lattice surgery tailored for the surface code~\cite{bourassa2021blueprint, brown2020MBQC_surface}. An array of homodyne detectors (bottom) are used for performing stabilizer measurements and measurement-based logical operations. By integrating components on a photonic chip, the proposed architecture provides a promising pathway for realizing MBQC with bosonic modes.}
    \label{fig:concatenation}
\end{figure}

With the experimental intricacies discussed in this chapter fresh in mind, we now offer a concise perspective on the scalability of GKP codes, particularly for applications in fault-tolerant quantum computing (FTQC), and briefly touch upon the challenges involved. For a deeper exploration of the theoretical and mathematical aspects of scaling up GKP codes using multimode codes, such as concatenated qubit codes, readers can refer to Chapter~\ref{sec:QEC-multimode}. For more in-depth discussions on FTQC with GKP codes, additional references, and further insights, please consult Chapter~\ref{sec:app_ftqc}.

Single-mode GKP codes exhibit a constant distance against translation errors and are limited by errors of the auxiliary qubit in the current superconducting circuit approach. Even though GKP qubits have set the record in QEC gain, a practical quantum memory necessitates significant improvements in error rates and qubit lifetimes by several orders of magnitude. To address this, a promising direction involves concatenating GKP codes with qubit codes (see Chapter~\ref{sec:QEC-multimode}), like surface codes which possess the highest-known threshold to-date~\cite{noh2020fault,noh2022low,Zhang2022GKPxzzx,terhal2020towards,vuillot2019toric}. Below the threshold value of the concatenated code, logical error rates can be decreased by increasing the size of the code; see Chapter~\ref{sec:app_ftqc} for further discussion. Furthermore, by using GKP qubits as the base layer of a concatenated code, we can achieve resource reduction in terms of the number of qubits required to achieve a target lifetime. Concatenated codes are a subset of a broader class of multimode GKP encoding introduced in Ref.~\cite{gkp2001}. We postpone the discussion of error correction with multimode encodings until the next section.

Scaling up GKP codes to realize multi-oscillator encodings will require the engineering of bosonic modes at scale, and the challenges faced will be different than small-scale experiments using one or two modes. It is likely that a co-design approach will be needed to realize a scalable bosonic architecture while accounting for trade-offs in physical lifetimes, stability, cross-talk, addressability, connectivity, and modularity. In Figure~\ref{fig:concatenation}, we show schematics for potential realizations of scalable fault-tolerant quantum computing architectures based on GKP codes in a superconducting architecture (Fig.~\ref{fig:concatenation}(a); see Ref.~\cite{puri2021rvw}) and a photonic chip (Fig.~\ref{fig:concatenation}(b); see Ref.~\cite{bourassa2021blueprint}). For superconducting circuit architectures, advances in the materials science and engineering of other types of superconducting qubits and codes, such as surface codes realized through arrays of planar transmons, could be adapted in the multimode bosonic code setting. For photonic architectures, the biggest hurdle is first generating an optical GKP state, which may require a large overhead due to the use of cryogenic components and mode multiplexing (see Chapter~\ref{sec:optical_gkp}). Furthermore, integrating all the necessary components for MBQC with GKP states on a chip is a grand challenge, especially due to the inline active operations (e.g., squeezing and displacements) that are typically required~\cite{bourassa2021blueprint}. Figure~\ref{fig:concatenation} shows the architecture for scaling up the GKP codes to a multi-oscillator code for both superconduting as well as photonic systems. Despite the engineering challenges ahead, we are optimistic about the future of scalable bosonic architectures, including those built for realizing GKP encodings, and we anticipate rapid progress in this direction in the coming years.



\section{QEC with Multimode GKP Qubit Codes}
 \label{sec:QEC-multimode}

There has been a recent surge of interest in understanding the properties and performance of multimode GKP qubit codes~\cite{harrington2001rates,baptiste2022multiGKP,conrad2022lattice,conrad2023good,lin2023surfaceGKP}. In Chapter~\ref{sec:gkp_lattice}, we mathematically demonstrated how to encode a qubit into a multimode ($2N$-dimensional) lattice; see Table~\ref{tab:lattice-code} for a quick overview. Nevertheless, we have yet to emphasize the potential benefits or QEC properties of such encodings. Intuitively, we anticipate that a multimode GKP qubit code will outperform a single-mode GKP qubit simply due to ``diffusion'' of logical information across a larger space---the $2N$-dimensional phase of $N$ bosonic modes---as opposed to the smaller 2-dimensional phase space for single mode codes; we make this concept more concrete in this section. Of course, it is also important to consider the trade-offs associated with multimode encoding, such as the potential increase in susceptibility to certain types of errors, e.g., crosstalk between neighboring modes.

In this chapter, we dive deeper into multimode GKP qubit codes, discuss why such encodings might be beneficial for QEC, and assess the QEC performance of some specific multimode codes---such as the tesseract and $D_4$ qubit codes. Additionally, we briefly discuss concatenating (inner layer) GKP qubit codes with (outer layer) discrete variable (DV) codes, which may prove useful for designing and analyzing fault-tolerant quantum computing architectures based on GKP codes.


\subsection{Why might a multimode GKP qubit be good for QEC?}
\label{ssec:multimodeGKP_better}


Theoretical studies~\cite{gkp2001,harrington2001rates,baptiste2022multiGKP,conrad2022lattice,conrad2023good,lin2023surfaceGKP} suggest that encoding a qubit into many modes via higher dimensional lattices is a robust and effective way to deal with noise. For instance, shortly after the conception of GKP states~\cite{gkp2001}, Harrington and Preskill~\cite{harrington2001rates} argued that high-dimensional ($N\rightarrow\infty$) GKP lattice codes can achieve the one-shot quantum capacity (optimized over Gaussian inputs) of the AGN channel [the lower bound on the AGN quantum capacity shown in Eq.~\eqref{eq:agn_bounds}]. These results have since been extended to thermal-loss channels~\cite{noh2019quantumcapacity}; see Chapter~\ref{subsec:apps_comms} for details. These findings suggest that multimode GKP codes might not only be good for QEC but are perhaps optimal (at least for Gaussian noise channels). Using the mathematical formalism established in this review, we provide some simple arguments as to why multimode GKP codes---and thus higher-dimensional lattices---might be better for QEC than single-mode encodings. 

We must first quantify the performance of a QEC code; for simplicity, we use the single-qubit error probability. We define the single-qubit error probability $p_{X\cup Y\cup Z}$ as the likelihood of any Pauli error occurring on the encoded qubit. To simplify the presentation, we take the union bound approximation,
\begin{equation}\label{eq:pe_defn}
    p_{X\cup Y\cup Z}\leq p_e\coloneqq \sum_{J\in\{X,Y,Z\}} p_J,
\end{equation}
which is accurate up to events with simultaneous errors that occur with probability $\sim\order{p_J^2}$. We emphasize that this is the probability of error induced by AGN on the GKP single-qubit codespace. Generically, this differs from the logical error rate $p_{L}$ that could be derived once an explicit decoding algorithm is given for the multimode code. Quite generally though, we have that $p_e \leq p_{L}$.

The task now is to identify lattices that yield good GKP codes. In general, quantifying what constitutes a ``good GKP code'' can be challenging. For example, given an iid AGN channel (a mathematical simplification of course), codes with larger Pauli distances are good quantum codes, at least indicated by the overly simplistic metric of $p_e$, as we show explicitly just below. However, this is not generically viable. One primary obstacle is constructing a good decoding algorithm and error correction procedure to infer the magnitude \emph{and} direction of the random shifts, a generally hard problem (more on this later). Furthermore, the encoding rate $R\coloneqq k/N$, where $k$ is the number of encoded logical qubits and $N$ is the number of oscillators or modes, likewise plays a crucial role in gauging performance. By definition, the encoding rate $R$ necessarily \textit{decreases} along with $p_e$\footnote{Of course, for $p_e$ to decrease, good lattices in higher dimensions must be considered.} as we go to higher dimensional codes, thereby indicating a tradeoff between a higher encoding rate and lower error probability. In follow-up sections, we briefly discuss decoding of GKP codes and highlight interesting subtleties arising in  practical realizations (e.g., sBs-type QEC protocols with a cQED architecture in mind). For the moment, we forget about these important subtleties in order to illuminate the potential benefits of scaling to many modes.

It turns out that, for iid random displacements $\bm e\sim\mathcal{N}(0,\sigma^2\bm I_{2N})$, we can compute the Pauli error probabilities $p_J$ exactly and thus estimate the total error probability $p_e$. Let $\bm j\in\{\bm x,\bm y, \bm z\}$ be the (shortest) Pauli displacement vector associated with a logical Pauli $J_L\in\{X_L,Y_L,Z_L\}$ in units $\ell=\sqrt{2\pi}$, such that $\norm{\bm j}$ is the $J$th Pauli distance of the code (see Chapter~\ref{sec:comp_lattice} for details). Then, the probability for a $J_L$ Pauli error to occur is simply,
\begin{equation}\label{eq:agn_pe}
    p_J\approx {\rm erfc}\left(\sqrt{\frac{\ell^2\norm{\bm j}^2}{8\sigma^2}}\right),
\end{equation}
where $\ell=\sqrt{2\pi}$. This can be found by integrating the multivariate Gaussian distribution of the AGN channel outside of the \emph{correctable region} $\mathscr{C}\coloneqq\left[-\frac{\ell\norm{\bm j}}{2},\frac{\ell\norm{\bm j}}{2}\right]$ along the direction $\hat{\bm\jmath}\cdot\bm e$. The approximation can be made more precise by considering higher-order correctable regions (corresponding to the likelihood of even numbers of Pauli errors, e.g. $X_L^{2k}$), however for small $\sigma$, the above estimate is very good.

Since $\rm{erfc}(x)$ is exponentially decreasing in $x$, we deduce that codes with fixed noise $\sigma$ but larger Pauli distances $\norm{\bm j}$ will exhibit substantially better code performance in terms of $p_e$. We saw several examples of extending Pauli lengths by going to higher dimensional GKP codes in Chapter~\ref{sec:gkp_lattice}. For instance, a two-dimensional (single-mode) square GKP qubit has $\norm{\bm x_\square}=\norm{\bm z_\square}=1/\sqrt{2}$ and $\norm{\bm y_\square}=1$. On the other hand, we can encode the qubit into a four-dimensional hypercube via the tesseract code [equivalent to a rotated four-dimensional hypercube; see Eq.~\eqref{eq:tess_M}] with $\norm{\bm x_{\widetilde{\square}^2}}=\norm{\bm z_{\widetilde{\square}^2}}=1/\sqrt[4]{2}$ and $\norm{\bm y_{\widetilde{\square}^2}}=\sqrt[4]{2}$. The logical Pauli's of the four-dimensional (two-mode) hypercube are $\sqrt[4]{2}$ times longer than the two-dimensional square qubit code. To illustrate the relative performance enhancement, we plot the error probability for square, tesseract, and $D_4$ qubit encodings in Figure~\ref{fig:pe_compare}(a) assuming ideal GKP states and ideal error correction for iid AGN.


\subsection{Assessing practical QEC of two-mode GKP qubit codes}\label{ssec:twomode_GKP_qec}



In the previous section, we argued that encoding a qubit into many modes and choosing ``better lattices'' can, in principle, lower single-qubit error probabilities and improve QEC performance. However, we made some implicit assumptions in arriving at these conclusions. In particular, we assumed \emph{perfect error correction}, in which case we can infer the magnitude \emph{and} direction of random displacements (up to logical Paulis) directly from syndrome measurements, in which case $p_e$ directly quantifies the logical error rate. This, in turn, led us to the result that the error probability [e.g., derived from Eq.~\eqref{eq:agn_pe}] is the defining metric and, consequently, that the performance of the code depends solely on the lengths of the shortest Pauli vectors, which can be relatively large (compared to, e.g., a square lattice) for multimode encodings. We discuss here why these assumptions are too idealistic.

For concreteness, we make a very specific comparison of two alternative two-mode (i.e., four-dimensional lattice) qubit codes: (i) the $D_4$ lattice code and (ii) the tesseract code; see Table~\ref{tab:2-mode-GKP} for an overview of the two codes. We also include the single-mode square code in our analyses for a baseline comparison. Our investigation focuses on the error correction capabilities of each code in two different scenarios: (1) by comparing the error probabilities [Eq.\eqref{eq:agn_pe}] for the different codes and (2) by evaluating the channel infidelities when auxiliary-qubit-based dissipative stabilization is employed for QEC. The results are succinctly summarized in Fig.~\ref{fig:pe_compare}. Given the assumptions described at the beginning of this section, the $D_4$ lattice should, in principle, have optimal performance in four dimensions due to longer code distance, however this is no longer the case when qubit-based stabilization is considered, as we elaborate further below. 

\begin{table}
\renewcommand{\arraystretch}{1.2}
    \centering
    \begin{tabular}{c | c  c}
    \hline\hline\\
        & Tesseract Code & $D_4$ Code  \\
        \hline\hline
        &&\\
         Generator Matrix & $\bm M(\widetilde{\square}^2)=\sqrt[4]{2}\begin{pmatrix}
             1&0&0&0\\
             0&\frac{1}{\sqrt{2}}&0&\frac{1}{\sqrt{2}}\\
             0&0&1&0\\
             0&\frac{1}{\sqrt{2}}&0&-\frac{1}{\sqrt{2}}
         \end{pmatrix}$& 
             \QZ{$\bm M({D_4})=\begin{pmatrix}
             1&1&0&1\\
             0&0&1&0\\
             1&0&-1&0\\
             0&-1&0&1
             \end{pmatrix}$}
         \\
         \hline
         &&\\
         Minimum Stabilizer length & $\sqrt[4]{2}$ & $\sqrt{2}$ \\
         \hline
         &&\\
         Logical Operators &$
             \begin{pmatrix}
                 \bm x^\top\\
                 \bm y^\top\\
                 \bm z^\top
             \end{pmatrix}=\sqrt[4]{2}\begin{pmatrix}
             \frac{1}{2} & 0 & \frac{1}{2} & 0\\
             \frac{1}{2} & \frac{1}{\sqrt{2}} & \frac{1}{2} & 0\\
             0 & \frac{1}{\sqrt{2}} & 0 & 0
             \end{pmatrix}$ &
             $\begin{pmatrix}
                 \bm x^\top\\
                 \bm y^\top \\
                 \bm z^\top
             \end{pmatrix}=\begin{pmatrix}
             \frac{1}{2}&\frac{1}{2}&\frac{1}{2}&\frac{1}{2}\\
             -\frac{1}{2}&\frac{1}{2}&\frac{1}{2}&\frac{1}{2}\\
             1&0&0&0
             \end{pmatrix}$ \\
         \hline
         &&\\
         Minimum Pauli length &$1/\sqrt[4]{2}$ & $1$ \\
         \hline
         &&\\
         Hadamard & $U_{\bm{R}(\pi/2)}^{\otimes 2} U_{\bm B(1/2)}$ &$U_{\bm{R}(\pi/4)}^{\otimes 2} U_{\bm B(1/2)}$\\
         \hline
         &&\\
         Phase Gate &$e^{-\frac{i}{\sqrt{2}}\hat{q}_A^2}\otimes I_A$ & $U_{\bm{R_A}(\pi/2)}\otimes I_A$\\
         \hline
         &&\\
         $CZ$ & $e^{-i\sqrt{2}\hat{q}_{A,1}\otimes \hat{q}_{B,1}}$&$e^{-i2 \hat{q}_{A,1}\otimes \hat{q}_{B,1}}$\\
         \hline
         &&\\
         Non-Clifford Gate &$\sqrt{H}=e^{i\frac{\pi}{32}\hat{F}^2}$&$\sqrt{H}=e^{i\frac{\pi}{8}\hat{n}_A^2}$\\
         \hline\hline
    \end{tabular}
    \caption{Two-mode GKP qubit code comparison: Tesseract code vs $D_4$ code. The table is inspired from Ref.~\cite{baptiste2022multiGKP} where the Tesseract code was first introduced and a practical set of universal gate operations for both codes were discussed. There are other options for gate operations, but the ones presented here are similar to (or better than) other available options in terms of the finite-energy GKP errors. $\bm R(\theta)$ is the single-mode phase-space rotation gate and $\bm B(T)$ is the beamsplitter interaction with transmission probability $T$; see Chapter~\ref{sec:gauss_unitaries}. Both these operations are envelope-preserving on the two modes. The Hadamard gate is also an envelope-preserving gate and thus does not suffer from envelope errors. The phase gate on the other hand is only envelope-preserving for the $D_4$ code. The two-mode entangling gate $CZ$ is the same for the both codes and equal to the corresponding entangling gate for single-mode GKP codes. Here $\hat{q}_{A,1}$ represents the $\hat{q}$ quadrature of the first mode (1) in the first qubit ($A$). Finally, for choices of non-Clifford gates, the Tesseract code can use quartic Hamiltonians like the Kerr-interaction to apply the square root of Hadamard, $\sqrt{H}$. Here $n_A=a_A^\dagger a_A$ is the number operator of the first mode while $F=\hat q_1^2+\hat p_1^2+\hat q_2^2+\hat p_2^2-2+\hat q_1\hat p_2-\hat p_1\hat q_2$. Justifications for these gates can be found in Ref.~\cite{baptiste2022multiGKP}. In lieu of the complicated code construction from classical lattices, ``easy'' gate implementations are available for multimode codes without increased difficulty (relative to single-mode codes).}
    \label{tab:2-mode-GKP}
\end{table}

\subsubsection{Comparison: oscillator errors}\label{sssec:osc-errors}

\begin{figure}
    \centering
    \includegraphics[width=.8\linewidth]{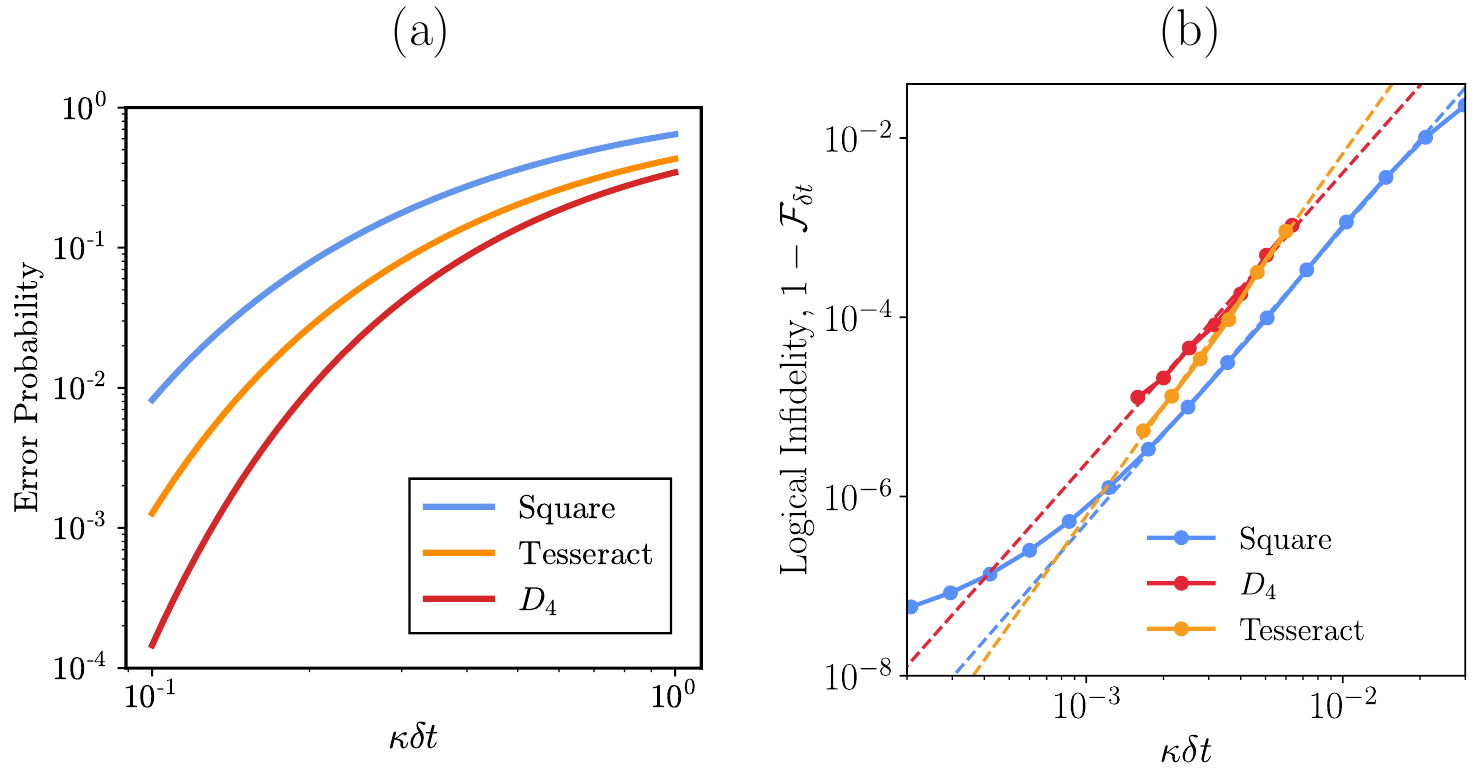}
    \caption{Comparison of multimode qubit codes. (a) Error probability $p_e\coloneqq \sum_{J\in\{X,Y,Z} p_J$ for infinite-energy GKP states, assuming noise conversion from loss (with loss rate $\kappa$) to AGN via pre-amplification. The AGN is related to the loss probability via $\sigma^2=1-e^{-\kappa\delta t}$. The plot was generated using Eq.~\eqref{eq:agn_pe}. Note that, although the two-mode Tesseract and D4 codes exhibit better performance as characterized by $p_e$, the encoding rate of these codes is half that of the single-mode square code. (b) Logical infidelity extracted from sBs type bit-wise decoding against amplitude damping (with rate $\kappa$) for different finite-energy GKP codes ($\bar n=5$ per mode). The channel fidelity is computed from $\mathcal{F}_t=(1+\sum_{\alpha}e^{-\gamma_\alpha t})/4$ where $\alpha\in\{X,Y,Z\}$, the rates $\gamma_\alpha$ are the decay rates of each Pauli eigenstate, and $\delta t$ corresponds to the time between each round of sBs. These rates are obtained by fitting an exponential decay to the time evolution of the associated logical Pauli operator after the projection measurement. Each logical damping rate is obtained by averaging over 200 trajectories. See Chapter~\ref{sec:qubit-dissipation} for a description of sBs. Figure 23(b) is adapted from Ref.~\cite{baptiste2022multiGKP}.}
    \label{fig:pe_compare}
\end{figure}

In Figure~\ref{fig:pe_compare}(a), we plot the error probability derived from Eq.~\eqref{eq:agn_pe} for the various codes assuming iid AGN. We see directly that the $D_4$ lattice code outperforms all other encodings by this metric. This is in line with the arguments presented in Chapter~\ref{ssec:multimodeGKP_better} in the context of densest lattice packings (i.e., longest Pauli vectors). However, the introduction of a practical error correction scheme brings about significant alterations to these results. 
In particular, the error probability (and thus the data plotted in Figure~\ref{fig:pe_compare}(a)) is a partial solution to the problem of GKP error correction since it assumes that the error vector $\bm e$---magnitude \textit{and} direction---can be inferred (modulo logical Paulis) from the $2N$ syndromes extracted from stabilizer measurements. However, in reality, the error channel displaces GKP states by a random vector in an unknown direction. This presents an added layer of complexity to the decoding problem.

It turns out that the Closest Vector Problem (CVP) is intimately related to the decoding problem for GKP codes. This problem---being NP-hard---has an exponential-time solution, the Micciancio-Voulgaris (MV) algorithm~\cite{micciancio2010deterministic}, which is $2^{O(n)}$ in space and time. The solution is based around the Voronoi cell of the lattice, which to recall is the correctable region of a GKP code. This feature makes the MV algorithm ideal for GKP error correction. Various different solutions to the CVP with comparable runtimes have been used in previous works to extract logical error rates of multimode codes~\cite{lin2023closest}. In contrast to the simple error probability metric, logical error rates from CVP-based decoding algorithms may exhibit a crossing point between curves of the same family, akin to a threshold-like behavior, as highlighted in Ref.~\cite{lin2023closest}. An example code family are the hypercubic lattices, e.g. square and Tesseract codes.\footnote{Note that the stabilizer matrix of a Tesseract code can be obtained from the concatenation of rectangular GKP codes with two-qubit repetition code. Thus, if we were to identify the family of codes with concatenation in mind, the generalization of this family would be as follows: $N=1$ (Rectangular GKP), $N=2$ (Two-qubit repetition code on rectangular GKP) as illustrated in~\cite{lin2023closest}. However, from the perspective of multimode codes, the Tessearct code (a 4-cube lattice) belongs to the family of hypercubic lattices whose generalization for $N=1$ case is a square GKP as used in this review.} It should be noted that such a crossing point cannot be strictly labeled as a critical point or threshold (as in the case of topological codes~\cite{dennis2002topological}) unless comparisons are made for a specific family of codes using a large number of modes to avoid any boundary effects. Finally, considering that the solution to CVP is exponential in time and space, for practical use cases, an approximate mapping of the problem to one with polynomial-time solution is desired, like minimum-weight perfect matching for topological codes~\cite{kolmogorov2009blossom}.

We now examine the scenario where sBs-type circuits (see section \ref{sec:qubit-dissipation}) are employed to stabilize the two-mode GKP codes; the corresponding results are plotted in Figure~\ref{fig:pe_compare}(b). We emphasize that the stabilization method here has been studied for photon loss errors in the absence of ancilla errors. Contrary to the behaviour of the error probability illustrated in Fig.~\ref{fig:pe_compare}(a), the $D_4$ lattice codes exhibits the poorest performance when sBs-type stabilization is used, as observed in Fig.~\ref{fig:pe_compare}(b). Intuitively, the longer stabilizers of $D_4$ code lead to lengthier QEC steps, resulting in an accumulation of errors throughout the process. One might expect that the longer logical Pauli vectors of the $D_4$ lattice could potentially mitigate this increased error accumulation. However, this expectation does not appear to manifest itself in practice. The reason is that sBs performs a bitwise error correction that alternates between two stabilizers; thus, the error correction will not be efficient unless the stabilizers are orthogonal. To see this, let us denote any two non-orthogonal stabilizers of the $D_4$ code by $S_1$ and $S_2$. During a bit-wise correction due to back action from the sBs map along $S_1$, the error along $S_2$ is affected in a random way unless explicitly taken care of in the process. Thus, such qubit-based dissipative stabilization, in its crudest sense, might not be the optimal error correction strategy for codes with non-orthogonal stabilizers like single-mode hexagonal codes and two-mode $D_4$ codes in comparison to hypercubic lattice codes, like the single-mode square code and two-mode Tesseract code. There might be cleverer tricks which could resolve this issue. We note that we could perform continuous error correction (as opposed to bit-wise extraction) by using a GKP measurement ancilla, such that the translation errors could be corrected along each stabilizer in a single step. Such error correction has been studied in terms of code concatenation~\cite{noh2020enhanced,noh2022low,Zhang2022GKPxzzx,terhal2020towards,vuillot2019toric}.   

Another interesting observation concerning Fig.~\ref{fig:pe_compare}(b) is that, under the error rates simulated in Ref.~\cite{baptiste2022multiGKP}, the Tesseract code performs less favorably than the single-mode square code. This may occur for the following reason: The GKP states utilized here contain an average of $5$ photons per mode, resulting in a total of $10$ photons for the two-mode codes compared to only $5$ photons in the single-mode code. Consequently, the two-mode code exhibits a higher total probability of photon loss than the single-mode code. However, it is worth noting that the curves for the simulated data are not parallel and might intersect at a lower photon loss rate $\kappa\delta t$. This tentatively implies the presence of a crossing point, akin the threshold behavior observed in qubit codes, below which two-mode codes may exhibit enhanced protection. Further investigations are required to explore this potentiality and generally understand its implications for the performance of multimode codes in a practical setting.



\subsubsection{Comparison: ancilla errors}
The authors of Ref.~\cite{baptiste2022multiGKP} noticed that an immediate advantage of two-mode codes can be seen in terms of protection against ancilla errors. The state-of-the-art beyond break-even GKP error correction results quoted in Chapter~\ref{sec:exp_arch} show that the logical error rate of GKP codes depend linearly on the ancilla decay rates, making ancilla errors the leading order contributor to GKP error rates. Chapter~\ref{sec:qubit-dissipation} presents an explanation for this issue and mentioned that Ref.~\cite{baptiste2022multiGKP} used the \emph{isthmus property} for the stabilization of two-mode codes to suppress these errors by some factor. Chapter~\ref{sssec:GKP-control} introduces the \emph{isthmus property} in the context of a single-mode GKP `state' with reference to Figure~\ref{fig:sBs}c. We revisit this technique here to explain the protection of a two-mode GKP `qubit'.
 
 For stabilization using sBs, if stabilizers are colinear to logical operators, the logical error probability of the code directly depends on ancilla decay. In dimensions higher than $2$, it is possible to choose stabilizers which have minimal overlap with the logical operators. For example, in the tesseract and $D_4$ lattice codes, it is possible to find a measurement circuit for each stabilizer such that probability of logical error due to an ancilla decay is limited to a point in phase space. Such paths are said to possess the \emph{isthumus} property. In this case, the effect of ancilla decay depends on the squeezing or envelope size of the GKP code.
 
 The isthumus property yields a degree of protection against single ancilla decay events without the use of biased-noise ancilla. However, this protection is limited to an improvement by an approximate factor of 10, determined by squeezing of the GKP state. This should be compared to the protection from using a biased-noise ancilla, such as a stabilized Kerr-cat~\cite{puri2019stabilized}, which could yield a possibly orders-of-magnitude improvement for GKP stabilization. Additionally, the isthmus property is only available for certain special lattices whereas the use of biased-noise ancilla is not constrained upon the type of lattice. So far, there has not been any experimental demonstration of either approach, but for these reasons, we believe a biased-noise ancilla is likely a more feasible option in the near-term to achieve a large improvement in GKP QEC~\cite{puri2019stabilized,grimm2020stabilization,terhal2020towards}.

\subsection{GKP codes concatenated with discrete-variable codes} \label{ssec:code-concatenation}


While our previous discussions have mainly focused on encoding a single qubit into a set of oscillators, scaling up to perform quantum computations with multiple qubits requires additional error protection. To achieve this, one approach is to combine a GKP qubit code---acting as the \emph{inner code}---with a discrete variable (DV) code---acting as the \emph{outer code}. Examples of DV codes include repetition codes and surface codes. In this context, we explore a straightforward method that addresses this challenge. Specifically, we discuss a few lattices recognized in Ref.~\cite{baptiste2022multiGKP} which result from concatenating DV quantum codes with an ensemble of GKP qubits organized in a scaled hypercubic lattice.

We take the inner code lattice to correspond to $n$ disjoint GKP qubits, each defined by the same (for the sake of simplicity) two-dimensional generator matrix $\bm M_2$, such that the overall generator matrix
\begin{equation}
    \bm M_{\rm inner}\coloneqq \bigoplus_{i=1}^n\bm M_2,
\end{equation} 
where $\det(\bm M_2^\top\bm\Omega \bm M_2)=2^2$ and thus $\det \bm M_{\rm inner}=2^n $. We then consider an outer DV code $\bm M_{\rm outer}\sim [[n,k,d]]$ that uses the $n$ physical GKP qubits of the inner code to encode $k$ logical qubits with distance $d$ (i.e., errors on less than $\frac{d-1}{2}$ physical qubits can be corrected). The $n-k$ stabilizers of the inner code can then be replaced by the stabilizers of the outer code. For the replacement to work properly, the $2n\times (n-k)$ stabilizer (parity-check) matrix of the qubit code $\bm M_{\rm outer}$, and the inner code matrix $\bm M_{\rm inner}$ should have the same ordering of the canonical variables. For instance, if the inner GKP code is composed of square GKP qubits with $(q_1,p_1,q_2,p_2,\dots)$ ordering, then the stabilizer matrix of the outer qubit code should have $(X_1,Z_1,X_2,Z_2,\dots)$ ordering.\footnote{We further note that the $n-k$ stabilizers of the $[[n,k,d]]$ code correspond to the columns of $\bm M_{\rm outer}$, as we use column convention for generator matrices.} Upon direct sum of $\bm M_{\rm outer}$ with $2\bm{I}_{n+k\times n+k}$, we get the $2n\times 2n$ stabilizer matrix of the outer code, $\bm T$. The corresponding multimode lattice is then described by,
\begin{equation}
 \bm M_{\rm final}=\bm L \bm T ,
\end{equation}
where,
\begin{equation}
    \bm L\coloneqq\frac{\bm M_{\rm inner}}{2} \qq{and}
 \bm T \coloneqq
 \left(
    \begin{array}{c | c c}
          &  & \bm 0_{(n-k)\times(n+k)} \\ \bm M_{\rm outer}\, & & \\
         & & 2\bm I_{(n+k)}
    \end{array}\right).
\end{equation}
By construction, $|\det\bm M_{\rm final}|=2^k$, hence the lattice encodes $k$ qubits as expected. The matrix $\bm L$ represents the logical Pauli matrix of the inner GKP code, i.e. with columns consisting of logical Pauli displacement vectors $\bm p$ for each GKP qubit. This construction is based on the assumption that the logical operators of the inner code are colinear with the  stabilizers. In Ref.~\cite{baptiste2022multiGKP}, the authors point out that further transformations by some unimodular matrix $\bm N$ could lead to improved error correction by providing generators of minimal length, such that
\begin{equation}
    \bm M_{\rm final}= \bm L \bm T \bm N,
\end{equation}
where $\bm N$ can be found using the Lenstra–Lenstra–Lovász (LLL) lattice basis reduction algorithm~\cite{lenstra1982factoring}. Examples of such codes can be found in Ref.~\cite{baptiste2022multiGKP}. In the next section, we study a special example of a two-mode code obtained from concatenating an outer two-bit repetition code with an inner rectangular GKP qubit code. 


With the aforementioned recipe, we can construct a valid stabilizer generator matrix $\bm M$ and place everything in the language of lattices. In the lattice-based formulation, it is clear that an $n$-mode GKP code requires only $2n$ stabilizer measurements, which are effectively the rows of $\bm M_{\rm final}$. This should be contrasted with ``standard approaches'' that concatenate an inner GKP code with an $[[n,k,d]]$ qubit code and decode in a layered fashion. The first decoding layer measures the $2n$ stabilizers of the inner GKP code---essentially the columns of $\bm M_{\rm inner}$; the second decoding layer then measures the $n-k$ stabilizers of the outer code---essentially the columns of the parity-check matrix $\bm M_{\rm outer}$---possibly augmented with analog information from the inner decoding layer~\cite{fukui2017PRLanalogQEC,fukui2018PRXftqc}. The former approach to decoding generically requires less measurements ($2n$ measurements compared to the standard approach requiring $3n-k$ measurements), but choosing between these two different decoding approaches may also depend on practical considerations as well as preference.

The lattice-inspired error correction procedure could be similar to what has already been described for GKP codes concatenated with surface codes or repetition codes~\cite{vuillot2019toric,noh2020enhanced,terhal2020towards, noh2022low,Zhang2022GKPxzzx,lin2023surfaceGKP, stafford2022GKPbiasedREP,hanggli2020pra}. For instance, in Refs.~\cite{vuillot2019toric,noh2020enhanced,terhal2020towards,noh2022low,Zhang2022GKPxzzx}, the stabilizers are measured using CV GKP ancillae (similar to the measurement approach described in Chapter~\ref{sec:syndromes}) followed by the mapping of syndromes to logical errors (or the identity) using decoding algorithms that solve the closest lattice vector problem, minimum-weight perfect matching etc. Taking a different approach, the authors of Ref.~\cite{puri2021rvw} also suggested a GKP surface code along with a hybrid system---where the ancillae are two-level Kerr-cat qubits with a biased-noise spectrum---to extract syndromes of GKP-surface codes fault-tolerantly (via similar methods described in Chapter~\ref{sec:qubit-dissipation}) followed by analysis of syndromes using decoding algorithms that map the syndromes to logical errors (or the identity). The claims for all-GKP surface code in these papers are supported by numerical simulations for high-distance surface code lattices which use physical GKP qubits as data qubits. 
The hybrid architecture requires a similar analysis to prove its utility against an all-GKP surface code or all-transmon surface code. In Chapter~\ref{sec:app_ftqc}, we provide a few more details about GKP surface codes, specifically focusing on their relevance to thresholds in fault-tolerant quantum computing architectures.

An alternative to the processes above could be to directly engineer the dissipators or use Hamiltonian-based-techniques for the multimode GKP manifold via methods as used in the previous section for Tesseract codes and D4 codes in~\cite{baptiste2022multiGKP}. In fact, the stabilizer matrices of Tesseract code and D4 code are the same as that of rectangular GKP code concatenated with two-qubit repetition code and square GKP code concatenated with two-qubit repetition code, respectively. The difference in this approach is that no classical post-processing is required to map the syndromes to logical errors. Such techniques could be meaningful to define local autonomous stabilization on smaller sections of larger codes, thus reducing the complexity of the decoding task.


\section{GKP Oscillators-to-Oscillators Codes} \label{sec:gkp_o2o_codes}

The previous chapter argued why multimode GKP qubit encodings may be beneficial for QEC in DV quantum information processing tasks. Here we argue in favor of multimode oscillators-to-oscillators (O2O) codes for CV quantum information processing. The main goal of an O2O code is to directly combat analog errors with CV resources, e.g. squeezing, allowing for protection of \emph{arbitrary} CV quantum states~\cite{noh2020o2o}. This is pertinent to a variety of CV quantum information processing tasks such as quantum sensing, target detection, and entanglement-assisted quantum communication to name a few. 
Early works along this line have explored possibility of protecting oscillators with multiple oscillators in limited scenarios~\cite{lloyd1998,braunstein1998,braunstein2003error}. Reference~\cite{noh2020o2o} proposed GKP oscillator-to-oscillators (O2O) codes that are capable of encoding an oscillator into oscillators for universal noise models. Reference~\cite{wu2022optimal} further generalizes to encoding multiple oscillators into more oscillators and provides optimal code design. 

In this chapter, we introduce the general formulation of GKP-O2O codes, including the encoding in Chapter~\ref{sec:general_encoding}, and decoding strategies in Chapter~\ref{sec:decoding}. In Chapter~\ref{sec:example_of_codes}, we use the GKP-two-mode-squeezing (GKP-TMS) code with square lattice~\cite{noh2020o2o} as a prototypical example. Then we present wide-ranging results on code reduction in Chapter~\ref{sec:code_redux}, which shows that all GKP-O2O codes can be reduced to general GKP-TMS codes. Finally, in Chapter~\ref{sec:no_threshold}, we address the non-existence of a threshold for generic O2O codes with finite squeezing, which establishes ultimate limits of performance on generic O2O codes.

\subsection{General encoding of qumodes}
\label{sec:general_encoding}

\begin{figure}[t]
    \centering
    \includegraphics[width=.75\linewidth]{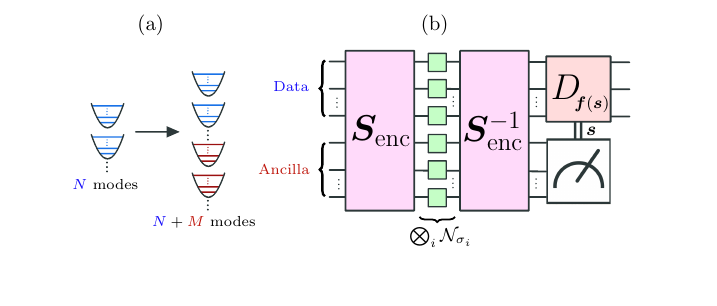}
    \caption{Oscillators-to-oscillators (O2O) code. (a) Illustration of encoding $N$ modes into $N+M$ modes. (b) O2O QEC circuit based on Gaussian encoding (symplectic matrix) $\bm S_{\rm enc}$ and syndrome-informed corrective operations (e.g., displacements $D_{\bm f(\bm s)}$).}
    \label{fig:code_general}
\end{figure}

A generic O2O code based on Gaussian operations encodes $N$ data modes into $K=N+M$ modes by entangling the data (by a Gaussian unitary) with $M$ ancilla modes that are prepared in some non-Gaussian resource state, e.g. a GKP lattice state; an intuitive picture is shown in Figure~\ref{fig:code_general}(a). In general, non-Gaussian states are required due to the no-go theorem of Gaussian error correction~\cite{niset2009nogo}. In this review, the non-Gaussian resources states are general GKP lattice states, $\ket{\mathcal{L}}$. Note that the Gaussian unitary $U_{\bm S_{\mathsf{enc}}}$ used for encoding can be described by the symplectic transform $\bm S_{\mathsf{enc}}$ (see Chapter~\ref{sec:gauss_evol} for discussion about Gaussian unitaries).

On the decoding side, the inverse transformation $\bm S_{\mathsf{enc}}^{-1}$ is applied to disentangle the data and the ancilla. However, such an operation correlates the noises of the data and ancilla, thus allowing for error correction via measurements on the ancilla. In particular, one measures the ancilla system to extract information about the data noise, which is encoded in an \textit{error syndrome}, $\bm s$. With this information in hand, error correcting displacement operations ${D}_{\bm f(\bm s)}$ are performed on the data, ideally ridding the data of noise. In our notation, $\bm f$ is an (vector) estimation function that takes the syndromes $\bm s$ as input and provides an estimate for the error displacements on the data. Due to the analog nature of the errors, error correction is never perfect, and there will be residual errors $\bm x_{\rm out}$ on the output data modes. The goal of error correction is to make these errors arbitrarily small so that they negligibly corrupt the data.

To quantify the error correction performance, we evaluate the output covariance matrix $\bm V_{\rm out}$ of the residual displacements $\bm x_{\rm out}$. As $\bm V_{\rm out}$ is a $2N\times2N$ symmetric matrix, we instead focus on single-valued quantities drawn from $\bm V_{\rm out}$, such as the geometric mean (GM) error
\begin{equation}
    \bar\sigma_{ \rm GM}^2\coloneqq\sqrt[2N]{\det\bm V_{\rm out}}.
    \label{eq:sigma_GM}
\end{equation}
We use the GM error as a figure of merit to benchmark code performance,
as it is invariant under symplectic operations on the data. Moreover, it can be shown that the GM error provides a lower bound on the quantum capacity of the input-output channel of the error correcting circuit, due to the fact that the channel is an additive non-Gaussian noise channel for GKP-O2O codes; see, e.g., Eq.~\eqref{eq:capacity_nonAGN}. As an alternative, one can also consider the root-mean-square (RMS) error, 
\begin{equation}
   \bar\sigma_{ \rm RMS}^2\coloneqq\frac{\Tr{\bm V_{\rm out}}}{2N}.
   \label{eq:sigma_RMS}
\end{equation}
The RMS error is often easier to evaluate since it requires only the diagonal elements of $\bm V_{\rm out}$ and provides an upper bound on the GM error, $\bar\sigma_{ \rm RMS}^2\ge  \bar\sigma_{ \rm GM}^2$.

\subsection{Lower bound on output error and break-even points}
\label{sec:output_bound}

Using quantum capacity arguments, a lower bound on the output error of a GKP-O2O code can be derived which only depends on the underlying AGN channel~\cite{noh2020o2o,wu2022optimal}. Consider a multimode GKP-O2O code to protect $N$ data modes with $M\geq N$ GKP ancilla. We assume non-identical AGN, such that $\bigotimes_{i=1}^{N+M}\mathcal{N}_{\sigma_i^2}$ is the error channel with variances $\sigma_i^2$. The output noise covariance matrix, $\bm V_{\rm out}$, such that $\bar{\sigma}^2_{\rm RMS}=\Tr{\bm V_{\rm out}}/2N$ and $\bar{\sigma}_{\rm GM}^2=\sqrt[2N]{\det\bm V_{\rm out}}$ are, respectively, the RMS and GM errors. It can then be shown that~\cite{wu2022optimal},
\begin{equation}\label{eq:lb_logicalnoise}
    \bar{\sigma}_{\rm RMS}\geq\bar{\sigma}_{\rm GM}\geq\sigma_{\rm LB}\coloneqq\frac{1}{\sqrt{e}}\sqrt[2N]{\left(\prod_{i=1}^{N+M}\frac{\sigma_i^2}{1-\sigma_i^2}\right)}.
\end{equation}
The key observation in proving this result is that the input-output quantum channel of the GKP-O2O code corresponds to non-Gaussian additive noise on the data, for which the lower bound of Eq.~\eqref{eq:capacity_nonAGN} applies. Moreover, since the GKP code is used to combat AGN, the non-Gaussian capacity is an achievable rate for the AGN channel, with the AGN channel having an upper bound given by Eq.~\eqref{eq:bounds_NmodeAGN}. Combining these upper and lower bounds, one arrives at the output error bound quoted above.

We can apply some data-processing arguments for the underlying AGN channel to find upper and lower bounds on the break-even point for GKP codes, resulting in $1/\sqrt{e}\leq\sigma^\star\leq 1/\sqrt{2}$. The upper bound comes from the upper bound on the quantum capacity for the AGN channel but can also be derived directly from Eq.~\eqref{eq:lb_logicalnoise}. The lower bound comes from a data-processing argument given that an achievable rate for the effective non-Gaussian additive noise channel [e.g., lower bound of Eq.~\eqref{eq:capacity_nonAGN}] of the GKP code is also an achievable rate for the underlying AGN channel. In Chapter~\ref{sec:code_opt_multi}, we find break-even points near $1/\sqrt{e}$ for multimode ($N=M=2$) GKP-O2O codes that use Minimum Mean Square Error (MMSE) estimation, whereas linear estimation leads a lower break-even point of $.558$~\cite{noh2020o2o}; see Chapter~\ref{sec:decoding} just below about decoding and estimation strategies.

Per Eq.~\eqref{eq:lb_logicalnoise}, if the number of ancilla modes is equal to the number of data modes ($M=N$), then error suppression is at best quadratic in $\sigma$. The performance can be further enhanced with concatenated codes ($M> N$) since ${\sigma_{\rm out}\sim \sigma^{1+\frac{M}{N}}}$, where ``out'' refers to RMS or GM error. This tells us how the output error should scale with the input error but does not tell us how to achieve such scaling in practice. In particular, it may require a large amount of physical resources to push the error to arbitrarily small values. In Chapter~\ref{sec:no_threshold}, we discuss these issues for general O2O codes, where finite squeezing becomes a limiting factor.

\subsection{Decoding strategies}
\label{sec:decoding}

For decoding, as shown in Figure~\ref{fig:code_general}, one needs to infer the actual error on the data from the error syndrome $\bm s$ obtained by stabilizer measurements on the ancilla modes (described in Chapter~\ref{sec:syndromes}), and then perform error correcting operations accordingly. In this section, we introduce two decoding strategies: linear estimation (see Chapter~\ref{sec:linear_est}) and minimum mean square error (MMSE) estimation (see Chapter~\ref{sec:MMSE}).

The first part of the decoding strategy is applying the Gaussian transformation $\bm S_{\mathsf{enc}}^{-1}$ that decorrelates the initial information in the data from the ancilla but correlates their additive noises; see Figure~\ref{fig:code_general}(b). The correlations between the additive noises of the data $\bx_d$ and ancilla $\bx_a$ are described by the covariance matrix 
\begin{align}
    \bm V_{x} = \bm{S}_{\rm{enc}}^{-1}\bm{V}_\xi\bm{S}_{\rm{enc}}^{-\top},
    \label{eq:Noise-CM-transform}
\end{align}
where $\bm V_{\xi}$ is the covariance matrix of the AGN channel (here we assume general form of noise, which may include correlated additive noise sources). Following the Gaussian transformation, one performs GKP-assisted homodyne measurements on the ancilla to extract the ancilla noise $\bm x_a$. This leads to an error syndrome $\bm s=\bm M^\top\bm\Omega\bm x_a \mod{\sqrt{2\pi}}$ [see Eq.~\eqref{eq:syndrome}], from which we can estimate the additive noise on the data $\bx_d$. To correct the data noise, we apply a syndrome-informed displacement on the data according to some estimation function $\bm f(\bs)$. The key part of the decoding strategy is to choose a good function $\bm f(\bs)$ to estimate the error. The information to make use of is the joint distribution of the data noise and the syndrome,
\begin{align}
    P(\bx_d,\bm s) 
 =  \sum_{\bm k} g(\bm V_d^{-1},\bm x_d+ \bm V^{-1}_d \bm V_{da}(\bm s-\bm k\sqrt{2\pi})) g(\bm V_{d|a}^{-1}, \bm s -\bm k\sqrt{2\pi}),
    \label{eq:xd-s-joint-distribution}
\end{align}
which can be solved from properties of the GKP lattice~\cite{wu2022optimal}. Here $g(\bm \Sigma, \bm x)$ is a multivariate Gaussian distribution, and the matrices above are defined implicitly via
\begin{align}
    \begin{pmatrix}
    \bm V_d & \bm V_{da}\\
    \bm V_{da}^\top & \bm V_a
    \end{pmatrix}^{-1} &\coloneqq 
    (\bI_{2N} \oplus {\bm M^\top \bm \Omega}) \bm V_x (\bI_{2N} \oplus {(\bm M^\top \bm \Omega)} ^{\top}),
    \label{Vd_Va_Vda}
\end{align}
and $\bm V_{d|a}=\bm V_a-\bm V_{da}^\top \bm V_d^{-1}\bm V_{da}$.

After the error correction, the output noise on the data is $\bx_{\rm{out}}=\bx_d-\bm f(\bs)$ and follows the joint probability density distribution (PDF)
\begin{equation}
    P(\bx_{\rm {out}})=\int_{\mathbb{R}^{2N}}\dd{\bx_d}\int_{\mathscr{I}^{2M}}\dd{\bm s}\;
    P(\bx_d, \bs)\delta\left(\bx_{\rm{out}}-\bx_d+\bm{f}(\bs)\right),
    \label{P_out_evaluation}
\end{equation}
where $\delta$ is the Dirac delta function and $\mathscr{I}\coloneqq[-\sqrt{\pi/2},\sqrt{\pi/2}]$. From the distribution, one can easily obtain the covariance matrix for the output error, $\bm V_{\rm out}$, which is just the second moments of $\bm x_{\rm out}$.

We note that the corresponding quantum channel $\widetilde{\mathcal{N}}_{\rm O2O}:\rho_{\rm in}\rightarrow\rho_{\rm out}$ of the O2O code for a generic input data state $\rho_{\rm in}$ is a non-Gaussian additive noise channel [see Eq.~\eqref{eq:ANGN_channel}] with displacement noise $\bm x_{\rm out}$ governed by the non-Gaussian PDF $P(\bm x_{\rm out})$.

\subsubsection{Linear estimation}
\label{sec:linear_est} 

Linear estimation refers an estimation of the noise which is linear in the syndrome---i.e., $\bm{f}(\bs)=\bm A \bs$, where $\bm A$ is some invertible matrix. Linear estimation is adopted in Ref.~\cite{noh2020o2o}, while Ref.~\cite{wu2022optimal} provided a general form. To obtain a reasonable choice of $\bm A$, we apply asymptotic analysis. Assume that the additive noises $\bx_d$ and $\bx_a$ are small, such that we can ignore the modular properties of the GKP lattice; then,
\begin{align}
    \bm s & = \bm M^\top \bm\Omega\bx_a \mod{\sqrt{2\pi}} \approx \bm M^\top \bm\Omega\bx_a.
\end{align}
In this case, the PDF of the data and the error syndrome, $P(\bx_d,\bm s)$ in Eq.~\eqref{eq:xd-s-joint-distribution}, is approximately a Gaussian distribution.
One can then show that the best choice is $\bm A =- \bm V_d^{-1} \bm V_{da}$, and thus,
\be 
\bm{f}_{\rm{Linear}}(\bs) = - \bm V_d^{-1} \bm V_{da} \bs.
\label{eq:f_linear_est}
\ee

\subsubsection{Minimum mean square error (MMSE) estimation} 
\label{sec:MMSE}

We consider minimum mean square error (MMSE) estimation, which is developed to minimize the RMS error of Eq.~\eqref{eq:sigma_RMS} and strictly performs better than linear estimation. We start with the joint PDF of the data and the error syndrome, $P(\bx_d,\bm s)$. The joint PDF is not a Gaussian distribution but, rather, is a sum of Gaussian distributions. The conditional distribution $P(\bx_d|\bm s)=P(\bx_d,\bm s)/P(\bm s)$, where $P(\bm s)$ is the marginal distribution for the syndrome $\bm s$, can be used to derive the MMSE estimator via $\bm f_{\rm MMSE}(\bm s)= \int_{\mathbb{R}^{2N}}\diff{\bx_d}\;\;  \bx_d P(\bx_d|\bm s)$. For GKP-O2O codes, the MMSE estimator can be derived in closed form (see Ref.~\cite{wu2022optimal} for a derivation):
\begin{theorem}\label{thm:fMMSE}
For a GKP-O2O code with GKP lattice state $\mathcal{L}$ described by generator matrix $\bm M$, the MMSE estimator for an error syndrome $\bm s$ is given by
\begin{align}
    &\bm f_{\rm MMSE}(\bm s)  = -\frac{\sum_{\bm n} \bm V_d^{-1}\bm V_{da}(\bm s-\bm n\sqrt{2\pi})g(\bm V_{d|a}^{-1},\bs-\bm n\sqrt{2\pi})}
    {\sum_{\bm m} g(\bm V_{d|a}^{-1},\bs-\bm m\sqrt{2\pi})},
    \label{eq:fMMSE_main}
\end{align}
where $g(\bm\Sigma, \bm x)$ is a multivariate Gaussian distribution and $\bm m, \bm n\in\mathbb{Z}^{2M}$. The matrices $\bm V_{da}$, $\bm V_d$ and $\bm V_a$ are defined through Eq.~\eqref{Vd_Va_Vda}
and $\bm V_{d|a}=\bm V_a-\bm V_{da}^\top \bm V_d^{-1}\bm V_{da}$.
\end{theorem}
This can be used to estimate the output error of O2O codes given for the MMSE estimation strategy.


\subsection{Example of codes and performances}
\label{sec:example_of_codes}

In this section, we review the performance of a few O2O code examples. In Ref.~\cite{noh2020o2o}, two codes are proposed based on the GKP square lattice, the GKP-TMS code and the GKP-squeezing-repetition code. We consider the simple two-mode case (one data mode and one GKP ancilla) and assume a two-mode, heterogenous AGN channel $\mathcal{N}_{\bm V_{\xi}}$, with noise covariance $\bm V_{\xi}={\rm diag}(\sigma_1^2,\sigma_1^2,\sigma_2^2,\sigma_2^2)$, which are analysed in more detail in Ref.~\cite{wu2021continuous}. 

\begin{figure}
\centering
\includegraphics[width=0.6\linewidth]{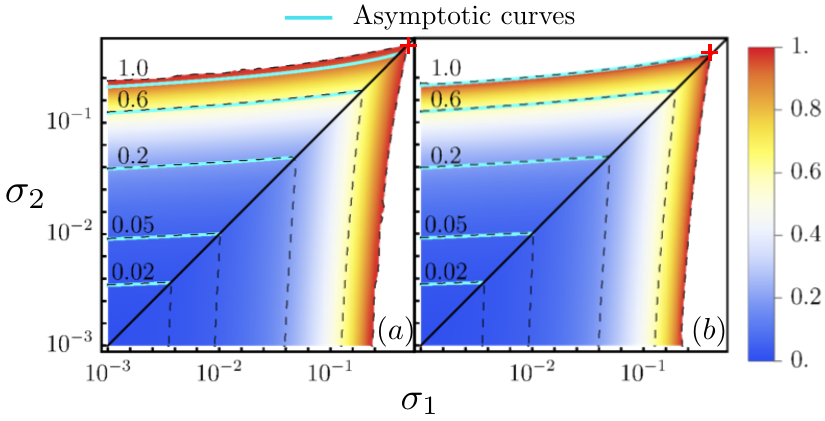}
\caption{Contours of the ratio $\bar{\sigma}_{\rm RMS}/\min[\sigma_1,\sigma_2]$ for 
(a) GKP-TMS code and
(b) GKP-squeezing-repetition code. Linear estimators are adopted for concreteness.
The end points marked by the red crosses are $\sigma_1=\sigma_2\simeq 0.56$ for (a) and $\sigma_1=\sigma_2\simeq 0.41$ for (b). Figure adopted from Ref.~\cite{wu2021continuous}.
\label{fig:contours}
}
\end{figure}

\begin{example}[GKP-two-mode squeezing code]
In a GKP-TMS-code with square GKP lattice, the symplectic transform of the encoding is $\bm S_{\rm enc}=\bm S_G$, where $\bm S_G$ is a TMS transformation of gain $G$ in Eq.~\eqref{eq:tms_matrix}. From Eq.~\eqref{eq:f_linear_est}, the linear estimator is explicitly
\begin{align}
    \bm f_{\rm{Linear}}(\bs) = -\bm V_d^{-1} \bm V_{da} \bs =  \tilde{\mu}\begin{pmatrix}
        0 & 1\\
        1 & 0
    \end{pmatrix} \bs,
\end{align}
where
$
\tilde{\mu} = \sqrt{G(G-1)}(\sigma_1^2+\sigma_2^2)/[(G-1)\sigma_1^2+G\sigma_2^2]$. This linear estimation scheme was adopted in Ref.~\cite{noh2020o2o}; essentially, it is optimized to reduce variance on the data in the Gaussian approximation (i.e., ignoring lattice effects arising from the GKP ancillae measurements) given syndrome information from the GKP ancilla. Likewise, from Eq.~\eqref{eq:fMMSE_main}, the MMSE estimator can be obtained as
\begin{align}
    \bm f_{\rm MMSE}(\bm s)  = \frac{\sum_{\bm n} \tilde{\mu} \begin{pmatrix}
        0 & 1\\
        1 & 0
    \end{pmatrix} (\bm s-\bm n\sqrt{2\pi})g(\sigma_G^2\bI,\bs-\bm n\sqrt{2\pi}) }
    {\sum_{\bm m} g(\sigma_G^2\bI,\bs-\bm m\sqrt{2\pi}) },
\end{align}
where $\sigma_G^2=G\sigma_2^2+(G-1)\sigma_1^2$. When $\sigma_1,\sigma_2\ll 1$, we obtain the asymptotic result from linear estimation,
\begin{align}
\label{eq:asympotic expressions_TMS}
\bar{\sigma}_{\rm RMS}^2 \approx
\frac{4 \bar{\sigma}^4}{\pi}\ln\left(\frac{\pi^{3/2}}{2\bar{\sigma}^4}\right),
\end{align}
where $\bar{\sigma}=\left(\sigma_1\sigma_2\right)^{1/2}$. The output error result aligns with broader findings on GKP O2O codes (albiet for iid AGN), as discussed heuristically in the chapter on no-thresholds for O2O codes (Chapter~\ref{sec:no_threshold}). Specifically, these arguments generally indicate a quadratic noise reduction---i.e., $\sigma\rightarrow \sigma_{\rm RMS}\sim \sigma^2$---when employing a single layer of error correction.
\end{example}

\begin{example}[Squeezing-repetition code]
    
The GKP-squeezing-repetition code (see also Ref.~\cite{zhuang2020distributed}) has the following encoding matrix for $N=2$ modes,
\begin{align}
    \label{eq:rep12}
    &
    \bm{S}^{[2]}_{\rm Sq-Rep} =
    \begin{pmatrix}
    {\kappa}/{\lambda} && 0 && 0 && 0\\
    0 && {\lambda}/{\kappa} && 0 && -\lambda\\
    \lambda && 0 && {\lambda}/{\kappa} && 0\\
    0 && 0 && 0 && {\kappa}/{\lambda}
    \end{pmatrix},
\end{align}
with $\lambda$ and $\kappa$ being tunable parameters. When $\sigma_1,\sigma_2\ll 1$, we obtain
\be 
\label{eq:asympotic expressions_SR}
\bar{\sigma}_{\rm RMS}^2 \approx \frac{4\bar{\sigma}^4}{\pi} \ln \left[\frac{\pi^{3/2}}{{2\bar{\sigma}^4}}\right]+\frac{4\bar{\sigma}^4}{\pi} \ln\left(\frac{\sigma_1^2+\sigma_2^2}{2\sigma_1^2}\right),
\ee 
which is identical to Eq.~\eqref{eq:asympotic expressions_TMS} in the leading-order, up to a next-order correction that disappears when $\sigma_1=\sigma_2$. For $\sigma_1\neq \sigma_2$, the GKP-SR code is asymmetric between the two channels. We plot contours of the ratio $\bar{\sigma}_{\rm RMS}/\min[\sigma_1,\sigma_2]$ in Figure~\ref{fig:contours}(b). The asymptotic results (cyan curves) agree well with the numerical results~\cite{wu2021continuous}. 
\end{example}

\subsection{General O2O code reduction to TMS codes}\label{sec:code_redux}

\begin{figure}
    \centering
    \includegraphics[width=\linewidth]{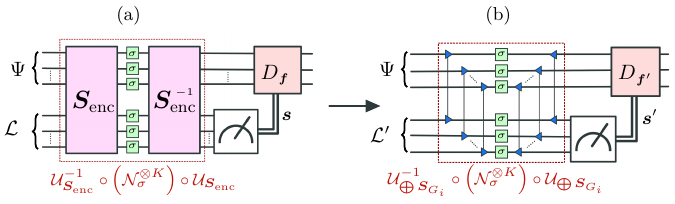}
    \caption{General reduction of a O2O code to a TMS code ($M=N$ here). (a) A general O2O code with encoding $\bm S_{\mathsf{enc}}$ and non-Gaussian ancilla $\ket{\mathcal{L}}$. The syndromes $\bm s$ are extracted from stabilizer measurements on the ancilla $\ket{\mathcal{L}}$ and inform the corrective operations (e.g., displacements) on the data, $D_{\bm f(\bm s)}$. (b) An equivalent TMS code, whereby the encoding is reduced to a set of TMS operations between the data and ancilla modes (Theorem~\ref{thm:code_redux}). The state $\ket{\mathcal{L}^\prime}$ is related to $\ket{\mathcal{L}}$ by a local symplectic transformation.}
    \label{fig:code_redux}
\end{figure}

To make progress towards describing generic features of GKP O2O codes, we focus on the simplifying case of independent and identically distributed (iid) AGN, with noise channel $\mathcal{N}_{\sigma}^{\otimes K}$, where $K=N+M$; here, $N$ is the number of data modes while $M$ is the number of ancillary modes. In Ref.~\cite{wu2022optimal}, it was proven that an arbitrary GKP-O2O code can be reduced to a generalized GKP-TMS code. A formal statement of the result is given just below in the form of a theorem; see also Figure~\ref{fig:code_redux}.

\begin{theorem}[GKP-O2O codes reduce to $\mathsf{TMS}^\otimes(\mathcal{L})$]\label{thm:code_redux}
For an iid AGN channel, the most general GKP-O2O code is formally equivalent to a product of $N$ GKP-TMS codes with an $M$-mode ancillary lattice state $\ket{\mathcal{L}}$. We refer to this general construction as $\mathsf{TMS}^\otimes(\mathcal{L})$.
\end{theorem}
This result is a consequence of the encoding/decoding structure of the code and follows from the modewise entanglement theorem (see Theorem~\ref{thm:modewise}). We sketch the proof here; see Ref.~\cite{wu2022optimal} for rigorous details and also Figure~\ref{fig:code_redux} for visual aid. Due to the modewise entanglement theorem, the multimode correlations between a sub-system $A$ of $N$ modes and sub-system $B$ of $M\geq N$ modes can be decomposed into $N$ pairwise TMS correlations up to local operations on $A$ and $B$. [This also assumes that the joint covariance matrix is of the form $\bm V_{AB}\propto\bm S\bm S^{\top}$.] We can apply this result to our current setting. Consider the sequence of operations $\bm S_{\rm enc}\rightarrow\mathcal{N}_{\sigma}^{\otimes K}\rightarrow\bm S_{\rm enc}^{-1}$ that transforms the iid AGN channel to a correlated AGN channel with covariance matrix $\bm Y_{\rm enc}=\sigma^2\bm S^{-1}_{\rm enc}\bm S^{-\top}_{\rm enc}$. By the modewise entanglement theorem, we can decompose the product $\bm S^{-1}_{\rm enc}\bm S^{-\top}_{\rm enc}$ into $N$ TMS operations between data and ancilla (with an identity on the remaining $M-N$ ancilla modes) up to local operations that can ultimately be absorbed into state preparation and the estimation strategy.

Given the TMS code reduction above, to optimize the code design, one needs to optimize over the $N$ gain parameters $G_i$ of the TMS operations, as well as the $M$-mode lattice state $\ket{\mathcal{L}}$. Interestingly, Theorem~\ref{thm:code_redux} does not actually require the non-Gaussian ancilla $\ket{\mathcal{L}}$ to be a GKP lattice state. In other words, our results hold for any codes based on Gaussian encoding with general non-Gaussian ancilla---not just GKP-O2O codes. This TMS construction therefore represents a general coding strategy for O2O codes based on Gaussian encoding. Moreover, considering that Gaussian operations supplied with non-Gaussian GKP ancilla are universal and sufficient for fault-tolerant quantum computation~\cite{baragiola2019GKPuniversal}, such codes appear to be generically sufficient for QEC as well. For single-mode data and ancilla, the optimal code design problem can be efficiently solved as discussed below. We provide concrete numerical examples of GKP-TMS codes in Chs.~\ref{sec:code_opt_single} and~\ref{sec:code_opt_multi}.

\subsubsection{Code optimization: two mode iid case}
\label{sec:code_opt_single}

Since all single-mode lattice states can be generated by local symplectic transformations on the canonical GKP state (see Proposition~\ref{prop:1mode_lattice}), Theorem~\ref{thm:code_redux} immediately implies the following:

\begin{theorem}\label{thm:tms_1mode}
For single-mode data and ancilla undergoing iid AGN, the TMS code $\mathsf{TMS}(\mathcal{L}_{\Lambda})$, with gain $G$ and ancilla lattice $\ket{\mathcal{L}_{\Lambda}}=U_{\bm\Lambda}\ket{\square}$, is the optimal GKP-O2O code.
\end{theorem}

Therefore, to find the best two-mode GKP-O2O code, we need to optimize the local Gaussian unitary $\bm\Lambda$ and the TMS gain $G$, as well as choose the best possible estimator $\bm f$. We are unaware of a way to derive an estimator $\bm f$ that minimizes the GM error.\footnote{Recall that the GM error has information theoretic roots [Eq.~\eqref{eq:capacity_nonAGN}], supporting its relevance.} On the other hand, since $\bar\sigma_{ \rm RMS}^2\ge  \bar\sigma_{ \rm GM}^2$, we can obtain an upper bound on the GM error from the RMS error, which we can minimize via MMSE estimation of Theorem~\ref{thm:fMMSE}.

As shown in Example~\ref{example:single_transform}, any single-mode Gaussian transformation has a decomposition $\bm R(\phi)\text{\bf Sq}(r)\bm R(\theta)$, where $\bm R(phi)$ is a $2\times2$ rotation matrix in Eq.~\eqref{eq:R_phi} and $\text{\bf Sq}(r)$ is single-mode squeezing of Eq.~\eqref{eq:Sq_r}. Due to the symmetry of the AGN, the last phase rotation $\bm R(\phi)$ does not alter performance; thus, we can ignore the last rotation and parameterize the transform as $\bm \Lambda =\text{\bf Sq}(r)\bm R(\theta)$ such that $\ket{\mathcal{L}}=U_{\bm \Lambda}\ket{\square}$. As examples, a rectangular GKP state (Figure~\ref{fig:lattice_schematic}c) is given by $\theta=0$ and $r>0$, and a hexagonal GKP state (Figure~\ref{fig:lattice_schematic}d) is given by $\theta=\pi/4$ and $r_{{\tiny\hexagon}}=\sqrt[4]{3}$.

\begin{figure}
    \centering
    \includegraphics[width=0.7\linewidth]{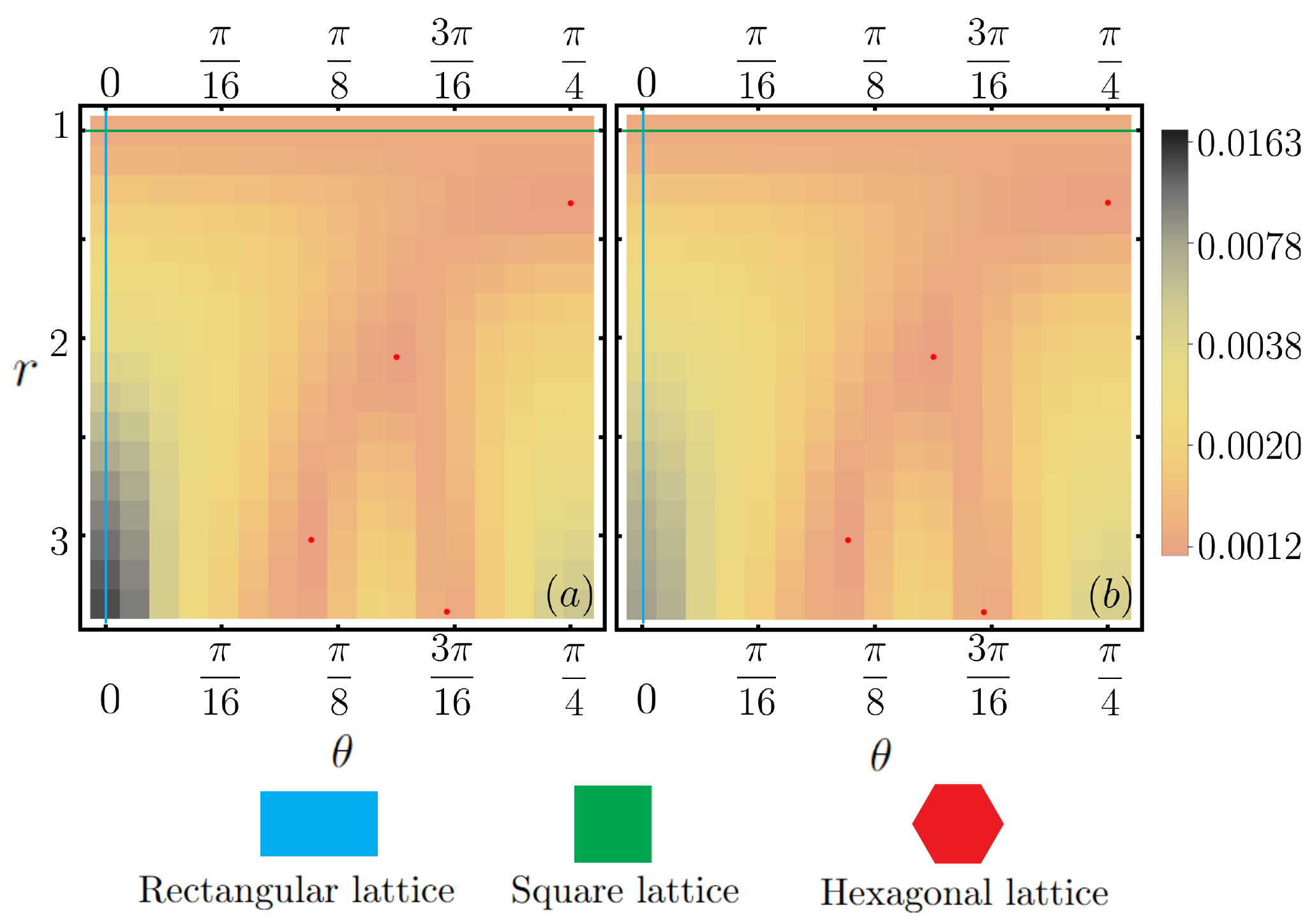}
    \caption{Output noise for a single-mode ($N=M=1$) GKP-O2O code. Input noise variance is $\sigma=10^{-2}$. We optimize the TMS gain $G$ for each point $(r,\theta)$. (a) RMS error $\bar\sigma_{ \rm RMS}^2$, (b) GM error $\bar\sigma_{ \rm GM}^2$. For the square lattice (green line), $\bar\sigma_{ \rm RMS}^2=1.25129(5)\times10^{-3}$and $\bar\sigma_{ \rm GM}^2=1.25129(5)\times10^{-3}$. The four hexagonal lattice points (red dots) have the same output noises of $\bar\sigma_{ \rm RMS}^2=1.15575(5)\times10^{-3}$ for RMS error and $\bar\sigma_{ \rm GM}^2 =1.15575(5)\times10^{-3}$ for GM error. Only the range $\theta\in[0,\pi/4]$ is considered due to symmetry; see Ref.~\cite{wu2022optimal}. This is a re-print of Figure~5 of Ref.~\cite{wu2022optimal}.}
    \label{fig:single_mode_lattice_contour}
\end{figure}

In Figure~\ref{fig:single_mode_lattice_contour}(a), we plot the contour of the RMS error $\bar\sigma_{ \rm RMS}^2$ for an MMSE decoder optimized over the TMS gain $G$ for each point $(r,\theta)$; note that each point $(r,\theta)$ corresponds to a different GKP ancilla lattice. We find four equivalent minima, corresponding to equivalent representations of the hexagonal lattice. The square lattice has $r=1$ with $\theta$ arbitrary (represented by the green line); the rectangular lattice has $\theta=0$ while $r$ squeezes the rectangle (represented by the blue line). The hexagonal lattice outperforms the square and rectangular lattices for all levels of AGN. In Figure~\ref{fig:single_mode_lattice_contour}(b), we plot the GM error $\bar\sigma_{ \rm GM}^2$ in $(r,\theta)$ parameter space for the same optimized gain values of Figure~\ref{fig:single_mode_lattice_contour}(a). The GM error and RMS error are almost equal, with some deviations at the left-bottom corner due to the large squeezing of a rectangular lattice. 

\subsubsection{Numerical results for multimode oscillator codes}
\label{sec:code_opt_multi}

In this section, we present recent numerical results for multimode ($N=M=2$) GKP-O2O codes~\cite{wu2022optimal}. These codes could be used to, e.g., protect a two-mode squeezed vacuum state. For simplicity we consider iid noise, as our focus is on comparing performance of various two-mode GKP lattices, $\ket{\mathcal{L}}$. Since GKP-TMS codes, $\mathsf{TMS}^{\otimes2}(\mathcal{L})$, represent a generic class of O2O codes via the code reduction theorem~\ref{thm:code_redux}, we focus on these codes here. The encoding (decoding) is given by $N=2$ TMS operations, with each TMS operation coupling one data mode to one ancilla mode; see Figure~\ref{fig:code_redux}. We numerically optimize the TMS gains to minimize the RMS error $\Bar{\sigma}_{\rm RMS}^2=\Tr{\bm V_{\rm out}}/4$ for MMSE estimation. To benchmark the results, we consider linear estimation with initial square GKP states, which was analyzed in the original work of Noh et al~\cite{noh2020o2o}. 

In Ref.~\cite{wu2022optimal}, three initial (canonical) GKP lattice states were considered: a direct product of square GKP states (i.e., a 4-dimensional hypercube), a direct product of hexagonal GKP states, and a canonical $D_4$ lattice that can be generated from a hypercube via a two-mode symplectic transformation [see Eq.~\eqref{eq:SD4}]. The canonical $D_4$ GKP state is entangled, and thus, the $D_4$ TMS code $\mathsf{TMS}^{\otimes2}(D_4)$ is a genuine multimode O2O code. For the direct product codes, since the TMS operations operate on the data modes independently and the additive noises are independent, the square and hexagonal TMS codes, $\mathsf{TMS}^{\otimes2}(\square)$ and $\mathsf{TMS}^{\otimes2}(\hexagon)$, produce equivalent results as their single-mode counterparts. 

\begin{figure}
    \centering
    \includegraphics[width=.9\linewidth]{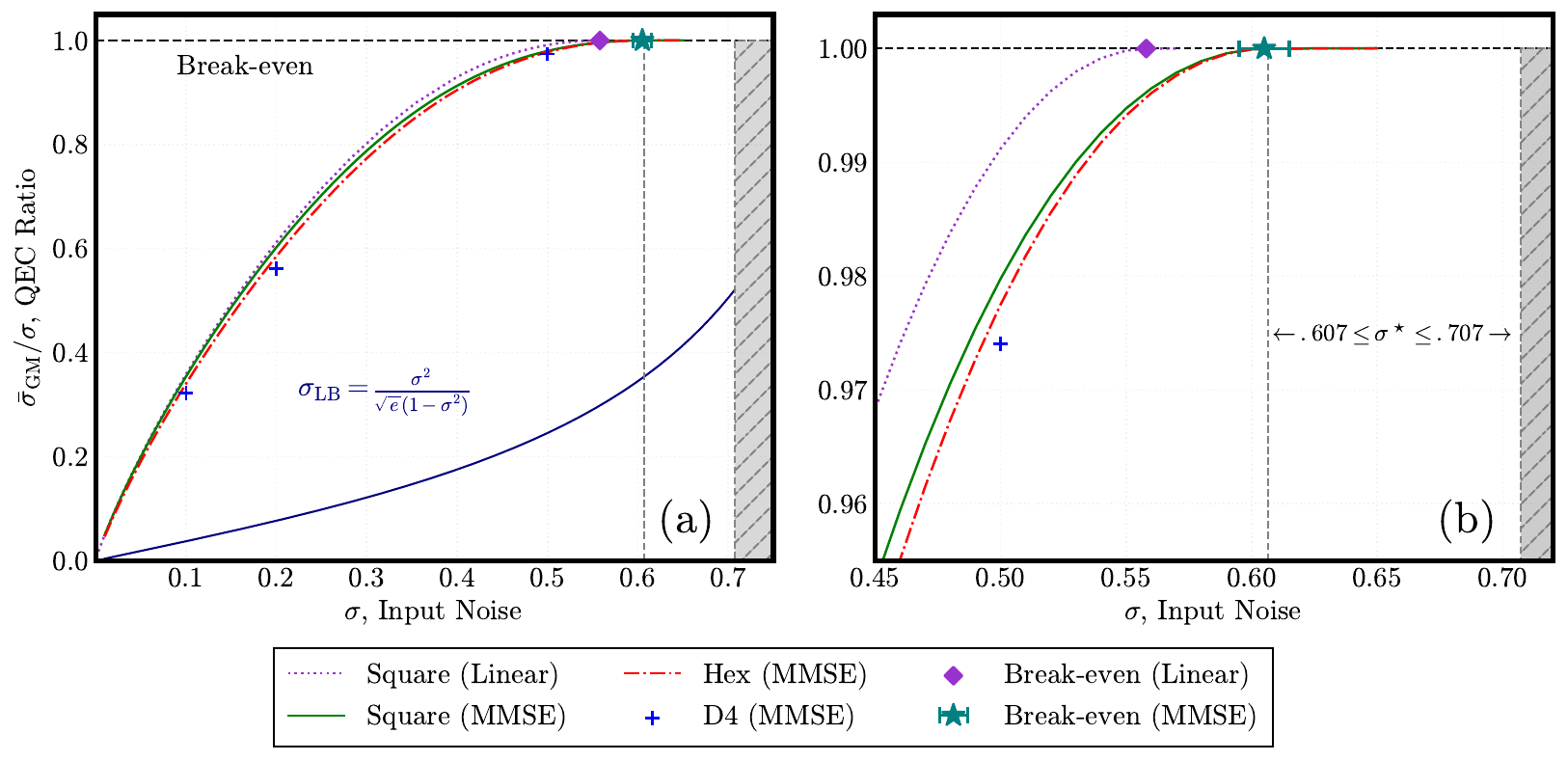}
    \caption{QEC gain between output noise and input noise of a multimode ($N=M=2$) GKP TMS code $\mathsf{TMS}^{\otimes2}(\mathcal{L})$ for different canonical lattices $\mathcal{L}$ (Square, Hexagonal, $D_4$) and estimation strategies (linear, MMSE). The square code with linear estimation (dotted purple) is presented as a benchmark~\cite{noh2020o2o}. Grey hatched region is forbidden by information theoretic arguments. The purple diamond and green star denote break-even points for linear and MMSE estimation, respectively, after which point no QEC gain is expected. Figure adapted from Ref.~\cite{wu2022optimal}.}
    \label{fig:output_noise}
\end{figure}

Results for the GM error $\Bar{\sigma}_{\rm GM}=\sqrt[4]{\det\bm V_{\rm out}}$ are presented in Fig~\ref{fig:output_noise}. The $D_4$ TMS code $\mathsf{TMS}^{\otimes2}(D_4)$ performs better than the square and hexagonal TMS codes $\mathsf{TMS}^{\otimes2}(\square)$ and $\mathsf{TMS}^{\otimes2}(\hexagon)$. This result is not too surprising, since the $D_4$ lattice has the densest sphere-packing in four dimensions. To quote some examples from the data, for $\sigma=.1$, the $\mathsf{TMS}^{\otimes2}(\hexagon)$ code outperforms the $\mathsf{TMS}^{\otimes2}(\square)$ code by a relative difference of about $3.95\%$, whereas $\mathsf{TMS}^\otimes(D_4)$ achieves a relative difference of about $9.04\%$. Hence, a $D_4$ lattice can improve code performance by roughly $10\%$ or better. The advantages become larger for lower $\sigma$.

Interestingly, from our numerical results, we observe that MMSE estimation leads to a break-even point $\sigma^\star_{\rm MMSE}\approx.605(5)$ (teal star in Figure~\ref{fig:output_noise}) irrespective of the lattice $\mathcal{L}$, whereas linear estimation leads to $\sigma^\star_{\rm lin}\approx.558$~\cite{noh2020o2o} (purple diamond). The value $\sigma^\star_{\rm MMSE}\approx.605(5)$ for MMSE agrees with the lower bound on the break-even point taken from capacity arguments ($.607\leq\sigma^\star\leq.707$) discussed in the Chapter~\ref{sec:output_bound}. A break-even point $\sigma^\star\approx .607$ was also observed in threshold behavior of GKP-surface codes~\cite{lin2023surfaceGKP}.

\subsubsection{No threshold for finite squeezing}
\label{sec:no_threshold}

Errors, such as additive Gaussian noise, are continuous errors, in contrast to the discrete Pauli errors in discrete variable quantum information processing. Analogously with error suppression in discrete variable systems, one begs the question: Can one shrink CV errors to arbitrarily small values? In Ref.~\cite{hanggli2021oscillator}, it was found that, for GKP-O2O codes relying on maximum likelihood decoding, AGN errors \textit{cannot} be made arbitrarily small with a finite amount of squeezing---no matter if we increase the number of modes---implying the existence of a no-threshold theorem for GKP-O2O. The authors of Ref.~\cite{wu2022optimal} then extended the no-threshold result to \textit{any} O2O code relying on Gaussian encoding, remarkably without reference to the non-Gaussian ancilla nor the estimation strategy used. In other words, we may generally state that O2O codes do not have a threshold. In some respects, the non-existence of a threshold is intuitive: We do not expect arbitrary lower error suppression of an analog/continuous error without consuming an arbitrarily high continuous-variable resource, such as squeezing. 

A simple proof of the no-threshold result follows from the code reduction of Theorem~\ref{thm:code_redux} and a classical data processing argument, which we now sketch. Consider a multimode O2O code with Gaussian encoding (decoding) $\bm S_{\rm enc}$ ($\bm S_{\rm enc}^{-1}$). Let $\bm x_d\in\mathbb{R}^{2N}$ and $\bm x_a\in\mathbb{R}^{2M}$ be the correlated data and ancilla noises, which are Gaussian distributed random variables with covariance matrix $\bm V_{x}=\sigma^2\bm S_{\rm enc}^{-1}\bm S_{\rm enc}^{-\top}$. By the Code Reduction Theorem~\ref{thm:code_redux}, the error matrix $\bm V_x$ can be decomposed into $N$ TMS blocks (each characterized by a TMS amplification gain $G_i$), with each data mode coupled to only one ancilla mode. Let $x_{a_i}$ be an element of $\bm x_a$ that is correlated with $ x_{d_i}$ of $\bm x_d$. Due to the structure of the TMS operation, there exists only $qq$ and $pp$ correlations---i.e., there are no cross correlations $qp$. Thus we can consider one data quadrature at a time. Furthermore, let $\Tilde{x}_{d_i}\coloneqq\Tilde{x}_{d_i}(x_{a_i})$ be the estimation of the data noise given information about the ancilla noise (practically extracted from syndrome measurements). A corollary of Theorem 8.6.6 in Ref.~\cite{cover2006elements} states that the estimation variance of a generic random variable $X$, given side information $Y$, is lower bounded via $\mathbb{E}[(X-\Tilde{X}(Y))^2]\geq\exp\left[2 S(X|Y)\right]/2\pi e$, where $S(X|Y)$ is the conditional differential entropy. For Gaussian random variables that are correlated via two-mode squeezing, it is easy to show that $S(x_{d_i}|x_{a_i}) =\ln(\frac{2\pi e\sigma^2}{2G_i-1})/2$. Therefore, 
\begin{equation}\label{eq:qestimation_variance}
    \mathbb{E}\left[\big(x_{d_i}-\tilde{x}_{d_i}\big)^2\right]\geq\frac{\sigma^2}{2G_i-1}.
\end{equation}
Summing over all quadratures and accounting for the double degeneracy of q/p variances, we find the following lower bound on the output RMS error,\footnote{A similar condition can be found for the GM error.}
\begin{equation}\label{eq:no_threshold}
\bar{\sigma}_{\rm RMS}^2\geq\frac{1}{N}\sum_{i=1}^{N}\frac{\sigma^2}{2G_i-1}.
\end{equation}

If we benchmark the error $\varepsilon\geq\bar{\sigma}_{\rm RMS}$, the average gain must scale as $G\sim\sigma^2/\varepsilon$ in order to be at or below the benchmark. Hence, without an infinite amount of squeezing, the error $\varepsilon$ cannot be made arbitrarily small, even with an infinite number of ancillary modes. This result is a consequence of the Gaussian encoding structure of O2O codes and, moreover, does not rely on the particular non-Gaussian ancilla nor the estimation strategy employed. No-threshold behavior thus seems to be a universal feature of O2O codes. Furthermore, we observe that the critical component for combating analog errors is the CV resource of squeezing, rather than the number of ancillary modes $M$, which does not even play a role in the bound. This contrasts with DV (qubit) codes, where increasing the number of ancilla qubits typically leads to vanishing logical error rates, given that the physical error rate is below a certain threshold.

Let us make a final observation before moving forward to applications of GKP codes. Equation~\eqref{eq:no_threshold} suggests that we can arbitrarily crank up the gain to attain a low output error, however this appears in contradiction with the bounded scaling $\sigma_{\rm out}\sim\sigma^{1+M/N}$ inferred from quantum capacity arguments [Eq.~\eqref{eq:lb_logicalnoise}] for GKP O2O codes. The apparent contradiction is fictitious and can be reconciled by the following heuristics. For GKP-O2O codes, the TMS gain $G$ cannot be made arbitrarily large as this would amplify the noisy displacements and harmful lattice effects would come into play. We are thus led to a crude constraint on the gain {$\sqrt{G}\xi\lesssim\ell$}, where $\xi\sim\mathcal{N}(0,\sigma^2)$ is a random displacement and $\ell$ is the lattice spacing, implying that the TMS gain must be bounded $G\sim\ell^2/\sigma^2$. The output error after one round of O2O QEC then scales as ${\sigma}_{\rm out, 1}\sim\sigma^2/\ell$. This argument provides a similar scaling (up to logarithmic corrections) as the more detailed analyses given in the original work of Ref.~\cite{noh2020o2o}; see also Eqs.~\eqref{eq:asympotic expressions_TMS} and~\eqref{eq:asympotic expressions_SR}. Going further, for $k=M/N$ rounds of error correction (or a concatenated code with $k$ levels), $\sigma_{\rm out, k}\sim \sigma^{1+k}/\ell^{k}$. Thus, one can shrink the error by increasing $k=M/N$, in accordance with the lower bound found in Eq.~\eqref{eq:lb_logicalnoise}. However, with decreasing error, the gain must correspondingly increase, in agreement with the no-threshold result~\eqref{eq:no_threshold}.

\section{Applications}
\label{sec:applications}

Bosonic QEC with GKP states has a wide range of useful quantum-specific applications, including fault-tolerant quantum computing with error-corrected bosonic modes, quantum communication, and error-correction enhanced quantum sensing. For example, GKP qubit codes have been shown to benefit from analog information at the CV level, enhancing the performance of outer DV codes that are useful in communication and computation. Additionally, it has been suggested that bosonic QEC is more resource efficient than typical DV approaches. Finally, it seems apparent that, in futuristic quantum networks, bosonic QEC will play a vital role for, e.g., long-distance quantum communication, distributed quantum information processing. The promising advantages of bosonic QEC have motivated extensive development of bosonic QEC codes, as well as a wide range of proposals for fault-tolerant quantum computers, quantum repeater designs, and QEC-enhanced sensors based on GKP qubit and O2O codes. In this section, we discuss some of these exciting avenues.

\subsection{Computing}
\label{sec:app_ftqc}

The race to build a fault-tolerant quantum computer (FTQC) is intensifying, although it may take some time before such a device becomes available. Despite this, GKP qubit codes have emerged as a promising solution for achieving this objective---offering several advantages over other qubit encoding schemes. Recent theoretical research has focused on optimizing the performance of GKP qubit codes and developing novel techniques for fault-tolerant quantum computing. On the experimental front, as we saw in previous sections, significant progress has been made in implementing GKP qubit codes using various physical systems, such as superconducting circuits and trapped ions, with numerous proposals for their use in optics. As we touched on briefly in Chapter~\ref{sec:scaling_up}, significant challenges still exist in scaling up these systems to construct large-scale fault-tolerant quantum computers, and further research is required to overcome these obstacles.

There have been a number of recent reviews on quantum computing with bosonic modes~\cite{terhal2020towards,puri2021rvw,cai2021bosonic,joshi2021quantum}. These reviews have provided valuable insights into the current state of the field, highlighting the progress made in developing and optimizing bosonic QEC codes for quantum computing and outlining the challenges that need to be addressed for the practical implementation of FTQC using GKP qubits. However, the field of bosonic QEC is progressing rapidly. Here we highlight some key aspects of quantum computing with GKP qubits and recent developments on both the theoretical and experimental fronts.

\subsubsection{Universality with Gaussian operations}\label{sec:gkp_universality}

Clifford gates (generated by Hadmard, S gate and CNOT) and measurements for GKP encoding consists of Gaussian operations and are therefore, in principle, relatively easy to implement. Remarkably, it has been shown in Ref.~\cite{baragiola2019GKPuniversal} that universal quantum computation with GKP qubits can be achieved without additional non-Gaussian elements beyond the GKP states themselves. This is nontrivial since non-Gaussian gates, such as the cubic phase gate, are typically necessary for universality; see Chapter~\ref{sec:NG_states} for a quick overview of non-Gaussianity. The driving force behind this \emph{GKP universality} is the following observation~\cite{baragiola2019GKPuniversal}: Applying GKP error correction (enabled by preparing an ancilla in a GKP state) can produce magic states on Gaussian inputs, in turn enabling the implementation of non-Clifford gates (magic gates). This result has important implications for the experimental realization of GKP-based quantum computation, as it suggests that the use of non-Gaussian operations is unnecessary, simplifying the experimental requirements. However, as described in Chapter~\ref{sssec:GKP-control}, the clifford equivalence of Gaussian operations is only correct for ideal GKP codes. For finite-energy GKP codes, the Gaussian operations need to be followed by a few stabilization rounds, since the finite-energy envelope induces correctable errors when ideal GKP operations are used with finite-energy GKP states. These errors can reduce the overall threshold when GKP qubit codes are concatenated with DV codes, as we elaborate in forthcoming sections.

\QZ{It is worthwhile to precisely pinpoint the resources leading to GKP universality, as such is relevant when consider the potential, classical simulatiability of quantum computation with GKP states. Recall that, in the DV domain, classical simulability is associated with Clifford operations and computational basis states/measurements. Whereas in the CV domain, the Gaussian nature of the quantum states and measurements (e.g., homodyne) is often associated with the simulability of CV circuits, as we discussed briefly in Section~\ref{sec:NG_states}. The amalgamation of DV-type computations with CV resources introduces novel challenges and avenues for identifying or quantifying simulatability in bosonic quantum information processing. While negativity of the Wigner function, a property that non-Gaussian states like GKP states possess, is necessary for classical non-simulatability~\cite{Chabaud2023BosonicQAdvantage}, it is not necessary and sufficient~\cite{alvarez2020prrSimulable_Negativity}. In fact, Refs.~\cite{alvarez2020prrSimulable_Negativity,Calcluth2022GKPeffSimulation_Quantum} provide a comprehensive analysis and establish conditions under which quantum circuits comprising \textit{only} GKP states, Gaussian unitaries, and homodyne detection can be classically simulatable.\footnote{This is due to the fact that one can track the GKP stabilizers through the Gaussian circuit and measurements, similar to stabilizer tracking in the DV domain.} Given these considerations in light of the GKP universality results discussed previously, it begs the question: Which physical resources actually lead to GKP universality? Remarkably, it turns out that vacuum or thermal states (albeit, of low occupation number) are the only additional elements necessary for universal quantum computation with GKP states~\cite{Calcluth2023PRA_Vacuum}, as such can be utilized to manifest GKP-magic~\cite{baragiola2019GKPuniversal}.}

\subsubsection{Fault tolerant quantum computing architectures}

\begin{figure}
    \centering
    \includegraphics[width=.8\linewidth]{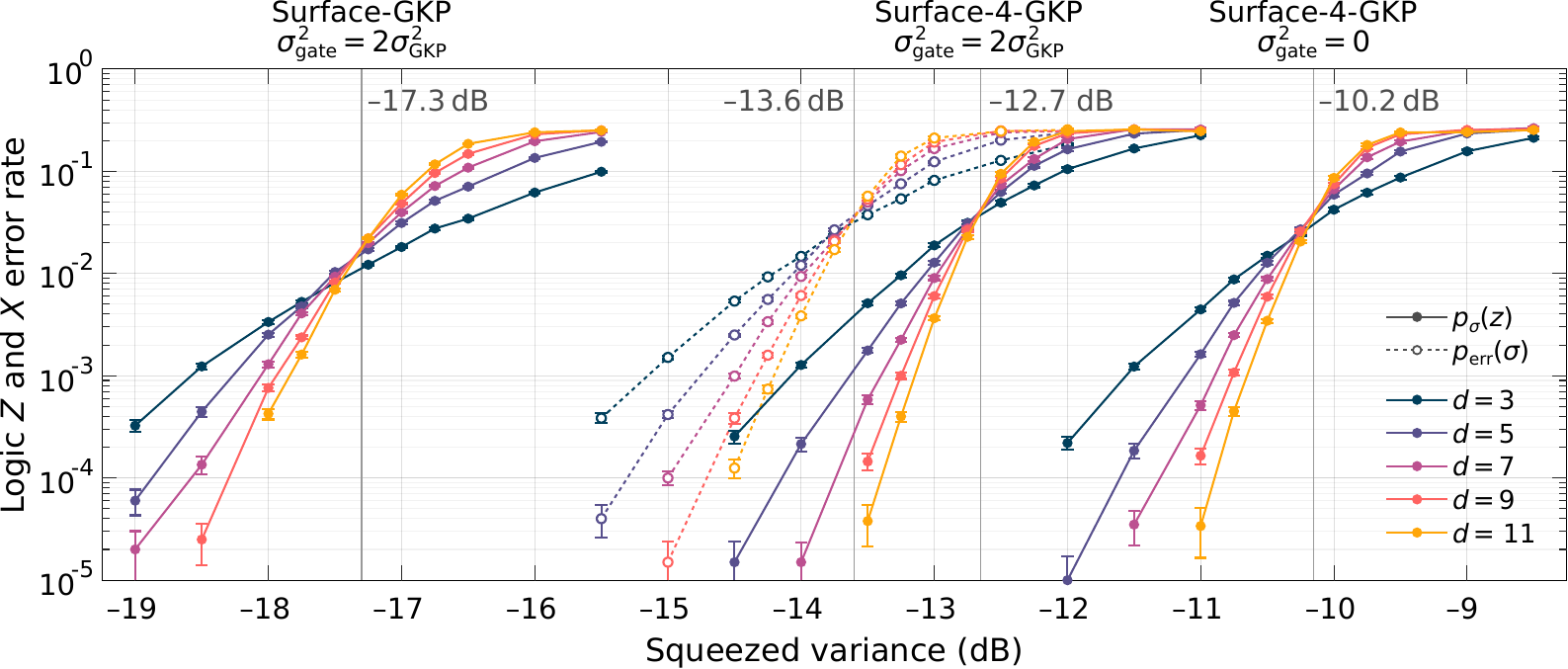}
    \caption{Threshold estimates for GKP surface codes in a MBQC architecture when finite-energy GKP noise and/or gate noise contribute to the overall noise budget. Threshold estimates vary from 10-17 dB (effective squeezing) depending on what code is utilized and if gate noise is present. Adapted from~\cite{larsen2021}.}
    \label{fig:threshold}
\end{figure}

A plethora of FTQC architectures based on GKP codes have been proposed in the last several years. These can generally be broken into two categories based on the platforms considered: cQED and ion-based platforms are well-suited for \emph{gate-based} quantum computing~\cite{fukui2018PRXftqc,noh2020fault,vuillot2019toric,noh2022low} because they have high fidelity gates and the ability to perform high-fidelity single-shot measurements. In contrast, photonic systems are considered good platforms for fault-tolerant measurement-based quantum computing (MBQC), which involves preparing a large entangled state, typically a cluster state, and then measuring the state in a particular pattern to perform quantum computations~\cite{larsen2021,bourassa2021blueprint,tzitrin2021staticFTcomp,bohan2020CVcomputing}. In either setting, finite-energy GKP noise and noise from faulty gates are limiting factors that need to be considered; see Figure~\ref{fig:threshold} for threshold estimates that include GKP noise and/or gate noise. We discuss both cQED- and optics-inspired approaches below.

In the context of FTQC with cQED, low-loss microwave modes are initialized in GKP qubits and gates and measurements are performed via Gaussian operations and, in principle, homodyne detection~\cite{terhal2020towards,puri2021rvw,cai2021bosonic,joshi2021quantum} or, in practice, via auxiliary qubit-based schemes like those presented in Chapter~\ref{sec:GKP_eng}. High-Q microwave cavities are a good platform for quantum information processing with GKP qubits as witnessed by the recent beyond break-even demonstration~\cite{sivak2022breakeven}. Furthermore, exquisite control of the quantum state of the microwave resonator is allowed by coupling the mode to a nearby transmon which can be used for single- and multimode quantum state engineering; see Chapter~\ref{sec:GKP_eng}.

Several architectures based on hybrid topological (e.g., surface) GKP codes have been considered as a viable option for FTQC in cQED~\cite{fukui2018PRXftqc,vuillot2019toric,terhal2020towards, noh2020fault,puri2021rvw,cai2021bosonic,joshi2021quantum,noh2022low, lin2023surfaceGKP}. Here, the GKP qubit code is used as inner code for analog QEC and a topological code is used as an outer code, in analogy with the code constructions presented in Chapter~\ref{sec:QEC-multimode}. Intuitively, the inner code provides a first layer of protection against Gaussian noise sources at the single-qubit level. The outer DV code then supplements the inner code by permitting single-qubit errors which can otherwise be corrected at the logical level. Importantly, since GKP states are the only non-Gaussian resource required for universality, logical operations can be implemented with Gaussian transformations on the GKP qubits. State-of-the-art results in numerical simulations of fault-tolerance thresholds for an all-GKP code is recorded at $\sigma\approx 0.602$~\cite{lin2023surfaceGKP}. In Ref.~\cite{hanggli2020pra,stafford2022GKPbiasedREP}, the authors study rectangular GKP codewords using XZZX codes and report threshold values around $\sigma\approx .6$ by using an asymmetric rectangular GKP lattice. Ref.~\cite{Zhang2022GKPxzzx} claim a threshold around $\sigma\approx .67$ is attainable via rectangular GKP concatenated with a XZZX code. It is to be noted that many such studies utilized ideal GKP operations and error correction with GKP ancillae. As illustrated in Chapter~\ref{sssec:GKP-control}, one can  achieve improved readout fidelities, state preparation, and error correction using single-qubit ancillae---which are pending analyses with respect to implementations of GKP-surface codes or concatenation with other qubit codes, like the low-density parity check (LDPC) codes, Color codes, etc.

In the optical domain, there have been recent proposals for MBQC with optical GKP qubits~\cite{larsen2021,bourassa2021blueprint,tzitrin2021staticFTcomp,bohan2020CVcomputing}. Similar to the microwave regime, a GKP qubit code is used as inner code to complement the outer qubit code. Gaussian operations on, e.g., GKP qubits generate an entangled cluster state. One then performs joint homodyne measurements on the GKP qubits to execute the computation. Arguably the most challenging obstacle to overcome in optical quantum computing architectures is GKP state generation. As discussed in Chapter~\ref{sec:optical_gkp}, many proposals to create optical GKP qubits rely on probabilistic methods, such as GBS devices that inject displaced squeezed vacua into a linear optical network and post-select on photon-number patterns from PNR detectors; see Ref.~\cite{konno2023propagatingGKP} for a recent demonstration. To achieve near-deterministic (but heralded) creation of GKP states, many GBS devices can be multiplexed. Additionally, high-quality GKP states are required since finite-squeezed GKP states introduce additional, effective noise in the architecture. Nevertheless, if reliable optical GKP states can be generated, cluster-state creation via Gaussian operations and computation via homodyne measurements can be straightforwardly achieved on-chip, making optical FTQC with optical GKP qubits an attractive prospect.

Numerical simulations assessing fault-tolerant thresholds based on GKP qubit codes concatenated with DV codes---in both cQED- and optics-inspired architectures---should be approached on a case-by-case basis due to the presence of caveats and underlying assumptions about noise modeling. However, several simulations targeting both platforms indicate a consistent trend. These simulations suggest fault-tolerant threshold behavior around 10-13 dB (equivalent squeezing), thus establishing a benchmark for the amount of squeezing required in the GKP qubits. A recent estimate using the GKP surface code~\cite{larsen2021}, as shown in Figure~\ref{fig:threshold}, exemplifies this trend. Universal FTQC is thus, in principle, possible with bosonic systems provided high-quality GKP states are available, and the error rates are within acceptable limits.

\subsection{Communication}\label{subsec:apps_comms}

The future of communication and networking will likely be encompassed in the so-called \textit{quantum internet}~\cite{kimble2008quantum,wehner2018quantum,kozlowski2019towards,quics2021prxQuantum}. The quantum internet, or more generally quantum networks, hypothetically consists of a set of quantum channels that link many quantum nodes---much like today's classical internet structure but with key distinctions. The channels of a quantum internet are ``quantum'' in the sense that they allow transmission of delicate quantum information (i.e., non-zero quantum capacity), while the nodes are ``quantum'' in the sense that the users at each node have (to varying degrees) some level of quantum information processing capabilities. The development of the quantum internet is currently in a gestational period, however the potential technological impact of futuristic quantum networks is apparent. Some specific applications that can be realized by a full-fledged quantum internet are, to name a few: distributed quantum computing and distributed quantum sensing over large-area networks, provably secure communication, and quantum clock synchronization (see Refs.~\cite{wehner2018quantum,quics2021prxQuantum} for more details and references)---with unforeseen applications and technological breakthroughs awaiting in our future.

Establishing a functional quantum network is not without its challenges. At the most primitive level, the primary task of any quantum network is to allow successful transmission and maintenance of quantum coherence and entanglement over large distances. This is generically a difficult task due to excessive noise in communication links and the fragility of quantum states. Similar to current classical communication networks, optical fibers and free-space channels are go-to approaches for fast and reliable quantum communication, but these lines of communication are, alas, noisy. At the theoretical level, these transmission media can be accurately modeled as thermal-loss channels described mathematically in Chapter~\ref{sec:gauss_evol}. It is thus vital to determine the best possible communication rates over these channels, explore practical avenues to achieve optimal communication rates, and construct approachable ways to connect network users in large-area quantum networks. 

For long-distance optical communications, we see GKP QEC codes playing a pivotal role in the future. In this section, we discuss how achievable rates with GKP codes approach the capacities of noisy Gaussian channels. This establishes GKP codes as a viable route to optimal quantum communication. We then discuss quantum repeaters based on GKP codes, which allows one to extend quantum communication links over, in principle, arbitrarily large distances. 

\subsubsection{Achievable Rates} 

In Chapter~\ref{sec:quantum_capacity}, we discussed, in quite general terms, the quantum capacities of Gaussian channels---placing upper bounds or providing achievable rates (lower bounds) in most cases. These bounds, however, are based on ``existence proofs'' with no specific route to achieve such in practice. GKP codes turn out to be a physically viable way of approaching such bounds. As a matter of fact, shortly after the conception of computational GKP states~\cite{gkp2001}, it was shown that multimode GKP codes can be used to achieve the one-shot quantum capacity of the AGN channel [lower bound in Eq.~\eqref{eq:agn_bounds}] by encoding the information into a high-dimensional GKP lattice~\cite{harrington2001rates}. The results from that work suggested that GKP codes may be useful for other types of Gaussian channels as well. This intuition has been proven correct. In fact, using loss-to-AGN conversion techniques presented in Chapter~\ref{sec:gauss_channels}, Noh et al~\cite{noh2019quantumcapacity} showed that GKP codes achieve the quantum capacity of thermal-loss channels [upper bound in Eq.~\eqref{eq:bounds_thermalLoss}] up to at most a constant factor gap. For the specific case of single-mode encoding, the authors demonstrated numerically that, starting from random initial codes, the hexagonal GKP code emerges as the optimal bosonic code for transmission across a thermal-loss channel; see also Fig.~\ref{fig:convergence2gkp} in Chapter~\ref{sec:noise_model} and surrounding discussions.

\subsubsection{Quantum Repeaters}

\begin{figure}[t]
    \centering
    \includegraphics[width=.75\linewidth]{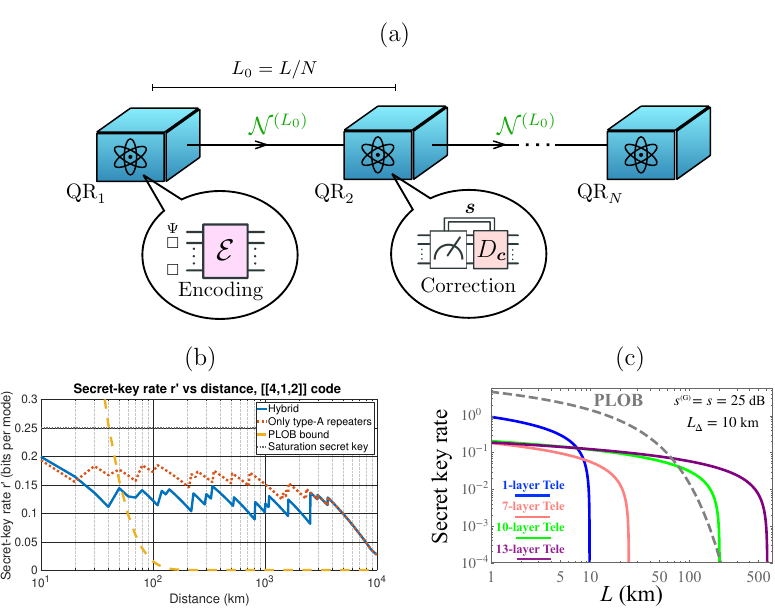}
    \caption{GKP quantum repeater. (a) High-level schematic of a one-way quantum repeater (QR) based on GKP qubits. A noisy quantum communication line of length $L$ is broken into $N$ segments of length $L_0=L/N$. At the first repeater station ${\rm QR}_1$, a GKP qubit $\Psi$ is encoded into a multi-qubit code (e.g., a [[4,1,2]] code) and then sent to the next repeater station, ${\rm QR}_2$. A QEC cycle is implemented to correct errors on the multi-qubit state (and re-encoded with fresh ancillae if necessary). The encoded information is sent down the line, with iterative QEC cycles implemented at intermediary repeater stations, until it reaches its destination at the final repeater station ${\rm QR}_N$, where the information is decoded and processed. (b) Secret-key rate for one-way GKP-QR architecture based on DV encoding (DV-QKD). Adapted from Ref.~\cite{rozpkedek2021quantum}. (c) Secret-key rate for GKP-QR architecture based on CV encoding and teleportation (CV-QKD; $L_0=10$ km). Adapted from Ref.~\cite{wu2022continuous}. PLOB bound refers to the best secret-key rate via repeaterless communication~\cite{plob2017}. In both designs, a critical distance of $\sim100$ km emerges, after which quantum repeaters appear to be an absolute necessity for reliable, high-rate, long-distance quantum communication. }
    \label{fig:gkp_qr}
\end{figure}

As we saw in our previous discussions, quantum capacities of Gaussian channels are (almost) achievable via GKP codes. However, these analyses only hold in regimes where the noise is not too large, meaning that the quantum capacity remains non-zero throughout the entire communication distance. For long-distance communication, this is no longer the case, and new avenues for successful quantum communication need to be explored. The key technology typically employed to overcome this pressing challenge is a quantum repeater.

Quantum repeaters are a crucial technology for long-distance quantum communication and are essential for the development of futuristic quantum networks~\cite{azuma2022QuRepeaterRvw}. They allow for the transmission of quantum information over long distances by breaking up the communication line into smaller segments and transmitting quantum information between adjacent segments via entanglement swapping or one-way QEC-assisted protocols; see Figure~\ref{fig:gkp_qr} for an illustration. Developing good QEC codes (and/or robust quantum memories) is thus critical to realizing effective quantum repeaters.

There has been growing interest in the use of GKP qubits for quantum repeater protocols~\cite{fukui2021allOptical,rozpkedek2021quantum,wu2022continuous,schmidt2023RepeaterGKPqudits,rozpedek2023allphotonic}, which utilize multi-layer QEC codes for the implementation of one- and two-way quantum repeaters; see Figure~\ref{fig:gkp_qr}. In particular, most proposed architectures employ an inner GKP code to provide a first-line defense against loss, followed by an outer layer qubit code---such as a $[[4,1,2]]$ or $[[7,1,3]]$ code---to protect against single-qubit errors that the inner code cannot handle on its own. Furthermore, these hybrid CV-DV architectures have demonstrated that, in principle, incorporating analog information for the inner CV code can lead to enhanced quantum repeater performance.\footnote{GKP encodings come with the additional benefit that Bell state measurements (a key ingredient for quantum teleportation) can be done deterministically, which is not the case for, e.g., ``all-optical'' implementations~\cite{lee2019AllOpticalFundamentals}.}

Apart from transmitting discrete information, transmitting genuinely analog information is essential for many quantum sensing and CV quantum communication tasks. Motivated by the need to transmit analog quantum information, there has been a recent proposal for an all-CV quantum repeater that utilizes GKP-O2O codes and CV quantum teleporation to achieve this task~\cite{wu2022continuous}. In particular, the authors of Ref.~\cite{wu2022continuous} showed that their CV repeater design can overcome attenuation in transmission and improve the performance of several physically motivated operational tasks such as entanglement-assisted communication, target detection, and CV quantum key distribution; see Figure~\ref{fig:gkp_qr}(c). Further research to optimize CV repeater architectures that, perhaps, leverage multimode GKP-O2O codes is warranted.

We point out that the resource requirements for the GKP-based (and other) quantum repeater architectures is significant because these designs typically require a large number of repeater stations that host quantum memories---such as high $Q$ microwave cavities---as well as high-fidelity quantum information processing capabilities in memory---such as the ability to couple multiple cavity modes. Additionally, transduction capabilities will likely be necessary for translating information between optical communication lines and memory~\cite{lauk2020PerspTransdux,quics2021prxQuantum,jiang2021OpticaTrdxRvw}. To bypass the transduction issue, one possibility is to directly link the optical channels to optical quantum information processors located at each repeater node.

\subsection{Sensing}

\begin{figure}[t]
    \centering
    \includegraphics[width=\linewidth]{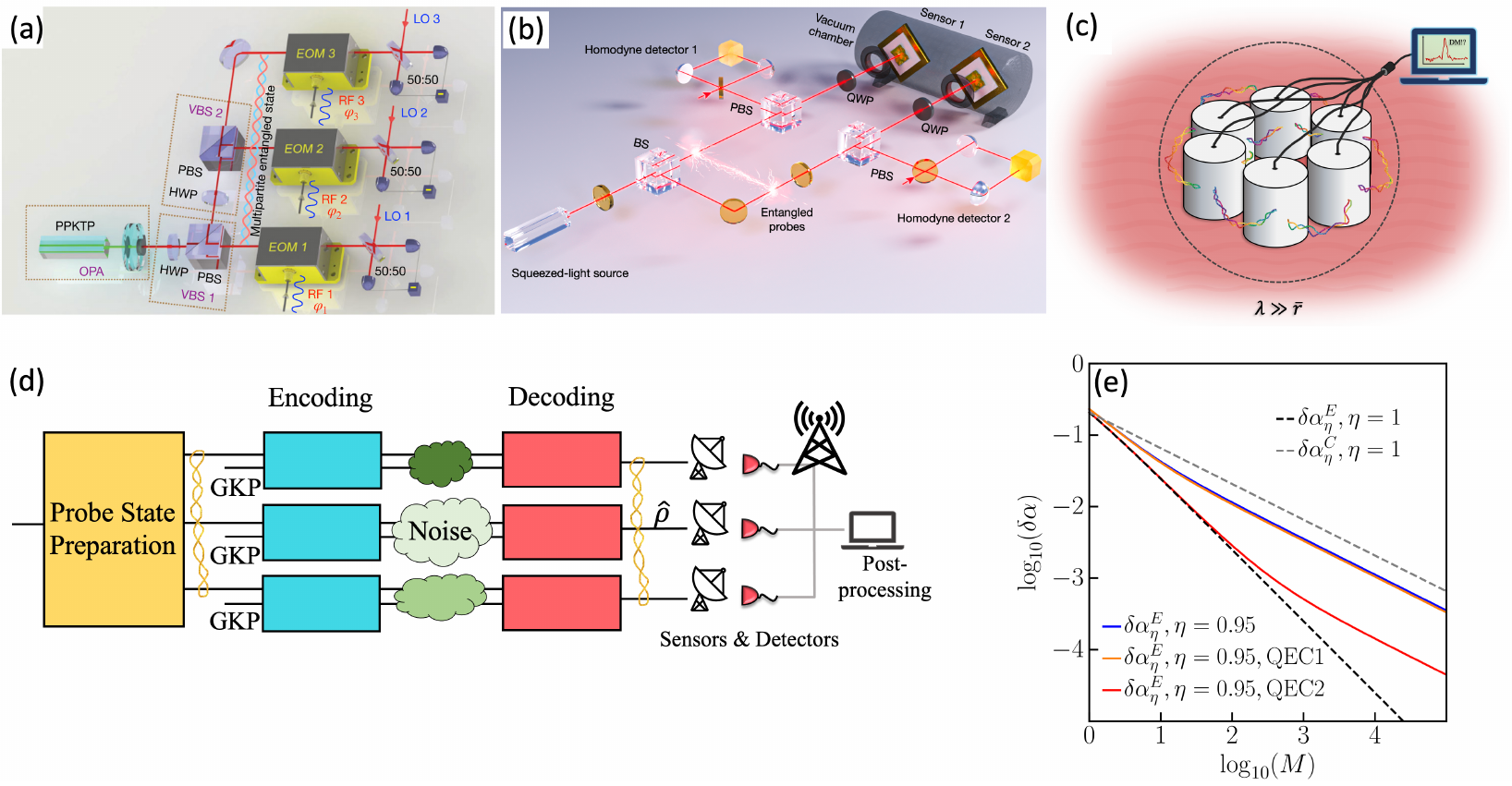}
    \caption{Schematic of distributed quantum sensing applied to (a) a radio-frequency photonic sensor array, (b) opto-mechanical sensors for force sensing and (c) microwave cavity sensors for dark matter detection. (d) Schematic of GKP-O2O codes applied to protect entanglement between sensors for distributed quantum sensing tasks. (e) Sensing accuracy $\delta\alpha$ versus number of sensors with and without GKP-O2O codes for protection. Black dashed: ideal entangled performance of Heisenberg scaling. Gray dashed: ideal classical performance. Blue: practical sensors enhanced by entanglement in presence of loss $1-\eta=0.05$. Red and orange are performance when GKP-O2O codes are used for QEC. Figures adopted from (a)~\cite{xia2020demonstration}, (b)~\cite{xia2023entanglement}, (c)~\cite{brady2022entangled}, (d)~\cite{zhou2022enhancing}, and (e)~\cite{zhuang2020distributed}.}
    \label{fig:dqs}
\end{figure}

Quantum sensing is a fascinating area of research that is poised to significantly impact society as a whole~\cite{degen2017quantum,pirandola2018RVWsensing,polino2020avsRVW,quics2021prxQuantum}. The key advantage of quantum sensing is that it can achieve higher sensitivity than classical sensing methods through the Heisenberg scaling of the measurement sensitivity $1/N$ (the fundamental limit on precision quantum measurements) which should be compared to the standard quantum limit $1/\sqrt{N}$ for classical sensing paradigms; here, $N$ is the number of sensors.

The \emph{in-principle} sensitivity advantage in the quantum sensing paradigm can be greatly beneficial for real-world applications---such as secure timing in clock networks~\cite{komar2014quantumClocks}, high-precision astronomy~\cite{gottesman2012longerBaseline,Tsang_2016}, fundamental physics applications like gravitational wave detection~\cite{tse2019quantumLIGO} and dark matter searches~\cite{backes2021quantum,dixit2021qubitDM,brady2022entangled,shi2023ultimate}, and data classification in supervised learning tasks~\cite{zhuang2019slaen}---to mention a few examples. However, the \emph{practical} advantage of quantum sensors is limited, due to the effects of decoherence and noise, which can depreciate the Heisenberg scaling advantage that is otherwise present in idealistic sensing scenarios.

This is where QEC comes in. By encoding quantum-sensor states (e.g., multi-partite entangled state) in a way that is robust against errors, QEC can improve the performance of quantum sensors, allowing them to recover the Heisenberg limit even in noisy environments~\cite{zhou2018achieving}. To this end, reducing the variance of estimation for a sensing task is paramount for enhanced sensitivity and reemergence of Heisenberg scaling. O2O codes---like those discussed in Chapter~\ref{sec:gkp_o2o_codes}---can substantially reduce the noise, as we have seen explicitly from the variance bound in Eq.~\eqref{eq:no_threshold}, and thus boost sensing performance. In this section, we discuss the QEC-enhanced sensing paradigm, emphasizing applications of GKP states and GKP codes for sensing in the CV domain.

\subsubsection{Error correction for entangled sensor networks}
Quantum sensing adopts different types of resources to boost the sensor performance in information acquisition. For instance, in the optical domain, sensors rely on squeezed vacua, two-mode squeezed vacua, and multi-partite continuous-variable entanglement to go beyond standard quantum limit dictated by vacuum fluctuations. The protection of these continuous-variable quantum resources against noise requires codes such as the GKP-O2O codes discussed in Chapter~\ref{sec:gkp_o2o_codes}. 

Recent works~\cite{zhuang2020distributed,zhou2022enhancing} on applying GKP-TMS codes and its concatenations have shown great promise in enhancing \emph{distributed quantum sensing}---a sensing paradigm where an entangled sensor-network extracts global information about local parameters with unprecedented precision~\cite{zhang2020distributed}. Distributed quantum sensing has a wide range of potential applications, including radio-frequency photonic sensor arrays~\cite{xia2020demonstration}, optomechanical sensor networks for force sensing~\cite{xia2023entanglement} and microwave dark matter detection~\cite{brady2022entangled}, as shown in Figure~\ref{fig:dqs}(a)-(c). In these applications, one does not have the capability to directly engineer the interactions between the sensors, which is otherwise required to implement the continuous-time control to recover Heisenberg scaling~\cite{zhou2018achieving}. Instead, to entangle the sensors, one must distribute an entangled state prepared at a central node to other sensor nodes via noisy quantum channels, as shown in Figure~\ref{fig:dqs}(d). 

In this regard, entanglement distribution loss is the major imperfection in the system, limiting distributed sensor networks to be local in the near-term. To mitigate the distribution loss, one can adopt the GKP-O2O code to encode each part being sent to individual sensors and decode at each individual sensor to recover the entangled state before participating in the final sensing process.
Ref.~\cite{zhuang2020distributed} performed initial analyses on the performance boost brought by the GKP-O2O codes in an optical displacement sensing scenario, which is the theoretical model behind the applications considered in, e.g., Refs.~\cite{xia2020demonstration,xia2023entanglement,brady2022entangled}. As shown in Figure~\ref{fig:dqs}(e), ideal error correction extends Heisenberg scaling in the entangled sensor network to a much larger number of sensors. Reference~\cite{zhou2022enhancing} further analyzed the remaining performance advantage with imperfect error correction.

\subsubsection{GKP state as a sensor}\label{subsec:gkp_sensor}

GKP grid states can themselves be helpful for quantum sensing purpose. 
Ref.~\cite{duivenvoorden2017} considers the single-shot measurement of complex displacement $D_\beta$. Suppose one has a single chance of measuring the real and imaginary part of the displacement. Due to uncertainty principle, $[\hat q,\hat p]=i$, the real and imaginary parts cannot be measured simultaneously with great precision. Indeed, states such as squeezed vacuum suppress the variance in a single quadrature at the cost of increasing the variance of the other quadrature. Superdense sensing circumvents this conundrum by introducing an entangled ancilla. Without entanglement, it is fundamentally impossible to violate uncertainty principle. However, despite the limitations imposed by the canonical commutation relations, the stabilizers in Eq.~\eqref{eq:canonical_stabilizers} commute, i.e. $[{\rm e}^{-i\sqrt{2\pi}\hat{p}},{\rm e}^{i\sqrt{2\pi}\hat{q}}]=0$ as discussed in more detail in Chapter~\ref{sec:gkp_lattice}. In other words, simultaneous precise estimation of $\hat p \mod \sqrt{2\pi}$ and $\hat q \mod \sqrt{2\pi}$ are possible. Suppose that one has prior knowledge that the displacements are guaranteed to be small, then the GKP state enables simultaneous precise measurement of both quadratures. Reference~\cite{zhuang2020distributed} further generalizes this idea to a sensor network. When one has a fairly precise prior knowledge about the weighted average of displacements, then one can achieve Heisenberg scaling of the estimation of weighted averages of both quadratures of displacement. However, without a precise prior, achieving simultaneous Heisenberg scaling will only be possible from entanglement-assisted superdense sensing protocols~\cite{zhuang2017entanglement}.

We emphasize that the advantages of GKP states as sensor states hold in single-shot measurements. If multiple measurements or continuous monitoring is possible, then a strategy of homodyne measuring a single quadrature at one time, boosted by squeezed vacuum, provides a better performance, considering there is no mod-$\sqrt{2\pi}$ ambiguity, and furthermore, the resources are much easier to engineer. For example, in dark matter detection with microwave cavities, the goal is to estimate thermal noise induced by dark matter, modeled as random displacements. Because continuous monitoring is possible, GKP states do not provide further advantage in improving the signal-to-noise ratio, and single-mode squeezed vacuum is the optimal source without entanglement assistance~\cite{shi2023ultimate,brady2022entangled}.

\section{Closing Remarks and Open Problems}\label{sec:discussion}

In this review, we have comprehensively covered recent theoretical and experimental advances in bosonic QEC using GKP codes, including proposals that employ multiple oscillators for safeguarding both discrete- and continuous-variable quantum information. Our aim has been to offer the reader both a pedagogical introduction to the field of bosonic QEC with GKP codes and a presentation of the latest state-of-the-art research results. With the development of quantum engineering in microwave cQED systems and the promise of scalable integrated photonic structures, multimode GKP qubit codes and GKP oscillators-to-oscillators codes provide many opportunities in quantum information processing such quantum computing, communication, and sensing. Yet many research directions have yet to be fully explored and open problems remain. Just below, we summarize a few open problems that we are aware of and that we think are worth addressing.

\paragraph{Concatenated GKP codes}  With advancements in universal control of single-mode GKP states, we would like to understand the required missing pieces to make the concatenation of GKP codes a feasible reality. We lay out a few open problems in different directions of GKP control and stabilization in order to improve the performance of GKP codes concatenated with qubits codes.

To further improve the lifetime of GKP memories in superconducting circuits, auxiliary qubit noise must be mitigated. One way is to directly engineer Hamiltonian protection such that no auxiliary qubit is required. Proposals towards this approach have been described in Chapter~\ref{sec:GKP_circuits}, however these proposals are challenging, requiring the combination of a number of recently developed and yet-to-be-developed technologies. For example, a suitable coherent quantum phase slip element has yet to be demonstrated, and it is likely that the best path forward to realizing this is by engineering emergent quantum phase slip dynamics with Josephson junction arrays. Advancements in the engineering and realization of other protected superconducting circuits, such as the fluxonium and $0-\pi$ qubit, will help towards realizing a fully protected GKP Hamiltonian through direct circuit-level engineering. Engineered continuous dissipation, as is discussed in Chapter~\ref{sec:dissipation_continuous}, is another promising candidate, with many challenges towards realization, including the need for high-impedance superconducting circuit elements. Any advancements that can reduce the requirements on the circuit impedance for these proposals will undoubtedly be beneficial in advancing the field.

An alternative path to mitigate auxiliary qubit noise is through the use of a biased-noise auxiliary qubit, such as a stabilized Kerr-cat \cite{puri2019stabilized}. This is a promising route forward that is currently being investigated experimentally. The heavy fluxonium is another promising candidate under investigation. Additionally, it might be fruitful to think of ways to implement conditional displacements using transmons in a fault-tolerant manner. Since conditional gates lie at the heart of most auxiliary-qubit based processes, if there were means to implement the conditional displacement gate in a path-independent way~\cite{ma2020path}, it could be used to mitigate errors. Such a method could potentially make use of the higher auxiliary qubit states.

Apart from stabilization, concatenation also requires Clifford gates. When using Gaussian operations to implement Clifford unitaries on GKP qubits, the finite-energy errors discussed in Chapter~\ref{sec:finite_GKP} and Chapter~\ref{sssec:GKP-control} can reduce the threshold of the concatenated implementation (also see Chapter~\ref{sec:app_ftqc}). Only a few attempts to resolve these errors by direct implementation of non-unitary finite-energy GKP logical operations have been made so far. Also, all of these methods use an auxiliary qubit or rely on stabilization rounds post infinite-energy operations. We thus need a rigorous study of the gain achieved in the presence of auxiliary qubit noise and also study its effect in threshold plots similar to Figure~\ref{fig:threshold}. On the experimental side, tunable superconducting circuits are to needed couple oscillators encoding GKP states without introducing spurious nonlinearities; this is an active topic of investigation. 

The non-Clifford resources for GKP codes, as proposed by Baragiola et al.~\cite{baragiola2019GKPuniversal}, are particularly appealing due to their reliance on logical Pauli states and Gaussian resources. Thorough analysis of this proposal is essential to assess its practicality in reducing the overhead of scalable fault-tolerant quantum computing. Firstly, the scheme relies on on how efficiently we can engineer state preparation, and thus SPAM errors become a major issue in realizing this scheme. Secondly, it is important to note that the scheme was designed for infinite-energy GKP codes and may not directly account for errors arising from the finite-energy envelope. A concrete analysis in this direction would be beneficial to gain insights into any potential overhead reduction in fault-tolerant quantum computing.

For readout, more research is needed to investigate the challenges faced in realistic superconducting circuit experiments. There has been only one proposal apart from the usual phase estimation approach to improve upon the readout fidelity of finite-energy GKP states~\cite{hastrup2021improved}. This scheme although good, does not solve all our problems. As mentioned in Chapter~\ref{sssec:GKP-control}, we need to focus on achieving high-fidelity end-of-the-line readout measurements in order to be able to use the GKP qubits in realistic quantum circuits where error correction will never fully correct the errors on the auxiliary qubit. Hence, a readout scheme closer to the square curve shown in Figure~\ref{fig:GKP_readout} can significantly increase the practical use-case of these states. Additionally, a readout scheme which mitigates auxiliary qubit noise would help to limit errors in experiment.

Finally, more research is needed to investigate the technical aspects of a hybrid concatenated code architecture in superconducting circuits. For example, linear oscillators with GKP states can be employed as the data qubits, while the measure qubits can be of a different variety, perhaps transmons or Kerr-cats. More analysis is needed to investigate noise propagation and thresholds for these hybrid architectures, and we believe they are a promising path forward in the near term. This was suggested in Ref.~\cite{puri2021rvw}, however to the best of our knowledge, a full analysis of noise in such a hybrid concatenated code architecture has not yet been published. Additionally, it was shown in Ref.~\cite{sivak2022breakeven} that postselection on outcome strings in sBs type error correction significantly increased the lifetime of the stabilized quantum memory, at the cost of a lower success probability. We anticipate that some classical post-processing on these output strings of sBs could yield better lifetimes even without post-selection. Furthermore, it would be intriguing to explore whether these measurement strings could be effectively employed for erasure conversions of GKP data qubit errors when combined with DV codes (e.g., surface codes), which then could be used to leverage the improved threshold with erasure noise. This feature has already been put to use for resource reduction of fault-tolerant quantum computing with surface code architectures realized in neutral atoms, trapped ions and superconducting circuits (via transmon and dual-rail codes) \cite{teoh2022dualrail,wu2022erasure,scholl2023erasure,kang2023quantum,kubica2022erasure,levine2023demonstrating}.

\paragraph{Optical GKP} Another critical engineering task involves the generation of optical GKP states, which play a pivotal role in enabling bosonic quantum error correction for quantum communication since optical photons serve as the primary carriers of quantum information in this context. \QZ{Although an initial demonstration of optically generated GKP states (albeit with low squeezing, 2.5 dB) shows promise~\cite{konno2023propagatingGKP}}, the absence of \textit{high-fidelity} optical GKP states poses challenges in photonic quantum computing, which otherwise offers the advantage of requiring less cooling compared to other quantum computing platforms. Chapter~\ref{sec:optical_gkp} presents several proposals in this area. Though, to achieve high-quality GKP states and photonic quantum information processing therefrom, a series of experimental challenges must be addressed. These challenges include developing high-quality photon number-resolving detectors, low-loss linear optical circuits, efficient microwave-optical transduction, and strong optical nonlinearity---as well as their integration onto novel quantum information processors, such as photonic chips. Overcoming these obstacles is essential to realize the full potential of GKP states. Exploring alternative proposals for generating optical GKP states may also hold promise in reducing the required engineering capabilities. As research further develops in this area, new and innovative approaches may emerge, potentially simplifying the experimental requirements and enhancing the feasibility of optical GKP state generation.
    
\paragraph{Multimode lattice codes}
While most strategies for scaling up GKP codes involve concatenation with surface codes, Chs.~\ref{sec:gkp_lattice} and~\ref{sec:QEC-multimode} have highlighted that this approach may only explore a limited subset of the available GKP encodings for multiple oscillators. Therefore, a more comprehensive understanding of the error correction properties of various high-dimensional GKP lattices is essential, along with assessing the costs associated with their implementation in terms of stabilization and control. In Chapter~\ref{sec:QEC-multimode}, an example of two-mode codes has been studied, presenting a practical option for QEC. However, considering the restrictions of practical QEC implementations, only certain lattices may demonstrate improvements over single-mode encoding in certain parameter regimes; hence the actual practical implementation of GKP codes is \emph{absolutely paramount} to consider when gauging performance of such codes. At the same time, we also need to explore other options for multimode GKP error correction that can achieve best results with \emph{optimal} lattices. We need to identify good GKP lattices~\cite{conrad2023good} and develop decoding strategies and implementations that match optimal recovery results~\cite{albert2018pra}. Investigating the feasibility of implementing multimode encoding using Hamiltonian engineering may offer a more viable path to scaling up GKP codes. Such investigations might lead to breakthroughs in developing practical quantum technologies that leverage the advantages of GKP codes and enable robust and efficient quantum error correction.

\paragraph{O2O codes} In terms of oscillator-to-oscillator encoding, many open directions are worth exploring. The search for optimal, higher-dimensional GKP lattice states capable of suppressing various noise sources remains open, despite some initial exploration on a few examples. Moreover, whether GKP states represent the best choice for oscillator-to-oscillator encoding is still an unresolved question. Employing numerical optimization techniques, akin to those used in Ref.~\cite{noh2019quantumcapacity}, could shed light on this matter, providing valuable insights into the most effective encoding schemes. Additionally, it is essential to consider the impact of better prior knowledge of errors on the code design, as such may lead to the development of more efficient and robust encoding strategies tailored to the specific error characteristics of the system.

\paragraph{Multitude of multimode codes} The phase-space of $N$ modes is a big place, and there is much room to explore. We have seen GKP codes concatenated with qubit codes for DV quantum information processing, GKP qubits encoded into ``typical'' classical lattices, and GKP O2O codes for protecting analog information. For instance, we have seen extensive studies on code-capacity thresholds for GKP qubit codes (e.g., GKP surface code) emerging over the past several years. However, more exhaustive comparisons and investigations are warranted, especially utilizing lattice GKP qubits, as discussed in Chs.~\ref{sec:QEC-multimode} and~\ref{sec:gkp_lattice}. Additionally, it would be intriguing to explore hybrid GKP qubit-O2O encodings, combining the strengths from both the CV and DV domains. One unique study in this regard was made in Ref.~\cite{xu2022qubito2o} where the authors compared a not-so-analog oscillator-to-oscillator encoding (see Chapter~\ref{sec:gkp_o2o_codes}) with GKP qubit codes---i.e., GKP as the inner layer and repetition codes or five-qubit codes as the outer layer.  In these preliminary toy examples, some qubit-O2O codes exhibited better performance than certain GKP qubit codes in specific parameter regimes. Studies like this showcase the extraordinary richness and potential of multimode GKP encodings.

\paragraph{GKP qudit encoding} Another direction of GKP-based research which has not been discussed extensively in this review is experimental advancement in GKP qudit encoding. We emphasize that for a single-mode GKP quantum memory, increasing the dimension $d$ of the qudit decreases the code distance for displacement errors. However, the development of qudit encodings can pave way towards reducing the circuit complexity of algorithms which demonstrate quantum advantage. In terms of error correction, Ref~\cite{gkp2001} further highlights the importance of increasing the code dimension of encodings in an oscillator. One major challenge in qudit encoding involves increased squeezing required to distinguish the logical states significantly with increasing code dimension $d$. Another pending analysis includes engineering low-error and easily implementable physical gates for GKP qudits. This could be a parallel next step towards improving the quantum volume using GKP encoding.

\medskip
\medskip
Bosonic QEC opens up a wide array of application spaces beyond the long-sought goal of fault-tolerant quantum computing, with quantum communication and sensing being prominent examples. For instance, the unique ability of bosonic QEC to protect quantum states of oscillators proves valuable in distributing CV multi-partite entanglement, offering advantages in quantum sensor networks. Moreover, the engineering of GKP states provides a means for fast and universal control of microwave cavities, enabling crucial tasks like entanglement distribution and distillation in bosonic quantum information processing. Additionally, unitary engineering holds the potential to open up opportunities in quantum simulation, offering a pathway to tackle fundamental problems crucial to condensed matter and high-energy physics.

To fully harness the potential capabilities of bosonic QEC with GKP codes, it is essential to address the open problems outlined earlier and actively search for new challenges to overcome. This pursuit will undoubtedly push the field forward and contribute to the development of fault-tolerant quantum technologies. As research continues in this rapidly evolving field, we eagerly anticipate unforeseen breakthroughs that will not only enhance our understanding of bosonic quantum error correction but also propel us toward our ultimate goal of achieving large-scale bosonic quantum information processing.

\section*{Acknowledgements}
The authors acknowledge Ben Barragiola, Philippe Campagne-Ibarcq, Jonathan Conrad, Michel Devoret, Steven M Girvin, Xanda Kolesnikow, Kyungjoo Noh, Baptiste Royer, and Volodymyr Sivak for helpful conversations and feedback. QZ, AJB and JW are supported by the DARPA Young Faculty Award (YFA) Grant No. N660012014029. QZ also acknowledges support from NSF CAREER Award CCF-2142882, NSF OMA-2326746 and ONR Grant No. N00014-23-1-2296. AE and SS acknowledge support by the Army Research Office (ARO) under grant number W911NF-23-1-0051. AE is also supported by the U.S. Department of Energy, Office of Science, National Quantum Information Science Research Centers, Co-design Center for Quantum Advantage (C2QA) under contract number DE-SC0012704. The views and conclusions contained in this document are those of the authors and should not be interpreted as representing the official policies, either expressed or implied, of the U.S. Government. The U.S. Government is authorized to reproduce and distribute reprints for Government purposes notwithstanding any copyright notation herein.
 


%

\end{document}